\documentclass[11pt, a4paper, oneside]{Thesis} 

\graphicspath{{Pictures/}} 

\usepackage{cite}
\usepackage{amsmath,latexsym} 
\usepackage{float}
\usepackage{commath}
\usepackage{relsize}
\usepackage{tikz}
\usetikzlibrary{positioning, shapes, calc}
\usepackage{mathtools}
\usepackage[all]{hypcap}
\usepackage{bm}
\usepackage{subfigure}
\usepackage{mathrsfs}
\usepackage{fancyhdr}
\usepackage{hyperref}
\hypersetup{colorlinks,citecolor=red}
\hypersetup{colorlinks=true,urlcolor=blue}

\usepackage{microtype}       
\allowdisplaybreaks[1]

\title{\ttitle} 
\DeclareUnicodeCharacter{2212}{-}
\DeclareUnicodeCharacter{2212}{+}

\begin{document}

\setstretch{1.3} 

\fancyhead{} 
\rhead{\thepage} 
\lhead{} 

%

\thesistitle{Late Time Cosmic Acceleration in Modified Gravity and Gauss--Bonnet Cosmology}
\documenttype{\textbf{THESIS}}
\supervisor{Prof. BIVUDUTTA MISHRA}
\supervisorposition{Professor}
\supervisorinstitute{BITS-Pilani, Hyderabad Campus}
\examiner{}
\degree{Ph.D. Research Scholar}
\coursecode{\textbf{DOCTOR OF PHILOSOPHY}}
\coursename{Thesis}
\authors{\textbf{LOHAKARE SANTOSH VIJAY}}
\IDNumber{\textbf{2020PHXF0052H}}
\addresses{}
\subject{}
\university{\texorpdfstring{\href{http://www.bits-pilani.ac.in/} 
                {Birla Institute of Technology and Science, Pilani}} 
                {Birla Institute of Technology and Science, Pilani}}
\UNIVERSITY{\texorpdfstring{\href{http://www.bits-pilani.ac.in/} 
                {\textbf{BIRLA INSTITUTE OF TECHNOLOGY AND SCIENCE, PILANI}}} 
                {BIRLA INSTITUTE OF TECHNOLOGY AND SCIENCE, PILANI}}


\department{\texorpdfstring{\href{http://www.bits-pilani.ac.in/pilani/Mathematics/Mathematics} 
                {Mathematics}} 
                {Mathematics}}
\DEPARTMENT{\texorpdfstring{\href{http://www.bits-pilani.ac.in/pilani/Mathematics/Mathematics} 
                {Mathematics}} 
                {Mathematics}}
\group{\texorpdfstring{\href{Research Group Web Site URL Here (include http://)}
                {Research Group Name}} 
                {Research Group Name}}
\GROUP{\texorpdfstring{\href{Research Group Web Site URL Here (include http://)}
                {RESEARCH GROUP NAME (IN BLOCK CAPITALS)}}
                {RESEARCH GROUP NAME (IN BLOCK CAPITALS)}}
\faculty{\texorpdfstring{\href{Faculty Web Site URL Here (include http://)}
                {Faculty Name}}
                {Faculty Name}}
\FACULTY{\texorpdfstring{\href{Faculty Web Site URL Here (include http://)}
                {FACULTY NAME (IN BLOCK CAPITALS)}}
                {FACULTY NAME (IN BLOCK CAPITALS)}}

\maketitle

\clearpage
\pagenumbering{roman} \setcounter{page}{1}
\Certificate
\Declaration
\setstretch{1.3} 

\pagestyle{empty} 

\Dedicatory{\bf \begin{LARGE}
\textit{\textbf{To}} 
\end{LARGE} 
\\
\vspace{1.5cm}
\begin{Large} \textit{\textbf{My loving family\\ \vspace{0.5cm} \& \vspace{0.5cm} \\inspiring teachers, with heartfelt gratitude}} \end{Large}\\
 \vspace{1cm}
 }

\addtocontents{toc}{\vspace{2em}}

\begin{acknowledgements}

It gives me great joy to have the chance to offer my sincere gratitude to everyone whose inspirational influences have supported me as I pursue my doctorate.

My supervisor, \textbf{Prof.~Bivudutta Mishra}, Professor, Department of Mathematics, BITS-Pilani, Hyderabad Campus, is to be sincerely thanked. His tenacity and unwavering excitement motivate me. The chance to collaborate with him has been a wonderful privilege. Along with his superior mathematical knowledge and teaching abilities, his humanism has served as a wonderful model for me. This dissertation would not exist without his committed help and direction. 

I would also like to thank my DAC members, \textbf{Prof.~Pradyumn Kumar Sahoo} and \textbf{Prof.~Rahul Nigam}, for their insightful questions, insightful comments, and constant feedback throughout this research.
 
I would like to extend my sincere gratitude to the Head of the Department, the DRC Convener, and the entire staff of the Department of Mathematics at BITS-Pilani Hyderabad Campus for their assistance, support, and encouragement in carrying out my research.

I am also grateful to the Associate Dean of AGRSD BITS-Pilani, Hyderabad Campus. 


I gratefully acknowledge BITS-Pilani, Hyderabad Campus, for providing me with the necessary facilities to carry out my research work.

I express my gratitude to the University Grants Commission (UGC Reference No. 191620116597) for its provision of support during the course of my research endeavors.

I would like to take this opportunity to thank my collaborators {\bf Prof.~Sunil Kumar Tripathy}, {\bf Prof.~Sunil Kumar Maurya} and {\bf Dr.~Ksh.~Newton Singh} for providing me with the opportunity to collaborate with them. I have gained a great deal of knowledge from them, for which I am quite grateful.

I would like to express my heartfelt gratitude to my research team: {\bf Dr.~Amar, Dr.~Siddheshwar, Lokesh, Shubham, Rahul, Kalpana, Shivam and Priyobarta.} I would like to thank {\bf Laxmipriya} for supporting and encouraging me throughout my PhD journey and helping me to clarify the fundamental concepts of my research work. Additionally, I want to acknowledge all my friends at the BITS-Pilani Hyderabad campus for their unwavering support and encouragement throughout this journey. While not all of them are named here, each has played a vital role in my success.

Most importantly, I would like to thank my parents, brothers, sisters, family members, and friends for their love, care, and support for my personal life.
\end{acknowledgements}

\clearpage

\begin{abstract}       
This thesis focuses on late-time cosmic acceleration within modified theories of gravity, using various observational data sets and statistical analysis. The Universe is assumed to be spatially homogeneous and isotropic and is described by the Friedmann Lema\^{i}tre Robertson Walker metric. The late-time acceleration of the Universe has posed a significant challenge to contemporary cosmology. General relativity addresses this by introducing the cosmological constant, forming the basis of the standard cosmological model ($\Lambda$CDM). However, this model has limitations, leading cosmologists to explore alternative explanations for late-time acceleration. These alternatives range from models involving a dynamic dark fluid known as dark energy, to large-scale modifications of gravitational interaction, known as modified gravity. The formulation of general relativity fundamentally changed our understanding of gravitation, redefining gravity as a manifestation of the curvature of spacetime rather than a force as described by Newton. Despite its success, general relativity has shown incompatibilities with observations, necessitating the introduction of dark matter and dark energy.

In the chapter \ref{Chapter1}, we provide an overview of background formulation, fundamental gravity theories, and cosmological observations. Chapters \ref{Chapter2}-\ref{Chapter5} delve into the dark sector of the Universe in modified gravity, utilizing Markov Chain Monte Carlo (MCMC) analysis and extensive data sets derived from measurements of the background expansion of the Universe. Our analysis includes the stability of the cosmological model through phase space analysis.

Chapter \ref{Chapter2} focuses on the acceleration of the Universe within the framework of curvature-based gravity with Gauss--Bonnet invariant. We explore the $f(R, \mathcal{G})$ cosmological model, demonstrating its compatibility with a radiation era, early deceleration and late-time acceleration. Our stability analysis and critical points behavior constrain the model parameter, ensuring the model aligns with observed cosmological data for matter and dark energy densities.

In chapter \ref{Chapter3}, we employ the parametrization approach in modified teleparallel Gauss--Bonnet gravity to recreate cosmological models. By integrating observational data, we examine parameters such as the Hubble, deceleration and equation of state parameters. Our results indicate a transition from deceleration to acceleration, with the equation of state parameter indicating a quintessence phase. Additionally, we explore violations of the strong energy condition and perform the $\text{Om}(z)$ diagnostic to determine the age of the Universe.

Chapter \ref{Chapter4} investigates the cosmological stability of $f(Q, B)$ gravity using a dynamical system approach and Bayesian statistical analysis by using a numerical approach. Our study reveals a stable critical point corresponding to the de Sitter phase, consistent with dark energy-dominated late-time accelerated expansion. The $f(Q, B)$ model demonstrates a smooth transition from deceleration to acceleration, presenting a viable alternative to the $\Lambda$CDM model.

In chapter \ref{Chapter5}, we present a numerical method for solving the Friedmann equations in modified $f(\mathcal{G})$ gravity, predicting the redshift behavior of the Hubble expansion rate. Using Bayesian MCMC techniques with late-time cosmic observations, we demonstrate that the $f(\mathcal{G})$ model aligns with the $\Lambda$CDM model at low redshifts but differs at high redshifts, lacking a standard matter-dominated epoch. The model effectively captures the evolution of energy components, supporting its validity as an alternative explanation for the accelerated expansion of the Universe. 

Chapter \ref{Chapter6} concludes with a summary of the key findings and a comprehensive discussion of the research presented throughout the thesis, with an outlook on potential directions for future studies and applications in the field.

\end{abstract}

\clearpage


\addtocontents{toc}{\vspace{1em}}
\tableofcontents 
\addtocontents{toc}{\vspace{1em}}
\lhead{\emph{List of Tables}}
\listoftables 
\addtocontents{toc}{\vspace{1em}}
\lhead{\emph{List of Figures}}
\listoffigures 
\addtocontents{toc}{\vspace{1em}}



\setstretch{1.5}
\lhead{\emph{List of Symbols and Abbreviations}}
\listofsymbols{ll}{
\begin{tabular}{cp{0.5\textwidth}}
$z$ &: Redshift\\
$\mathcal{L}_{\text{m}}$ & : Matter Lagrangian\\
$g_{ij}$ &: Lorentzian metric\\
$g$ &: Determinant of $g_{ij}$\\ 
$G_{ij}$ &: Einstein tensor\\
$\Lambda$ &: Cosmological constant\\
$\tilde{\Gamma}^k{}_{ij}$ &: General affine connection \\ 
$\Gamma^k{}_{ij}$ &: Levi--Civita connection\\ 
$\hat{\Gamma}^k{}_{ij}$ &: Weitzenb\"{o}ck connection\\
$\nabla_{i}$ &: Covariant derivative \\ 
$R^k{}_{\sigma ij}$ &: Riemann tensor \\
$R_{ij}$ &: Ricci tensor \\
$R$ &: Ricci scalar \\
$T$ &: Torsion \\
$Q$ &: Nonmetricity scalar\\
$S_{M}$ &: Matter action\\
$T_{ij}$ &: Energy-momentum tensor\\
$\mathcal{T}$ &: Trace of the energy-momentum tensor \\
$H$ &: Hubble parameter\\
$q$ &: Deceleration parameter\\
$s$ &: Snap parameter\\
$\mathrm{j}$ &: Jerk parameter\\
\end{tabular}
}

\clearpage

\noindent
\begin{center}
\begin{tabular}{cp{0.5\textwidth}}
$L^k{}_{ij}$ &: Disformation tensor \\ 
$P^k{}_{ij}$ &: Super potential \\
$\chi^2$ &: Chi-square\\
$\Omega$ &: Density parameter\\
$G$ &: Newton's gravitational constant\\
$\square$ &: D'Alembert operator\\
\textbf{GR} &: \textbf{G}eneral \textbf{R}elativity\\
\textbf{TEGR} &: \textbf{T}eleparallel \textbf{E}quivalent to \textbf{G}eneral \textbf{R}elativity\\
\textbf{STEGR} &: \textbf{S}ymmetric \textbf{T}eleparallel \textbf{E}quivalent to \textbf{G}eneral \textbf{R}elativity\\
\textbf{$\Lambda$CDM} &: $\Lambda$ \textbf{C}old \textbf{D}ark \textbf{M}atter\\
\textbf{SNe Ia} &: Type \textbf{I}a \textbf{S}uper\textbf{n}ovae\\
\textbf{EoS} &: \textbf{E}quation \textbf{o}f \textbf{S}tate\\
\textbf{CC} &: \textbf{C}osmic \textbf{C}hronometers\\
\textbf{CMB} &: \textbf{C}osmic \textbf{M}icrowave \textbf{B}ackground\\
\textbf{CMBR} &: \textbf{C}osmic \textbf{M}icrowave \textbf{B}ackground \textbf{R}adiation\\
\textbf{BAO} &: \textbf{B}aryon \textbf{A}coustic \textbf{O}scillations\\
\textbf{MCMC} &: \textbf{M}arkov \textbf{C}hain \textbf{M}onte \textbf{C}arlo\\
\textbf{DE} &: \textbf{D}ark \textbf{E}nergy\\
\textbf{FLRW} &: \textbf{F}riedmann \textbf{L}ema\^{i}tre \textbf{R}obertson \textbf{W}alker
\end{tabular}
\end{center}


%
%


\clearpage 

\lhead{\emph{Glossary}} 

 


\mainmatter 
\setstretch{1.2}
\pagestyle{fancy} 



\chapter{Introduction} 
\label{Chapter1}

\lhead{Chapter 1. \emph{Introduction}} 

    Cosmology is a branch of astrophysics that explores the origin, evolution, and large-scale structure of the Universe. Over the past two decades, remarkable progress in theoretical and observational studies has deepened our understanding of the formation and evolution of the Universe. Einstein’s GR extends Newtonian gravity, which is valid only in weak gravitational fields, to describe strong-field regimes such as those near neutron stars, black holes and white dwarfs \cite{Moffat_2021_2021_02, Vilhena_2023_2023_04}. GR predicts black holes and gravitational waves, with both concepts verified by groundbreaking observations \cite{Abbott_2016_116}. In 1917, Einstein derived the first exact cosmological solution to his field equations, proposing a static Universe \cite{Einstein_1917_142} model by introducing the cosmological constant ($\Lambda$) to counteract gravitational collapse. Einstein's General Relativity (GR) revolutionized our understanding of gravity and laid the theoretical foundation for modern cosmology. However, his initial model, which described a static Universe, was found to be unstable. In contrast, Alexander Friedmann derived solutions to Einstein’s equations that predicted a dynamically expanding Universe, originating from an initial high-density state approximately 13.8 billion years ago \cite{Friedman_1922_10}. This insight led to the development of the Big Bang theory, which posits that the Universe evolved from a singular point of extreme density and temperature. Recognizing this, Einstein abandoned his cosmological constant, though the concept would later be revived to explain the accelerated expansion of the Universe.
    
    The standard cosmological model is based on the cosmological principle, which states that the Universe is homogeneous and isotropic on large scales, meaning it has a uniform distribution of matter and looks the same in all directions when averaged over vast distances. Although galaxies and galaxy clusters create inhomogeneities on scales smaller than approximately 350 Mpc,\footnote{1 Mpc = 106 pc; 1 pc $= 3.09 \times 1016 \text{m} = 3.26$ light-years (ly).} the Universe remains isotropic on larger scales, as evidenced by the uniformity of the CMB \cite{Yadav_2005_364, Clowes_2013_429}. 
    Additionally, CMBR \cite{Hu_2002_40} shows that anisotropies in the Universe are remarkably small, measuring less than roughly $0.001\%$. Einstein’s GR provides the foundation for the standard cosmological model, serving as the primary framework for describing gravitational interactions at cosmological scales. The early $20^\text{th}$ century saw the development of powerful telescopes and spectroscopic techniques that provided evidence for the existence of galaxies beyond our own, which are receding from us \cite{Slipher_1917_56}.  Subsequently, Friedmann \cite{Friedman_1924_21, Friedman_1922_10}, Lemaître \cite{Lemaitre_1927_49}, Robertson \cite{Robertson_1935_82} and Walker \cite{Walker_1937_42} introduced new solutions to Einstein’s field equations for an expanding Universe, leading to the development of the FLRW metric, a cornerstone of the standard cosmological model.

    In 1929 \cite{Hubble_1929_15}, Edwin Hubble discovered that galaxies recede from us at speeds proportional to their distances, leading to the introduction of the Hubble constant $H_{0}$ and establishing the concept of cosmic expansion \cite{Hubble_1931_74}. The observed expansion was well-explained by the FLRW metric, derived from Einstein’s field equations. The discovery of the CMB provided further evidence for the Big Bang model, which was already supported by Hubble’s observations. The CMB, anticipated through the FLRW metric, was experimentally confirmed by Arno Penzias and Robert Wilson in 1965. In the following decades, both theoretical and observational advancements reshaped our understanding of cosmic evolution. While the Big Bang model explains the early expansion of the Universe, it does not account for the observed large-scale uniformity of the CMB or the formation of cosmic structures. These challenges led to the development of inflationary theory. In the late $20^\text{th}$ century, the theory of cosmic inflation, proposed by Starobinsky \cite{Starobinsky_1979_30, Starobinsky_1982_117}, Guth \cite{Guth_1981_23} and Linde \cite{Linde_1982_108} provided a compelling explanation for the uniformity of the CMB and the formation of large-scale structures. This theory describes an exponential expansion in the early Universe and resolves fundamental puzzles such as the horizon, flatness and monopole problems. Guth's early inflationary models \cite{Guth_1981_23} predicted that quantum fluctuations during inflation seeded the large-scale structures observed today. While inflation explains the early Universe, observations of SNe Ia in the late 1990s provided strong evidence that the expansion of the Universe is not just continuing but accelerating over time. The discovery of the accelerating expansion of the Universe, supported by observations of SNe Ia, was independently confirmed by the High-Z Supernova Search Team \cite{Riess_1998_116} and the Supernova Cosmology Project \cite{Perlmutter_1998_517}. This unexpected phenomenon challenged our understanding of gravity and cosmic composition. Subsequent data from the Planck satellite \cite{Ade_2016_594a} refined our understanding of the Universe, showing that ordinary matter accounts for only 4--5\% of the total energy content, while dark matter contributes 25\% and dark energy makes up the remaining 70\%. These discoveries led to the formulation of the $\Lambda$CDM model in which dark matter governs the formation of cosmic structures and DE represented by the cosmological constant drives the accelerated expansion at late times. Despite its success, the cause of this acceleration remains one of the most perplexing open questions in cosmology. In response, various theoretical models have been proposed to explain late-time acceleration, either by introducing an additional field known as dark energy or by modifying gravity itself \cite{Sahni_2006_15, COPELAND_2006_15, Sami_2016_25, Brax_2018_81}. However, no universally accepted model has emerged, leaving the nature of cosmic acceleration as one of the central mysteries in modern physics.

    This chapter provides an overview of the standard cosmological model, discussing its theoretical foundations, observational evidence and challenges. In particular, it highlights the late-time accelerated expansion of the Universe, a phenomenon that has led to the development of modified gravity theories as possible explanations. The following sections explore the mathematical framework of cosmology, beginning with the cosmological principle and its implications for large-scale structure formation. For further insights into modern cosmology, one may refer to standard literature \cite{Weinberg_1972, Raychaudhuri_1979, Padmanabhan_2000, Rich_2009, Liddle_2015}. The following sections delve into the mathematical foundations of cosmology, beginning with the homogeneous and isotropic Universe.
\section{Homogeneous and Isotropic Universe}
    A homogeneous and isotropic Universe, described by the cosmological principle, is fundamental to cosmology. It means the Universe is uniform in composition and structure (homogeneity) and looks the same in all directions (isotropy) on large scales. Observational evidence supports this principle. For instance, the CMBR, which is the afterglow of the Big Bang, appears uniform in all directions, indicating isotropy. Additionally, large-scale surveys of galaxies show that their distribution is uniform when viewed over vast distances, supporting homogeneity. These assumptions simplify the mathematical models used in cosmology, allowing scientists to describe the large-scale structure of the Universe and evolution with greater precision. The cosmological principle underpins the standard model of cosmology, including the Big Bang theory and the expansion of the Universe. It also provides a foundation for understanding how matter and energy are distributed across the cosmos and how they have evolved over time.
\subsection{Scale Factor and Hubble’s Law}
    The scale factor and Hubble’s law are fundamental concepts in cosmology that describe the expansion of the Universe. The scale factor denoted as $a(t)$, measures how distances in the Universe change over time, providing a way to quantify the expansion. Hubble’s law proposed by Edwin Hubble, states that the recessional velocity of galaxies is directly proportional to their distance from us, expressed as 
    \begin{equation}
        v = H_0 \times D \, ,
    \end{equation}
    where $v$ is the velocity, $H_0$ is the Hubble constant and $D$ is the distance. This law provides strong evidence for the expansion of the Universe, indicating that galaxies move away from each other at a speed proportional to their separation. The scale factor is crucial in this context as it evolves over time, influencing the rate of expansion described by Hubble’s law.

    The rate of change of the distance from the relation $\vec{r} = a(t)\,\vec{x}$, where $x$ is the comoving distance and $\vec{r}$ can be defined as
        \begin{equation}
            \dot{\vec{r}} = \dot{a}(t) \vec{x} = \dot{a}(t) \frac{\vec{r}}{a(t)},
        \end{equation}
    in the form of Hubble parameter 
        \begin{equation}
            \dot{\vec{r}}= \frac{\dot{a}(t)}{a(t)} \vec{r} = H(t)\,\vec{r}\, . 
        \end{equation}
    Throughout this thesis, the notation ‘dot’ is utilized to indicate a derivative with respect to cosmic time $t$ and the Hubble parameter quantifies the rate of expansion of the Universe
        \begin{equation}
            \label{Hub}
            H(t) = \frac{\dot{a}(t)}{a(t)} \,\, .
        \end{equation}
\section{FLRW Metric}
    The fundamental assumption of modern cosmology is that the Universe exhibits symmetry on large scales, which significantly simplifies its mathematical description. The two key symmetries that shape the structure of spacetime in cosmology are
    \begin{itemize}
        \item \textbf{Translational symmetry:} The Universe appears the same regardless of position, meaning the laws of physics and the large-scale structure remain unchanged if we shift an observer to another location.
        \item \textbf{Rotational symmetry:} The Universe looks the same in all directions when viewed from any fixed point, implying that there are no preferred directions in space.
    \end{itemize}
    These symmetries are mathematically encoded in the FLRW metric, which describes an expanding Universe that is homogeneous (due to translational symmetry) and isotropic (due to rotational symmetry). When the FLRW metric \eqref{gen_flrw_metric} is substituted into Einstein field equations, it yields the Friedmann equations, which describe the evolution of the Universe expansion.
\subsection{Mathematical Formulation}
    The Robertson-Walker metric provides a mathematical framework for modeling a spacetime with translational and rotational symmetry. Due to rotational symmetry, the metric must not contain any preferred spatial directions, meaning the mixed components must vanish
        \begin{eqnarray}
            g_{i0} = g_{0i} = 0 \hspace{0.8cm} \text{for} \hspace{0.8cm} i = 1, 2, 3.
        \end{eqnarray}
    The general line element for a symmetric, expanding spacetime is then written as
        \begin{eqnarray}
            ds^2 &=& g_{\mu \nu} dx^\mu dx^\nu \nonumber\\
            &=& g_{00} dx^0 dx^0 + g_{ij} dx^i dx^j + 2g_{0i} dx^0 dx^i \nonumber\\
            &=& -N^2(t) \ c^2 dt^2 + a^2(t) \gamma_{ij} dx^i dx^j \, .
        \end{eqnarray}
    where $N^2(t) \equiv -g_{00}$ is a lapse function, often set to $N(t) = 1$ in standard cosmological conventions. The constant $c$ represents the speed of light and for simplicity, we adopt natural units, setting $c = 1$, which is common in cosmology and GR.
\subsection{Spatial Geometry and Curvature}
    The spatial part of the metric must be compatible with the assumption of rotational symmetry, meaning distances between points should depend only on their separation, not their direction. The most general form satisfying this condition is
        \begin{eqnarray}
            \gamma_{ij} = \frac{\delta_{ij}}{ \left( 1 + \frac{k}{4} x^n x^m \delta_{nm} \right)^2} \, .
        \end{eqnarray}
    where the dimensionless constant $k$ represents spatial curvature and takes the following values
    \begin{itemize}
        \item $k = 0$ \hspace{0.6cm} $\Rightarrow$ \hspace{0.3cm} \textbf{Flat (Euclidean) geometry}
        \item $k = +1$ \hspace{0.3cm} $\Rightarrow$ \hspace{0.3cm} \textbf{Closed (spherical) geometry}
        \item $k = -1$ \hspace{0.3cm} $\Rightarrow$ \hspace{0.3cm} \textbf{Open (hyperbolic) geometry}
    \end{itemize}
    Thus, the FLRW line element is
        \begin{eqnarray}
            ds^2 = -dt^2 + a^2(t) \frac{\delta_{ij} dx^i dx^j}{\left( 1 + \frac{k}{4} (x^2 + y^2 + z^2) \right)^2} \, .
        \end{eqnarray}
    Alternatively, in spherical coordinates ($r, \theta, \phi$), the FLRW metric takes the form
        \begin{eqnarray} \label{gen_flrw_metric}
            ds^2 = -dt^2 + a^2(t) \left( \frac{dr^2}{1 - k r^2} + r^2 d\theta^2 + r^2 \sin^2 \theta \, d\phi^2 \right) \, .
        \end{eqnarray}
    The metric tensor representation, which encodes distances, angles and causal relationships in spacetime, is given by
        \begin{eqnarray} \label{metric_diagonal_form}
            g_{\mu \nu} = \text{diag} \left( -1, \,\, \frac{a^2(t)}{1 - k r^2}, \,\, a^2(t)r^2, \,\, a^2(t)r^2 \sin^2 \theta \right) \, .
        \end{eqnarray}
    This metric serves as the foundation of modern cosmology, providing a framework for understanding the evolution of the Universe and its large-scale structure.
\section{Cosmological Redshift}
    The observation that nearly all celestial objects appear to be receding from us and that their apparent velocity increases with distance is fundamental evidence for cosmic expansion. As photons travel through the expanding Universe, the stretching of space increases their wavelengths, causing a loss of energy. If these photons originate in the visible spectrum, they start with a shorter wavelength (blue) and gradually shift toward longer wavelengths (red) as they travel. This phenomenon is known as cosmological redshift. Since galaxies are moving away from us, the redshift $z$ is defined as
        \begin{equation} \label{eq:redshift_def}
            z = \frac{\lambda_{\text{obs}} - \lambda_{\text{em}}}{\lambda_{\text{em}}},
        \end{equation}
    where $\lambda_{\text{obs}}$ and $\lambda_{\text{em}}$ are the observed and emitted wavelengths, respectively.

    The redshift of spectral lines strongly supports an expanding Universe and can be directly related to the scale factor $a(t)$. To derive this relationship, consider a photon travelling between two nearby points separated by an infinitesimal distance $dr$. Though depicted as galaxies, these points represent any two locations moving with the Hubble flow.

    According to Hubble's law, the relative velocity $dv$ between these points is given by
        \begin{equation}
            dv = H \, dr = \frac{\dot{a}}{a} \, dr,
        \end{equation}
    where $ H = \dot{a}/a $ is the Hubble parameter. Since the points are sufficiently close, we can apply the Doppler effect to relate the fractional change in wavelength to the velocity
        \begin{equation}
            \frac{d\lambda}{\lambda_{\text{em}}} = \frac{dv}{c}.
        \end{equation}
    Since the separation changes with cosmic expansion, the time required for light to traverse the distance $dr$ is $dt = dr/c$. Combining these equations gives
        \begin{equation}
            \frac{d\lambda}{\lambda_{\text{em}}} = \frac{\dot{a}}{a} dt = \frac{da}{a}.
        \end{equation}
    Integrating both sides results in
        \begin{equation}
            \ln \lambda = \ln a + C,
        \end{equation}
    or equivalently,
        \begin{equation}
            \lambda \propto a.
        \end{equation}
    This relation implies that as the Universe expands, photon wavelengths stretch in direct proportion to the scale factor. If a photon's wavelength has doubled, the Universe was half its present size when it was emitted.

    The redshift $z$, as defined in equation \eqref{eq:redshift_def}, is directly related to the scale factor
        \begin{equation} \label{eq:redshift_scale_factor}
            1 + z = \frac{\lambda_{\text{obs}}}{\lambda_{\text{em}}} = \frac{a(t_0)}{a(t_{\text{em}})},
        \end{equation}
    where $t_0$ is the present cosmic time and $t_{\text{em}}$ is the time of emission. The redshift parameter is widely used in observational cosmology to describe the history of expansion in the Universe.   
\section{Cosmological Distances}
    Distance measurements are fundamental to cosmology, serving as a bridge between theory and observation. In contrast to flat, static Euclidean space, defining distances in an expanding Universe requires careful treatment of relativistic effects and the dynamical nature of spacetime. Various cosmological distance measures such as comoving distance, angular diameter distance and luminosity distance are crucial for interpreting astronomical observations and understanding the large-scale structure and evolution of the Universe.
\subsection{Hubble Distance and Comoving Distance}
    The Hubble distance establishes a fundamental length scale in cosmology, setting a natural benchmark for cosmic distances. It is given by
        \begin{eqnarray}
            d_H = \frac{c}{H_0} \approx 3000 \, h^{-1} \, \text{Mpc} \approx 9.26 \times 10^{25} \, h^{-1} \, \text{m}.
        \end{eqnarray}
    Here, $c$ is the speed of light and $H_0$ is the Hubble parameter at the present epoch. The dimensionless Hubble parameter $h$ is given by
        \begin{eqnarray}
            h = \frac{H_0}{100 \, \text{km} \, \text{Mpc}^{-1} \, \text{s}^{-1}}.
        \end{eqnarray}
    In an expanding Universe, the proper separation between objects changes over time. The comoving distance $d_C(z)$ accounts for this expansion, representing the comoving separation (present-day separation) of two objects moving with the Hubble flow. It is defined as
        \begin{eqnarray}
            d_C(z) = d_H \int_0^z \frac{\mathrm{d}z^\prime}{E(z^\prime)},
        \end{eqnarray}
    where the dimensionless expansion rate $E(z)$ is given by
        \begin{eqnarray}
            E(z) = \frac{H(z)}{H_0}.
        \end{eqnarray}
\subsection{Transverse Comoving Distance}
    The transverse comoving distance $d_P(z)$ accounts for spatial curvature of space and depends on the geometry of the Universe. It is defined as
        \begin{eqnarray}
            d_P(z) =
            \begin{cases} 
            \frac{d_H}{\sqrt{\Omega_k}} \sinh\left(\sqrt{\Omega_k} \frac{d_C(z)}{d_H}\right)\, , & \Omega_k > 0 \, , \\
            d_C(z)\, , & \Omega_k = 0\, , \\
            \frac{d_H}{\sqrt{|\Omega_k|}} \sin\left(\sqrt{|\Omega_k|} \frac{d_C(z)}{d_H}\right)\, , & \Omega_k < 0\, .
            \end{cases}
        \end{eqnarray}
    Here, $\Omega_k$ is the curvature density parameter, which determines whether the Universe is open ($\Omega_k > 0$), flat ($\Omega_k = 0$), or closed ($\Omega_k < 0$).
\subsection{Angular Diameter Distance}
    The angular diameter distance $d_A(z)$ relates an object's physical size $D$ to its observed angular size $\delta\theta$ through  
        \begin{eqnarray}
            d_A(z) = \frac{d_P(z)}{1 + z} \, .
        \end{eqnarray}  
    At low redshifts ($z \ll 1$), $d_A \approx d_P$, but at higher redshifts, $d_A$ decreases due to the cumulative effect of cosmic expansion. This leads to a non-monotonic behavior, where objects at very high redshifts may appear larger than expected, despite being farther away. The observed angular size $\delta\theta$ of an object follows the relation  
        \begin{eqnarray}
            \delta\theta = \frac{D}{d_A(z)} \, .
        \end{eqnarray}  
    This inverse relationship implies that the object appears smaller in the sky as the angular diameter distance increases. Angular diameter distance plays a crucial role in interpreting the apparent sizes of galaxies, galaxy clusters and features in the CMB, helping to constrain cosmological parameters and the expansion history of the Universe.
\subsection{Luminosity Distance}
    The luminosity distance $d_L(z)$ is crucial for determining the brightness of astronomical objects. It relates the intrinsic luminosity $L$ of a source to its observed flux $F$, given by 
        \begin{eqnarray}
            F = \frac{L}{4 \pi d_L^2} \, ,
        \end{eqnarray}
    and also it is given as 
        \begin{eqnarray}
            d_L(z) = (1 + z) d_P(z).
        \end{eqnarray}
    The additional factor of $(1+z)$ in the luminosity distance arises due to two key relativistic effects: (i) the redshift of photons, which decreases their energy and (ii) time dilation, which alters the observed arrival rate of photons from distant sources.

    For a flat Universe, the luminosity distance simplifies to
        \begin{eqnarray}
            d_L(z) = (1 + z) d_C(z).
        \end{eqnarray}
\subsection{Observational Applications: Distance Modulus and Cosmological Constraints}
    SNe Ia act as ``standard candles," because their well-known intrinsic luminosities allow for precise distance measurements. These supernovae provide a robust method for constraining cosmological parameters. This allows astronomers to measure luminosity distances directly. The observed magnitude $m$ of a supernova at redshift $z$ is related to its absolute magnitude $M$ and luminosity distance by the distance modulus
        \begin{eqnarray}
            \mu = m - M = 5 \log_{10}\left(\frac{d_L(z)}{(1\ \text{Mpc})}\right) + 25.
        \end{eqnarray}
    This relation is fundamental for reconstructing the expansion history of the Universe.

    Comparing observed supernova luminosity distances with theoretical predictions enables tight constraints on fundamental cosmological parameters, including the Hubble constant ($H_0$), matter density ($\Omega_\text{m}$) and DE density ($\Omega_\Lambda$). These constraints play a crucial role in distinguishing between different cosmological models. Precise cosmological distance measurements provide key insights into the nature of gravity at cosmic scales. Although GR remains the leading framework, recent observations question its completeness, prompting the exploration of alternative theories of gravity. These alternative formulations, which involve curvature, torsion and nonmetricity, provide different perspectives on the interaction between spacetime and matter.
\section{Geometrical Foundations of Gravity}
    Gravity can be geometrically formulated in three equivalent ways, each providing a distinct perspective on the interaction between spacetime and matter-energy. These formulations are based on different geometric properties of spacetime: curvature, torsion and nonmetricity. These formulations are characterized by three fundamental scalars: the Ricci scalar ($R$), the torsion scalar ($T$) and the nonmetricity scalar ($Q$), collectively known as the geometric trinity of gravity.
\subsection{Curvature} 
    In GR, spacetime is described as a smooth, differentiable manifold with a metric tensor $g_{\mu\nu}$. The curvature of this manifold is captured by the scalar curvature $R$, derived from the Riemann curvature tensor and the corresponding connection used is the Levi--Civita connection. The Levi--Civita connection is a torsion-free and metric-compatible connection, meaning
    \begin{itemize}
        \item $\nabla_\lambda g_{\mu\nu} = 0$ (metric-compatibility),
        \item $\Gamma^\lambda{}_{\mu\nu} = \Gamma^\lambda{}_{\nu\mu}$ (no torsion).
    \end{itemize}
    The Levi--Civita connection $\Gamma^\lambda{}_{\mu\nu}$ is given explicitly by
        \begin{eqnarray}
            \Gamma^\lambda{}_{\mu\nu} = \frac{1}{2} g^{\lambda\rho} \left( \partial_\mu g_{\nu\rho} + \partial_\nu g_{\mu\rho} - \partial_\rho g_{\mu\nu} \right)\, .
        \end{eqnarray}
    Using this connection, we can construct the Riemann curvature tensor $R^\rho_{\ \sigma\mu\nu}$, which measures how vectors are transported around infinitesimal loops in spacetime
    \begin{equation}
        R^\rho{}_{\sigma\mu\nu} = \partial_\mu \Gamma^\rho{}_{\nu\sigma} - \partial_\nu \Gamma^\rho{}_{\mu\sigma} + \Gamma^\rho{}_{\mu\lambda} \Gamma^\lambda{}_{\nu\sigma} - \Gamma^\rho{}_{\nu\lambda} \Gamma^\lambda{}_{\mu\sigma} \, .
    \end{equation}
    From the Riemann tensor, we obtain the Ricci tensor $R_{\mu\nu}$ by contracting the first and third indices
    \begin{equation}
        R_{\mu\nu} = R^\lambda{}_{\mu\lambda\nu} \, .
    \end{equation}
    Finally, the scalar curvature $R$ is the trace of the Ricci tensor
    \begin{equation}
        R = g^{\mu\nu} R_{\mu\nu} \, .
    \end{equation}
    The Levi--Civita connection plays a key role in defining how vectors change along curves in spacetime, ensuring that the geometry is torsion-free and governed by curvature alone. The Einstein--Hilbert action in GR is
        \begin{equation}
            S=\int d^{4} x \sqrt{-g} \left(\frac{R}{16 \pi G}+ \mathcal{L}_\text{m}\right) \, ,
        \end{equation}
    The extension of GR to $f(R)$ gravity is a standard modification where the Einstein--Hilbert action is generalized by replacing the Ricci scalar $R$ with an arbitrary function $f(R)$. This modification introduces higher-order curvature terms, enabling explanations for cosmic acceleration and other phenomena without invoking DE.
\subsection{Torsion} 
    In teleparallel gravity, the gravitational interaction is described using torsion rather than curvature. The key object here is the torsion scalar $T$ and the connection used is the Weitzenb\"{o}ck connection, which has zero curvature but nonzero torsion. The Weitzenb\"{o}ck connection $\hat{\Gamma}^\lambda{}_{\mu\nu}$ is chosen such that spacetime is flat (i.e., the Riemann curvature tensor vanishes), but torsion remains. This connection is defined using the tetrad (or vierbein) fields $e^a{}_{\mu}$, which relate the spacetime metric to an orthonormal frame
    \begin{equation}
        g_{\mu\nu} = \eta_{ab} e^a{}_{\mu} e^b{}_{\nu} \, ,
    \end{equation} 
    where $\eta_{ab}$ is the Minkowski metric. The Weitzenb\"{o}ck connection is then given by
    \begin{equation}
    \hat{\Gamma}^\lambda{}_{\mu\nu} = e^\lambda{}_{a} \, \partial_\mu \, e^a{}_{\nu} \, ,
    \end{equation}
    This connection has torsion but no curvature.

    The torsion tensor $T^\lambda{}_{\mu\nu}$ is defined as the antisymmetric part of the Weitzenb\"{o}ck connection
    \begin{equation} 
    T^\lambda{}_{\mu\nu} = \hat{\Gamma}^\lambda{}_{\nu\mu} - \hat{\Gamma}^\lambda{}_{\mu\nu} \, .
    \end{equation}

    The torsion scalar $T$ is constructed from contractions of the torsion tensor
    \begin{equation} 
    T = \frac{1}{4} T^\lambda{}_{\mu\nu} T_\lambda{}^{\mu\nu} + \frac{1}{2} T^\lambda{}_{\mu\nu} T_\lambda{}^{\nu\mu} - T^\lambda{}_{\lambda\mu} T_\nu{}^{\mu\nu} \, .
    \end{equation}
    In teleparallel gravity, the torsion scalar $T$ replaces the scalar $R$ in describing the gravitational field. The field equations in teleparallel gravity are derived from the action involving $T$, analogous to the Einstein--Hilbert action in GR
        \begin{equation}
            S=\int d^{4} x \ e \left(\frac{T}{16 \pi G}+ \mathcal{L}_\text{m}\right)  \, ,
        \end{equation}
    where $e = \det(e^a{}_{\mu})$ is the determinant of the tetrad.

    Teleparallel gravity is a reformulation of GR where torsion, rather than curvature, mediates the gravitational interaction. This approach is used in theories like $f(T)$ gravity \cite{Duchaniya_2022_82, Capozziello_2011_84, Bamba_2011_01, Cai_2016_79, Myrzakulov_2011_71, Baojiu_2011_83, Tamanini_2012_86, Anagnostopoulos_2019_100, Debenedictis_2022_105_084020, Nair_2022_105}, where the gravitational Lagrangian is a general function of $T$.
\subsection{Nonmetricity} 
    In symmetric teleparallel gravity, also known as $f(Q)$ gravity, the geometric description of gravity is based on nonmetricity rather than curvature or torsion. The key scalar is the nonmetricity scalar $Q$, and the corresponding connection is a general affine connection that is flat (zero curvature) and torsion-free but allows nonmetricity. In symmetric teleparallel gravity, the nonmetricity tensor $Q_{\lambda\mu\nu}$ measures how the metric changes under parallel transport
    \begin{equation}
    Q_{\lambda\mu\nu} = \nabla_\lambda g_{\mu\nu} \, . 
    \end{equation}
    For a metric-compatible connection like the Levi--Civita connection, this term vanishes ($\nabla_\lambda g_{\mu\nu} = 0$), but in symmetric teleparallel gravity, nonmetricity is allowed, so the connection does not preserve the metric.

    The nature of the geometry is determined by the affine connection, which is represented by $\tilde{\Gamma}^\zeta{}_{\mu\nu}$. The three independent components can be expressed in the general form of an affine connection as follows
        \begin{equation}\label{ch4_eq: affine connection}
            \tilde{\Gamma}^\sigma{}_{\mu \nu}=\Gamma^\sigma{}_{\mu\nu}+K^\sigma{}_{\mu\nu}+ L^\sigma{}_{\mu\nu} \,,
        \end{equation}
    where $\Gamma^\sigma{}_{\mu\nu}$, $L^\sigma{}_{\mu\nu}$, and $K^\sigma{}_{\mu\nu}$ respectively represent the Levi--Civita connection, the deformation tensor, and the contortion tensor. These are defined as
        \begin{eqnarray} \label{ch4_connections formulae}
            \Gamma^\sigma{}_{\mu\nu}&=&\frac{1}{2}g^{ \sigma\zeta}\left(\partial_\mu g_{\zeta \nu}+\partial_ \nu g_{\zeta \mu}-\partial_\zeta g_{\mu \nu}\right), \\
            L^\sigma{}_{\mu \nu}&=&\frac{1}{2}g^{\sigma \zeta} (-Q_{\mu \zeta \nu} - Q_{\nu \zeta \mu} + Q_{\zeta \mu \nu}), \\ K^\sigma{}_{\mu\nu}&=&\frac{1}{2}g^{\sigma \zeta} (T_{\mu \zeta \nu} + T_{\nu \zeta \mu} + T_{\zeta \mu \nu}).
        \end{eqnarray}

    The general affine connection $\tilde{\Gamma}^\lambda{}_{\mu\nu}$ in this framework is chosen to satisfy the following conditions:
    \begin{itemize}
        \item Zero curvature: $R^\lambda{}_{\mu\nu\rho} = 0$,
        \item Zero torsion: $T^\lambda{}_{\mu\nu} = 0$,
        \item Non-zero nonmetricity: $Q_{\lambda\mu\nu} \neq 0$.
    \end{itemize}
    The nonmetricity scalar $Q$ is constructed from the nonmetricity tensor, using contractions similar to how $R$ and $T$ are constructed
    \begin{equation}
    Q = g^{\mu\nu} \left( L^\lambda{}_{\mu\nu} L^\rho{}_{\lambda\rho} - L^\lambda{}_{\mu\rho} L^\rho{}_{\lambda\nu} \right) \, , 
    \end{equation}
    where $L^\lambda{}_{\mu\nu}$ is the distortion tensor related to the nonmetricity tensor.

    In symmetric teleparallel gravity, the nonmetricity scalar $Q$ plays a role similar to $R$ in GR and $T$ in teleparallel gravity, describing the gravitational field through nonmetricity instead of curvature or torsion. The field equations are derived from an action based on $Q$
        \begin{equation}
            S=\int d^{4} x \ \sqrt{-g} \left(\frac{Q}{16 \pi G}+ \mathcal{L}_\text{m}\right)  \, .
        \end{equation}
    Symmetric teleparallel gravity provides a third geometric description of gravity, focusing on how lengths and angles change under parallel transport rather than spacetime curvature or torsion. It leads to modifications of GR in theories like $f(Q)$ gravity, offering alternative ways to address cosmological and astrophysical problems.

    While GR has been remarkably successful, recent observations such as the accelerated cosmic expansion suggest the need for modifications or extensions to its framework. The Geometric Trinity of Gravity provides three distinct but equivalent approaches to describing gravitational interactions, offering potential avenues for addressing these open questions.
\section{Theory of General Relativity}
    In Newtonian physics, space and time are treated as absolute, unchanging backgrounds where objects move under the influence of gravity, which is described as a force acting at a distance. However, this perspective was fundamentally transformed in the early $20^\text{th}$ century when Einstein presented a radically different interpretation of gravity, describing it not as a force but as the curvature of spacetime. Building upon the principles of Special Relativity, Einstein proposed that the laws of physics should appear identical to all freely falling observers, irrespective of their position in a gravitational field. This profound insight, known as the equivalence principle, became the cornerstone of his new theory of gravity. This insight enabled Einstein to connect gravity to the fundamental nature of spacetime, transforming our understanding of gravitational phenomena. In GR, the curvature of spacetime is described using two essential mathematical structures. The first is the metric tensor $g_{\mu\nu}$, which quantifies distances between points and distinguishes spatial from temporal directions, effectively defining the geometry of spacetime. Its components vary with position, reflecting the local curvature. The second structure is the affine connection $\tilde{\Gamma}^\alpha{}_{\mu\nu}$, which defines parallel transport and enables the comparison of tensors at different spacetime points. 
    
    The curvature associated with this Levi--Civita connection, ${R}{}^\lambda{}_{\mu\nu\alpha}$, is fundamental in explaining gravitational interactions within the framework of GR. To understand how GR works, we first need to understand its theoretical foundations. So, let us start with the Einstein--Hilbert action
        \begin{equation} \label{ch1_gr_action}
            S=\frac{1}{2 \kappa^2} \int d^{4} x \sqrt{-g}(R-2 \Lambda)+\int d^{4} x \sqrt{-g}\, \mathcal{L}_\text{m} [g_{\mu \nu}, \Psi^{i}] \, ,
        \end{equation}
    where $g$ is a metric determinant, $\mathcal{L}_\text{m}$ describes matter Lagrangian describing the motion of matter fields $\Psi^{i}$, $\kappa^2=8\pi G$, $G$ is the Newton's gravitational constant and $g_{\mu\nu}$ represents the metric tensor, $R$ denotes the Ricci scalar and $\Lambda$ is the cosmological constant. The term $d^4x \sqrt{-g}$ refers to the volume element expressed in terms of the determinant of the metric $g_{\mu\nu}$. To derive the Einstein equations, we employ the principle of least action to equation \eqref{ch1_gr_action}. By introducing a small perturbation to the metric, such that $g_{\mu\nu} \rightarrow g_{\mu\nu} + \delta g_{\mu\nu}$, the action \eqref{ch1_gr_action} transforms accordingly
        \begin{equation}
            \delta S=\int d^{4} x\left[\frac{1}{16 \pi G}\left(\frac{\delta R}{\delta g^{\mu \nu}}+\frac{(R-2 \Lambda)}{\sqrt{-g}} \frac{\delta \sqrt{-g}}{\delta g^{\mu \nu}}\right)+\frac{1}{\sqrt{-g}} \frac{\delta\left(\sqrt{-g} \mathcal{L}_\text{m} \right)}{\delta g^{\mu \nu}}\right] \sqrt{-g} \delta g^{\mu \nu} \, .
        \end{equation}
    The variation of the metric determinant can be expressed as
        \begin{equation}
            \delta \sqrt{-g}=-\frac{1}{2} \sqrt{-g} g_{\mu \nu} \delta g^{\mu \nu}  \, ,
        \end{equation}
    and as a metric variation $\delta g^{\mu \nu}$ smoothly vanishes when approaching the boundary of spacetime, in conjunction with the tensor identities in \cite{Carroll-2019}, we can derive the Einstein field equations
        \begin{equation} \label{ch1_einstein field eq}
            G_{\mu \nu} \equiv R_{\mu \nu}-\frac{1}{2} R g_{\mu \nu}+\Lambda g_{\mu \nu} = \kappa^2 T_{\mu \nu} \, .
        \end{equation}
    The energy-momentum tensor $T_{\mu \nu}$ describes the distribution of energy and momentum in spacetime and is defined as  
        \begin{equation}
            T_{\mu \nu} \equiv \frac{-2}{\sqrt{-g}} \frac{\delta\left(\sqrt{-g} \mathcal{L}_\text{m} \left[g_{\mu \nu}, \Psi^{i}\right]\right)}{\delta g^{\mu \nu}} \, .
        \end{equation}  
    Its conservation follows from the Bianchi identity, which arises due to the diffeomorphism invariance of GR. Specifically, the Einstein tensor $G_{\mu\nu}$ satisfies  
        \begin{equation}
            \nabla_{\mu} G^{\mu \nu}=0 \, ,
        \end{equation}  
    where $\nabla_{\mu}$ denotes the covariant derivative associated with the metric $g_{\mu\nu}$. Through the Einstein field equations \eqref{ch1_einstein field eq}, this identity directly implies the conservation of the energy-momentum tensor
        \begin{equation} \label{ch1_energymomentum_tensor_zero}
            \nabla_{\mu} T^{\mu \nu}=0 \, .
        \end{equation}  
    Expanding this equation in terms of the Christoffel symbols, we obtain  
        \begin{equation}
            \nabla_{\mu} \ T^{\mu \nu} \equiv \partial_{\mu} \ T^{\mu \nu} + \Gamma^\nu{}_{\mu \rho} \ T^{\mu \rho} + \Gamma^\mu{}_{\mu \rho} \ T^{\rho \nu} = 0 \, .
        \end{equation}  
    This equation expresses the local conservation of energy and momentum, ensuring that no energy is created or destroyed in an isolated system. The conservation law plays a fundamental role in GR, governing the dynamics of matter and energy in curved spacetime. 
\subsection{Geodesics}
    In the study of particle motion, the equation of motion is derived by varying the action with respect to the coordinates of a point particle. The separation between two events in spacetime is quantified by the line element $ds^{2} = g_{\mu \nu} dx^{\mu} dx^{\nu}$, where $g_{\mu \nu}$ is the metric tensor. Proper time $\tau$, is defined through the relation $d\tau^2 = -ds^2$. A freely falling particle follows a trajectory that maximizes its proper time, which corresponds to the shortest path in curved spacetime known as a geodesic.

    To describe this motion, we formulate the action for a point particle of mass $m$, parameterized by its worldline $x^\mu = x^\mu(\tau)$
        \begin{eqnarray}
            S = m \int \sqrt{-g_{\mu \nu} \frac{dx^{\mu}}{d\tau} \frac{dx^{\nu}}{d\tau}} \, d\tau.
        \end{eqnarray}
    Varying the action yields the geodesic equation, which describes the motion of a freely falling particle in curved spacetime. In a more compact and general form, it can be written using the inverse metric $g^{\mu \sigma}$ as
        \begin{eqnarray} \label{ch1_geodesic equation}
            \frac{d^{2} x^{\mu}}{d\tau^{2}} = -\frac{1}{2} g^{\mu \sigma} \left( \partial_{\alpha} g_{\sigma \beta} + \partial_{\beta} g_{\sigma \alpha} - \partial_{\sigma} g_{\alpha \beta} \right) \frac{dx^{\alpha}}{d\tau} \frac{dx^{\beta}}{d\tau}.
        \end{eqnarray}
    This is the geodesic equation, which governs the trajectory $x^{\mu}(\tau)$ of a particle or light ray in a gravitational field. By solving equation \eqref{ch1_geodesic equation}, we find the path of the particle in spacetime, which reflects the geometry determined by the metric $g_{\mu \nu}$. The geodesic equation is fundamental to understanding how gravity shapes the motion of objects in curved spacetime. The next section derives the Friedmann equations, which describe the expansion history under these alternative gravitational models.
\subsection{Field Equations in General Relativity}
    The cosmological framework established so far remains general and does not depend on any specific gravitational theory. It applies not only to GR but also to modified theories of gravity. In this section, we derive the Friedmann equations, which govern the dynamical evolution of the Universe. These equations arise from Einstein’s field equations when applied to the homogeneous and isotropic FLRW metric. These equations are a set of two differential equations in time involving the scale factor $a(t)$, the density $\rho(t)$ and the pressure $p(t)$ of the fluid. Along with the continuity equation, these equations form a set of three equations, two of which are independent, with three unknown dynamic variables, viz., $a(t), \, \rho(t),\, \text{and}\,\, p(t)$, fully describing our system once solved.

    The FLRW metric in comoving coordinates is given by
        \begin{eqnarray}
            ds^2 = -dt^2 + a^2(t) \left( \frac{dr^2}{1 - k r^2} + r^2 d\Omega^2 \right),
        \end{eqnarray}
    where $a(t)$ is the scale factor, $k$ is the curvature parameter and $d\Omega^2 = d\theta^2 + \sin^2\theta \, d\phi^2$ is the angular part of the metric. The stress-energy tensor $T_{\mu\nu}$ for a perfect fluid, the stress-energy tensor is
        \begin{eqnarray}
            T_{\mu\nu} = (\rho + p) u_\mu u_\nu + p \, g_{\mu\nu},
        \end{eqnarray}
    where $ \rho $ is the energy density, $ p $ is the pressure, and $ u_\mu = (1,0,0,0) $ is the four-velocity of the fluid. The contraction of $T_{\mu \nu}$ gives $T_\nu{}^{\mu} = \text{diag}(-\rho, p, p, p)$, with trace $\mathcal{T} =  -\rho + 3p$.
\subsubsection{Friedmann Equations for Non-Zero Curvature (\texorpdfstring{$ k \neq 0 $}{})}
    Substituting the FLRW metric into the Einstein field equations yields the Friedmann equations
        \begin{eqnarray}
            \left( \frac{\dot{a}}{a} \right)^2 = \frac{8\pi G}{3} \rho - \frac{k}{a^2}, \quad \text{(First Friedmann equation)}  
        \end{eqnarray}
        and
        \begin{eqnarray}
            \frac{\ddot{a}}{a} = -\frac{4\pi G}{3} (\rho + 3p). \quad \text{(Second Friedmann equation)}
            \end{eqnarray}
\subsubsection{Friedmann Equations for Flat Universe (\texorpdfstring{$ k = 0 $}{})}
    In the special case of a spatially flat Universe ($ k = 0 $), the Friedmann equations take a particularly simple form, reflecting a balance between the expansion rate and the matter-energy content
        \begin{eqnarray} \label{first Friedmann eq}
            \left( \frac{\dot{a}}{a} \right)^2 = \frac{8\pi G}{3} \rho \, , 
        \end{eqnarray} 
        \begin{eqnarray} \label{second Friedmann eq}
            \frac{\ddot{a}}{a} = -\frac{4\pi G}{3} (\rho + 3p) \, .
        \end{eqnarray}   
    The conservation equation expresses the fundamental relationship between these quantities
        \begin{equation} \label{general continuity eq}
            \dot{\rho}+3 H \rho (1+\omega)=0 \, ,
        \end{equation}
    where $\omega = \frac{p}{\rho}$. The critical density, denoted as $\rho_c$, is defined as the density of a substance when the curvature of the Universe ($k$) is equal to zero
        \begin{equation}
            \rho_{\text{c}}(t)=\frac{3 H^{2}}{8 \pi G} \, .
        \end{equation}
    The energy content of the Universe is typically described using dimensionless density parameters, each representing the contribution of a different component to the total energy density of the Universe. These parameters evolve with redshift according to their respective equations of state. The given information enables the definition of a density parameter, denoted as $\Omega$
        \begin{equation}
            \Omega_{\text{m}} \equiv \frac{\rho_\text{m}}{\rho_{\text{c}}} \, , \hspace{0.5cm} \Omega_{\text{r}} \equiv \frac{\rho_\text{r}}{\rho_{\text{c}}} \hspace{0.7cm} \text{and} \hspace{0.5cm} \Omega_\Lambda \equiv \frac{\rho_\Lambda}{\rho_{\text{c}}}
        \end{equation}
    In the same way, we can establish $\Omega_k$ to represent the curvature of the Universe. The evolution of the expansion of the Universe is governed by the combined effects of these parameters
        \begin{equation}
            \Omega_{\text{m}} + \Omega_{\text{r}} + \Omega_{k} + \Omega_{\Lambda}=1 \, .
        \end{equation}
    This can then be combined with equations \eqref{first Friedmann eq} and \eqref{general continuity eq} to obtain
        \begin{equation} \label{general_LCDM_eq}
            H^{2}(z)=H_{0}^{2}\left[\Omega_{\text{r}0}(1+z)^{4}+\Omega_{\text{m}0}(1+z)^{3}+\Omega_{k0}(1+z)^{2}+\Omega_{\Lambda 0}\right]  \, .
        \end{equation}
    We define the density parameters at the present epoch ($z = 0$) as $\Omega_{\text{m}0}$ (matter), $\Omega_{\text{r}0}$ (radiation), $\Omega_{k0}$ (spatial curvature), and $\Omega_{\Lambda 0}$ (DE). The following expressions are derived using the standard normalization of the scale factor, where $a_0 = 1$ at the present epoch. The evolution of these parameters with redshift $z$ is described by
    \begin{itemize}
        \item \textbf{Pressureless matter:} $\Omega_\text{m}(z) = 3 H_0^2 \Omega_{\text{m}0} (1 + z)^3$,
        \item \textbf{Radiation:} $\Omega_\text{r}(z) = 3 H_0^2\Omega_{\text{r}0} (1 + z)^4$,
        \item \textbf{Spatial curvature:} $\Omega_k(z) = 3 H_0^2 \Omega_{k0} (1 + z)^2\, .$ 
    \end{itemize}
    The parameters in equation \eqref{general_LCDM_eq} were precisely measured by the Planck collaboration in 2018 \cite{Aghanim_2020_641}. The energy density scaling with the scale factor for different components of the Universe is summarized in Table \ref{gen_energy_density_table_first}.
    \begin{table}[H]
        \centering
        \begin{tabular}{|c|c|c|}
        \hline Fluid type & EoS & $\rho(a)$ \\
        \hline \hline Cold matter & $w=0$ & $\propto a^{-3}$ \\
        \hline Radiation & $w=1 / 3$ & $\propto a^{-4}$ \\
        \hline Cosmological constant & $w=-1$ & $\propto$ constant \\
        \hline Curvature & $w=-1 / 3$ & $\propto a^{-2}$ \\
        \hline
    \end{tabular}
        \caption{Energy density scaling with the scale factor for different components of the Universe.}
        \label{gen_energy_density_table_first}
    \end{table}
    The Friedmann equations provide the foundation for understanding cosmic expansion, which depends on the dominant energy components at different epochs: matter, radiation and DE. Each component behaves as a perfect fluid with an EoS $p = \omega \rho$, where $\omega$ determines its properties. Matter ($\omega = 0$) represents non-relativistic particles, such as galaxies, while radiation ($\omega = \frac{1}{3}$) describes ultra-relativistic particles like neutrinos and photons. DE ($\omega \approx -1$) drives the accelerated expansion of the Universe, exhibiting a repulsive gravitational effect. These fluid-like behaviors shape the large-scale cosmic evolution, linking thermodynamics to the dynamic history of the Universe.
\subsection{Theoretical Foundations and Observational Challenges in General Relativity}
    GR is a foundational theory in modern physics, offering a geometrical description of gravitation. It postulates that spacetime is a dynamic, curved entity influenced by the presence of matter and energy. This subsection provides an in-depth exploration of the mathematical underpinnings of GR and the observational challenges associated with testing its predictions.
\subsubsection{Limitations of General Relativity: A Case for Modified Gravity}
    Despite its remarkable success in describing gravitational interactions, GR struggles to account for several key cosmological observations without invoking unseen components, such as dark matter and DE.
    \begin{enumerate}
        \item \textbf{The Dark Matter Problem:} Several independent observations, such as galaxy rotation curves, gravitational lensing in galaxy clusters, and anisotropies in the cosmic microwave background, provide strong evidence for the existence of an invisible mass component known as dark matter. Within the $\Lambda$CDM model, dark matter contributes a density parameter $\Omega_{\text{m}} = 0.265$, suggesting it comprises approximately five times the amount of baryonic matter. Despite its inferred gravitational effects, the fundamental nature of dark matter remains unknown.
        \item \textbf{The DE and Cosmic Acceleration Problem:} DE represents a still-unknown component of the Universe responsible for accelerating cosmic expansion. Unlike matter, DE exhibits negative pressure, counteracting gravitational attraction on cosmic scales. Observations indicate that DE behaves like a perfect barotropic fluid, characterized by the EoS $p = \omega \rho$ with $\omega = -1.03 \pm 0.03$ \cite{Aghanim_2020_641}. While the exact nature of DE remains unknown, this value aligns with the $\Lambda$CDM model, which posits a cosmological constant $\Lambda$ (where $\omega = -1$). The $\Lambda$CDM model is the prevailing cosmological framework aligning with most observational data. This model assumes that GR accurately describes gravitational interactions at cosmological distances. As suggested by its name, the $\Lambda$CDM model posits that dark matter and DE represented by the cosmological constant, primarily influence the overall energy composition of the Universe. The standard model of cosmology has achieved significant success, but it comes with certain limitations. Atomic matter constitutes less than $5\%$ of the Universe, and the nature of DE and matter remains unknown. These components have only been inferred through their gravitational effects and have not been directly detected. Furthermore, the $\Lambda$CDM model encounters various theoretical and observational challenges. 

        This thesis examines the phenomenon of late-time cosmic acceleration within the context of extensions to GR. By analyzing alternative gravitational models and their predictions, we aim to explore potential explanations for the observed accelerated expansion of the Universe and evaluate how well these models align with cosmological data.
        \item \textbf{The Hubble Tension:} A major challenge to the $\Lambda$CDM paradigm is the Hubble tension, which refers to a persistent discrepancy between the value of the Hubble constant inferred from early Universe measurements, such as CMB data and those obtained from late-time observations, such as supernova distances. The Planck collaboration finds $H_0 = 67.4 \pm 0.5$ $\text{Km} \, \text{s}^{-1} \, \text{Mpc}^{-1}$, while late-time observations suggest a higher value around $H_0 = 73.52 \pm 1.62$ $\text{Km} \, \text{s}^{-1} \, \text{Mpc}^{-1}$. This discrepancy known as the Hubble tension, may suggest limitations in GR or the applicability of the $\Lambda$CDM model at different cosmic epochs. As previously noted, the $\Lambda$CDM model aligns well with cosmological observations, including CMB data and the late Universe observations, such as late-time acceleration. However, recent findings indicate notable tensions within the $\Lambda$CDM framework between these two categories of observations \cite{Verde_2019_3}. Specifically, the estimated value of $H_0$ derived from late Universe observations exhibits a $5\sigma$ tension with the $H_0$ value obtained from CMB data \cite{Riess_2021_934}. Estimates of $H_0$ from the early Universe tend to be higher than those derived from local Universe observations. As observational precision continues to improve, it has been suggested that exploring alternatives to the $\Lambda$CDM model may be necessary to resolve the $H_0$ tension. Another significant cosmological tension arises from the observations of large-scale structures and the corresponding parameter values derived from CMB data. This 3$\sigma$ tension is evident in the constraints on the matter-energy density parameter $(\Omega_\text{m})$ and the amplitude $(\sigma_8)$ of matter fluctuations, as estimated from both local Universe measurements and CMB observations \cite{Macaulay_2013_111, Asgari_2021_645}. Numerous alternative theories of gravity have been suggested to tackle these challenges, encompassing modified gravity models incorporating geometric adjustments to scalar-tensor theories \cite{Clifton_2012_513, Nojiri_2017_692, COPELAND_2006_15, DiValentino_2021_38}. While these models can explain the observed late-time evolution of the Universe, most are not favored over the $\Lambda$CDM model based on current cosmological observations. Therefore, it is crucial to develop alternatives to the $\Lambda$CDM model that better aligns with observational data.
    \end{enumerate}
    Beyond observational challenges, GR also faces deep theoretical issues:  
    \begin{itemize}  
        \item \textbf{Singularity Problem:} The theory predicts singularities (e.g., in black holes and the Big Bang), where curvature diverges to infinity, signaling the breakdown of classical GR.  
        \item \textbf{Cosmological Constant Problem:} The observed value of the cosmological constant is many orders of magnitude smaller than predicted by quantum field theory.  
        \item \textbf{Quantum Gravity Incompatibility:} GR is non-renormalizable, meaning it cannot be consistently quantized like other fundamental interactions in the Standard Model.  
    \end{itemize}  
    These open problems motivate modified gravity theories, which aim to address cosmic acceleration, structure formation, and quantum gravity within a broader theoretical framework.  
\subsubsection{Theoretical Motivations for Modified Gravity Theories}
    To address the limitations of GR, various modified gravity theories have been proposed, altering equations of GR or introducing new fields:
    \begin{itemize}
        \item \textbf{Nonlinear Modifications:} These models extend the Einstein--Hilbert action by incorporating nonlinear functions of the Ricci scalar $f(R)$, allowing for self-accelerating solutions that may explain cosmic acceleration without requiring a cosmological constant.
        \item \textbf{Scalar-Tensor Theories:} In these models, a dynamical scalar field is coupled to gravity, effectively modifying the gravitational constant. Examples include Brans--Dicke gravity, where a scalar field replaces Newton’s constant and Horndeski theories, which provide self-consistent modifications that preserve second-order field equations.
        \item \textbf{Higher-Dimensional Models:} Braneworld models explore extra dimensions that could influence gravity at large scales, offering potential resolutions to the DE and dark matter problems within a unified framework.
    \end{itemize}
        These theories aim to provide viable alternatives that could reduce or eliminate the need for dark components, particularly by modifying gravitational dynamics over large distances.
\section{Modified Gravity}
    Since Einstein’s formulation of GR, numerous alternative theories have been proposed to address its limitations and extend gravitational dynamics to different regimes. While many models fail due to complexity or inconsistency with observations, several viable frameworks have emerged that offer testable predictions. Despite this, researchers have continued to propose modifications. There are several reasons to consider altering gravity. From a mathematical standpoint, there is no inherent requirement to adopt the Einstein--Hilbert action (linear in $R$ ) as the fundamental action of gravity. However, suppose we aim to derive field equations that are, at most, second-order partial differential equations with respect to the metric tensor. In that case, Einstein’s equations are the only ones that meet this criterion.\footnote{This holds true exclusively for metric theories, where the metric tensor is the only independent field in the action. However, this does not apply to Palatini and Metric-Affine theories of gravity, where the connection is independent of the metric tensor.}

    The Gauss--Bonnet invariant arises naturally in higher-dimensional gravity and string-inspired modifications of GR. In four dimensions, it contributes as a topological invariant, meaning that it does not affect the equations of motion when included linearly. However, when coupled to a dynamical scalar field or included in $f(\mathcal{G})$ modification, it introduces novel effects that can influence cosmic evolution. It is expressed as a specific combination of curvature invariants 
        \begin{equation}
        \mathcal{G} = R^2 - 4R_{\mu\nu}R^{\mu\nu} + R_{\mu\nu\rho\sigma}R^{\mu\nu\rho\sigma}, \end{equation}
    where $R$ is the Ricci scalar, $R_{\mu\nu}$ is the Ricci tensor, and $R_{\mu\nu\rho\sigma}$ is the Riemann tensor. The Gauss--Bonnet term has a unique property in four dimensions: its integral contributes as a topological invariant.

    Modified gravity theories incorporate the Gauss--Bonnet invariant to introduce higher-order curvature corrections while avoiding ghost-like instabilities. This arises in low-energy string theory, where such terms naturally emerge in the effective action, motivating their inclusion in four-dimensional cosmological models. This stems from string theory and quantum gravity, where such terms naturally arise in the low-energy effective action. In 1971, Lovelock's theorem \cite{Lovelock_1971_12} provided a formal foundation, showing that the Gauss--Bonnet term is the only second-order scalar curvature invariant that avoids these issues, making it a central candidate for constructing viable higher-dimensional or modified gravity models. In the context of four-dimensional cosmology, the Gauss--Bonnet invariant gained attention as a mechanism for addressing fundamental issues in the standard model of cosmology, such as DE and cosmic acceleration. By coupling the Gauss--Bonnet term to a function $f(\mathcal{G})$, researchers introduced a modification of Einstein's GR that could potentially explain these phenomena without invoking a cosmological constant. This modification extends the standard Einstein--Hilbert action, resulting in the so-called $f(\mathcal{G})$ gravity models.

    These models aim to provide a dynamic explanation for DE and avoid the fine-tuning and coincidence problems associated with $\Lambda$CDM. The Gauss--Bonnet invariant has proven particularly useful in achieving late-time acceleration by generating effective energy density and pressure terms that modify the Friedmann equations. Moreover, the inclusion of $\mathcal{G}$ helps address issues related to the early and late-time behavior of the Universe, offering an alternative approach to understanding cosmic evolution. Consequently, the Gauss--Bonnet invariant has become a cornerstone in many modified gravity theories, including scalar-tensor, Horndeski, and braneworld models.

    However, in higher-dimensional spacetimes or when incorporated into modified gravity theories like the $f(R, \mathcal{G})$ gravity or Einstein--Gauss--Bonnet gravity, the Gauss--Bonnet term introduces significant modifications to the dynamics. Such theories extend GR by adding corrections to the Einstein--Hilbert action, typically of the form
        \begin{eqnarray}
            S = \int d^4x \sqrt{-g} \left[\frac{1}{2\kappa^2} (R + \alpha \mathcal{G}) + \mathcal{L}_{\text{m}}\right] \, ,
        \end{eqnarray}
    where $\alpha$ is a coupling constant, and $\mathcal{L}_\text{m}$ is the matter Lagrangian. These modifications are motivated by the need to address cosmological issues, particularly the phenomena of late-time cosmic acceleration and the dynamics of early cosmic inflation. Additionally, they aim to illuminate the constraints of string theory in the low-energy regime. The presence of the Gauss--Bonnet term can influence the equations governing the cosmic expansion, potentially explaining DE or yielding non-trivial black hole solutions.
\subsection{How to Modify General Relativity?}
    Modifying GR involves extending its geometric and physical principles to address limitations such as the accelerated expansion of the Universe, dark matter, and quantum gravity. Common approaches include adding higher-order curvature terms like in $f(R)$ gravity, introducing new fields as in scalar-tensor theories, or modifying the geometric description of gravity itself using torsion or nonmetricity, as seen in teleparallel and symmetric teleparallel gravity. Other modifications investigate the inclusion of extra dimensions, as in braneworld scenarios, or the incorporation of nonlocal terms in action, enabling gravity to exhibit distinct behavior on large or small scales. These alternatives provide diverse frameworks for exploring gravitational phenomena beyond the scope of GR while maintaining consistency with observations.
\subsection{The Gravitational Action}
    In this thesis, we aim to investigate late-time cosmic acceleration within the frameworks of modified gravity theory and Gauss--Bonnet cosmology.\footnote{A comprehensive description of these gravity theories will be given in the upcoming chapters.}
\subsubsection{\texorpdfstring{$f(R)$}{} Gravity}
    The most general action for $f(R)$ gravity is \cite{De_Felice_2010_13}
    \begin{equation}
        S = \int d^4x \ \sqrt{-g} \left[ \frac{1}{2 \kappa^2} f(R)  + \mathcal{L}_\text{m} \right]\, .
    \end{equation}
    Throughout this thesis, we use the convention $\kappa^2 = 8 \pi G = 1$, where $G$ is the Newton's gravitational constant, and $\mathcal{L}_\text{m}$ represents the matter Lagrangian. 
\subsubsection{\texorpdfstring{$f(\mathcal{G})$}{} Gravity}
    We consider an action that encompasses GR and a functional dependent on the Gauss--Bonnet term \citep{Nojiri_2005_631, Nojiri_2006_39}
    \begin{equation}
        S = \int d^4x \ \sqrt{-g} \left[ \frac{1}{2 \kappa^2} R + f(\mathcal{G}) + \mathcal{L}_\text{m} \right]\, .
    \end{equation}
\subsubsection{\texorpdfstring{$f(R, \mathcal{G})$}{} Gravity}
    The action of $f(R, \mathcal{G})$ gravity, a modification of GR \cite{Laurentis_2015_91, Wu_2015_92, Santos_da_Costa_2018_35, ODINTSOV_2019_938_935, Kumar_Sanyal_2020_37, Brout_2022_938} is,
    \begin{equation}\label{frg_action}
        S=\int d^{4}x \ \sqrt{-g}\left[\frac{1}{2\kappa^2}f(R,\mathcal{G})+\mathcal{L}_\text{m}\right]  \, .
    \end{equation}  
\subsubsection{\texorpdfstring{$f(T, T_\mathcal{G})$}{} Gravity}
    In $f(T, T_\mathcal{G})$ gravity, the total modified gravitational action has the following form \cite{Kofinas_2014_90_084044, Kofinas_2014_90_084045}
        \begin{equation}\label{eq.fttg action}
            S= \int d^{4}x\, e\,\left[ \frac{1}{2\kappa^2} f(T, T_\mathcal{G}) +\mathcal{L}_\text{m}\right] \, ,
        \end{equation}
    which is based on the torsion scalar $T$ and the Gauss--Bonnet invariant term, $T_\mathcal{G}$.
\subsubsection{\texorpdfstring{$f(Q, B)$}{} Gravity}
    A recent extension of the $f(Q, B)$ theory \cite{Capozziello_2023_83, De_2024_2024_50, Palianthanasis_2024_43} incorporates a boundary term into the gravitational action integral. This generalization includes the gravitational action integral in the following manner
    \begin{equation} \label{eq: fqb action}
        S = \int d^{4}x \ \sqrt{-g} \left[\frac{1}{2 \kappa^2} f(Q,B) +\mathcal{L}_\text{m} \right]  \, .
    \end{equation}
\subsection{Geometric Trinity}
    The three main formulations of GR that make up the geometric trinity of gravity are (i) the standard formulation, which explains gravity in relation to spacetime curvature, (ii) TEGR \cite{Ferraro_2007_75, Bengochea_2009_79, Linder_2010_81, Arcos_2004_13, Maluf_2013_525, Aldrovandi_2013_TG_INTRO}, which describes gravity in terms of spacetime torsion, and (iii) STEGR, which explains gravity in terms of spacetime nonmetricity.

    The two mathematical frameworks yield identical field equations, despite initially appearing to be entirely unrelated theories. Additionally, there is the TEGR, which is based on the torsion scalar $T$ and the related boundary term $\Tilde{B}$ (with $B \neq \Tilde{B}$). TEGR represents another equivalent formulation of GR \cite{Capozziello_2022_82, Aldrovandi_2013_TG_INTRO, Cai_2016_79}. These three theories, with dynamics governed by the Ricci scalar $R$ for GR, the torsion scalar $T$ for TEGR, and the nonmetricity scalar $Q$ for STEGR, make up the Geometric Trinity of Gravity \cite{Capozziello_2022_82, Jimenez_2019_5}.
    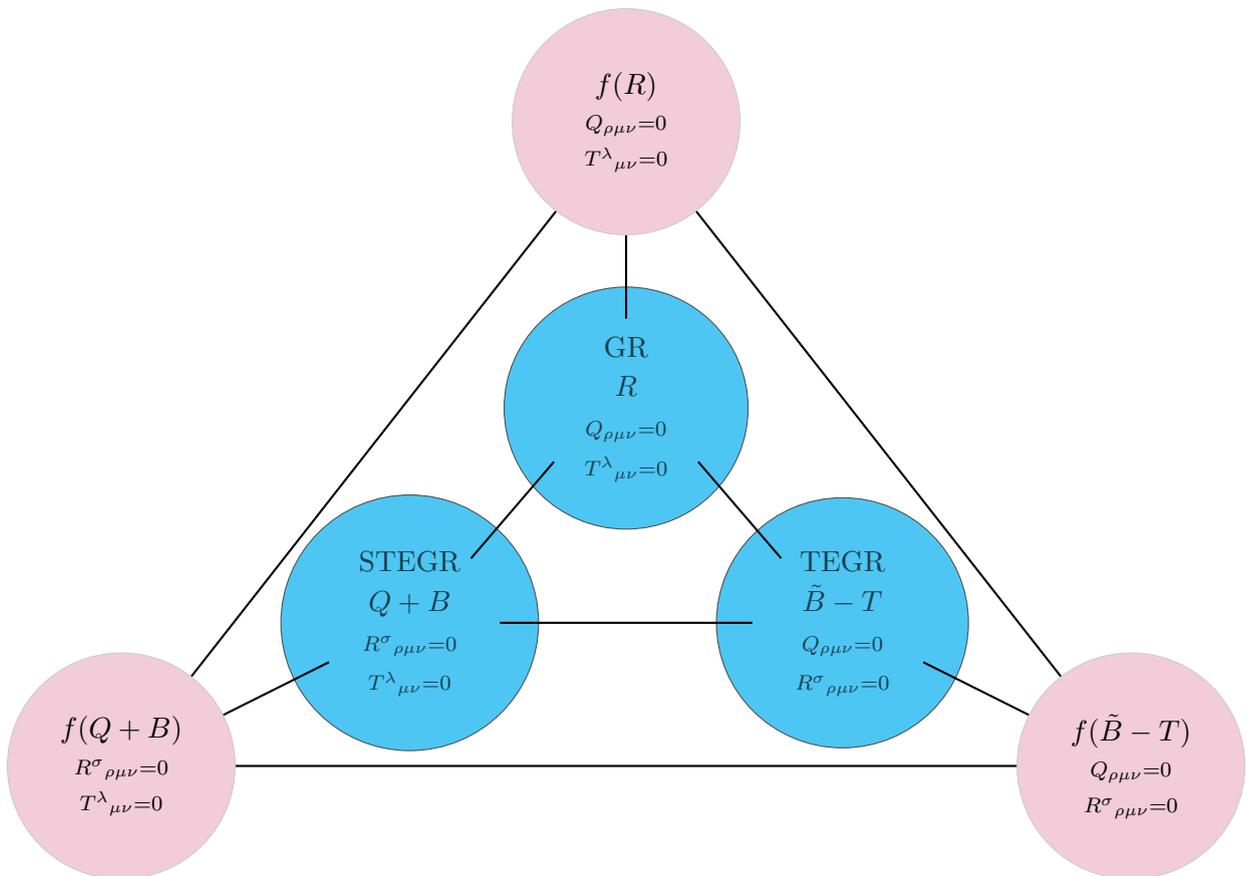
\begin{figure}[!htbp]
    \centering
    \begin{tikzpicture}[scale=0.95]
    \tikzset{venn circle A/.style={draw,circle,minimum width=3cm,fill=purple,opacity=0.2}}
    \tikzset{venn circle B/.style={draw,circle,minimum width=3cm,fill=purple,opacity=0.2}}
    \tikzset{venn circle C/.style={draw,circle,minimum width=3cm,fill=purple,opacity=0.2}}

    \node [venn circle A] (A) at (0,8) {};
    \node [venn circle B] (B) at (7,-1) {};
    \node [venn circle C] (C) at (-7,-1) {};

    \node at (barycentric cs:A=1,B=1,C=-0.2) {\begin{tabular}{c} 
    \end{tabular}};
    \node at (barycentric cs:A=1,C=1,B=-0.2) {\begin{tabular}{c} 
    \end{tabular}};
    \node at (barycentric cs:B=1,C=1,A=0) {\begin{tabular}{c} 
    \end{tabular}};

    \node at (A) [anchor=center, align=center] {$f(R)$ \\ $\scriptstyle{Q_{\rho\mu\nu}=0}$\\ $\scriptstyle{T^\lambda{}_{\mu\nu}=0}$};
    \node at (B) [anchor=center, align=center] {$f(\Tilde{B}-T)$ \\ $\scriptstyle{Q_{\rho\mu\nu}=0}$ \\ $\scriptstyle{R^\sigma{}_{\rho\mu\nu}=0}$};
    \node at (C) [anchor=center, align=center] {$f(Q+B)$ \\ $\scriptstyle{R^\sigma{}_{\rho\mu\nu}=0}$ \\ $\scriptstyle{T^\lambda{}_{\mu\nu}=0}$};

    \draw[thick] (A) -- (B);
    \draw[thick] (B) -- (C);
    \draw[thick] (C) -- (A);

    \node [draw,circle,minimum width=2cm,fill=cyan,opacity=0.7] at (0,4) {\begin{tabular}{c} GR\\ $R$ \\ $\scriptstyle{Q_{\rho\mu\nu}=0}$\\ $\scriptstyle{T^\lambda{}_{\mu\nu}=0}$ \end{tabular}};
    \node [draw,circle,minimum width=2cm,fill=cyan,opacity=0.7] at (3,1) {\begin{tabular}{c} TEGR\\ $\Tilde{B}-T$ \\ $\scriptstyle{Q_{\rho\mu\nu}=0}$ \\ $\scriptstyle{R^\sigma{}_{\rho\mu\nu}=0}$ \end{tabular}};
    \node [draw,circle,minimum width=2cm,fill=cyan,opacity=0.7] at (-3,1) {\begin{tabular}{c} STEGR\\ $Q+B$ \\ $\scriptstyle{R^\sigma{}_{\rho\mu\nu}=0}$ \\ $\scriptstyle{T^\lambda{}_{\mu\nu}=0}$ \end{tabular}};

    \draw[thick] (0,5.25) -- (A);
    \draw[thick] (4.12,0.45) -- (B);
    \draw[thick] (-4.12,0.45) -- (C);

    \draw[thick] (1,3.25) -- (2.15,1.9);
    \draw[thick] (1.75,1) -- (-1.75,1);
    \draw[thick] (-2.15,1.9) -- (-1,3.25);

    \end{tikzpicture}  
    \caption{Geometric trinity in gravity (inside circle) and Extended geometric trinity of gravity (outside circle)}
    \label{fig:trinity_gr}
    \end{figure}
    The geometric trinity of gravity consists of three dynamically equivalent reformulations of Einstein's GR, each based on a different fundamental geometric quantity:
    \begin{itemize}
        \item \textbf{Curvature-based gravity (GR):} Governed by the Ricci scalar $R$.
        \item \textbf{Teleparallel gravity (TEGR):} Described by the torsion scalar $T$.
        \item \textbf{Symmetric teleparallel gravity (STEGR):} Defined by the nonmetricity scalar $Q$.
    \end{itemize}
    While these formulations produce identical field equations in their simplest forms, their extended versions such as $f(R)$, $f(T)$, and $f(Q)$ gravity are not fundamentally equivalent. Each introduces distinct modifications to gravitational dynamics, leading to different physical implications, particularly in the context of cosmic acceleration.

    It can be demonstrated that these Lagrangians are equivalent to a boundary term, which differs in GR, TEGR, and STEGR. This equivalence implies that TEGR and STEGR yield the same field equations as GR. When transitioning to extended theories such as $f(R)$, $f(T)$, and $f(Q)$ gravities, a challenge arises due to their lack of dynamic equivalence. Specifically, $f(R)$ gravity, in metric formalism, is a fourth-order theory, while $f(T)$ and $f(Q)$ are second-order theories. However, it is feasible to reestablish equivalence among these extended theories by introducing a suitable boundary term. In the general frameworks $f(T, \Tilde{B})$ and $f(Q, B)$ (where $\Tilde{B} \neq B$), it can be observed that $f(\Tilde{B} - T)$, $f(Q + B)$ are dynamically equivalent to $f(R)$ gravity. These three theories form what we may refer to as the extended geometric trinity of gravity (see figure \ref{fig:trinity_gr}). While $R$, $\Tilde{B} - T$, $Q + B$ represent a geometric trinity of gravity of second-order, $f(R)$, $f(\Tilde{B} - T)$, $f(Q + B)$ configure a geometric trinity of gravity of fourth-order. In brief, including boundary terms enhances the degrees of freedom by functioning as effective scalar fields. This approach can be further concluded to encompass higher-order metric-affine theories expressed within metric, teleparallel, and symmetric teleparallel formalisms. Having introduced several extensions of GR, we now examine how these modifications affect the cosmological evolution of the Universe. 
\section{Age of the Universe}
    The age of the Universe is a fundamental parameter in cosmology, offering insights into the evolution and dynamics of the cosmos. It is typically estimated within the framework of the $\Lambda$CDM model, which incorporates both DE (represented by the cosmological constant, $\Lambda$) and cold dark matter (CDM). The age of the Universe $t_0$, can be calculated by integrating the Hubble parameter $H(t)$, which governs the rate of expansion of the Universe. The age of the Universe is then obtained by integrating the inverse of the Hubble parameter from the time of the Big Bang ($z \to \infty$) to the present day ($z = 0$)
        \begin{eqnarray}
            t_0 = \int_0^\infty \frac{dz}{(1+z)H(z)} \, .
        \end{eqnarray}
    Recent observations, such as those of the CMB by the Planck satellite, indicate that the age of the Universe is estimated to be around $t_0 \approx 13.8$ billion years. This result is consistent with independent measurements from distant supernovae and BAO, providing robust constraints on the cosmological parameters that define our understanding of the expansion history of the Universe. The most precise measurements come from the Planck satellite, which analyzed the CMB and provided an age of $13.787 \pm 0.020$ billion years. These findings are consistent with other methods, such as studying the oldest star clusters and the distribution of galaxies. Understanding the age of the Universe helps cosmologists trace the history of cosmic evolution and the formation of structures within it. However, to test these theoretical predictions, we rely on observational data from various sources, such as the CMB, Supernovae, and BAO. The next section explores the statistical methods used to constrain cosmological models.
\section{Observational constraints and statistical analysis}
    Cosmology is an observational science in which theoretical models must be tested against real data. To achieve this, statistical methods such as chi-squared minimization, likelihood analysis, and Markov Chain Monte Carlo (MCMC) techniques are used to extract cosmological parameters and assess model viability. This section discusses key techniques such as $\chi^2$ minimization, maximum likelihood estimation, and MCMC analysis. These methods enable us to determine best-fit parameter values, explore parameter spaces, and establish robust constraints on cosmological models.
\subsection{\texorpdfstring{$\chi^2$}{} Minimization}
    The $\chi^2$ minimization method is a standard technique for statistical inference in cosmology. It quantifies the goodness-of-fit between observed data and theoretical models by computing the sum of squared deviations, weighted by uncertainties. The $\chi^2$ function quantifies the difference between data points $y_i$ and model predictions $f(x_i, \{\theta\})$, weighted by observational uncertainties $\sigma_i$ as
        \begin{equation}
            \chi^2 = \sum_{i=1}^{N} \frac{(y_i - f(x_i, \{\theta\}))^2}{\sigma_i^2} \, ,
        \end{equation}
    where $\{\theta\}$ represents the model parameters, and $x_i$ often corresponds to redshift $z$ in cosmological analyses. 
    
    For data sets with correlated uncertainties, the $\chi^2$ function generalizes to
        \begin{equation}
            \chi^2 = \sum_{i,j} [y_i - f(x_i, \{\theta\})] \  \text{Cov}_{ij}^{-1} \ [y_j - f(x_j, \{\theta\})] \, ,
        \end{equation}
    where $\text{Cov}_{ij}$ is the covariance matrix. The combined $\chi^2$ for multiple data sets is the sum of individual $\chi^2$ values i.e.,
        \begin{equation}
            \chi_{\text{tot}}^2 = \sum_{m} \chi_{m}^{2} \, .
        \end{equation}
    Minimizing $\chi^2$ provides the best-fit parameter values, while confidence intervals are determined using $\Delta \chi^2 = \chi^2 - \chi_{\min}^2$ (e.g., $1\sigma$ or $2\sigma$ confidence levels).
\subsection{Maximum Likelihood Analysis}
    Maximum likelihood estimation provides an alternative approach to parameter estimation by determining the set of parameters most likely to produce the observed data. The likelihood function $\mathcal{L}(\{\theta\})$ measures the probability of observing the data $D$ given the parameters $\{\theta\}$
        \begin{equation}
            \mathcal{L}(\{\theta\}) = \exp \left(-\frac{\chi^2}{2} \right) \, .
        \end{equation}
    Maximizing the likelihood is equivalent to minimizing $\chi^2$. In Bayesian terms, the posterior probability distribution $p(\{\theta\} \mid D)$ is given by Bayes' theorem
        \begin{equation}
            p(\{\theta\} \mid D) \propto \mathcal{L}(\{\theta\}) \cdot p(\{\theta\}) \, ,
        \end{equation}
    where $p(\{\theta\})$ is the prior distribution, encoding existing knowledge about the parameters before observing the data.
\subsection{Priors in Cosmology}
    In Bayesian cosmology, priors encode prior knowledge or theoretical constraints on model parameters, influencing the resulting probability distribution. Common types of priors include:
    \begin{itemize}
        \item \textbf{Uniform Priors:} Used when little prior information is available. For example, a uniform prior on the Hubble constant $H_0$ might span a broad range: $50 < H_0 < 100$ $\text{Km} \ \text{s}^{-1} \ \text{Mpc}^{-1}$.

        \item \textbf{Gaussian Priors:} Incorporate precise measurements from previous experiments. For instance, the matter density $\Omega_\text{m}$ might have a Gaussian prior centered on $0.315 \pm 0.007$ (from Planck data).
    
        \item \textbf{Log-Uniform Priors:} Suitable for scale parameters (e.g., dark matter particle mass), ensuring positivity.
    
        \item \textbf{Physical Priors:} Enforce theoretical constraints, such as $\Omega_\text{m} > 0$ or $w \geq -1$ for the DE EoS.
    \end{itemize}
    Choosing appropriate priors is critical, as they can significantly influence the posterior distribution, especially in cases of weak data or parameter degeneracies.
\subsection{MCMC Analysis}
    MCMC methods allow for efficient sampling of high-dimensional parameter spaces, making them indispensable in cosmological data analysis. MCMC techniques generate a sequence of probable parameter sets that converge to the posterior distribution, enabling robust uncertainty estimation. Key steps include:
    \begin{enumerate}
        \item \textbf{Proposal:} Suggest new parameter values based on a proposal distribution.
        \item \textbf{Acceptance/Rejection:} Use the Metropolis-Hastings criterion to decide whether to accept or reject the new values.
        \item \textbf{Convergence:} Run the chain until it converges to the posterior distribution.
    \end{enumerate}
    MCMC is particularly useful in cosmology for:
    \begin{itemize}
        \item \textbf{Parameter Estimation:} Constraining cosmological parameters from data (e.g., CMB, supernovae, large-scale structure).
        \item \textbf{Marginalization:} Integrating over nuisance parameters to obtain constraints on parameters of interest.
        \item \textbf{Model Comparison:} Computing Bayesian evidence to compare competing models.
    \end{itemize}
    We employ the \texttt{emcee}\footnote{ \url{https://github.com/dfm/emcee}} Python implementation of the collective sampler for MCMC, developed by Foreman--Mackey et al. \cite{Foreman_Mackey_2013_emcee}. Two-dimensional confidence contours are visualized using the \texttt{GetDist}\footnote{ \url{https://github.com/cmbant/getdist}} module in Python, created by Lewis \cite{Lewis_getdist}. Additionally, we utilized the \texttt{ChainConsumer}\footnote{ \url{https://github.com/Samreay/ChainConsumer}} Python gallery for our analysis. By combining these statistical methods, we can extract meaningful insights from cosmological data and advance our understanding of the Universe.
\subsection{Observational Data sets}
    Cosmology relies on a diverse set of observational probes to trace the evolution of the Universe, test gravitational theories, and constrain cosmological parameters. The following subsections provide a summary of key observational data sets and highlight major discoveries that have shaped our understanding of the Universe.
\subsubsection{Cosmic Chronometers}
    CC measurement provides a direct way to measure the Hubble parameter $H(z)$ at different redshifts, allowing us to reconstruct the expansion history of the Universe. Direct measurements of $H(z)$ at various redshifts, derived from the ages of the most massive and passively evolving galaxies, serve as a crucial cosmological probe. These $H(z)$ measurements are obtained through two primary methods: the galaxy differential age technique (also known as the CC) and the radial BAO size method \cite{Yu_2018_856}.

    In their studies, present 13 $H(z)$ values derived from the BC03 \cite{Moresco_2012_2012_006, Moresco_2016_2016_014} and MaStro SPS models \cite{Bruzual_2003_344}, referred to as the CCB and CCM compilations, respectively. Additionally, contribute 5 $H(z)$ values using the BC03 model, which has been incorporated into the CCB compilation \cite{Zhang_2014_14, Maraston_2011_418}. Moresco provides combined MaStro/BC03 values for 2 $H(z)$ measurements \cite{Moresco_2015_450}. Furthermore, \cite{Stern_2010_2010_008} introduces an alternative SPS model, distinct from MaStro and BC03, comprising 11 $H(z)$ values, known as the CCH compilation, along with 26 BAO-assessed points \cite{Sharov_2018_01}. Our analysis utilizes 32 objects spanning the redshift range $0.07 \leq z \leq 1.965$ \cite{Moresco_2022_25, Lohakare_2023_40_CQG}.
\subsubsection{Type Ia Supernovae}
    SNe Ia act as standard candles, providing precise measurements of luminosity distances over cosmic time. Their apparent brightness helps constrain cosmic acceleration and investigate the properties of DE. This explosion occurs when the white dwarf approaches the Chandrasekhar mass limit by accumulating mass from a companion star. Consequently, SNe Ia serves as standard candles for measuring luminosity distance \cite{Riess_1998_116, Perlmutter_1998_517}. In 1998, Riess et al. \cite{Riess_1998_116} discovered the accelerated expansion of the Universe using observations of 16 distant and 34 nearby SNe Ia from the Hubble telescope. This finding was confirmed in 1999 by Perlmutter et al. \cite{Perlmutter_1998_517}, who analyzed 18 nearby supernovae from the Calan-Tololo sample and 42 high-redshift SNe. Numerous research groups have since focused on this area, including the Sloan Digital Sky Survey (SDSS) SNe Survey \cite{Holtzman_2008_136, Kessler_2009_185}, the Lick Observatory Supernova Search (LOSS) \cite{Leaman_2011_412, Li_2011_412}, the Nearby Supernova Factory (NSF) \cite{Copin_2007_50, Scalzo_2010_713}, the Supernova Legacy Survey (SNLS) \cite{Astier_2006_447, Baumont_2008_491}, and the Higher-Z Team \cite{Riess_2004_607, Riess_2007_659}, among others. Recently, the Union 2.1 SNe Ia data set, comprising 580 SNe Ia, was released \cite{Visser_2004_21}.

    The Pantheon$^+$ compilation is one of the most comprehensive data sets on SNe Ia, containing 1048 data points within the redshift range $(0.01 < z < 2.26)$ \cite{Scolnic_2018_859}. We will also utilize the Pantheon SNe Ia data compilation, which encompasses 1701 measurements of relative luminosity distances for SNe Ia, spanning a redshift range from $0.00122 < z < 2.2613$ \cite{Brout_2022_938}. This comprehensive data set includes distance moduli derived from 1701 light curves of 1550 spectroscopically confirmed SNe Ia, gathered from 18 distinct surveys. Significantly, 77 of these light curves are associated with galaxies that contain Cepheid variables. The Pantheon$^+$ data set is particularly valuable as it not only helps constrain the Hubble constant $H_0$ but also refines other model parameters.
\subsubsection{Baryon Acoustic Oscillations}
    BAO originate from sound waves propagating in the hot plasma of the early Universe. These oscillations imprint a characteristic scale on the large-scale distribution of galaxies, serving as a standard ruler for measuring cosmic expansion \cite{Peebles_1970_162, Peebles_2003_75}. Like SNe Ia, BAO acts as a standard ruler in cosmology, helping us understand the expansion of the Universe. BAO creates a unique pattern on the matter power spectrum, which can be detected through galaxy cluster surveys at low redshifts $(z < 1)$ \cite{Alam_2004_2004_008}. Furthermore, BAO scales can be observed through reionization emissions, offering insights into the early Universe at high redshifts $(z > 2)$ \cite{Slosar_2013_04_026, Green_2020_2020_12_050, Villaescusa_2017_466}. The Hubble parameter and angular diameter distance can be calculated from the apparent magnitude of BAO observed in astronomical data. Many studies, such as the Two-degree-Field Galaxy Redshift Survey (2dFGRS) \cite{Colless_2003_arXiv} and the Sloan Digital Sky Survey (SDSS) \cite{York_2000_120, Tegmark_2006_74}, have focused on BAO measurements, with SDSS being particularly successful, continuously releasing its eighth data set (SDSS DR8) in 2011 (\href{https://www.sdss3.org/dr8/}{www.sdss3.org/dr8/}). The upcoming chapters will apply the modified gravity theories discussed earlier to address specific challenges presented by this observational data.

    The most recent observational data, including the Dark Energy Spectroscopic Instrument (DESI) surveys \cite{Adame_2024_DESI_collaboration}, SNe Ia \cite{Riess_1998_116, Perlmutter_1998_517}, Wilkinson microwave anisotropy probe experiment (WMAP) \cite{Spergel_2003_148}, CMB \cite{Hinshaw_2013_208}, Baryon oscillation spectroscopic survey (BOSS) \cite{Alam_2017_470} and the BAO data sets \cite{Eisenstein_2005_633} have prompted researchers to consider modifications and expansions to the principles of GR in theories such as $f(R)$ \cite{Sotiriou_2010_82}, $f(T)$ \cite{Ferraro_2007_75}, and $f(Q)$ \cite{Jimenez_2018_98, Heisenberg_2023_1066_Review}. These alternative theories aim to accommodate better and explain the new observational data. One of the most straightforward extensions to Einstein’s gravity is the so-called $f(R)$ gravity, where $f$ is an arbitrary function of the Ricci scalar $R$ \cite{Nojiri_2007_04}. Even in this relatively simple case, constructing viable $f(R)$ models consistent with cosmological and local gravity constraints is not straightforward. This complexity arises because $f(R)$ gravity introduces a strong coupling between DE and non-relativistic matter in the Einstein frame \cite{capozziello2008extended}. Extensions of GR, including the Gauss--Bonnet invariant in the gravitational action, have generated significant interest \cite{Nojiri_2005_631, Li_2007_76, Elizalde_2010_27, Felice_2009_80_063516, Maurya_2021_81, Maurya_2022_925, Nojiri_2017_692, ODINTSOV_2019_938_935, Lohakare_2023_40_CQG, Nojiri_2008_172_81}. One such theory that has garnered significant interest is $f(\mathcal{G})$ gravity. 

    The insights gained from CCs, Supernovae, and BAO have significantly shaped our understanding of cosmic acceleration and large-scale structure formation. However, the observed tensions and unexplained phenomena in these data sets suggest possible deviations from Einstein’s gravity, motivating the exploration of alternative gravitational theories.
\section{Dynamical System Analysis}
    Dynamical systems theory provides a powerful mathematical framework for understanding long-term behavior in complex systems, including cosmology, fluid mechanics, and molecular biology. In cosmology, the evolution of the Universe can be modeled as a dynamical system, where the scale factor, density parameters, and curvature evolve according to autonomous differential equations. It is an essential tool for analyzing the long-term behavior of systems. In cosmology, dynamical systems techniques help classify different phases of the evolution of the Universe, such as early inflationary states, radiation-dominated eras, and late time acceleration. By identifying fixed points (equilibrium solutions) and analyzing their stability, we can determine whether the Universe evolves toward a stable attractor or undergoes transitions between different expansion phases. Dynamical systems theory provides a framework for characterizing this behavior. For comprehensive reviews on dynamical systems in cosmology, one may refer \cite{Coley_1999, Bahamonde_2018_775}.

    This section will delve into the mathematical techniques related to dynamical systems that will be utilized throughout this thesis. Specifically, we will analyze fixed points in four dimensions and the behavior of trajectories in phase space. We will briefly introduce dynamical systems and fixed points, followed by an outline of linear stability theory. For additional reading on these topics, see \cite{Strogatz_2018_book, Glendinning_1994_book, Wiggins_2003_book} or \cite{Wainwright_2005_book} for applications in cosmology. We will then progress to more advanced treatments of dynamical systems, examining non-hyperbolic fixed points and extending beyond linear stability theory. In the subsequent sections, these techniques will be applied to cosmological models framed as sets of autonomous differential equations, offering valuable insights into the dynamics and stability of each model. 
\subsection{Introduction to Dynamical Systems}
    Dynamical systems are broadly classified into two types: differential equations and iterated maps. Differential equations describe systems where time is continuous, which is typically the scenario in cosmology. Conversely, iterated maps consider time as a discrete variable and will not be the focus of our discussion. The differential equations of primary interest to us involve a single independent variable with time being a prominent example. These are referred to as ordinary differential equations.

    To begin, we define the set of variables $x_1, \ldots, x_n \in X \subseteq \mathbb{R}^n$, representing coordinates in an $n$ - dimensional phase space. Additionally, we introduce the independent variable $t \in \mathbb{R}$, which may not necessarily correspond to time. Within this framework, we can describe a dynamical system using ordinary differential equations, as referenced in \cite{Strogatz_2018_book}
    \begin{eqnarray} \label{ch4_CMT_system}
        \dot{x}_1 &= f_1(x_1, \ldots, x_n) \, , \nonumber\\
        &\vdots \label{general_system_Intro} \\ 
        \dot{x}_n &= f_n(x_1, \ldots, x_n), \nonumber
    \end{eqnarray}
   where over-dot notation represents the derivative with respect to $t$, such that $\dot{x}_i \equiv \frac{dx_i}{dt}$. The function $f_i : X \to X$. The system is described as autonomous if it does not explicitly depend on the independent variable $t$. To express equation \eqref{general_system_Intro} more concisely but equivalently, we can rewrite it as follows
    \begin{eqnarray} \label{general_signle_system}
        \dot{\textbf{x}} = \textbf{f}(\textbf{x}) \, .
    \end{eqnarray}
    This represents a system of differential equations where each $\dot{\textbf{x}}$ (the derivative of $x_i$ with respect to time) is a function of all the variables $x_1$ to $x_n$. The function $\textbf{f(x)} = \left(f_1 (\textbf{x}), . . . , f_n (\textbf{x}) \right)$ can be interpreted as a vector field on $\mathbb{R}^n$. We will focus on scenarios where $f(x)$ is both smooth and real-valued. In regions of the phase space where these conditions are not met, the methods discussed will not be applicable. In our cosmological examples presented in later sections, there are instances where $f(x)$ exhibits divergences for certain values of $x$, and these cases require careful handling. Any specific solution to equation \eqref{general_signle_system} for a given initial condition $\textbf{x}_0$ corresponds to a point tracing a curve in the phase space, denoted as $\psi(t)$. This curve, or solution $\psi(t)$, is referred to as the trajectory or orbit. Consequently, the phase space is populated with trajectories originating from various initial conditions. By analyzing these geometric representations, we can derive insights about the system through the examination of trajectory flows within the phase space.
\subsubsection{Equilibrium Points and Stability}
    \textbf{Definition:} \textit{Equilibrium point or Critical point or Fixed point}: A fixed (also known as critical or equilibrium) point at $\textbf{x} = \textbf{x}_*$ exists if and only if it meets the condition $\textbf{f}(\textbf{x}_*) = 0$ for an autonomous system described by equation \eqref{general_signle_system} \cite{Wiggins_2003_book}.

    If a dynamical system, represented by a set of autonomous equations, has a fixed point as defined in the above definition, then any trajectories that originate exactly from this point will remain stationary and unchanged over time. To analyze the impact of small disturbances around this point or the behavior of trajectories that pass nearby, it is essential to define the stability of the fixed point.\\
    \textbf{Definition:} \textit{Lyapunov Stable Fixed Point or Stable Fixed Point}: A fixed point $\textbf{x}_*$ is considered stable (or Lyapunov stable) if, for every small $\epsilon > 0$, there exists a $\delta > 0$ such that $||\psi(t_0) - \textbf{x}_*|| < \delta$. This ensures that the solution $\psi(t)$ remains within $||\psi(t) - \textbf{x}_*|| < \epsilon$ for all future times $t \geq t_0$ \cite{Bahamonde_2018_775}. 

    A stable fixed point means that if the system starts slightly perturbed from equilibrium, it will remain bounded within a small neighbourhood over time. In cosmology, stable fixed points correspond to self-regulating cosmic phases, such as the late-time de Sitter expansion, where the Universe remains in an accelerating phase indefinitely (as $t \to \infty$). However, points within this radius are not necessarily required to converge to the stable fixed point. A stronger definition is needed to satisfy that criterion.\\
    \textbf{Definition:} \textit{Asymptotically Stable Fixed Point}: A fixed point $\textbf{x}_*$ is considered asymptotically stable if it is stable and there exists a $\delta$ such that $||\psi(t_0) - \textbf{x}_*|| < \delta$. Under these conditions, the solution $\psi(t)$ satisfies $\lim\limits_{t \to \infty} \psi(t) = \textbf{x}_*$.

    If a fixed point is asymptotically stable, as defined in the above definition, trajectories that come sufficiently close to it will eventually converge to that point. This type of equilibrium is particularly significant in cosmology, where most stable fixed points are also asymptotically stable. It is important to note, however, that the definition does not specify the duration required for a trajectory to converge to the asymptotically stable fixed point. Lastly, an unstable fixed point is an equilibrium point that lacks stability.

    Let us highlight some key aspects of the phase space in an autonomous dynamical system and its fixed points. Excluding periodic orbits, distinct trajectories within the phase space cannot intersect. This implies that the solutions to the autonomous ODE are unique, provided that $f(x)$ is smooth. An orbit connecting two fixed points is known as a heteroclinic orbit, which will be particularly significant.

    If all solutions $\psi(t)$ within a well-defined subspace $S$ of the full phase space $S \subset X \subseteq \mathbb{R}^n$ remain within that subspace for all $t \in \mathbb{R}$, then the set of points $x \in S$ is called the invariant set, and the subspace $S$ is known as the invariant manifold. Essentially, any orbits isolate the invariant manifold from the rest of the phase space. A related concept is the invariant sub-manifold, an invariant manifold with a dimension one or less than the phase space. Invariant sub-manifolds partition the phase space into smaller, independent sections that are not connected by any orbits.
\subsection{Linear Stability Theory}
    We employ linear stability theory to analyze the dynamics of trajectories near a critical point. We can approximate the non-linear dynamics of complex systems, still represented by $\dot{\textbf{x}} = \textbf{f(x)}$, by linearizing around a critical point $\textbf{x}_*$. This approximation is valid if we assume $\textbf{f(x)}$ is sufficiently regular. By performing a Taylor expansion of $\textbf{f(x)}$ around the critical point $\textbf{x}_*$, we obtain
    \begin{eqnarray}
        \textbf{f(x)} = \textbf{f}(\textbf{x}_*) + (\textbf{x} - \textbf{x}_*) \left. \frac{\partial \textbf{f}}{\partial \textbf{x}} \right|_{\textbf{x} = \textbf{x}_*} + \ldots \, ,
    \end{eqnarray}
    where only the first partial derivatives need to be considered \cite{Strogatz_2018_book}. By definition, $\textbf{f}(\textbf{x}_*) = 0$, so the evolution of the points $(\textbf{x} - \textbf{x}_*)$ is governed by the Jacobian matrix evaluated at the critical points
    \begin{eqnarray} \label{general Jacobian matrix}
        J|_{\textbf{x} = \textbf{x}_*} = \left. \frac{\partial \textbf{f}}{\partial \textbf{x}} \right|_{\textbf{x} = \textbf{x}_*} = 
        \begin{bmatrix}
        \left. \frac{\partial f_1}{\partial \textbf{x}_1} \right|_{\textbf{x} = \textbf{x}_*} & \cdots & \left. \frac{\partial f_1}{\partial \textbf{x}_n} \right|_{\textbf{x} = \textbf{x}_*} \\
        \vdots & \ddots & \vdots \\
        \left. \frac{\partial f_n}{\partial \textbf{x}_1} \right|_{\textbf{x} = \textbf{x}_*} & \cdots & \left. \frac{\partial f_n}{\partial \textbf{x}_n} \right|_{\textbf{x} = \textbf{x}_*} 
    \end{bmatrix} \, .
    \end{eqnarray}
    The Jacobian matrix also known as the stability matrix, provides information about the stability of the critical points $\textbf{x}_*$ through its eigenvalues. These eigenvalues can be determined manually, as demonstrated in the upcoming example, or through computational methods for more complex systems of equations. The same approach applies to identifying the fixed points of the system.

    Let us briefly examine the classification of fixed points for linear systems, which are determined by the eigenvalues of the Jacobian matrix at the critical point \eqref{general Jacobian matrix}. If all eigenvalues have positive real parts, the point is considered unstable (or a repeller), causing trajectories to diverge from it. Conversely, if all eigenvalues have negative real parts, the point is stable (or an attractor), drawing trajectories towards it. When at least two eigenvalues have opposite signs, the point is classified as a saddle-node, with trajectories being attracted in some directions and repelled in others. These three classifications cover most fixed points found in cosmological systems. Additionally, spirals may occur in two dimensions when eigenvalues have non-zero imaginary parts. These spirals can be either stable or unstable, depending on the real parts of the eigenvalues. While a broader range of critical points can be classified \cite{Strogatz_2018_book, Glendinning_1994_book}, only these will be relevant for this work. Linear stability analysis is effective when all eigenvalues have nonzero real parts. However, in many cosmological models, non-hyperbolic fixed points arise when one or more eigenvalues vanish, such as in bouncing cosmologies or the transition between radiation and matter domination. In these cases, a more advanced approach is needed since the stability of the critical point is not solely determined by the linear terms \cite{Wiggins_2003_book}.\\
    \textbf{Definition:} \textit{Hyperbolic Point}: A fixed point $\textbf{x}_*$ of the system $\dot{\textbf{x}} = \textbf{f(x)}$ is considered hyperbolic if all eigenvalues of its Jacobian matrix have non-zero real parts. Otherwise, the point is classified as non-hyperbolic.
\subsection{Center Manifold Theory} \label{Intro: CMT}
    Center manifold theory extends stability analysis beyond hyperbolic points, allowing us to study critical points where linear stability theory fails. This is particularly useful for understanding early Universe transitions, emergent Universe, or higher-order gravitational effects in cosmology. As described by Perko \cite{Perko_2013}, when zero eigenvalues appear, linear analysis becomes inconclusive, meaning the fate of the system cannot be determined using standard stability methods. Center manifold theory addresses this by reducing the dimensionality of the system and capturing the essential slow dynamics near the fixed point. This theory simplifies the analysis by reducing the dimensionality of the system near critical points. When a system approaches a critical point, it follows an invariant local center manifold, denoted as $W^c$. This manifold is associated with eigenvalues with zero real parts, encapsulating the essential dynamics of the system near equilibrium.

    Consider a function $f$ in $C^r(E)$, where $E$ is an open subset of $\mathbb{R}^n$ that includes the origin, and $r \geq 1$. Assume $f(0) = 0$ and that the derivative $Df(0)$ has $c$ eigenvalues with zero real parts and $s$ eigenvalues with negative real parts, where $c + s = n$. Typically, the system can be reformulated as follows
    \begin{equation}\label{CMT_system}
        \begin{aligned}
        \dot{x} = A x + F(x, y)  \hspace{0.5cm} \text{and} \hspace{0.5cm} 
        \dot{y} = B y + G(x, y) \, ,
        \end{aligned}
    \end{equation}
    where $A$ is a square matrix with $c$ eigenvalues having zero real parts, and $B$ is a square matrix with $s$ eigenvalues having negative real parts, with $(x, y) \in \mathbb{R}^c \times \mathbb{R}^s$. The functions $F$ and $G$ satisfy $F(0) = G(0) = 0$ and their derivatives at zero are also zero. Additionally, there exists a small positive value $\epsilon > 0$ and a function $g(x)$ in $C^r(N_\epsilon(0))$, which defines the local center manifold and satisfies certain conditions
    \begin{equation} \label{local_center_manifold}
        \operatorname{Dg}(x)[A x + F(x, g(x))] - B g(x) - G(x, g(x)) = \mathcal{N}(g(x)) = 0 \, ,
    \end{equation}
    for $|x| < \epsilon$. The center manifold can be derived using the system of differential equations
    \begin{equation}
        \dot{x} = A x + F(x, g(x))\, ,
    \end{equation}
    for all $x \in \mathbb{R}^c$ with $|x| < \epsilon$.
\chapter{Observational Constrained \texorpdfstring{$f(R,\mathcal{G})$}{} Gravity Cosmological Model and the Dynamical System Analysis} 

\label{Chapter2} 

\lhead{Chapter 2.~\emph{Observational Constrained $f(R,\mathcal{G})$ Gravity Cosmological Model and the Dynamical System Analysis}} 

\vspace{10.5 cm}
* The work in this chapter is covered by the following publication: 

\textbf{Santosh V. Lohakare}, K. Rathore, and B. Mishra, ``Observational Constrained $f(R, \mathcal{G})$ Gravity Cosmological Model and the Dynamical System Analysis", \href{https://iopscience.iop.org/article/10.1088/1361-6382/acfc0f/meta}{\color{blue}\textit{{Classical and Quantum Gravity}} \textbf{40} (2023) 215009.}

 \clearpage
 
\section{Introduction}
    The Einstein field equations can be modified to fit the matter-energy content of the observable Universe by changing the geometrical sector. On the explanation of the evolution of the Universe, various modified gravity theories have been proposed \cite{Capozziello_2011_509, Faraoni_2010_170, Nojiri_2011_505, Carroll_2004_70, Nojiri_2007_04, Nojiri_2017_692}. One of the important findings is that it is possible to define early inflation with different coupling parameters and describe the late-time DE-dominated era with precision \cite{Starobinsky_1980_91}. In $f(R)$ gravity \cite{Carroll_2004_70, Nojiri_2011_505}, the gravitational action generalizes the Einstein--Hilbert action by introducing a generic function of the Ricci scalar curvature $R$ and GR can be restored by assuming $f(R) = R$. The general relativistic gravitational Lagrangian may be modified to include a broader range of curvature invariants, such as $R$, $R_{i\,j}R^{i\,j}$ and $R_{i\,j\,k\,l}R^{i\,j\,k\,l}$ among others. The $f(R, \mathcal{G})$ model acts as a viable alternative to DE \cite{Martino_2020_102, Benetti_2018_27, ODINTSOV_2019_938_935, Elizalde_2010_27}. The gravitational Lagrangian in Gauss--Bonnet gravity theories is a function $f(R, \mathcal{G})$, where the Gauss--Bonnet invariant $\mathcal{G}$ is defined as $\mathcal{G} \equiv R^2-4R^{i\,j} R_{i\,j}+R^{i\,j\,k\,l}R_{i\,j\,k\,l}$. In differential geometry and topology, the Gauss--Bonnet invariant modifies the Einstein--Hilbert action that governs the dynamics of gravity. In Refs. \cite{Baojiu_2007_76, Lattimer_2014_784, De_Felice_2009_675, Nojiri_2005_631, Cognola_2006_73}, the $f(R,\mathcal{G})$ gravity was proposed to incorporate $R$ and $\mathcal{G}$ into a bivariate function that supports the double inflationary scenario \cite{Laurentis_2015_91} and are also strongly supported by observations \cite{Capozziello_2014_29}. Besides its stability, the $f(R,\mathcal{G})$ theory is well-suited to describe the crossing of the phantom divide line and the transformation between an accelerating and decelerating state of celestial bodies.

    In scalar-tensor gravity, phase space is vibrant due to the fourth-order contributions of the Gauss--Bonnet invariant and the second-order contributions of the scalar field \cite{Konstantinos_2022_2211.06076}. Several invariant structures in phase space are necessary for the theory to be valid and viable in describing the evolution of the Universe \cite{Chatzarakis_2020_419}. Using dynamical system analysis, Shah et al. \cite{Shah_2019_79} analyzed the stability properties and acceleration phase of the Universe under various circumstances. The combined study of the data $H(z)$ and $f\sigma_8(z)$ shows that for $n=2$, the Starobinsky model of $f(R)$ fits well with the observational data and is a feasible alternative to the $\Lambda$CDM model \cite{Bessa_2022_82}. Using the dynamical system approach and constraining observational data, Bayarsaikhan et al. \cite{Bayarsaikhan_2023_83} have examined regularized Einstein-Gauss--Bonnet gravity in four dimensions with a non-minimal scalar coupling function.

    In cosmological observation, the CC approach can be used to determine the age and expansion rate of the Universe. The CC technique consists of three basic components: i) the definition of a sample of optimal CC tracers, ii) the determination of the differential age, and iii) the assessment of systematic effects \cite{Moresco_2022_25}. The value of the Hubble parameter $H(z)$ is instrumental in determining the energy content of the Universe and its acceleration mechanism. The estimation of $H(z)$ is carried out mainly at $z=0$. However, there are methods to determine $H(z)$ such as the detection of BAO signal in the clustering of galaxies and quasars, analyzing SNe Ia data, Ref. \cite{Riess_2018_853, Riess_2021_908, Font_Ribera_2014_2014_027, Raichoor_2020_500, Hou_2020_500}) and so on. Pantheon$^+$ is the successor to the original Pantheon analysis \cite{Scolnic_2018_859} and expands the original Pantheon analysis framework to combine an even larger number of SNe Ia samples to understand the complete expansion history. Here, we have used the CC sample, Pantheon$^+$, and BAO data sets to investigate the expansion history of the Universe, as well as the behavior of other geometrical parameters. 

    Noether symmetry analysis revealed that one such symmetry is admissible for $f(R, \mathcal{G})=\alpha R^n \mathcal{G}^{1-n}$ \cite{Capozziello_2014_29}. Viable cosmological solutions that include stability criteria for such a form have also been explored \cite{Santos_da_Costa_2018_35}. However, the absence of a linear term in the Ricci scalar is a serious concern, while other forms may also be possible from symmetry analysis. We investigate the geometric and dynamical characteristics of the cosmological model defined by $\mathcal{F}(R, \mathcal{G}) = \alpha R^2 \mathcal{G}^\beta$, where $R$ is the Ricci scalar and $\mathcal{G}$ refers to the Gauss--Bonnet invariant. We constrain the model parameters using observational cosmological data. Our results demonstrate the viability of a radiation-dominated era, transitioning to an early deceleration followed by late-time acceleration in the matter-dominated phase. A phase-space analysis has been conducted to evaluate stability, leading to restrictions on the parameter $\beta$, specifically excluding $\beta = -1$ \cite{Santos_da_Costa_2018_35}. Furthermore, we examine the stability of the model by analyzing the behavior of critical points, ultimately determining the current values of the density parameters for both matter and DE components, which align with those derived from the cosmological data sets. This chapter is organized as follows: In section \ref{CH2_SEC-II}, we present the mathematical formalism of $f(R, \mathcal{G})$ gravity. Section \ref{CH2_SEC-III} discusses and uses the observational data sets derived from the CC sample, Pantheon$^+$ samples and the BAO. The geometrical and dynamical parameters are also constrained by using these data sets.  Dynamical system analysis has been performed for the model in section \ref{CH2_SEC-IV}. Finally, we summarize our results in section \ref{CH2_SEC-V} with the conclusion.

\section{Basic Formalism of \texorpdfstring{$f(R, \mathcal{G})$}{} Gravity and Cosmology} \label{CH2_SEC-II}
    The action of $f(R, \mathcal{G})$ gravity, a modification of GR \cite{Laurentis_2015_91, Wu_2015_92, Santos_da_Costa_2018_35, ODINTSOV_2019_938_935, Kumar_Sanyal_2020_37, Brout_2022_938} is,
        \begin{equation}\label{ch2_1}
            S=\int d^{4}x \ \sqrt{-g}\left[\frac{1}{2\kappa^2}f(R,\mathcal{G})+\mathcal{L}_\text{m}\right] \, ,
        \end{equation}
    The Gauss--Bonnet invariant is defined as
        \begin{equation}\label{ch2_2}
            \mathcal{G} \equiv R^2-4 R^{i\,j} R_{i\,j} + R^{i\,j\,k\,l} R_{i\,j\,k\,l},
        \end{equation}
    with the Ricci tensor and Riemann tensor, respectively, denoted by $R^{i\,j}$ and $R^{i\,j\,k\,l}$. The definition of $\mathcal{G}$ in differential geometry is
        \begin{equation}\label{ch2_3}
            \int_{\mathcal{M}} \mathcal{G} d^{n}x=\chi (\mathcal{M}),
        \end{equation}
    in 4-D, $\mathcal{G}=R^{i\,j\,k\,l} R_{i\,j\,k\,l}=\chi (\mathcal{M})$, which is metric independent, and so a topologically invariant Euler number. Consequently, $\int \mathcal{G} \sqrt{-g} d^{4}x $ yields a surface term. Thus, the Gauss--Bonnet term contributes only either through dynamical coupling or considering non-linear terms. Here, we consider curvature scalar coupled non-linear Gauss--Bonnet term. By varying the action \eqref{ch2_1} with respect to the metric tensor $g_{i\,j}$, the field equations of $f(R,\mathcal{G})$ gravity can be written as,
        \begin{eqnarray} \label{ch2_4}
            \nonumber f_R{G}_{i\,j}&=&\kappa^2{T}_{i\,j}+\frac{1}{2}g_{i\,j}[f(R,\mathcal{G})-R f_{R}]+\nabla_{i} \nabla_{j} f_{R} - g_{i\,j} \Box f_{R} + f_{\mathcal{G}}\Big({-2R}{R_{i\,j}} +4R_{i\,k}R^{k}_{j} \nonumber\\& & -2R^{k\,l\,m}_{i}R_{j\,k\,l m} + 4g^{k\,l} g^{m\,n} R_{i\,k\,j\,m} R_{l\,n}\Big) +2(\nabla_{i}\nabla_{j}f_{\mathcal{G}})R - 2g_{i\,j}(\Box f_{\mathcal{G}})R + 4(\Box f_{\mathcal{G}})R_{i\,j} \nonumber\\& & - 4(\nabla_{k} \nabla_{i} f_{\mathcal{G}})R^{k}_{j}-4(\nabla_{k} \nabla_{j} f_{\mathcal{G}})R^{k}_{i} + 4g_{i\,j}(\nabla_{k} \nabla_{l} f_{\mathcal{G}})R^{kl}-4(\nabla_{l} \nabla_{n} f_{\mathcal{G}})g^{kl}g^{mn}R_{i\,k\,j\,m}\, ,\nonumber\\
        \end{eqnarray}
    where $G_{i\,j}$ represents the Einstein tensor, $\nabla_{i}$ describes the covariant derivative operator associated with $g_{i\,j}$, $\Box \equiv g^{i\,j}\nabla_{i}\nabla_{j}$ represents the covariant d'Alembert operator, and ${T}_{i\,j}$ represents the energy-momentum tensor. Additionally, the following quantities have been specified.
        \begin{equation*} 
            f_R\equiv \frac{\partial f(R,\mathcal{G})}{\partial R},\hspace{1cm} f_\mathcal{G}\equiv \frac{\partial f(R,\mathcal{G})}{\partial \mathcal{G}}.
    \end{equation*}
    The spacetime for the flat FLRW metric can be given as
        \begin{equation} \label{ch2_flrw metric}
            ds^{2}=-dt^{2}+a^{2}(t)(dx^{2}+dy^{2}+dz^{2}),
        \end{equation}
    where $a(t)$ is the scale factor and the Hubble parameter, $H\equiv\frac{\dot{a}(t)}{a(t)}$. Subsequently, the Ricci scalar and the Gauss--Bonnet invariant respectively, becomes
        \begin{equation} \label{ch2_6}
            R=6(\dot{H}+2H^{2}), \hspace{1cm} \mathcal{G}=24H^{2}(\dot{H}+H^{2}) \, .
        \end{equation}
    By substituting equation \eqref{ch2_flrw metric} and \eqref{ch2_6} into the gravitational field equation \eqref{ch2_4}, we obtain the field equations of $f(R,\mathcal{G})$ gravity as,
        \begin{eqnarray} 
            3H^{2} f_{R}&=&{\kappa^{2}} \rho+\frac{1}{2}\left[R {f_{R}}+\mathcal{G} {f_{\mathcal{G}}} - {f(R,\mathcal{G})}\right]-3H \dot{F}_{R} -12H^{3} \dot{F}_{\mathcal{G}},\label{ch2_8}\\
            (2\dot{H}+3H^{2}) f_{R} &=& -\kappa^{2} p +\frac{1}{2}\left[R f_{R}+\mathcal{G} f_{\mathcal{G}} - f(R,\mathcal{G})\right] -2H\dot{F}_{R}-\ddot{F}_{R}\nonumber\\& &-8H\left(\dot{H}+H^2\right) \dot{F}_{\mathcal{G}}-4H^{2} \ddot{F}_{\mathcal{G}}.\label{ch2_9}
        \end{eqnarray}
    Background cosmology can be simplified by rewriting these equations as effective fluids, embodying additional terms due to higher-order curvature terms incorporated into the expression. We consider the mapping, $f(R,\mathcal{G}) \longrightarrow R+\mathcal{F}(R,\mathcal{G})$ \cite{Marco_2020_135}. The motivation behind considering this form is its consistency with the concordance $\Lambda$CDM model. For  $\mathcal{F}=-2\Lambda$, with $\Lambda$ being the cosmological constant, it corresponds to $\Lambda$CDM paradigm. Accordingly, equation \eqref{ch2_8} and \eqref{ch2_9} reduce to,
        \begin{eqnarray}
            3 H^2 &=& \kappa^2 (\rho_\text{m}+\rho_\text{r} + \rho_{\text{DE}}) = \kappa^2 \rho_{\text{eff}}, \label{ch2_10}\\
            3 H^2+2 \dot{H} &=& -\kappa^2 (p_\text{m}+p_\text{r}+p_{\text{DE}}) = - \kappa^2 p_{\text{eff}}, \label{ch2_11}
        \end{eqnarray}
    and resulting in the following identities as
        \begin{eqnarray}
            \kappa^2 \rho_{\text{DE}} &=& -3 H^2 \mathcal{F}_R+\frac{1}{2}\big(R \mathcal{F}_R + \mathcal{G} \mathcal{F}_\mathcal{G}-\mathcal{F}(R,\mathcal{G}) - 6 H \dot{\mathcal{F}}_R-24 H^3 \dot{\mathcal{F}}_\mathcal{G}\big),  \label{ch2_12}\\
            \kappa^2 p_{\text{DE}} &=& (2 \dot{H}+3 H^2) \mathcal{F}_R-\frac{1}{2} \big(R \mathcal{F}_R+\mathcal{G} \mathcal{F}_{\mathcal{G}}-\mathcal{F}(R,\mathcal{G}) - 4 H \dot{\mathcal{F}}_R\nonumber\\& & - 2\ddot{\mathcal{F}}_R-8 H^2 \ddot{\mathcal{F}}_{\mathcal{G}}-16 H \dot{H} \dot{\mathcal{F}}_{\mathcal{G}}-16 H^3 \dot{\mathcal{F}}_{\mathcal{G}}\big). \label{ch2_13}
        \end{eqnarray}
    To solve the system equations \eqref{ch2_12}-\eqref{ch2_13}, some viable form of $\mathcal{F}(R, \mathcal{G})$ would be required. Hence, we consider
        \begin{eqnarray} \label{ch2_14}
            \mathcal{F}(R,\mathcal{G})=\alpha R^2 \mathcal{G}^\beta ,
        \end{eqnarray}
    where $\alpha$ and $\beta$ are arbitrary constants, $\beta \neq 1$, and study cosmological consequence and the stability criteria, following the work \cite{Capozziello_2014_29}. It is a double inflationary scenario connected to the existence of Noether symmetries. The form of the function $\mathcal{F}(R, \mathcal{G})$ in modified gravity models can have intriguing cosmological implications. It can lead to modified field equations that govern the dynamics of the Universe, affecting the expansion rate, the evolution of cosmic structures, and the behavior of matter and energy. Studying the cosmological consequences of this form of $\mathcal{F}(R, \mathcal{G})$ allows researchers to explore new scenarios, such as inflationary models and DE models, and compare them with the cosmological observations. In the case of a positive second derivative of $f(R, \mathcal{G})$ with respect to $R$, the model is free from instability within the context of Dolgov-Kawasaki instability \cite{Dolgov_2003_576}, and accordingly, the limits on the model parameters are $\alpha > 0$, and $\beta$ is even. We rewrite $R$ and $\mathcal{G}$ in the redshift parameter to get the expansion rate $(1 + z)H(z) = -\frac{dz}{dt}$ as
        \begin{eqnarray}
            R&=&6\left(2 H_0^2 E(z)-\frac{H_0^2 (1+z) E'(z)}{2}\right), \nonumber\\ \mathcal{G}&=&24H_0^2 E(z) \left(H_0^2 E(z)-\frac{H_0^2 (1+z) E'(z)}{2}\right), \label{ch2_15}
        \end{eqnarray}
    where $H^2 (z)=H_0^2 E(z)$, $H_0$ represents the present value of the Hubble parameter, and throughout this thesis, the notation ‘prime’ is utilized to indicate a derivative with respect to redshift $ z $. We use the following functional form for $E(z)$ \cite{Lemos_2018_483},
        \begin{eqnarray} \label{ch2_16}
            E (z) = A\, (1+z)^3 + B + C\, z + D\, \text{ln}(1+z) ,
        \end{eqnarray}
    where $A$, $B$, $C$, and $D$ are free parameters. The $A(1+z)^3$ term accounts for the main effect of matter, as its energy density dilutes with the expansion of the Universe.  The terms $B$, $C\,z$  and $D\, \text{ln}(1+z)$ in the above expression are associated with the contribution from DE, which drives the accelerated expansion of the Universe. The flexible $E(z)$ allows us to explore alternative cosmological scenarios beyond the standard $\Lambda$CDM model. Changing the parameters, one can examine the impact of different components and modifications of gravity on the expansion of the Universe.
\section{Observational Constraints} \label{CH2_SEC-III}
    In cosmology, the Hubble and Pantheon$^+$ data sets are important to study the expansion history of the Universe and the properties of DE. Here, we shall use the early-type galaxies expansion rate data such as the $H(z)$, Pantheon$^+$ data, BAO and CMB distance priors. Since $H(z)$ provides the basic information about the energy content and the main physical mechanisms driving the present acceleration of the Universe; therefore the accurate determination of the expansion rate of the Universe has become important. In CC measurement, the expansion rate of the Universe is directly and cosmology-independently estimated without any assumptions about the origin of the Universe. There is no direct correlation between the observations and cosmological models. Therefore, these data sets serve as an independent tool to estimate the parameters of the cosmological models.
        \begin{figure} [H]
        \centering
            \includegraphics[width=120mm]{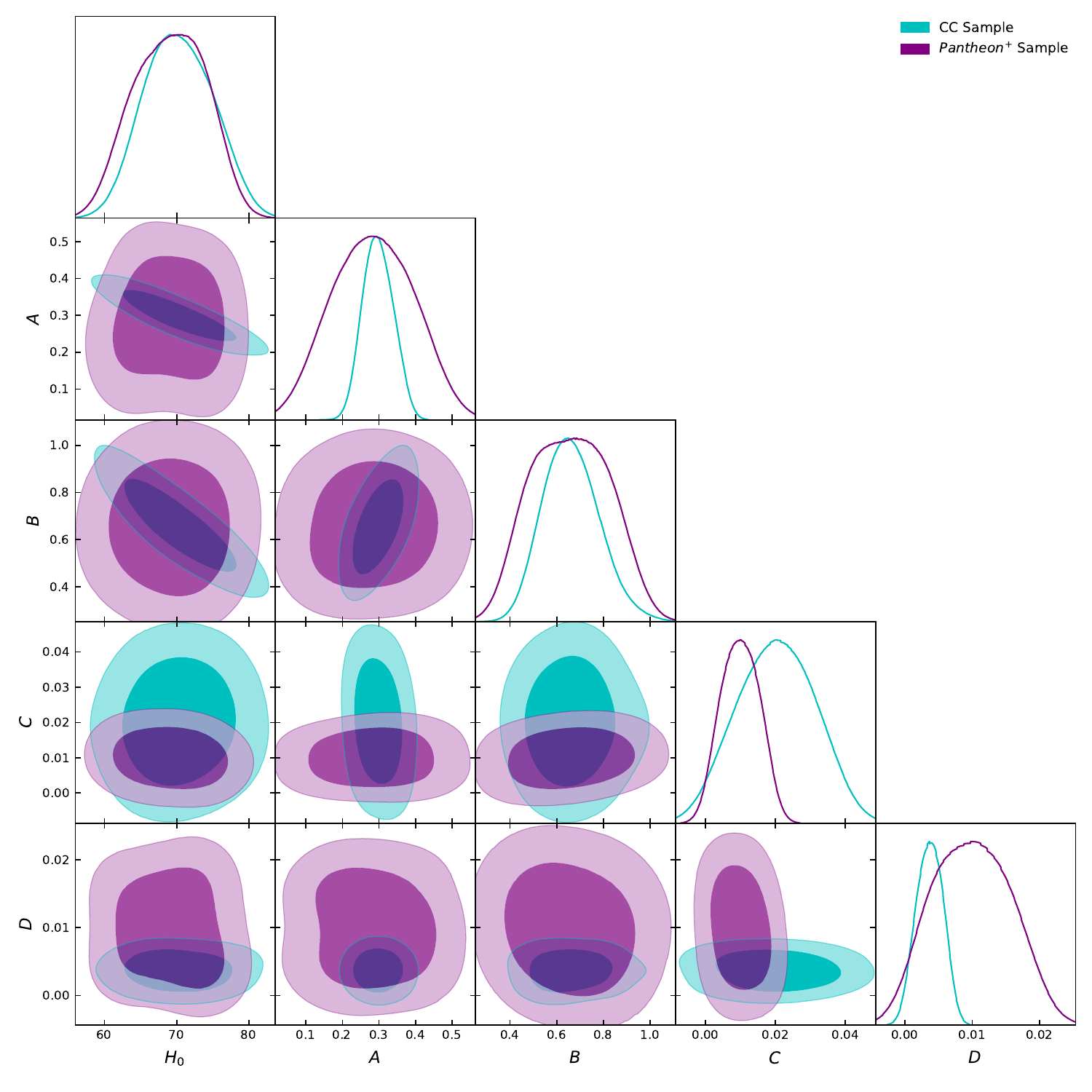}
            \caption{Contour plots of $H_0$, $A$, $B$, $C$, and $D$ with $1\sigma$ and $2\sigma$ errors.}
        \label{CH2_FIG2}
        \end{figure}
        \begin{figure} [H]
        \centering
            \includegraphics[width=120mm]{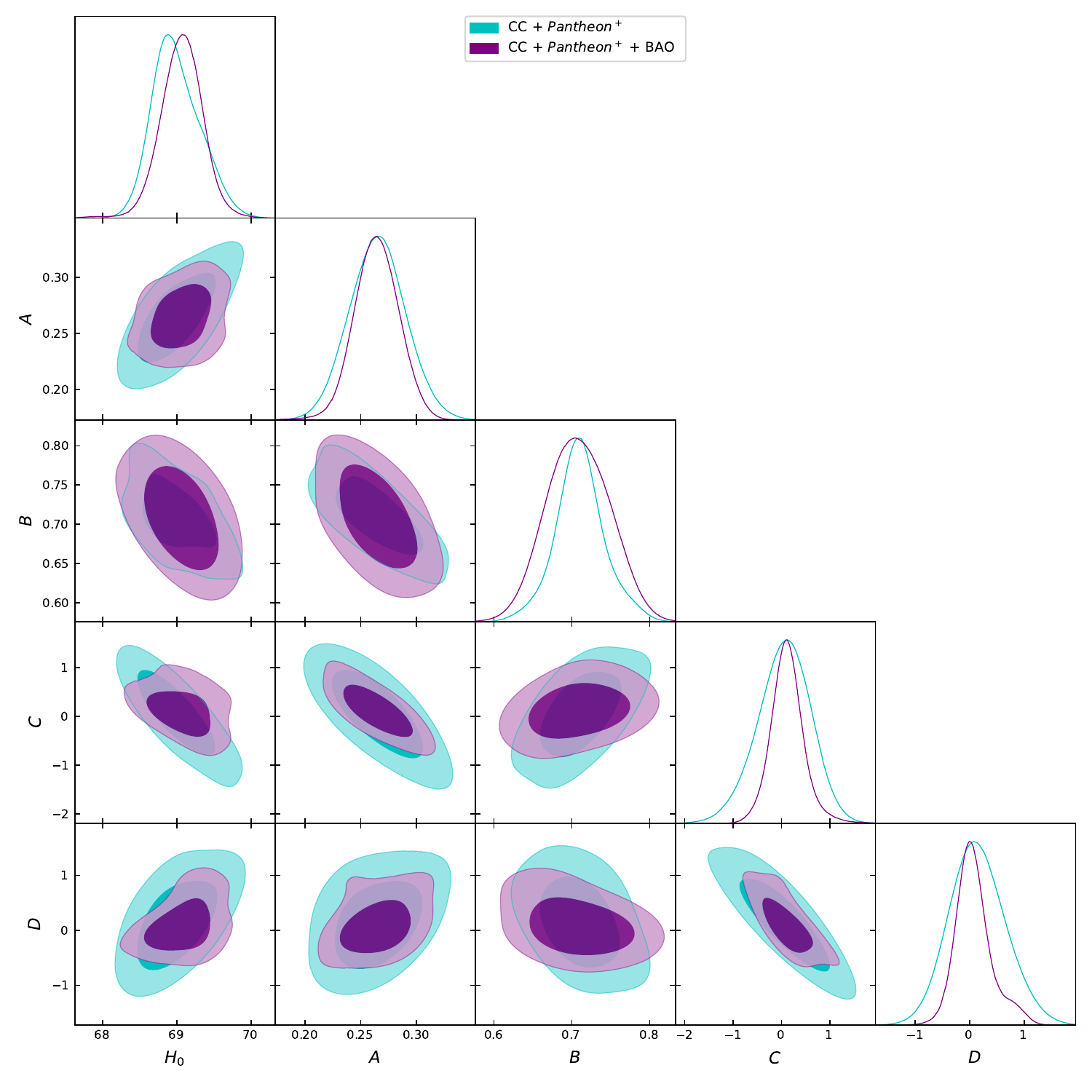}
            \caption{Contour plots of $H_0$, $A$, $B$, $C$, and $D$ with $1\sigma$ and $2\sigma$ errors for combined data sets.}
        \label{CH2_FIG3}
        \end{figure}
    \begin{table*} [!htb]
        \centering
        \scalebox{0.81}{
        \begin{tabular}{|*{5}{c|}}\hline
    \parbox[c][0.8cm]{2cm}{\textbf{Coefficients}} & \textbf{CC Sample}  & \textbf{Pantheon$^+$} & \textbf{CC + Pantheon$^+$} & CC + \textbf{Pantheon$^+$ + BAO}\\ [0.5ex]
        \hline \hline
    \parbox[c][0.7cm]{2cm}{$H_0$} & 70.2 $\pm$ 4.6 & 69.1 $\pm$ 4.8 & $68.69_{-0.59}^{+0.67}$ & $69.26_{-0.53}^{+0.57}$ \\
        \hline
    \parbox[c][0.7cm]{2cm}{$A$} & 0.297 $\pm$ 0.04 & 0.28 $\pm$ 0.11 & $0.285^{+0.050}_{-0.048}$ & $0.264^{+0.039}_{-0.036}$ \\
        \hline
    \parbox[c][0.7cm]{2cm}{$B$} & 0.66$^{+0.11}_{-0.13}$ & 0.64 $\pm$ 0.16 & $0.689^{+0.071}_{-0.067}$ & $0.698^{+0.070}_{-0.071}$ \\
        \hline
    \parbox[c][0.7cm]{2cm}{$C$} & 0.0099 $\pm$ 0.0053 & 0.02 $\pm$ 0.011 & $0.012^{+0.98}_{-1.1}$ & $0.012 \pm 0.71$ \\[0.5ex] 
        \hline 
    \parbox[c][0.7cm]{2cm}{$D$} & 0.0037 $\pm$ 0.0019 & 0.0099 $\pm$ 0.056 & $0.014^{+1.1}_{-0.99}$ & $0.0025^{+0.81}_{-0.61}$ \\[0.5ex]
        \hline
        \end{tabular}}
        \caption{Constrained values of $H(z)$ model parameter based on the CC, Pantheon$^+$ samples, and BAO data sets.}
        \label{CH2_TABLE I}
    \end{table*}

\subsection{Cosmic Chronometers}
    The Hubble parameter $H(z)$ can be estimated at certain redshifts $z$ using the following formula
    \begin{equation}
        H(z) = \frac{\dot{a}}{a} = -\frac{1}{1+z}\frac{dz}{dt} \approx -\frac{1}{1+z}\frac{\Delta z}{\Delta t} \, ,  \label{ch5_eq: 20}
    \end{equation}
        where $\dot{a}$ is the derivative of the scale factor $a$ with respect to time $t$, and $\Delta z$ and $\Delta t$ are the differences in redshift and time, respectively, between two objects. The value of $\Delta z$ can be determined by a spectroscopic survey, while the differential ages $\Delta t$ of passively evolving galaxies can be used to estimate the value of $H(z)$. Compiling such observations can be regarded as a CC sample. We use 32 objects spanning the redshift range $0.07 \leq z \leq 1.965$ \cite{Moresco_2022_25}. For these measurements, one can construct a $\chi^2_{\text{CC}}$ estimator as follows
    \begin{eqnarray}
        \chi^2_{\text{CC}} = \sum_{i=1}^{32} \frac{[H_{\text{th}}(z_i) - H_{\text{obs}}(z_i)]^2}{\sigma^2_{H}(z_i)} \, , \label{ch5_eq: 21}
    \end{eqnarray}
        Here, $H_{\text{obs}}$ and $H_{\text{th}}$ represent the observational and theoretical values of the Hubble parameter, respectively, with $\sigma_H$ being the error in the observational value.

\subsection{Pantheon\texorpdfstring{$^+$}{} Sample}
    We will also take into account the Pantheon$^+$ SNe Ia data set, which includes 1701 measurements of the relative luminosity distance of SNe Ia spanning the redshift range of $0.00122 < z < 2.2613$ \cite{Brout_2022_938}. The Pantheon$^+$ compilation consists of distance moduli derived from 1701 light curves of 1550 spectroscopically confirmed SNe Ia within the redshift range of $0.00122 < z < 2.2613$, collected from 18 different surveys. It is worth noting that 77 of the 1701 light curves are associated with galaxies containing Cepheids. The Pantheon$^+$ data set is advantageous in that can also be used to limit the value of $H_0$ besides the model parameters. To estimate the model parameter from the Pantheon$^+$ samples, we minimize the $\chi^2$ function. To calculate the chi-square $(\chi^2_{\text{SNe}})$ value using the Pantheon compilation of 1701 supernovae data points, we use the following formula
   \begin{equation}
        \chi^2_{\text{SNe}}= \Delta\mu^T (C_{\text{Sys} + \text{Stat}}^{-1})\Delta\mu \, , \label{ch5_eq: 22}
    \end{equation}
        where inverse covariance matrix, $C_{\text{Sys} + \text{Stat}}^{-1}$, associated with the Pantheon$^+$ data set, incorporates both systematic and statistical uncertainties. The term $\Delta \mu$, defined below, signifies the distance residual
    \begin{equation}
        \Delta\mu = \mu_{\text{th}}(z_i,\theta)-\mu_{\text{obs}}(z_i) \, . \label{ch5_eq: 23}
    \end{equation}
        The distance modulus is specifically defined as the difference between an observed apparent magnitude ($m$) of the object and its absolute magnitude ($M$), which quantifies its intrinsic brightness. At a given redshift $z_i$, the distance modulus is expressed as follows
        \begin{equation}
        \mu_{\text{th}}(z_i,\theta)=5\log_{10}\left(d_L(z,\theta) \right) + 25 = m - M \, , \label{ch5_eq: 24}
    \end{equation}
        where $d_L$ denotes the luminosity distance in megaparsec (Mpc), contingent upon the specific model, which is
        \begin{equation}
        d_L(z,\theta)=\frac{c(1+z)}{H_0}\int_0^z \frac{d\zeta}{E(\zeta)}\, , \label{ch5_eq: 25}
    \end{equation}
        where $E(z) = \frac{H(z)}{H_0}$, with $c$ representing the speed of light. Furthermore, the residual distance is indicated by
    \begin{equation}
        \Delta\Bar{\mu}=\begin{cases}
            \mu_k-\mu_k^{cd}, & \text{if $k$ is in Cepheid hosts}\\
            \mu_k-\mu_{\text{th}}(z_k), & \text{otherwise}
        \end{cases} \label{ch5_eq: 26}
    \end{equation}
        where $\mu_k^{cd}$ represents the Cepheid host-galaxy distance as determined by SH0ES. This covariance matrix can be integrated with the SNe Ia covariance matrix to form the covariance matrix for the Cepheid host galaxy. Incorporating both statistical and systematic uncertainties from the Pantheon$^+$ data set, the combined covariance matrix is expressed as $C^{\text{SNe}}_{\text{Sys} + \text{Stat}} + C^{cd}_{\text{Sys} + \text{Stat}}$. This formulation defines the $\chi^2$ function for the combined covariance matrix, which is utilized to constrain cosmological models in the analysis
    \begin{equation}
        \chi^2_{\text{SNe}^+}= \Delta\Bar{\mu} (C^{\text{SNe}}_{\text{Sys}+\text{Stat}}+C^{cd}_{\text{Sys}+\text{Stat}})^{-1}\Delta\Bar{\mu}^T \, . \label{ch5_eq: 27}
    \end{equation}

\subsection{Baryon Acoustic Oscillations}
    Early Universe is being studied by analyzing BAO. There are three types of BAO measurements namely: High-resolution Sloan Digital Sky Surveys (SDSS), Six Degree Field Galaxy Surveys (6dFGS), and BOSS \cite{Percival_2010_401}. We present results from SDSS, 6dFGS, and BOSS-DR12 based on available BAO data. The following expressions for measurable quantities are used to obtain BAO constraints
        \begin{eqnarray}
            d_A (z)&=&\int_{0}^{z} \frac{dz^*}{H(z^*)}, \label{ch2_19}\\
            D_v (z)&=&\left( \frac{d_A (z)^2 z}{H(z)}\right)^{\frac{1}{3}}, \label{ch2_20}
        \end{eqnarray}
    and
        \begin{eqnarray} \label{ch2_21}
            \chi^2_{\text{BAO}}=X^T C^{-1} X,
        \end{eqnarray}
    where the angular diameter distance and the dilation scale are represented by $d_A(z)$, $D_V(z)$, respectively, and $C$ represents the covariance matrix \cite{Giostri_2012_2012_027}.
    \begin{figure} [H]
        \centering
        \includegraphics[width=7.78cm,height=5.5cm]{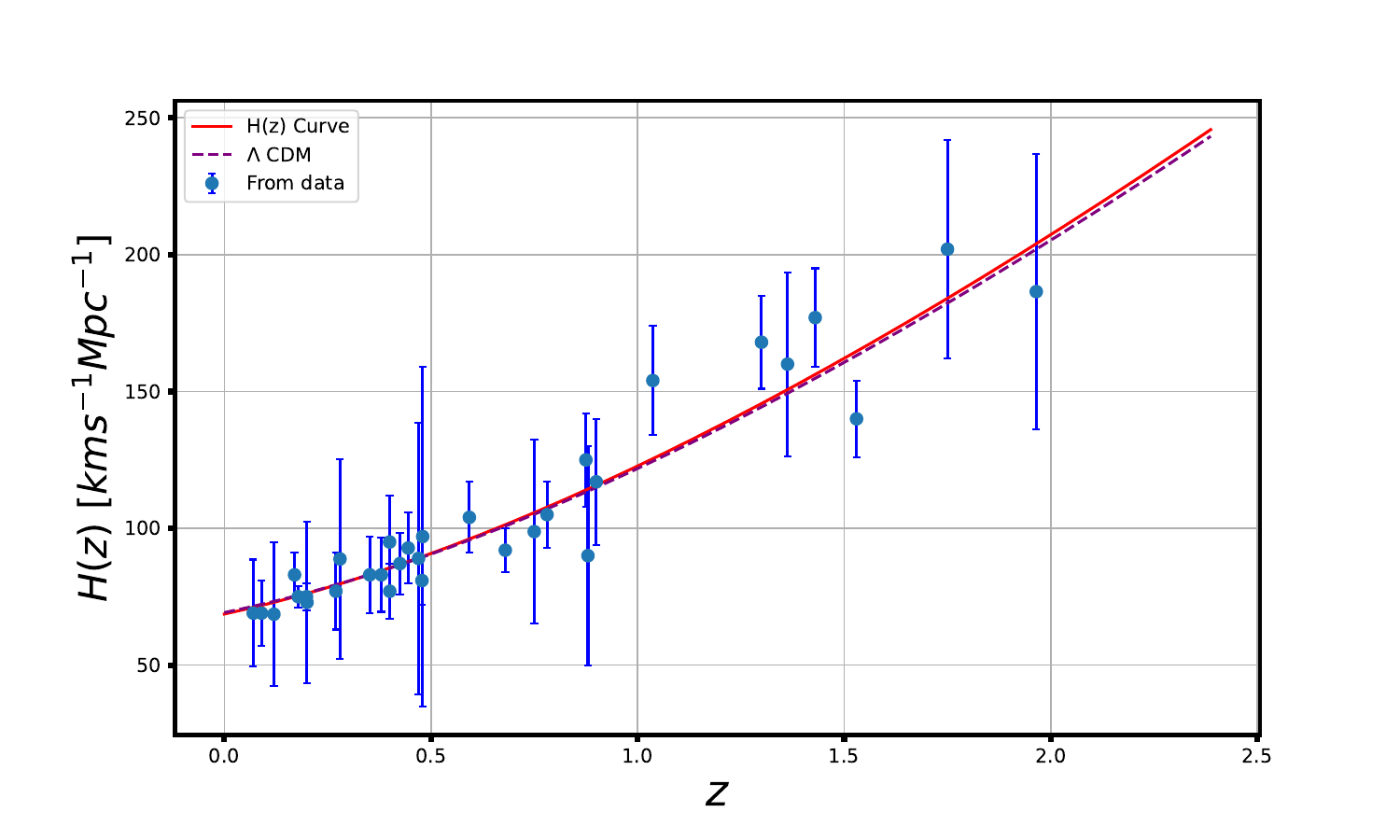}
        \includegraphics[width=7.78cm,height=5.5cm]{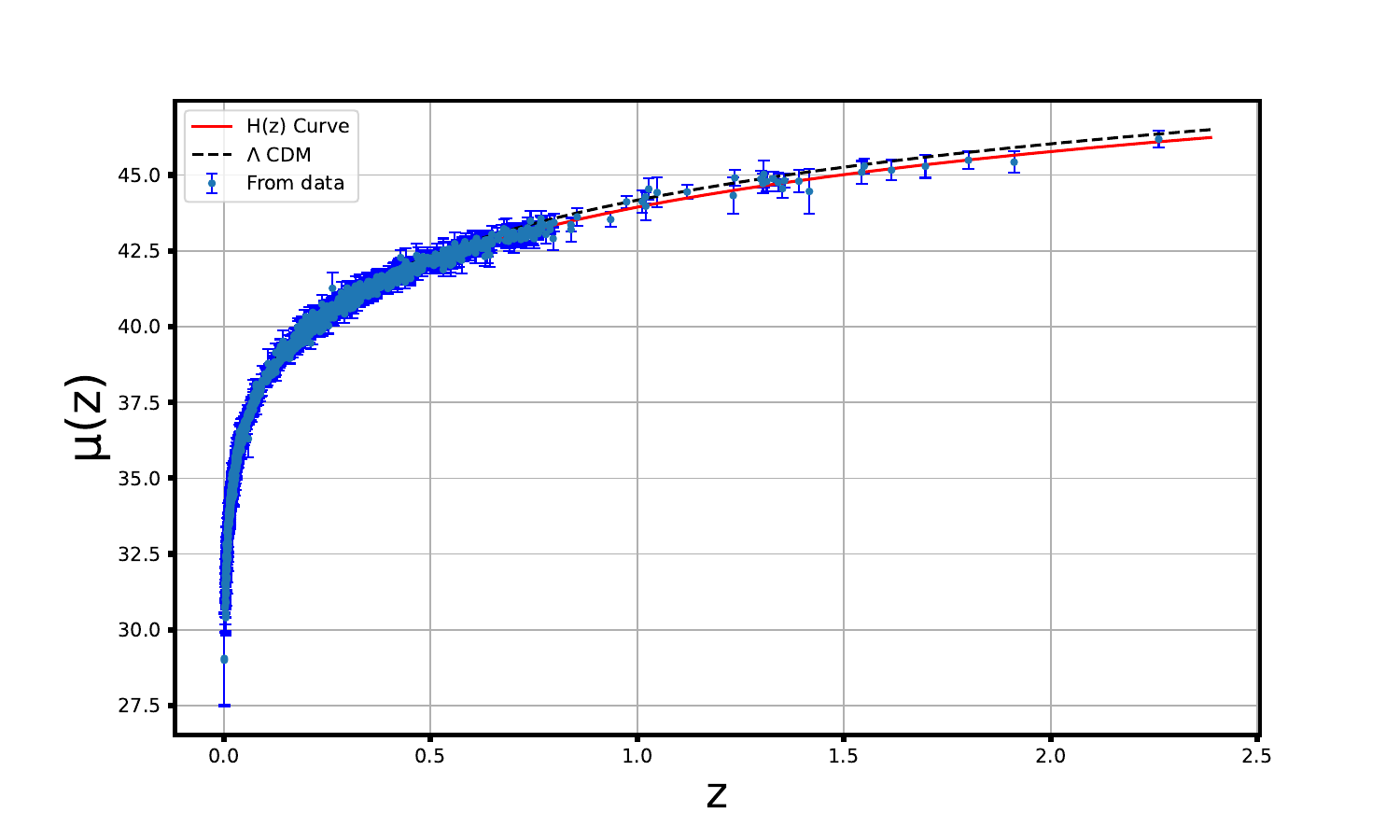}
        \caption{Graphical behavior of error bars are from the 32 points of the CC sample, the solid red line is of the model (left panel). In (right panel), the red line is the plot of the model's distance modulus $\mu(z)$ versus $z$, which exhibits a better fit to the 1701 points of the Pantheon$^+$ data sets along with its error bars and the broken black line is for the $\Lambda$CDM.}
        \label{CH2_FIG1}
    \end{figure}
    \begin{figure} [H]
        \centering
        \includegraphics[scale=0.44]{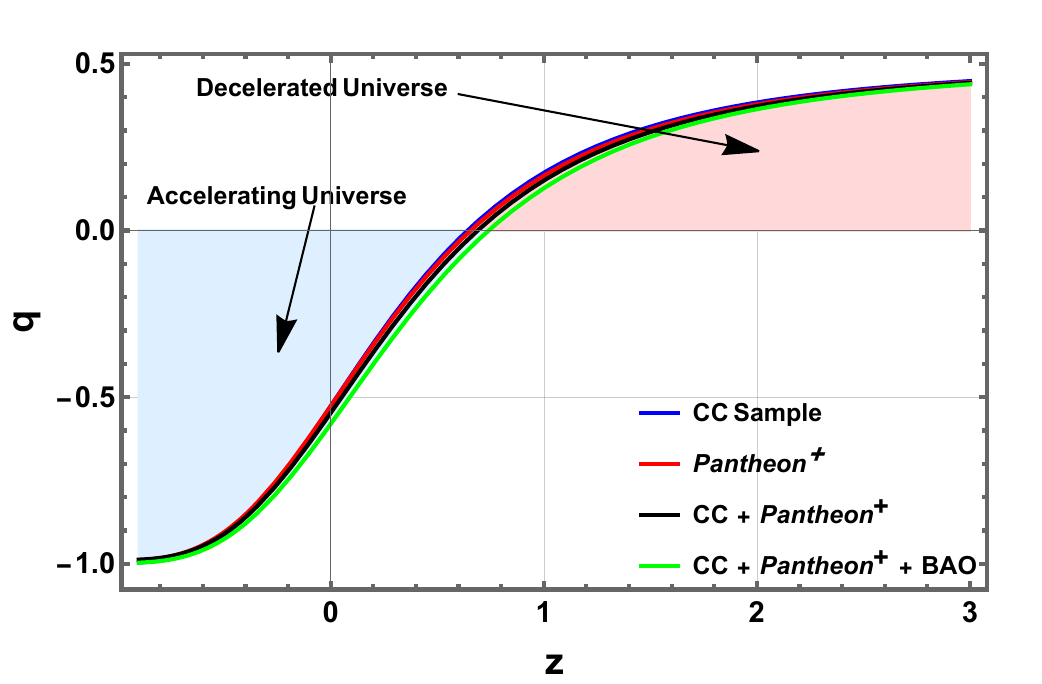}
        \includegraphics[scale=0.44]{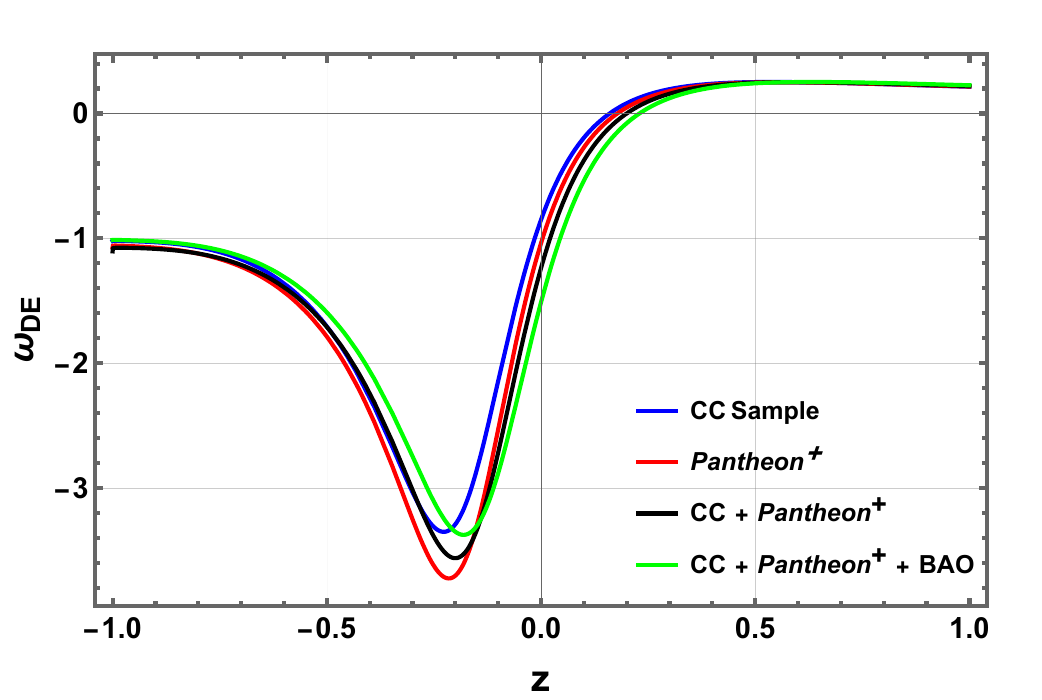}
        \caption{Graphical behavior of deceleration and EoS parameter with CC, Pantheon$^+$ and BAO data sets for the parameters $\alpha=1.1$, $\beta=4$.}
        \label{CH2_FIG4}
    \end{figure}
    \begin{table*}[!htb]
        \centering
        \scalebox{0.7}{
    \begin{tabular}{|*{5}{c|}}\hline
        \parbox[c][0.7cm]{2cm}{\textbf{Parameters}} & \textbf{CC Sample} & \textbf{Pantheon$^+$} & \textbf{CC + Pantheon$^+$} & \textbf{CC + Pantheon$^+$ + BAO}\\ [0.5ex]
    \hline \hline
        \parbox[c][0.7cm]{2cm}{$\hspace{.85cm} q$} & -0.526 ($z_t \approx 0.636$) & -0.529 ($z_t \approx 0.656$) & -0.548 ($z_t \approx 0.691$) & -0.579 ($z_t \approx 0.74$) \\
    \hline
        \parbox[c][0.7cm]{2cm}{\hspace{.7cm}$\omega_{\text{DE}}$} & -0.8478 & -1.02 & -1.224 & -1.47 \\
    \hline
        \parbox[c][0.7cm]{2cm}{\hspace{.7cm}$\omega_{\text{eff}}$} & -0.684 & -0.686 & -0.7 & -0.72 \\
    \hline
    \end{tabular}}
        \caption{Present value of deceleration and EoS parameters based on the CC samples, Pantheon$^+$ samples, and BAO data sets.} 
        \label{CH2_TABLE II}
    \end{table*}
    The contour plots with $1\sigma$ and $2\sigma$ errors are given in figure \ref{CH2_FIG2} for the CC and Pantheon$^+$ sample data sets, whereas in figure \ref{CH2_FIG3} for the CC + Pantheon$^+$ and CC + Pantheon$^+$ + BAO data sets. In figure \ref{CH2_FIG1}, one can observe that the $H(z)$ curve is lying well within the error bars. The blue error bars are from the 32 points of the CC sample, the solid red line is of the model, and the broken black line is for the $\Lambda$CDM (left panel). In (right panel), the red line is the plot of the model's distance modulus $\mu(z)$ versus $z$, which exhibits a better fit to the 1701 points of the Pantheon$^+$ data sets along with its error bars. All the values obtained for the parameters are listed in Tables \ref{CH2_TABLE I} and \ref{CH2_TABLE II}.
    The deceleration parameter $q=-1-\frac{\dot{H}}{H^2}$ describes the rate of acceleration of the Universe, where a positive $q$ indicates that the Universe is in a decelerated phase, while a negative $q$ indicates that the Universe is in an accelerated phase. The constrained values of model parameters in the Hubble, Pantheon$^+$, and BAO data sets resulted in $q$ changing from a positive value in the past, suggesting an early slowdown, to a negative value in the present, indicating an acceleration at present, as seen in figure \ref{CH2_FIG4}. In the current cosmic epoch, Hubble and Pantheon data are relatively consistent with the range $q_0=-0.528^{+0.092}_{-0.088}$ determined by recent observations \cite{Christine_2014_89} and a redshift from deceleration to acceleration occurs at $z_t=0.8596^{+0.2886}_{-0.2722}$, $z_t=0.65^{+0.19}_{-0.01}$ \cite{Yang_2020_2020_059, Capoziello_2008_664, Capozziello_2014_90_044016}. The deceleration parameter $q_0= -0.526$, $q_0= -0.529$, $q_0=-0.548$ and $q_0=-0.579$ at the current cosmic epoch and our derived model shows a smooth transition from a deceleration phase of expansion to an acceleration phase, at $z_t = 0.636$, $z_t = 0.656$, $z_t = 0.691$ and $z_t = 0.74$ for CC, Pantheon$^+$, CC + Pantheon$^+$ and CC + Pantheon$^+$ + BAO data sets, respectively. The recovered transition redshift value $z_t$ is consistent with certain current constraints based on 11 $H(z)$ observations reported by Busca et al. \cite{Busca_2013_552} between the redshifts $0.2 \leq z \leq 2.3$, $z_t = 0.74 \pm 0.5$ from Farooq et al. \cite{Farooq_2013_766}, $z_t = 0.7679^{+0.1831}_{-0.1829}$ by Capozziello et al. \cite{Capozziello_2014_90_044016} and $z_t = 0.60^{+0.21}_{-0.12}$ by Yang et al. \cite{Yang_2020_2020_059}

    Among the parameters that define the behavior of the Universe is the deceleration parameter, which determines whether the Universe continuously decelerates or accelerates constantly, has a single phase of transition or several, etc. Energy sources play a similar role in the evolution of the Universe according to the EoS parameter $\big(\omega_{\text{DE}}=\frac{p_{\text{DE}}}{\rho_{\text{DE}}}\big)$. Calculating the related energy density and pressure of DE, as illustrated in figure \ref{CH2_FIG4}, allows us to see the variations in the effective EoS of DE with respect to the redshift variable. The present value of EoS for DE $\omega_{\text{DE}}(z = 0)$ respectively obtained as, $-0.8478$, $-1.02$, $-1.224$ and $-1.47$ for CC, Pantheon$^+$, CC + Pantheon$^+$ and CC + Pantheon$^+$ + BAO data sets. It shows the phantom behavior (at $z \leq -0.015$) and its approach to $-1$ at late times. Whereas the present value of effective EoS ($\omega_{\text{eff}}$) parameter respectively obtain as $-0.684, \, -0.686, \, -0.7, \, -0.72$ [Table \ref{CH2_TABLE II}]. The numerical value of the EoS parameter has also been restricted by several cosmological investigations, including the Supernovae Cosmology Project $\omega_{\text{DE}}=-1.035^{+0.055}_{-0.059}$ \cite{Amanullah_2010_716}, Planck 2018, $\omega_{\text{DE}}=-1.03\pm 0.03$ \cite{Aghanim_2020_641} and WAMP+CMB, $\omega_{\text{DE}}=-1.079^{+0.090}_{-0.089}$ \cite{Hinshaw_2013_208}.
\section{Dynamical System Analysis} \label{CH2_SEC-IV}
    The $f(R, \mathcal{G})$ gravity model has been able to address some of the key issues of the early and late Universe, and it is always good to know its general phase space structure. Among higher-order theories of gravity, $f(R, \mathcal{G})$ gravity has one of the most complicated field equations, and dynamical system analysis has been important in understanding its physical behavior. The prime represents the derivative with respect to $N = \text{ln}\, a$. This method can generate the general form of the dynamical system for the modified FLRW equations, which are defined by equation \eqref{ch2_8}. As an autonomous system, the set of cosmological equations of the model is written with the following dimensionless variables \cite{Santos_da_Costa_2018_35}
        \begin{eqnarray} \label{ch2_22}
            u_1=\frac{\dot{F}_R}{H f_R},\hspace{0.5cm} u_2=\frac{F}{6 H^2 f_R},\hspace{0.5cm} u_3=\frac{R}{6 H^2}, u_4=\frac{\mathcal{G} f_{\mathcal{G}}}{6 H^2 f_R},\hspace{0.5cm} u_5=\frac{4 H \dot{F}_\mathcal{G}}{f_R},
        \end{eqnarray}
    with the energy density parameters
        \begin{eqnarray} \label{ch2_23}
            u_6=\Omega_\text{r}= \frac{\kappa^2 \rho_\text{r}}{3 H^2 f_R},\hspace{1cm} u_7=\Omega_\text{m}= \frac{\kappa^2 \rho_\text{m}}{3 H^2 f_R} \, ,
        \end{eqnarray}
    Thus, we have the algebraic identity
        \begin{equation} \label{ch2_24}
            1=-u_1-u_2+u_3+u_4-u_5+\Omega_\text{r}+\Omega_\text{m} \, .\\
        \end{equation}
    The dynamical system is
    \begin{eqnarray}
        \frac{du_1}{dN}&=& \frac{\ddot{F}_{R}}{f_{R}H^{2}} -u_1^{2}-u_1\frac{\dot{H}}{H^{2}}, \label{ch2_25}\\
        \frac{du_2}{dN}&=& \frac{\dot{F}}{6f_{R}H^{3}}-u_1u_2-2u_2\frac{\dot{H}}{H^{2}}, \label{ch2_26}\\
        \frac{du_3}{dN}&=& \frac{\dot{R}}{6H^{3}}-2u_3 \frac{\dot{H}}{H^{2}}, \label{ch2_27}\\
        \frac{du_4}{dN}&=& \frac{\dot{\mathcal{G}}}{\mathcal{G}H} u_4+\frac{\mathcal{G}}{24H^{4}} u_5-u_1u_4-2u_4(u_3-2), \label{ch2_28}\\
        \frac{du_5}{dN}&=& u_5 \frac{\dot{H}}{H^{2}}+4\frac{\ddot{f_{\mathcal{G}}}}{f_{R}}-u_1u_5, \label{ch2_29}\\
        \frac{du_6}{dN}&=& -2u_3 u_6-u_1 u_6, \label{ch2_30}\\
        \frac{du_7}{dN}&=& -u_7 \left(3+u_1+2 \frac{\dot{H}}{H^{2}}\right). \label{ch2_31}
    \end{eqnarray}
    To close the system, all terms on the right-hand side of the above equations must be expressed in terms of variables specified in equation \eqref{ch2_14}. Thus, we find
        \begin{eqnarray}
             \frac{\dot{H}}{H^{2}}&=&u_3-2, \label{ch2_32}\\
             \frac{\dot{F}}{6f_{R}H^{3}}&=&{-u_1 u_3}, \label{ch2_33}\\
             \frac{\dot{R}}{6H^{3}}&=&{u_1 u_3}, \label{ch2_34}\\
             \frac{\mathcal{G}}{24H^{4}}&=&u_3-1, \label{ch2_35}\\
             \frac{\dot{\mathcal{G}}}{\mathcal{G}H}&=&\frac{1}{u_3-1}\left[{u_1 u_3}+2(u_3-2)^{2}\right]. \label{ch2_36}
        \end{eqnarray}
    \begin{table*} [!htb]
    \centering 
    \scalebox{0.85}{
    \begin{tabular}{|*{6}{c|}}\hline
    \parbox[c][0.8cm]{0.85cm}{\textbf{C.P.}} & \textbf{$u_3$} & \textbf{$u_4$} & \textbf{$u_6$} & \textbf{$u_7$} & \textbf{Exists for} \\ [0.5ex] 
    \hline\hline 
    \parbox[c][0.8cm]{0.85cm}{$\mathcal{P}_{1}$} & 0 & 0 & 1 & 0 & always \\
     \hline
    \parbox[c][0.8cm]{0.85cm}{$\mathcal{P}_{2}$} & 0 & $\frac{1-u_6}{5}$ & $u_6$ & 0 & $3 + 2 u_6 \neq 0$, $\beta=\frac{1}{2}$ \\
     \hline
    \parbox[c][0.8cm]{0.85cm}{$\mathcal{P}_{3}$} & 2 & -2 & 0 & 0 & $-1+4\beta \neq 0$ \\
   \hline
    \parbox[c][0.8cm]{0.85cm}{$\mathcal{P}_{4}$} & 0 & $u_4$ & 0 & 0 & $-1+u_4 \neq 0, -1+2 u_4 \neq 0, \beta =\frac{-3-u_4}{8(-1+u_4)}$ \\
    \hline
    \parbox[c][0.8cm]{0.85cm}{$\mathcal{P}_{5}$} & $u_3$ & $\frac{1}{2}(-6+u_3)$ & 0 & 0 & $-1+u_3 \neq 0, -2+u_3 \neq 0, 14-12 u_3+3 u_3^2 \neq 0, \beta =0$ \\
    \hline
    \end{tabular}}
     \caption{Critical points of the dynamical system with coordinates: ($u_3, u_4, u_6, u_7$).}
    \label{CH2_TABLE-III}
    \end{table*}
    A theory specified by $\Gamma=\frac{\ddot{F}_{R}}{f_{R}H^{2}}$ is used. It can be inferred that the system can only be considered complete once it is expressed in terms of dynamical variables (\ref{ch2_22}), (\ref{ch2_23}). From equations (\ref{ch2_14}), (\ref{ch2_22}) and equation (\ref{ch2_29}), we can get
        \begin{eqnarray}
            u_3&=&2u_2, \label{ch2_37}\\
            u_5&=&\frac{u_4}{u_3-1}\left[{2u_1}+\frac{\beta-1}{u_3-1}\left[2(u_3-3)^{2}+{u_1u_3}\right]\right]. \label{ch2_38}
        \end{eqnarray}
    Using these relations and the constraint [equation (\ref{ch2_24})], the system can be reduced to a set of four equations as
        \begin{eqnarray}
            \frac{du_3}{dN}&=&u_1u_3-2u_3(u_3-2), \label{ch2_39}\\
            \frac{du_4}{dN}&=&\frac{\beta u_4}{u_3-1}\left[2(u_3-3)^{2}+{u_1u_3}\right]+u_1u_4-2u_4(u_3-2), \label{ch2_40}\\
            \frac{du_6}{dN}&=&-2u_3 u_6-u_1 u_6, \label{ch2_41}\\
            \frac{du_7}{dN}&=&-u_7(2u_3+u_1-1), \label{ch2_42}
        \end{eqnarray}
    where
        \begin{equation} \label{ch2_43}
            u_1 = \frac{-1+\frac{3}{2}u_3+u_6+u_7+u_4-2(\beta-1)\frac{(u_3-2)^{2}}{(u_3-1)^{2}}u_4}{1+\frac{u_4}{(u_3-1)}\left[2+u_3\frac{(\beta-1)}{(u_3-1)}\right]},
        \end{equation}
    and
        \begin{eqnarray} \label{ch2_omega_total}
            \omega_{\text{eff}}=-1-\frac{2\dot{H}}{3H^2}=-1-\frac{2}{3}u_3.
        \end{eqnarray}
    \begin{table} [H]
        \centering 
        \begin{tabular}{|*{6}{c|}}\hline
    \parbox[c][0.6cm]{0.85cm}{\bf{C.P.}} & \bf{$\Omega_\text{m}$} & \bf{$\Omega_\text{r}$} & \bf{$\Omega_{\text{DE}}$} & \bf{$q$} & \bf{$\omega_{\text{eff}}$} \\ [0.5ex] 
        \hline\hline 
    \parbox[c][0.6cm]{0.5cm}{$\mathcal{P}_{1}$} & 0 & 1 & 0 & 1 & $\frac{1}{3}$ \\
        \hline
    \parbox[c][0.6cm]{0.5cm}{$\mathcal{P}_{2}$} & 0 & $u_6$ & $1-u_6$ & 1 & $\frac{1}{3}$ \\
        \hline
    \parbox[c][0.6cm]{0.5cm}{$\mathcal{P}_{3}$} & 0 & 0 & 1 & -1 & $-1$ \\
        \hline
    \parbox[c][0.6cm]{0.5cm}{$\mathcal{P}_{4}$} & 0 & 0 & 1 & 1 & $\frac{1}{3}$ \\
        \hline
    \parbox[c][0.6cm]{0.5cm}{$\mathcal{P}_{5}$} & 0 & 0 & 1 & $1-u_3$ & $\frac{1}{3} (1-2u_3)$ \\
        \hline
    \end{tabular}
     \caption{Deceleration, EoS and density parameters for the critical points.} 
    \label{CH2_TABLE-IV}
    \end{table}

    \begin{table} [H]
        \centering
        \begin{tabular}{|*{2}{c|}}\hline
    \parbox[c][0.5cm]{0.85cm}{\textbf{C.P.}} & \parbox[c][1cm]{12cm}{\textbf{Eigenvalues}}\\\hline \hline
    \parbox[c][0.5cm]{0.5cm}{$\mathcal{P}_1$} & \parbox[c][1cm]{12cm}{$\big\{4, -1, 1, -4(-1+2 \beta)\big\}$}\\\hline
    \parbox[c][0.5cm]{0.5cm}{$\mathcal{P}_2$} & \parbox[c][1cm]{12cm}{$\big\{0, \frac{-5 (-1+2 u_6)}{3+2 u_6}, 1, 4\big\}$}\\\hline
    \parbox[c][0.5cm]{0.5cm}{$\mathcal{P}_3$} & \parbox[c][1cm]{12cm}{$\big\{-4, -3, \frac{3-12\beta-\sqrt{9-136 \beta +400 \beta^2}}{2(-1+4 \beta)}, \frac{3-12 \beta +\sqrt{9-136 \beta +400 \beta^2}}{2(-1+4 \beta)}\big\}$}\\\hline
    \parbox[c][0.5cm]{0.5cm}{$\mathcal{P}_4$} & \parbox[c][1cm]{12cm}{$\big\{\frac{4u_4}{(-1+u_4)(-1+2 u_4)}, \frac{-3+u_4}{-1+u_4}, \frac{2(-1+3 u_4)}{-1+u_4}, \frac{1-7 u_4+10 u_4^2}{(-1+u_4)(-1+2 u_4)} \big\}$}\\\hline
    \parbox[c][0.5cm]{0.5cm}{$\mathcal{P}_5$} & \parbox[c][1cm]{12cm}{$\big\{0, \frac{-6(1-2 u_3+u_3^2)}{14-12 u_3+3 u_3^2}, -4(-1+u_3), (5-4u_3) \big\}$}\\\hline
    \end{tabular}
    \caption{Eigenvalues for fixed points.}
    \label{CH2_TABLE-VI}
    \end{table}
    Tables \ref{CH2_TABLE-III} and \ref{CH2_TABLE-IV} show the conditions under which the critical points of these systems exist and the eigenvalues of these systems. The critical points can be calculated to analyze their features and behavior. Table \ref{CH2_TABLE-IV} represents the cosmological parameters corresponding to the critical points. Below, we will discuss the properties of each critical point and their potential connection with different evolutionary eras of the Universe, which are divided into five critical points.
\subsection{Visualization of Phase Portraits}
    For a complete understanding of the distinguishing features of each critical point, it is crucial to describe its behavior in proper diagrams. The phase portraits for each critical point are presented in this section, along with the critical steps involved in their derivation and whether they are compatible with the analysis of Tables \ref{CH2_TABLE-III} and \ref{CH2_TABLE-VI}. The properties of each of the five critical points are separately discussed, and their possible connections to the eras of the evolution of the Universe are explored.
\begin{itemize}
    \item {\bf{Point $\mathcal{P}_1$:}} In a radiation-dominated Universe, the first critical point $\mathcal{P}_1$ occurs. Table \ref{CH2_TABLE-III} shows that the critical point exists for all values of the free parameters. This critical point applies to any free model parameter based on Table \ref{CH2_TABLE-IV}, $\Omega_\text{r}=1$. The EoS parameter  $\omega_{\text{tot}}=\frac{1}{3}$ and deceleration parameter $q=1$ demonstrate that the background level does not experience late-time acceleration in this solution. Table \ref{CH2_TABLE-III} shows that our critical point is a saddle hyperbolic. Point $\mathcal{P}_1$ possesses a 2D local unstable manifold with boundaries defined only within the neighbourhood of the critical point, whereas the description of local indicates that these boundaries are determined only within the neighbourhood of the critical point.
\end{itemize}
\begin{figure} [H]
    \centering
    \includegraphics[width=58mm]{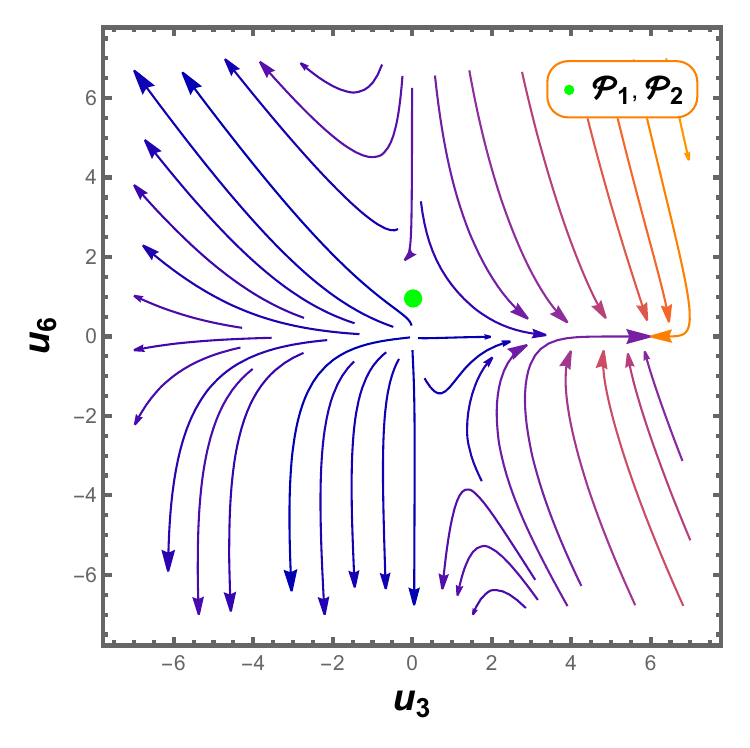}
    \includegraphics[width=58mm]{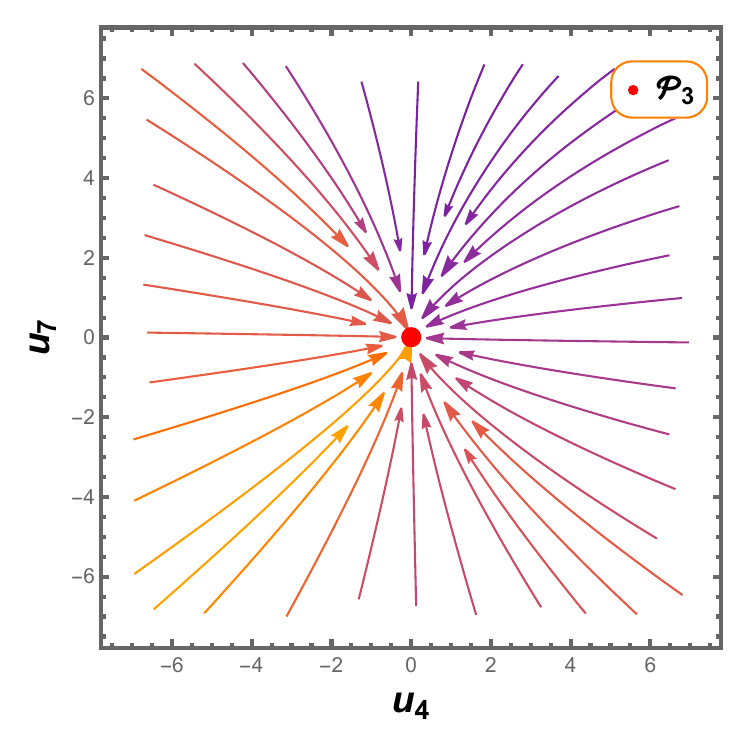}\\
    \includegraphics[width=58mm]{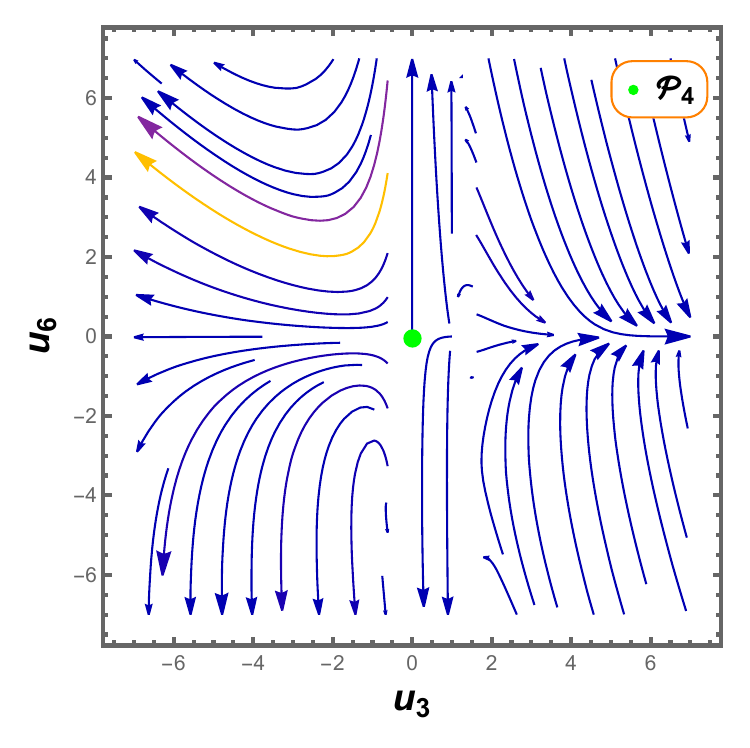}
    \includegraphics[width=58mm]{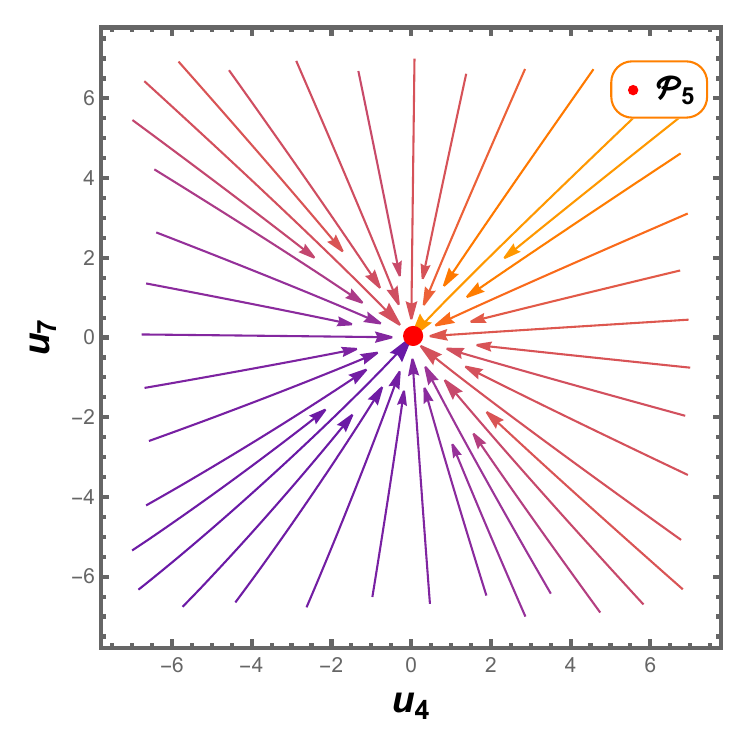}
    \caption{$2D$ phase portrait for the dynamical system.} \label{CH2_FIG5}
\end{figure}
\begin{table*} [!htb]
    \centering 
    \scalebox{0.87}{
    \begin{tabular}{|*{4}{c|}}\hline
    \parbox[c][0.6cm]{0.75cm}{\textbf{C.P.}} & \textbf{Acceleration equation} & \textbf{Phase of the Universe} & \textbf{Stability condition} \\ [0.5ex] 
    \hline\hline 
    \parbox[c][0.6cm]{0.5cm}{$\mathcal{P}_{1}$} & $\dot{H}=-2 H^2$ & $a(t)= t_{0} (2 t+c_{1})^\frac{1}{2}$ & Unstable  \\
     \hline
    \parbox[c][0.6cm]{0.5cm}{$\mathcal{P}_{2}$} & $\dot{H}=-2 H^2$ & $a(t)= t_{0} (2 t+c_{1})^\frac{1}{2}$ & Unstable  \\
     \hline
    \parbox[c][0.6cm]{0.5cm}{$\mathcal{P}_{3}$} & $\dot{H}=0$ & $a(t)=t_0 e^{c_1 t}$ & Stable \\
   \hline
    \parbox[c][0.6cm]{0.5cm}{$\mathcal{P}_{4}$} & $\dot{H}=-2 H^2$ & $a(t)= t_{0} (2 t+c_{1})^\frac{1}{2}$ & Unstable \\
    \hline
    \parbox[c][0.6cm]{0.5cm}{$\mathcal{P}_{5}$} & $\dot{H}=(-2+u_3) H^2$ & $a(t)=t_0 \left((2-u_3)t+c_1\right)^\frac{1}{2-u_3}$ & Stable \\
    \hline
    \end{tabular}}
     \caption{Acceleration equation, phase of the Universe with stability conditions.} 
    \label{CH2_TABLE-V}
\end{table*}
    \begin{itemize}
        \item {\textbf{Point $\mathcal{P}_2$:}} Table \ref{CH2_TABLE-III} shows that the second critical point $\mathcal{P}_2$ exists for $3 + 2 u_6 \neq 0$ and $\beta=\frac{1}{2}$. The Universe is in a radiation-dominated phase with $\Omega_\text{r}=u_6$, $\Omega_\text{m}=0$, and $\Omega_{\text{DE}}=1-u_6$. This is further evidenced by the EoS parameter ($\omega_{\text{tot}}$) being equal to $\frac{1}{3}$ and the deceleration parameter $q$ having a value of 1. The Jacobian matrices associated with these critical points have real positive and negative parts and zero eigenvalues, indicating that it has an unstable saddle behavior.\\
        \item {\textbf{Point $\mathcal{P}_3$:}} Under the conditions in Table \ref{CH2_TABLE-III}, this point $\mathcal{P}_3$ corresponds to a Universe dominated by DE. Since it is stable under the conditions shown, it can be considered a late-time phase of the Universe. Interestingly, under conditions with $-1+4 \beta \neq 0$, the EoS parameter ($\omega_{\text{tot}}$) equals the value of the cosmological constant $-1$ at this critical point, where $\Omega_{\text{DE}} = 1$, $\omega_{\text{tot}} = -1$. The deceleration parameter $q = -1$. Since these features are compatible with observations, they are a great advantage of the scenario under consideration; furthermore, they can only be obtained by using $f(R,\mathcal{G})$ gravity without explicitly including a cosmological constant or a canonical or phantom scalar field. It is stable when $0<\beta\leq \frac{9}{100}$. The corresponding eigenvalue is\\ $\big\{-4, -3, \frac{3-12\beta-\sqrt{9-136 \beta +400 \beta^2}}{2(-1+4 \beta)}, \frac{3-12 \beta +\sqrt{9-136 \beta +400 \beta^2}}{2(-1+4 \beta)}\big\}$.\\
        \item {\textbf{Point $\mathcal{P}_4$:}} This critical point exists in a radiation-dominated Universe for $-1 + u_4 \neq 0$, $-1 + 2 u_4 \neq 0$ and $\beta =\frac{-3 - u_4}{8(-1 + u_4)}$, leading to a decelerating phase of the Universe with an EoS parameter $\omega_{\text{tot}} =\frac{1}{3}$ and deceleration parameter $q = 1$. The corresponding density parameters are $\Omega_\text{r}=0$, $\Omega_\text{r}=0$, and $\Omega_\text{r}=1$. The eigenvalues associated with this critical point reveal positive and negative signs by taking some restrictions on $u_4$, indicating that it is an unstable node.\\
        \item {\textbf{Point $\mathcal{P}_5$:}} At late times, point $\mathcal{P}_5$ could attract the Universe due to its stability under the conditions presented in Table \ref{CH2_TABLE-III}. There are similarities between this point and $\mathcal{P}_3$, but there are differences in parameter regions. In particular, it suggests an accelerating Universe dominated by DE. The negative value of the deceleration parameter indicates the accelerating phase of the Universe, and $\omega_{\text{tot}} = -1$ behaves as a cosmological constant at this critical point, where $\Omega_{\text{DE}} = 1$ and $q = -1$ at the background. The deceleration parameter shows the accelerating behavior when $1<u_3$. It is stable when $u_3 > \frac{5}{4}$. The corresponding eigenvalue is\\ $\big\{0, \frac{- 6(1 - 2 u_3 + u_3^2)}{14 - 12 u_3 + 3 u_3^2}, -4(-1 + u_3), (5 - 4 u_3) \big\}$.
    \end{itemize}
    Table \ref{CH2_TABLE-V} summarises all the results and the corresponding scale factor.
        \begin{figure}[H]
            \centering
            \includegraphics[width=80mm]{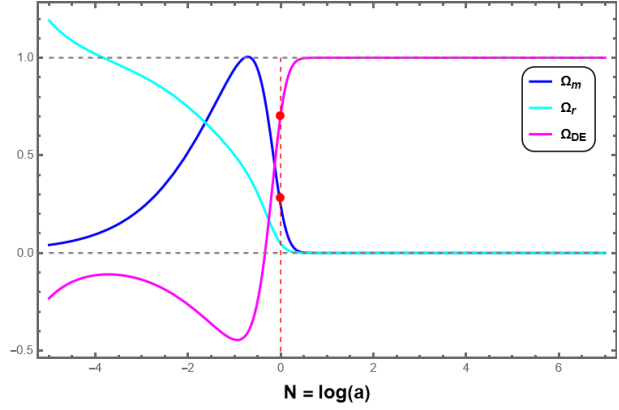}
            \caption{Evolution of density parameters DE (magenta), matter (blue) and radiation (cyan).} 
        \label{CH2_FIG6}
        \end{figure}
    Figure \ref{CH2_FIG6}  shows the cosmic evolution of the density parameter for matter, radiation, and DE for the model (\ref{ch2_14}) with the initial conditions $u_3=10^{-9.45}$, $u_4=0.01$, $u_6=1.28999$ and $u_7=0.448 \times 10^{-1.2}$. The behavior is consistent with recent cosmic observations on the evolution of density parameters. To obtain the current densities, $\Omega_\text{m} \approx 0.28$, $\Omega_{\text{DE}} \approx 0.679$, and $\Omega_\text{r} \approx 0.047$ are calculated. Radiation dominance is shown in figure \ref{CH2_FIG6} at the beginning, followed by a brief phase of matter dominance and, at the end, the de Sitter phase.

\section{Conclusion} \label{CH2_SEC-V}
    In this chapter, we have investigated the cosmological behavior of a modified Gauss--Bonnet gravity model by describing the gravitational action involving the Ricci scalar and Gauss--Bonnet invariant. We parameterized the Hubble and other geometrical parameters, constraining the coefficients using the CC sample, the extensive Pantheon$^+$, and BAO data sets. The best-fit values of these coefficients are detailed in Table \ref{CH2_TABLE I}. Additionally, we constrained the deceleration parameter and the EoS parameter, with the best-fit values presented in Table \ref{CH2_TABLE II}. Our model demonstrates a smooth transition from a decelerating phase to an accelerated expansion phase. The transition redshifts for CC, Pantheon$^+$, CC + Pantheon$^+$, and CC + Pantheon$^+$ + BAO data are $z_t=0.636$, $z_t=0.656$, $z_t=0.691$, and $z_t=0.74$, respectively. The DE EoS parameter indicates that the expansion of the Universe has accelerated, remaining within the phantom region for $z \leq -0.015$. At $z = 0$, the DE EoS parameter values are $-0.8478$, $-1.02$, $-1.224$, and $-1.47$ for CC, Pantheon$^+$, CC + Pantheon$^+$, and CC + Pantheon$^+$ + BAO data sets, aligning with recent cosmological observations. The present effective EoS ($\omega_{\text{eff}}$) parameter values are $-0.684$, $-0.686$, $-0.7$, and $-0.72$ for the respective data sets.

    In the second phase of the analysis, we conducted a dynamical system analysis focusing on the $f(R, \mathcal{G})$ function type. This analysis allowed us to examine the global behavior and stability of the cosmological model. Preliminary findings for the finite phase space of a power-law class of fourth-order gravity models $\mathcal{F}(R, \mathcal{G})=\alpha R^2 \mathcal{G}^\beta$ are presented. Equations (\ref{ch2_39})-(\ref{ch2_42}) describe the dynamical system for the mixed power law $\mathcal{F}(R, \mathcal{G})$ gravity model. Critical points and existing conditions for the model are provided in Table \ref{CH2_TABLE-III}, while Table \ref{CH2_TABLE-IV} presents values for the deceleration, EoS, and density parameters. We identified five critical points, with two ($\mathcal{P}_3, \mathcal{P}_5$) being stable and three ($\mathcal{P}_1, \mathcal{P}_2, \mathcal{P}_4$) unstable. Stable critical points emerged during the de Sitter phase, while unstable behavior was observed during the radiation-dominated phase. The eigenvalues' signature and phase-space portrait support the critical points' behavior. The trajectory behavior indicates that unstable critical points act as release points, while stable ones serve as attractor points (Figure \ref{CH2_FIG5}).

    The accelerating behavior of the model is confirmed by the EoS ($\omega_{\text{tot}}=-1$) and deceleration parameter ($q=-1$) values. The density parameters are $\Omega_\text{m} \approx 0.28$, $\Omega_{\text{DE}} \approx 0.679$, and $\Omega_\text{r} \approx 0.047$. Figure \ref{CH2_FIG6} illustrates the transition from radiation dominance to matter dominance and finally to de Sitter dominance. Using cosmological data sets, we determined the present values of matter density and DE density parameters to be $\Omega_\text{m}\approx 0.28$ and $\Omega_{\text{DE}}\approx 0.68$. Using initial conditions for the dynamical variables, the dynamical system analysis shows present values of $\approx 0.28$ for matter density and $\approx 0.679$ for DE density parameters. Both approaches confirm the alignment of density parameter values and the stable accelerating behavior of the model.


\chapter{Analyzing the Geometrical and Dynamical Parameters of Modified Teleparallel--Gauss--Bonnet Model} 

\label{Chapter3} 

\lhead{Chapter 3. \emph{Analyzing the Geometrical and Dynamical Parameters of Modified Teleparallel--Gauss--Bonnet Model}} 

\vspace{10.5 cm}
* The work in this chapter is covered by the following publication:

\textbf{Santosh V. Lohakare}, B. Mishra, S. K. Maurya, and Ksh. N. Singh, ``Analyzing the Geometrical and Dynamical Parameters of Modified Teleparallel--Gauss--Bonnet Model" \href{https://doi.org/10.1016/j.dark.2022.101164}{\color{blue}\textit{Physics of the Dark Universe} \textbf{39} (2023) 101164.}

\clearpage

\section{Introduction} 
    Late-time accelerated expansion can also be studied by modified gravitational theory without using the DE model. However, this approach is impossible in GR \cite{Lue_2006_423, Nojiri_2007_04}, and one can extend the geometrical part of the Einstein--Hilbert action to address the cosmic expansion issue. In the teleparallel formulation of gravity, higher curvature corrections can be introduced such as the Gauss--Bonnet combination $\mathcal{G}$, so the action would involve higher-torsion modifications \cite{Boulware_1985_55, Wheeler_1986_268, Antoniadis_1994_415, Nojiri_2005_71}. The torsion invariant $T_\mathcal{G}$ has been extracted without imposing the Weitzenb$\ddot{o}$ck connection, equivalent to the Gauss--Bonnet term $\mathcal{G}$ \cite{Kofinas_2014_90_084044}. This has led to another interesting class of modified theories of gravity, known as $f(T, T_\mathcal{G})$ gravity \cite{Kofinas_2014_90_084044, Kofinas_2014_90_084045}. Another modified gravity formulated with the torsion scalar $T$ is coupled with the trace of energy-momentum tensor $\mathcal{T}$. In the cosmological applications, the unified description of the inflationary phase, matter-dominated expansion, and late time acceleration can be realized \cite{Harko_2014_2014_021}. Also, the extension of $f(T)$ gravity can be obtained by including the non-minimal torsion-matter coupling in the action of $f(T)$ gravity. This has been successful in getting the DE sector of the Universe \cite{Harko_2014_89}. This study explores a gravitational action composed of the torsion scalar and the Gauss--Bonnet component, leading to the $f(T, T_\mathcal{G})$ theories. These have been extensively studied in various contexts (Ref. \cite{Kofinas_2014_90_084045, Chattopadhyay_2014_353, Capozziello_2016_76_629}), yielding exciting results on multiple scales. We focus on the cosmological dynamics of a subclass of $f(T, T_\mathcal{G})$ models chosen based on symmetry considerations. Our goal is to use late-time cosmic observations to test the viability of such a scenario and determine whether it could be a viable alternative to the standard cosmological paradigm.

    Motivated by the successful cosmological results of the extension of $f(T)$ gravity, we shall study the cosmological scenario in $f(T, T_\mathcal{G})$ gravity in this chapter. In particular, we shall focus on the behavior of the Universe at the late time of its evolution. This chapter is organized as follows: In section \ref{CH3_SEC II}, we have set up the field equations of $f(T, T_\mathcal{G})$ gravity. In section \ref{CH3_SEC III}, we have performed observational constraints using Hubble and Pantheon data. The $f(T, T_\mathcal{G})$ model is suggested to obtain the solutions to the field equations, including the behavior of cosmological parameters such as deceleration parameter and EoS parameter and also discussed energy conditions in section \ref{CH3_SEC IV}. In section \ref{CH3_SEC V}, we present the $\text{Om}(z)$ diagnostic and the age of the Universe. Finally, section \ref{CH3_SEC VI}, we have presented the results and conclusions. 
\section{\texorpdfstring{$f(T,T_\mathcal{G})$}{} Gravity  Field Equations and Dynamical Parameters} \label{CH3_SEC II}
    In $f(T, T_\mathcal{G})$ gravity, the total modified gravitational action has the following form \cite{Kofinas_2014_90_084045}
        \begin{equation}\label{CH3_eq.1}
            S=\frac{1}{2\kappa^2} \int d^{4}x\,\, e\,\, f(T, T_\mathcal{G}),
        \end{equation}
    which is based on the torsion scalar $T$ and the Gauss--Bonnet invariant term, $T_\mathcal{G}$. In torsion-based gravity, $f(T, T_{\mathcal{G}})$ gravity, the invariant term $T_\mathcal{G}$ can be  defined as,
\begin{eqnarray}\label{CH3_eq.2}
T_\mathcal{G}=\Big(\mathcal{K}^{a_1}_{~~e a} \mathcal{K}^{e a_2}_{~~b} \mathcal{K}^{a_3}_{~~f c} \mathcal{K}^{f a_4}_{~~d}-2 \mathcal{K}^{a_1 a_2	}_{~~~~a} \mathcal{K}^{a_3}_{~~e b} \mathcal{K}^{e}_{~~f c} \mathcal{K}^{f a_4}_{~~d}+2\mathcal{K}^{a_1 a_2	}_{~~~~a} \nonumber\\\times \mathcal{K}^{a_3}_{~~e b} \mathcal{K}^{e a_4}_{~~f} \mathcal{K}^{f}_{~c d}   +2\mathcal{K}^{a_1 a_2	}_{~~~~a} \mathcal{K}^{a_3}_{~~e b} \mathcal{K}^{e a_4}_{~~c,d} \Big)\delta^{a b c d}_{a_1 a_2 a_3 a_4}\, .~~~~~~~
\end{eqnarray}
The following gravitational field equations obtained by modifying the action \eqref{CH3_eq.1} about vierbein
\begin{eqnarray}\label{CH3_eq.3}
&& 2(H^{[ac]b}+H^{[ac]b})_{,c} +2(H^{[ac]b}+H^{[ba]c}-H^{[cb]a}) \mathcal{C}^{d}{}_{dc} +(2H^{[ac]d}+H^{dca}) \mathcal{C}^{b}{}_{cd} \nonumber\\ && +4 H^{[db]c} \mathcal{C}_{(dc)}{}^{a}+T^{a}{}_{cd} H^{cdb}-h^{ab} +(F-Tf_T-T_\mathcal{G} f_{T_\mathcal{G}}) \eta^{ab}=0 \, ,
\end{eqnarray}
with 
\begin{eqnarray}
H^{abc}&=&f_T (\eta^{ac} \mathcal{K}^{bd}{}_{d}-\mathcal{K}^{bca}+f_{T_\mathcal{G}}\Big[\epsilon^{cprt}(2\epsilon^{a}{}_{dkf} \mathcal{K}^{bk}{}_{p}\mathcal{K}^{d}{}_{qr} +\epsilon_{qdkf} \mathcal{K}^{ak}{}_{p} \mathcal{K}^{bd}{}_{r} \nonumber \\&&\hspace{-0.3cm} + \epsilon^{ab}{}_{kf} \mathcal{K}^{k}{}_{dp} \mathcal{K}^{d}{}_{qr}) \mathcal{K}^{qf}{}_{t} + \epsilon^{cprt} \times \epsilon^{ab}{}_{kd} \mathcal{K}^{fd}{}_{p} (\mathcal{K}^{k}{}_{fr,t}-\frac{1}{2} \mathcal{K}^{k}{}_{fq} \mathcal{C}^{q}{}_{tr}) \nonumber \\ && \hspace{-0.3cm} +\epsilon^{cprt}\epsilon^{ak}{}_{df} \mathcal{K}^{df}{}_{p} \times (\mathcal{K}^{b}{}_{kr,t}-\frac{1}{2} \mathcal{K}^{b}{}_{kq} \mathcal{C}^{q}{}_{tr})\Big]+\epsilon^{cprt} \epsilon^{a}{}_{kdf}\big[(f_{T_\mathcal{G}} \mathcal{K}^{bk}{}_{p} \mathcal{K}^{df}{}_{r})_{,t}\nonumber \\&&\hspace{-0.3cm}+f_{T_\mathcal{G}} \mathcal{C}^{q}{}_{pt} \mathcal{K}^{bk}{}_{[q} \mathcal{K}^{df}{}_{r]}\big] \, ,
\end{eqnarray}
and
\begin{equation*}
    h^{ab}=f_T \epsilon^{a}{}_{kcd} \epsilon^{bpqd} \mathcal{K}^{k}{}_{fp} \mathcal{K}^{fc}{}_{q} \, ,
\end{equation*}
where $f_T$ and $f_{T_\mathcal{G}}$ respectively denote the partial derivative with respect to the torsion scalar $T$ and Gauss--Bonnet invariant $T_\mathcal{G}$.
To derive the field equations of $f(T,T_{\mathcal{G}})$, we consider an isotropic and homogeneous FLRW space-time \eqref{ch2_flrw metric}. For such spacetime, the diagonal vierbein is,
\begin{equation}\label{CH3_eq.5}
e^{a}{}_{\mu}=\text{diag}(1,\, a(t),\,\, a(t),\,\, a(t)) \, ,
\end{equation} 
and its determinant is $e=a^3(t)$, where its dual is represented as 
\begin{equation} \label{CH3_eq.6}
e_{\mu}{}^{a}=\text{diag}(1,\, a^{-1}(t),\,\, a^{-1}(t),\,\, a^{-1}(t)) \, .
\end{equation}
Now, the torsion scalar and Gauss--Bonnet invariant term can be expressed respectively in Hubble term as $T=6H^{2}$ and $ T_{\mathcal{G}}=24H^{2}(\dot{H}+H^{2})$. In addition, we consider a matter action $S_\text{m}$, which is equivalent to an energy-momentum tensor $T^{\mu \nu}$, with a particular emphasis on the case of a perfect fluid with energy density $\rho$ and pressure $p$. 

Varying the total action $S + S_\text{m}$, the following equations are produced in FLRW geometry \cite{Kofinas_2014_90_084044, Kofinas_2014_90_084045}
\begin{eqnarray} 
f-12 H^2 f_{T} - T_{\mathcal{G}} f_{T_{\mathcal{G}}}+24 H^3 \dot{f}_{T_{\mathcal{G}}} & = & 2 \kappa^2 \rho,  \label{CH3_eq.7} \\
f-4(\dot{H}+3 H^2)f_{T}-4 H \dot{f}_{T}-T_{\mathcal{G}} f_{T_{\mathcal{G}}}+\frac{2}{3H} T_{\mathcal{G}} \dot{f}_{T_{\mathcal{G}}}+8 H^2 \ddot{f}_{T_{\mathcal{G}}} & = & -2 \kappa^2 p\, ,   \label{CH3_eq.8}
\end{eqnarray} 
with
\begin{eqnarray} \label{CH3_eq.9}
f_T \equiv \frac{\partial f(T, T_\mathcal{G})}{\partial T},\,\,\,\,\,\,\,\,\,\,\,\,\, f_{T_\mathcal{G}} \equiv \frac{\partial f(T, T_\mathcal{G})}{\partial T_\mathcal{G}} \, ,
\end{eqnarray}
For brevity, we represent $F\equiv f(T, T_{\mathcal{G}})$ and to frame the cosmological model, we calculate the pressure and energy density for a physically acceptable form of $f(T, T_\mathcal{G})$. Now, the derivative of $F$ can be obtained as,
\begin{eqnarray}
&&\hspace{-0cm}\dot{f}_T=f_{TT}\dot{T}+f_{TT_{\mathcal{G}}}\dot{T}_{\mathcal{G}} \, , \nonumber\\
&&\hspace{-0cm}\dot{f}_{T_{\mathcal{G}}}=f_{T T_\mathcal{G}}\dot{T}+f_{T_\mathcal{G} T_\mathcal{G}}\dot{T}_\mathcal{G} \, , \nonumber\\ &&\hspace{-0cm}\ddot{f}_{T_\mathcal{G}}=f_{TTT_{\mathcal{G}}}\dot{T}^2+2f_{{T T_\mathcal{G}} T_\mathcal{G}}\dot{T}\dot{T}_\mathcal{G}+f_{T_\mathcal{G} T_\mathcal{G} T_\mathcal{G}}\dot{T}_{\mathcal{G}}^2+f_{T T_{\mathcal{G}}} \ddot{T}+f_{T_\mathcal{G} T_\mathcal{G}}\ddot{T}_{\mathcal{G}} ,\nonumber   
\end{eqnarray}
where $f_{TT}$, $f_{TT_\mathcal{G}}$,...are the mathematical expressions used to indicate several partial differentiations of $f(T, T_\mathcal{G})$ over $T$, $T_\mathcal{G}$.

\section{Observational Constraints} \label{CH3_SEC III}
    We use the MCMC sampling method and the Python emcee \cite{Foreman_Mackey_2013_emcee} package to explore the parameter space. Note that the normalizing constant will not be computed to estimate the parameters. However, the prior and likelihood estimates can be used to calculate the posterior parameter distributions. In this analysis, we have used the Pantheon data set, which contains 1048 SNe Ia experiment findings from surveys such as the Low-z, SDSS, SNLS, Pan-STARRS1(PS1) Medium Deep Survey, and HST \cite{Scolnic_2018_859}, in the redshift range $z\in(0.01,2.26)$. With an emphasis on the evidence relevant to the expansion history of the Universe, such as the distance-redshift connection, two separate current observational data sets are employed to limit the model under consideration.  More importantly, new studies investigating the roles of $H(z)$ and SNe Ia data in cosmological constraints have found that both can restrict cosmic parameters. The parameters for this model are $\alpha$, $\beta$, $\zeta,$ and $\Omega_{\text{m}0}$. To determine the expansion rate $(1+z) H(z)=-\frac{dz}{dt}$, we rewrite $T$ and $T_\mathcal{G}$ in redshift parameter as,
\begin{small}
\begin{eqnarray}
 \label{CH3_eq.10}
    T=6 H_0^2 E(z),~~\,\,\,\,\,\,\,\, T_\mathcal{G}=24 H_0^2 E(z) \left(- \frac{H_0^2 (1+z) E'(z)}{2}+H_0^2 E(z) \right),~~~~ 
\end{eqnarray}
\end{small}
where $H^2 (z) = H_0^2 E(z)$ and $H_0 = 67 \pm 4~\text{km}\,\text{s}^{-1} \text{Mpc}^{-1}$\cite{Yu_2018_856} be the late-time Hubble parameter and the prime denotes the derivative to the redshift parameter. In addition, we have considered $H_0=70.7$ $\text{km s}^{-1} \text{Mpc}^{-1}$ for our analysis. We use the following functional form for $E(z)$ [Sahni et al.\cite{Sahni_2003_77_201}],
\begin{equation}\label{CH3_eq.11}
    E(z)=\Omega_{\text{m}0} (1+z)^3 + \zeta (1+z)^2 + \beta (1+z)+ \alpha,
\end{equation}
    where the constants $\Omega_{\text{m}0}$, $\zeta$, $\beta$, and $\alpha$ are determined by fitting the experimental data and their measurements. Additionally, the restriction $E(z=0)=1$ constrains the relationship between these coefficients, as $\alpha + \beta + \zeta =1-\Omega_{\text{m}0}$.
\subsection{Hubble Data}
    To recreate the cosmological models, we employed the parametrization approach. Some interesting works have utilized the parametrization approach to investigate cosmological models \cite{Saini_2000_85, Capozziello_2005_71}. The real benefit of using this approach is that observational data may be used to evaluate cosmological theories. However, it is challenging to depict the precise history of $H$ due to the integration in its formula $H(z)$. As a result, the $H(z)$ data can indicate the fine structure of the expansion history of the Universe. Sharov et al. \cite{Sharov_2018_01} present the entire list of data sets. We estimate the model parameters using the $\chi^2$ test using MCMC simulation. We use 55 Hubble data points, as listed in the \hyperref[Appendices]{Appendices}, to compute the chi-square value $\chi_{\text{Hubble}}^2$ for the observational Hubble parameter data.
\subsection{Pantheon Data}
    The chi-square function for a sample of 1048 SNe Ia from the Pantheon study is used to compare theoretical and observed distance moduli, where the standard error in the observed Hubble parameter is denoted by $\sigma_{i}^{'}$ \cite{Scolnic_2018_859}. The theoretical distance modulus $\mu_{\text{th}}$ is defined by $\mu_{\text{th}}^i=\mu(D_{L})=m-M=5 log_{10}D_L(z)+\mu_0$, where the apparent and absolute magnitudes are denoted by $m$ and $M$, respectively, and $\mu_0$ is specified as $\mu_0=5log(\frac{H_0^{-1}}{Mpc})+25$. The formula for luminosity distance $D_L$ is given by $D_L(z)=(1+z) H_0 \int \frac{1}{H(z^*)} dz^*$. The series of $H(z)$ is constrained to the tenth term and approximately integrates the constrained series to get the luminosity distance.
\begin{figure} [!htb]
\centering
\includegraphics[width=115mm]{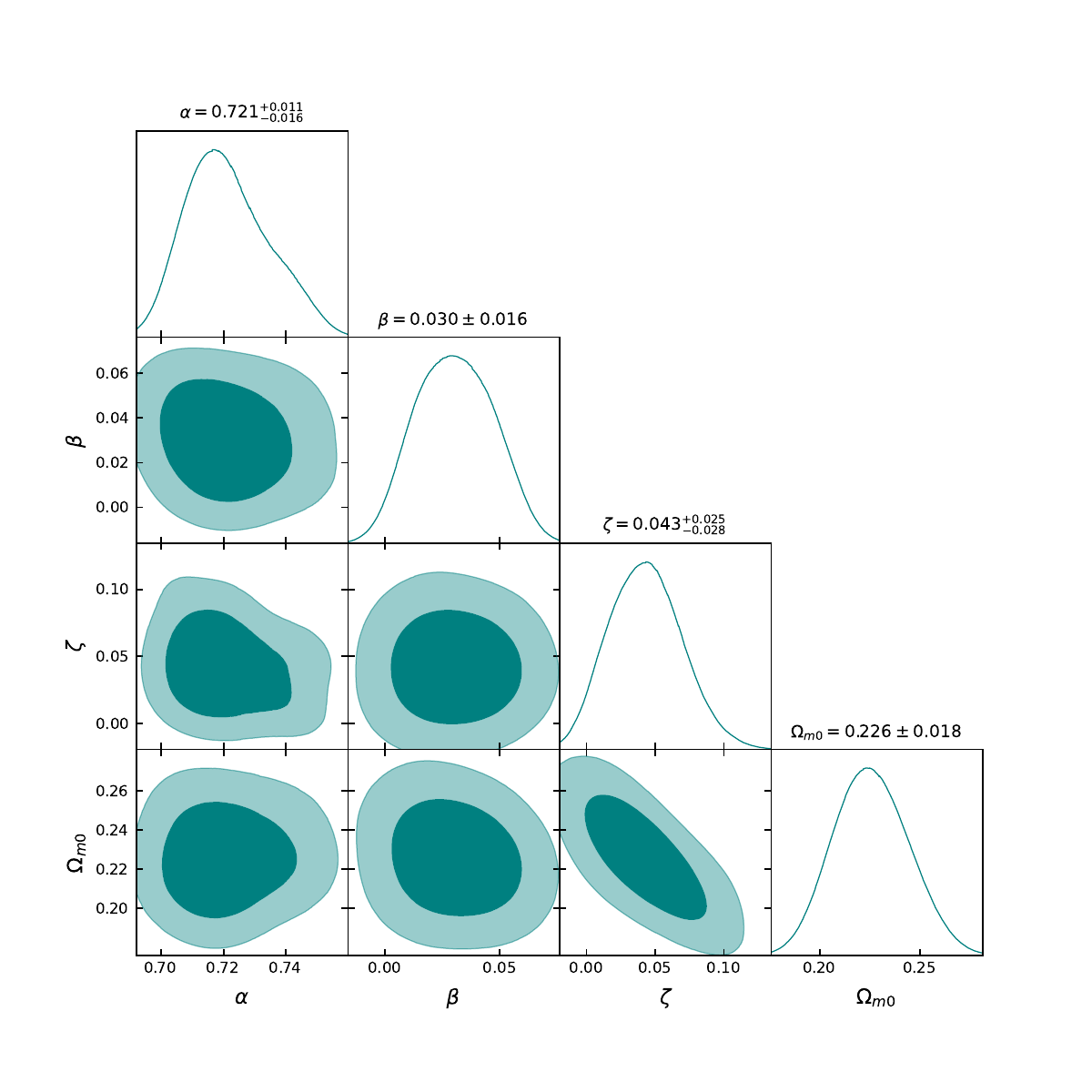}
\caption{The contour plots with $1\sigma$ and $2\sigma$ errors for the parameters $\alpha$, $\beta$, $\zeta$, and $\Omega_{\text{m}0}$. Additionally, it contains the parameter values that better match the 55-point Hubble data set defined in Table \ref{CH3_table: Table II}.}
\label{CH3_FIG2}
\end{figure}
\begin{figure} [!htb]
\centering
\includegraphics[width=115mm]{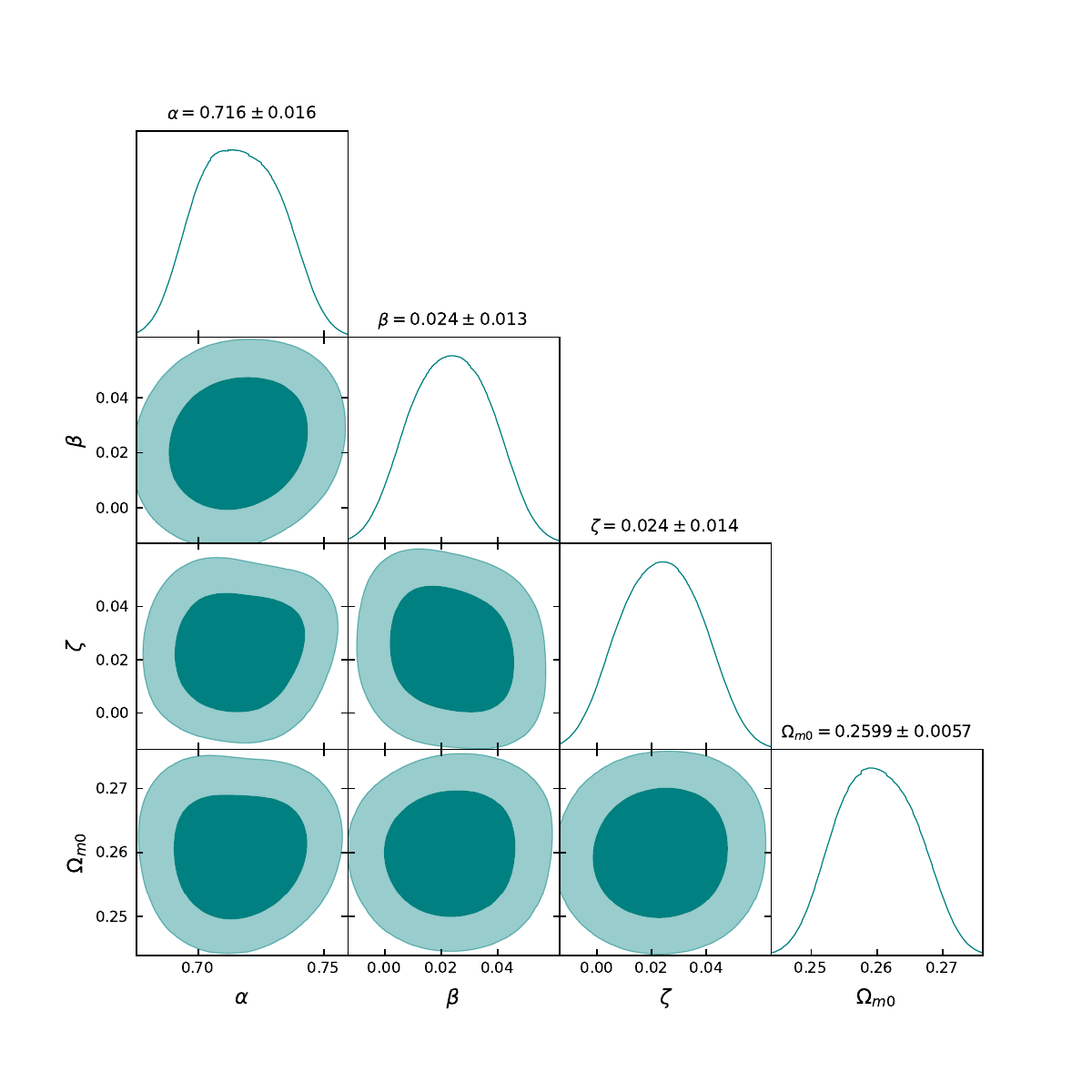}
\caption{The contour plots with $1\sigma$ and $2\sigma$ errors for the parameters $\alpha$, $\beta$, $\zeta$, and $\Omega_{\text{m}0}$. It also contains parameter values that better match the 1048-point Pantheon sample.}
\label{CH3_FIG3}
\end{figure}

\begin{figure} [!htb]
\centering
\includegraphics[width=8.9cm,height=6cm]{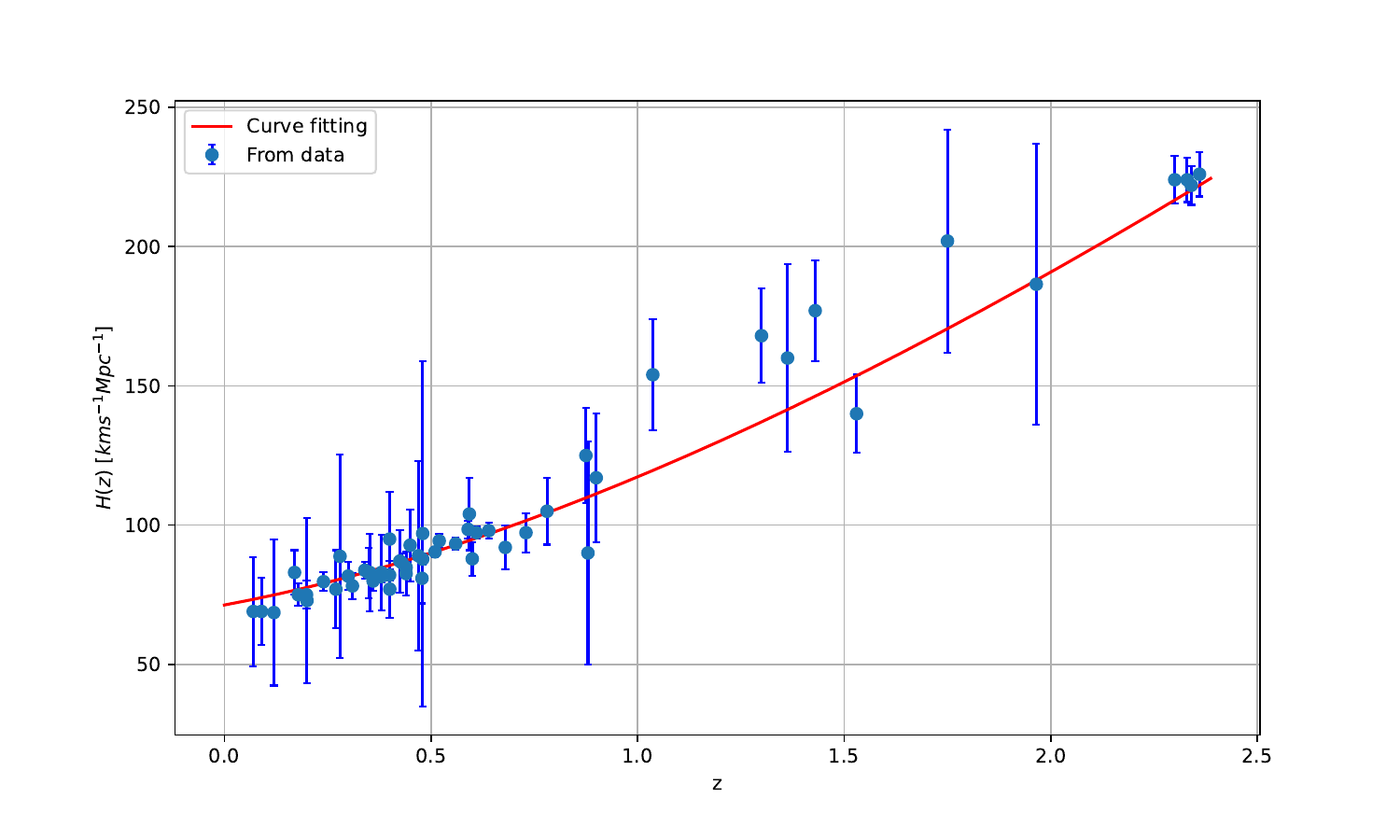} \includegraphics[width=8.9cm,height=6cm]{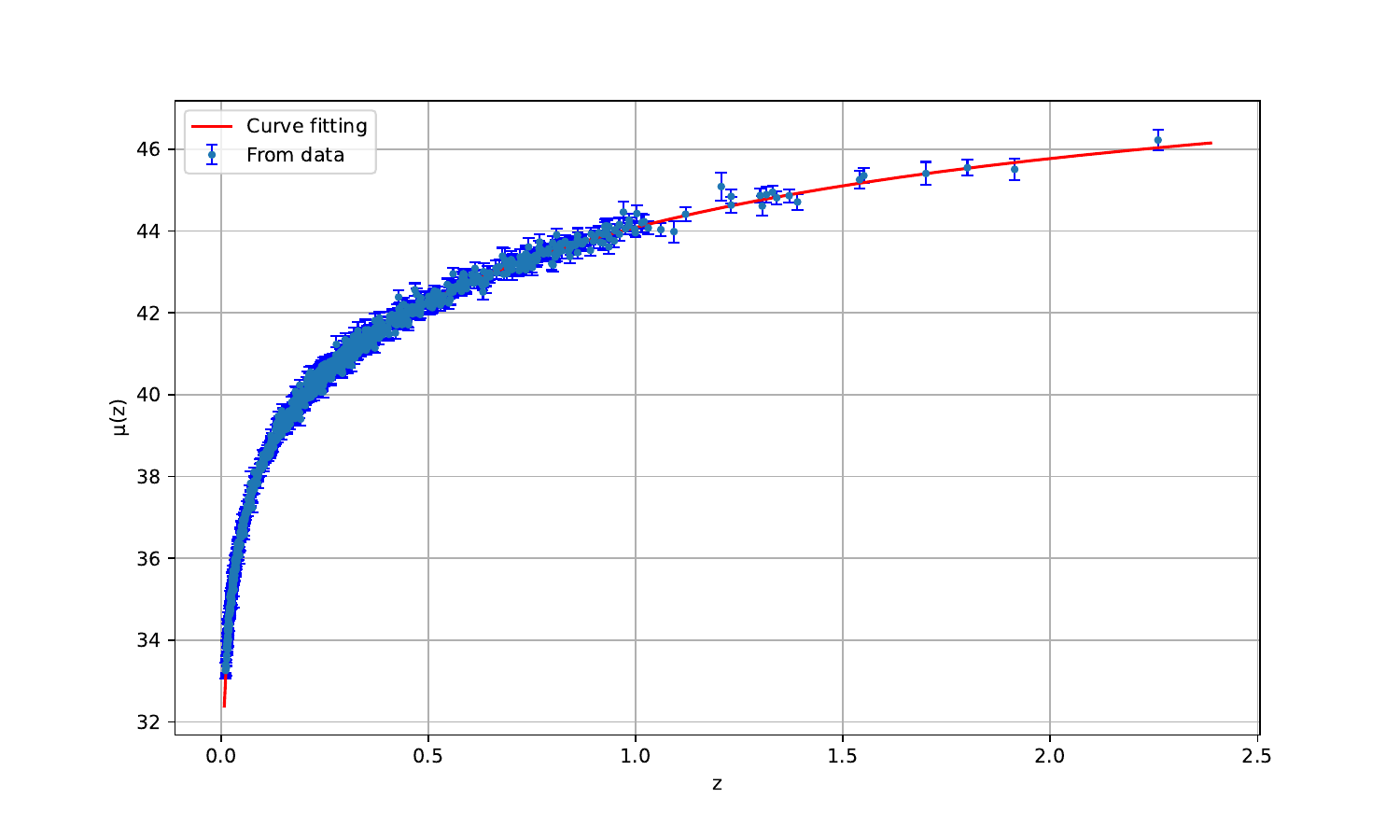}
\caption{Graphical behavior of $H(z)$ and $\mu (z)$ for Hubble and Pantheon data sets with its error bars.}
\label{CH3_FIG1}
\end{figure}

Figure \ref{CH3_FIG1} shows graphical behavior of the provided model (solid red line) has a better fit to the $H(z)$ data sets for $\alpha=0.721$, $\beta=0.030$, $\zeta=0.043$ and $\Omega_{\text{m}0}=0.226$, which is shown in the upper panel plot along with the 55 points of the $H(z)$ data sets (blue dots) and accompanying error bars (see Table \ref{CH3_table: Table II}). In the lower panel, The red line is the plot of our model's distance modulus $\mu(z)$ vs $z$, which exhibits a better fit to the 1048 points of the Pantheon data sets along with its error bars for $\alpha=0.716$, $\beta=0.024$, $\zeta=0.024$ and $\Omega_{\text{m}0}=0.2599$. In figure \ref{CH3_FIG2} and figure \ref{CH3_FIG3}, we see the marginalized distribution for the parameters $\alpha$, $\beta$, $\zeta$ and $\Omega_{\text{m}0}$ which has been displayed with the triangle plots. The contour shows where the $1\sigma$ and $2\sigma$ confidence intervals are located.

\begin{table} [!htb]
\centering 
\begin{tabular}{|c|c|c|} 
\hline 
Coefficients & Hubble data set  & Pantheon data set\\ [0.5ex] 
\hline \hline
$\alpha$ & $0.721^{+0.011}_{-0.016}$ & 0.716 $\pm$ 0.016 \\
\hline
$\beta$ & 0.030 $\pm$ 0.016 & 0.024 $\pm$ 0.013 \\
\hline
$\zeta$ & $0.043_{-0.028}^{+0.025}$ & 0.024 $\pm$ 0.014 \\
\hline
$\Omega_{\text{m}0}$ & 0.226 $\pm$ 0.018  & 0.2599$\pm$0.0057 \\[0.5ex] 
\hline 
\end{tabular}
\caption{Constrained values of $H(z)$ model parameters based on the Hubble and Pantheon data sets.} 
\label{CH3_table:nonlin} 
\label{CH3_TABLE I}
\end{table}
Figure \ref{CH3_FIG2} and figure \ref{CH3_FIG3}  exhibit the $1\sigma$ and $2\sigma$ confidence regions that have been illustrated in our constraint findings. These are retrieved by the respective contour analyses of $\chi^2$ in the parameter space. Table \ref{CH3_TABLE I} also summarizes the best-fit value of the parameters and their associated uncertainty.

\section{The Functional \texorpdfstring{$f(T,T_\mathcal{G})$}{}} \label{CH3_SEC IV}
    The above analysis requires the specification of the $f(T, T_\mathcal{G})$ form. The corrections of $T$-powers are included first in conventional $f(T)$ gravity. However, in the present context, $T_\mathcal{G}$ is in the same order as $T^2$ because it includes quartic torsion components. Because $T$ and $\sqrt{T^2 +\lambda_2 T_\mathcal{G}}$ have the same order, both should be employed in a modified theory. As a result, the most fundamental non-trivial model, which is distinct from GR and does not introduce a new mass scale into the problem, is $f(T, T_\mathcal{G})=-T + \lambda_1 \sqrt{T^2+\lambda_2 T_\mathcal{G}}$. The couplings $\lambda_1, \lambda_2$ are dimensionless, and the model is predicted to be essential in late times. Although straightforward, this model can produce remarkable cosmic behavior demonstrating the advantages, possibilities, and novel aspects of $f(T, T_\mathcal{G})$ cosmology. We note here that this scenario simplifies to TEGR, or GR, with simply a rescaled Newton's constant, whose dynamical analysis has been carried out in detail in the literature \cite{Copeland_1998_57, Ferreira_1997_79, Chen_2009_2009_04} when $\lambda_2 = 0$. Therefore, in the following sections, we restricted our study to the condition $\lambda_2 \neq 0$.

    Now plugging the Hubble parameter $H(z)$ and $f(T, T_\mathcal{G})$ model into the field equations \eqref{CH3_eq.7} and \eqref{CH3_eq.8} the following set of field equations are obtained, 
\begin{eqnarray}
\rho&=& -\frac{T}{2}+\frac{\lambda_1  \left(2 T^2+\lambda_2 T_\mathcal{G} \right)}{4 \sqrt{T^2+\lambda_2 T_\mathcal{G}}}+H^2 \left(6-\frac{6 \lambda_1  T}{\sqrt{T^2+\lambda_2 T_\mathcal{G}}}\right)-\frac{3 \lambda_1  \lambda_2  H^3 \left(2 T \dot{T}+\lambda_2  \dot{T}_\mathcal{G}\right)}{\left(T^2+\lambda_2 T_\mathcal{G}\right)^{3/2}} \label{CH3_eq.15}\\
p&=& \frac{T}{2}-\frac{\lambda_1  \sqrt{T^2+\lambda_2 T_\mathcal{G}}}{2}+\frac{\lambda_1  \lambda_2 T_\mathcal{G}}{4 \sqrt{T^2+\lambda_2 T_\mathcal{G}}}+2 \left(\dot{H}+3 H^2\right) \left(\frac{\lambda_1  T}{\sqrt{T^2+\lambda_2 T_\mathcal{G}}}-1\right)\nonumber\\& &-\frac{\lambda_1  \lambda_2  H^2 \left(3 \left(2 T \dot{T}+\lambda_2  \dot{T}_\mathcal{G}\right)^2-2 \left(T^2+\lambda_2 T_\mathcal{G}\right) \left(2 T \ddot{T}+2 \dot{T}^2+\lambda_2  \ddot{T}_\mathcal{G}\right)\right)}{2 \left(T^2+\lambda_2 T_\mathcal{G}\right)^{5/2}}\nonumber\\& &-\frac{ \lambda_1  \lambda_2  H \left(T \dot{T}_\mathcal{G}-2 T_\mathcal{G} \dot{T}\right)}{\left(T^2+\lambda_2  T_\mathcal{G}\right)^{3/2}}+\frac{\lambda_1 \lambda_2 T_\mathcal{G} \left(2 T \dot{T}+\lambda_2 \dot{T}_\mathcal{G}\right)}{12 H \left(T^2+\lambda_2 T_\mathcal{G}\right)^{3/2}}  \, , \label{CH3_eq.16}
\end{eqnarray}
where, $T=6 H^2 ,\hspace{0.2cm} T_{\mathcal{G}}=24H^{2}(\dot{H}+H^{2}),\hspace{0.2cm} \dot{T}=12 H \dot{H}, \hspace{0.2cm} \dot{T}_\mathcal{G}= 24 H \left(H \ddot{H}+2 \dot{H} (\dot{H}+2 H^2)\right), \hspace{0.2cm} \ddot{T}=12 H \ddot{H}+12 \dot{H}^2,\\ \ddot{T}_\mathcal{G}= 24 \left(4 H^3 \ddot{H}+2 \dot{H}^3+H^2 (\dot{\ddot{H}}+12 \dot{H}^2)+6 H \dot{H} \ddot{H}\right)$ and by using $(1+z) H(z)=-\frac{dz}{dt}$, we obtained $\dot{H}, \ddot{H}$ and $\dot{\ddot{H}}$ over redshift as 
    \begin{eqnarray}
&&\hspace{-0.3cm} \dot{H}=-\frac{1}{2} H_0^2 (z+1) \left(\beta+z (2 \zeta+3 \Omega_{\text{m}0} z + 6 \Omega_{\text{m}0})+2 \zeta+3 \Omega_{\text{m}0}\right) \, ,\\
&&\hspace{-0.3cm} \ddot{H}= \frac{1}{2} H_0^3 (z+1) \left(\beta+z (4 \zeta+9 \Omega_{\text{m}0} z+18 \Omega_{\text{m}0})+4 \zeta+9 \Omega_{\text{m}0}\right) \nonumber\\&& \hspace{0.6cm} \times \sqrt{\alpha+\beta (z+1)+\zeta (z+1)^2+\Omega_{\text{m}0}(z+1)^3},\\
&&\hspace{-0.3cm} \dot{\ddot{H}}=-\frac{1}{4} H_0^4 (z+1) \Bigg[z \Big(z \Big\{6 \Omega_{\text{m}0} (9 \alpha+34 \beta+100 \zeta)+z \Big[68 \beta \Omega_{\text{m}0}+24 \zeta^2+400 \zeta \Omega_{\text{m}0} +810 \Omega_{\text{m}0}^2 \nonumber \\&&
  \hspace{0.6cm} +\Omega_{\text{m}0} z (100 \zeta+81 \Omega_{\text{m}0} z+405 \Omega_{\text{m}0}) \Big]+24 \zeta (\beta+3 \zeta)+810 \Omega_{\text{m}0}^2 \Big\}+4 \Omega_{\text{m}0} (27 \alpha+51 \beta+100 \zeta)\nonumber \\ && \hspace{0.6cm} +16 \alpha \zeta+3 \beta^2 +48 \beta \zeta +72 \zeta^2 +405 \Omega_{\text{m}0}^2 \Big)+2 \alpha (\beta+8 \zeta+27 \Omega_{\text{m}0})+3 \beta^2+24 \beta \zeta+68 \beta \Omega_{\text{m}0}  \nonumber\\&&
  \hspace{0.6cm} + 24 \zeta^2+100 \zeta \Omega_{\text{m}0}+81 \Omega_{\text{m}0}^2 \Bigg]\, .
\end{eqnarray}
\begin{figure} [!htp]
\centering
\minipage{1\textwidth}
\includegraphics[width=\textwidth]{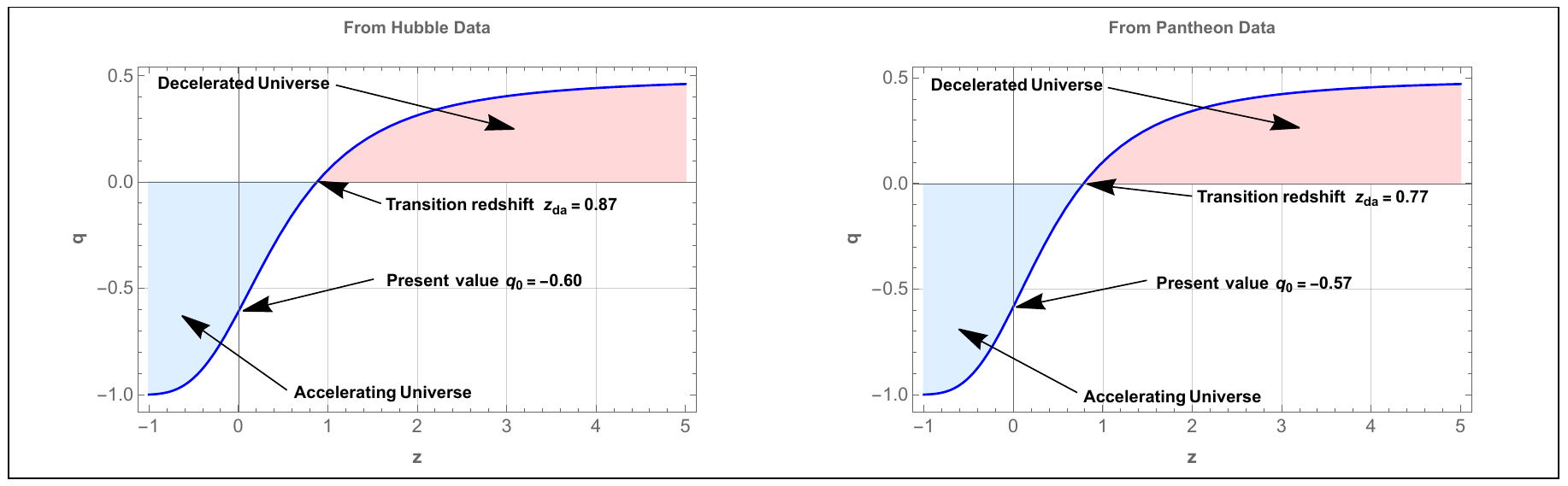}
\endminipage
\caption{Graphical behavior of the deceleration parameter versus redshift with the constraint values of the coefficients obtained from figure \ref{CH3_FIG2} and \ref{CH3_FIG3} (The parameter scheme: Mean of parameter values).}
\label{CH3_FIG4}
\end{figure}

\subsection{Deceleration and Equation of State Parameter}
The deceleration parameter $q=-1-\frac{\dot{H}}{H^2}$ is a function of the Hubble parameter that describes the rate of acceleration of the Universe. For positive $q$, the Universe is in a decelerated phase; for negative $q$, the accelerated phase can be realized. The model parameters $\alpha, \beta, \zeta,$ and $\Omega_{\text{m}0}$ are used to calculate the deceleration parameter $q$. The graph shown in figure \ref{CH3_FIG4} explains the expansion from the past to the present by depicting how $q$ behaves for redshift $z$. In figure \ref{CH3_FIG2} and figure \ref{CH3_FIG3}, the restricted values of model parameters from the examined Hubble and Pantheon data sets cause $q$ to transit from positive in the past, indicating early deceleration to negative in the present, indicating current acceleration. The deceleration parameter $q_0=-0.60$, $q_0=-0.57$ for Hubble and Pantheon data respectively, at the current cosmic epoch, is relatively consistent with the range $q_0=-0.528^{+0.092}_{- 0.088}$ as determined by a recent observation \cite{Christine_2014_89}.

The Universe makes a smooth transition from a decelerated phase of expansion to an accelerated phase in our derived model, with a deceleration-acceleration redshift of $z_{\text{da}} = 0.87$, $z_{\text{da}}=0.77$ for Hubble and Pantheon data respectively shown in figure \ref{CH3_FIG4}. The recovered value of the deceleration-acceleration redshift $z_{\text{da}}=0.82 \pm 0.08$ is consistent with certain current constraints, based on 11 $H(z)$ observations made by Busca et al. \cite{Busca_2013_552} between redshifts $0.2 \leq z \leq 2.3$, $z_{\text{da}}=0.74\pm 0.05$ of Farooq et al. \cite{Farooq_2013_766}, $z_{\text{da}}=0.69^{+23}_{-12}$ of Lu et al. \cite{Lu_2011_699}, $z_{\text{da}}=0.7679^{+0.1831}_{-0.1829}$ of Capozziello et al. \cite{Capozziello_2014_90_044016}, and $z_{\text{da}}=0.60^{+0.21}_{-0.12}$ of Yang et al. \cite{Yang_2020_2020_059}.

The kinematic variables are significant in the analysis of cosmological models. The deceleration parameter, for instance, defines the behavior of the Universe, including whether it is always decelerating, constantly accelerating, has single or several transition phases, etc. The EoS parameter similarly defines the physical significance of energy sources in the evolution of the Universe. The EoS parameter $(\omega)$ is,
\begin{equation}\label{CH3_eq.18}
\omega= \frac{p}{\rho}.
\end{equation}
In the dust phase, the EoS parameter, $\omega= 0$, whereas in the radiation-dominated phase, $\omega = \frac{1}{3}$. The vacuum energy or the $\Lambda$CDM model is represented by $\omega=-1$. In addition, for the accelerating phase of the Universe, e.g. in the quintessence phase $(-1 < \omega < 0)$ and in phantom regime $(\omega < -1)$.
\begin{figure} [!htp]
\centering
\minipage{1\textwidth}
\includegraphics[width=\textwidth]{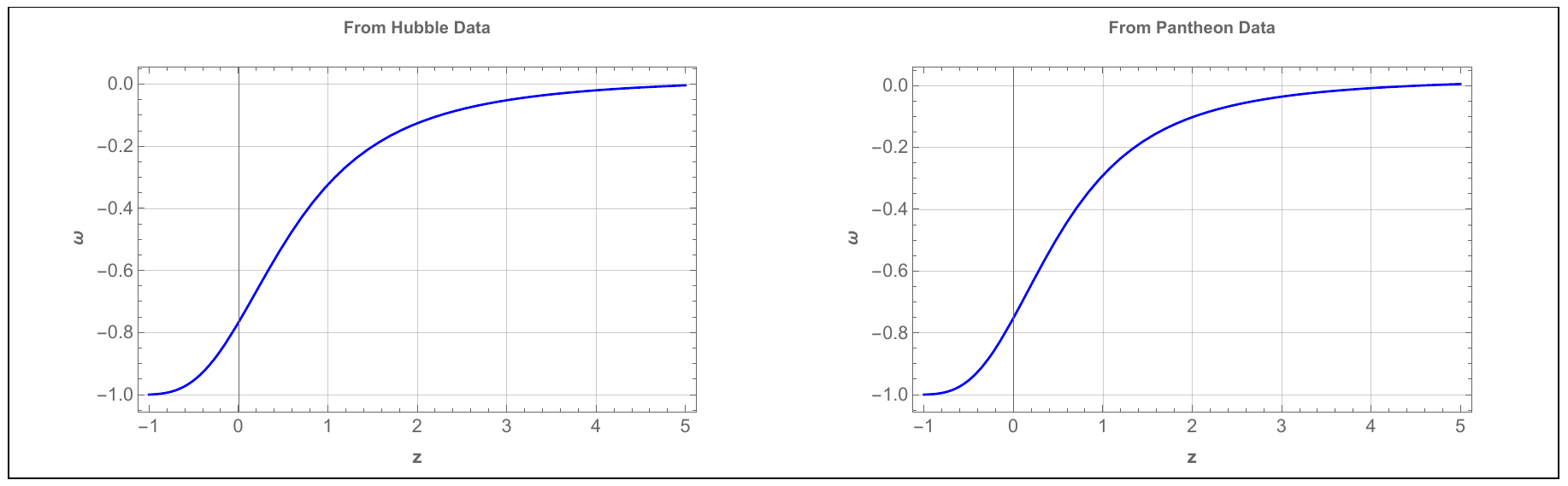}
\endminipage
\caption{Graphical behavior of the EoS parameter versus redshift with the constraint values of the coefficients obtained from figure \ref{CH3_FIG2} and \ref{CH3_FIG3} (The parameter scheme: Mean of parameter values).}
\label{CH3_FIG5}
\end{figure}
    We may visualize the variations in EoS of DE equation \eqref{CH3_eq.18} in terms of the redshift variable by calculating the associated energy density and pressure of DE, as shown in figure \ref{CH3_FIG5}. This diagram represents the quintessence-like behavior and its approach to $-1$ at late times so that the current value of EoS $(z = 0)$ equals $-0.77$, $-0.755$ for Hubble and Pantheon data, respectively, for the values of model parameter $\lambda_1=0.3, \lambda_2=0.36$. As a result, we conclude that the Universe is expanding faster, which is compatible with the cosmological data provided by Amanullah et al. \cite{Amanullah_2010_716}.

    The Pantheon study constrained the parameter space, a newly proposed observational data set. The $2\sigma$ limitations for the parameters in our study are $\alpha =0.716 \pm 0.016, \beta =0.024 \pm 0.013$, $\zeta=0.024 \pm 0.014$ and $\Omega_{\text{m}0} = 0.2599 \pm 0.0057$. The 1048 Pantheon samples and our model taking into account $H_0 = 70.7$ kms$^{-1}$ Mpc$^{-1}$, show a good fit to the observational findings in the error bar plot. Valentino et al. \cite{Valentino_2016_761} have performed the combined analysis of the Planck and $R16$ results in an extended parameter space. In place of the usual six cosmological parameters, twelve parameters were simultaneously varying and obtained the phantom-like DE component, with $\omega=-1.29_{-0.12}^{+0.15}$ at $68\%$ of confidence interval Some other experiments on this parameter suggests the range for EoS parameter as $\omega\approx -1.3$ \cite{Vagnozzi_2020_102}. In addition, Efstathiou and Gratton \cite{Efstathiou_2020_496} have obtained the range of the curvature density parameter $\Omega_k=0.0004\pm0.0018$, which is in agreement with the Planck 2018 result. Further Vagnozzi et al. \cite{Vagnozzi_2021_908} obtained $\Omega_k=0.0054\pm0.0055$, which is consistent with the spatially flat Universe by combining Planck 2018 CMB temperature and polarization data with the latest CC measurements. 
\subsection{Energy Conditions}
    The Raychaudhuri equation, a fundamental tool in gravitational theory, plays a crucial role in defining energy conditions. It provides a robust framework for analyzing the nature of attractive gravity and its implications in spacetime geometry. The Raychaudhuri equation indicates that \cite{Santos_2007_76, Kar_2007_69},
        \begin{equation} \label{CH3_eq.19}
            \frac{d\theta}{d\tau}=-\frac{1}{2} \theta^2-\sigma_{ab} \sigma^{ab}+w_{ab} w^{ab}-R_{ab} k^a k^b,
        \end{equation}
    where the expansion scalar is $\theta$, the shear and vorticity tensors are $\sigma_{ab}$ and $w_{ab}$, respectively. Also, $k^a$ is a null vector field. The Raychaudhuri equation avoids any reference to gravitational field equations, which is essential to establish. Instead, it is viewed as a purely geometric statement. If we consider any orthogonal congruence hypersurface $(w_{ab}=0)$. Then, as a result of $\frac{d\theta}{d\tau}<0$, we can formulate the criteria for attractive gravity as $R_{ab} k^a k^b \geq 0$ because the shear tensor's spatial nature is $\sigma^2 =\sigma^{ab} \sigma_{a b}\geq 0$. The previous condition, known as the null energy condition, can be written in terms of the stress-energy tensor in the context of Einstein's relativistic field equations as $\mathcal{T}_{ab}k^a k^b \geq 0$, where $k^a$ is any null vector. More precisely, the weak energy condition indicates that $\mathcal{T}_{ab}u^a u^b \geq 0$, where $u^a$ denotes the time-like vector and assumes a positive local energy density.
\begin{figure} [!htb]
\centering
\minipage{1\textwidth}
\includegraphics[width=\textwidth]{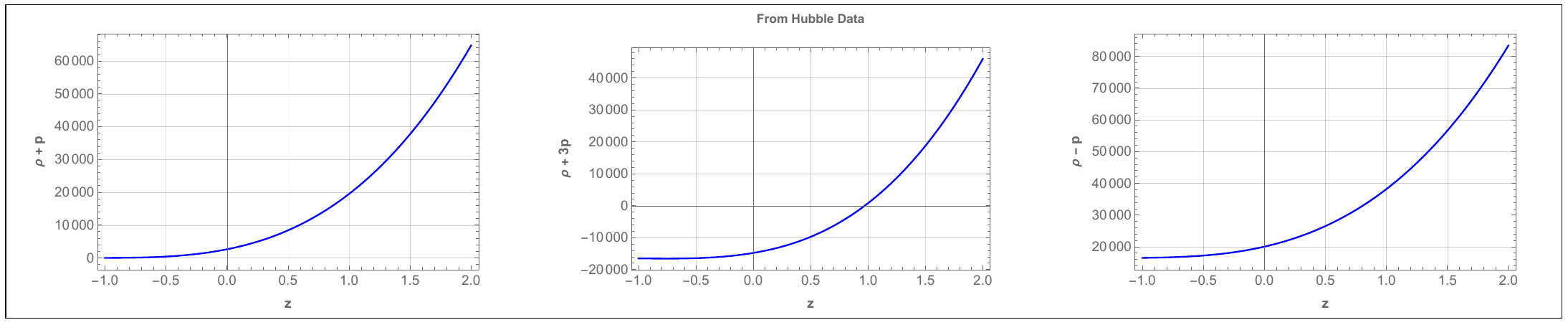}
\endminipage
\caption{Graphical behavior of the energy conditions versus redshift with the constraint values of the coefficients obtained from figure \ref{CH3_FIG2} and \ref{CH3_FIG3} (The parameter scheme: Mean of parameter values).}
\label{CH3_FIG6}
\end{figure}
\begin{figure} [!htb]
\centering
\minipage{1\textwidth}
\includegraphics[width=\textwidth]{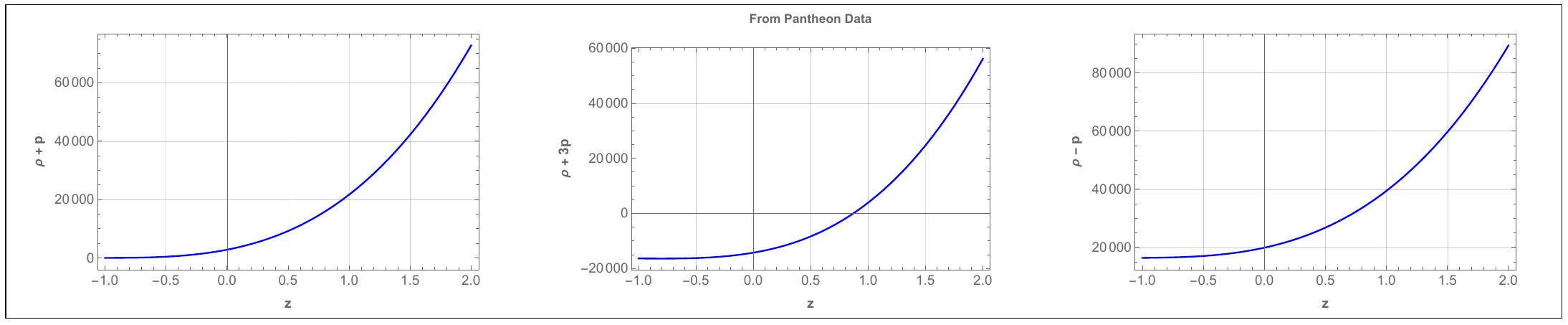}
\endminipage
\caption{Graphical behavior of the energy conditions versus redshift.}
\label{CH3_FIG7}
\end{figure}
The energy conditions are essentially boundary conditions for maintaining a positive energy density \cite{Hawking_1973_book, Poisson_2004_book}. Hence, we present here, Null Energy Condition (NEC): $ \rho+p \geq 0$, Weak Energy Condition (WEC): $\rho \geq 0$ and $\rho + p \geq 0$, Strong Energy Condition (SEC): $ \rho+3p \geq 0$ and $ \rho+p \geq 0$, Dominant Energy Condition (DEC): $\rho \geq 0$ and $\rho \pm p \geq 0$. The NEC violation suggests that none of the energy conditions specified are valid. The SEC is now the topic of significant discussion because of the current accelerated expansion of the Universe \cite{Barcelo_2002_11}. SEC must be violated in cosmological situations throughout the inflationary expansion and now \cite{Visser_1997_56}. The graph of the energy conditions is shown in figure \ref{CH3_FIG6} and \ref{CH3_FIG7}. We check if the NEC and DEC hold, but the SEC violates the model, which directly points to the accelerated expansion of the Universe.
\begin{figure} [!htb]
\centering
\minipage{1\textwidth}
\includegraphics[width=\textwidth]{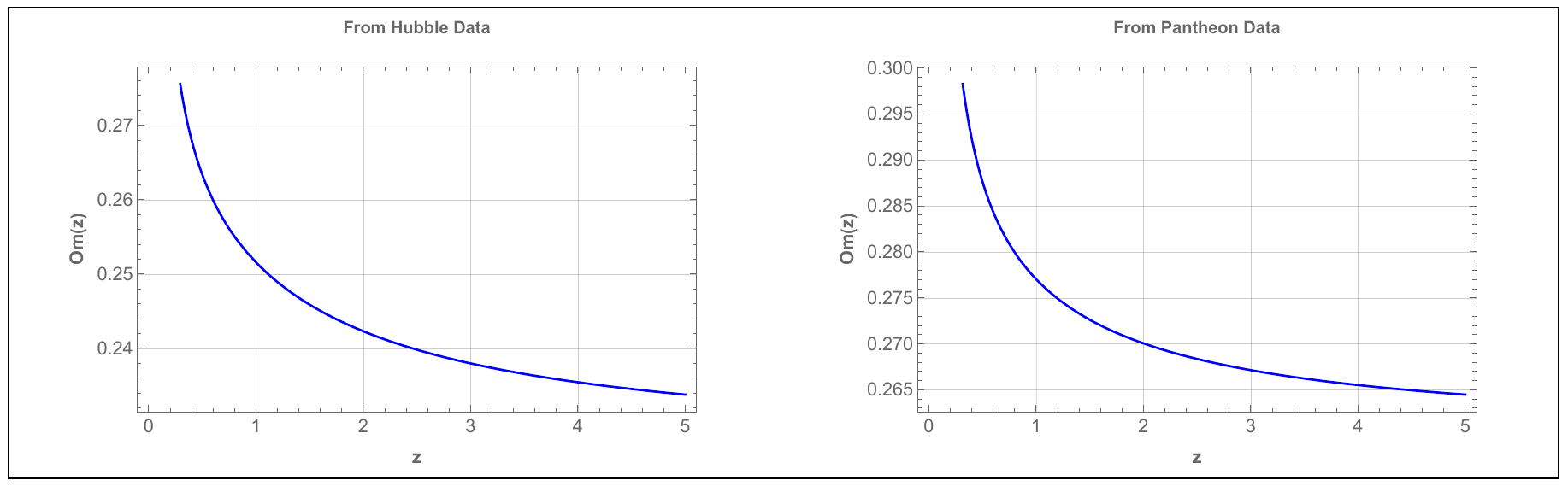}
\endminipage
\caption{Graphical behavior of the $\text{Om}(z)$ versus redshift with the constraint values of the coefficients obtained from figure \ref{CH3_FIG2} and \ref{CH3_FIG3} (The parameter scheme: Mean of parameter values).}
\label{CH3_FIG8}
\end{figure}
Figure \ref{CH3_FIG6} and \ref{CH3_FIG7} illustrate that the WEC is positive from the early time to the late time phase. Since our model exhibits quintessential behavior, we can predict how satisfied DEC and NEC are at the late stages of evolution. At the same time, the SEC started a violation from $z \approx 0.972$, $z \approx 0.879$, and was previously satisfied for both data sets. Simultaneously, the SEC was violated at the late time from $(z \approx 0.9)$ and satisfied at the early time. In particular, a detailed analysis of these energy conditions may be accomplished when the cosmic dynamics are fixed up by a calculated or assumed Hubble rate.

\section{\texorpdfstring{$\text{Om}(z)$}{} Diagnostic and Age of the Universe} \label{CH3_SEC V}
In this section, we are interested in how the model responds to the $\text{Om}(z)$ diagnostic. For some DE theories, the $\text{Om}(z)$ parameter is considered another effective diagnostic tool \cite{Sahni_2008_78_103502, Sahni_2014_L40} and which is defined as,
\begin{equation} \label{CH3_eq.20}
\text{Om}(z)=\frac{E(z)-1}{(1+z)^3-1} \, ,
\end{equation}
where, $E(z)=\frac{H^2(z)}{H^2_0}$ is dimensionless parameter and $H_0$ is the Hubble rate of the present epoch. The two-point difference diagnostic is
\begin{equation}\label{CH3_eq.21}
    \text{Om}(z_1, z_2)= \text{Om}(z_1) - \text{Om}(z_2) \, ,
\end{equation}
Alternatively for quintessence, $\text{Om}(z_1,z_2) > 0$, while for phantom  $\text{Om}(z_1,z_2)<0$, ($z_1 < z_2$). For the $\Lambda$CDM model, the $\text{\text{Om}}(z)$ diagnostic provides a null test \cite{Sahni_2008_78_103502}, and more data was subsequently gained on its sensitivity with the EoS parameter \cite{Ding_2015_803_L22, Zheng_2016_825_17, Qi_2018_18_066}. The DE concept will form a cosmological constant if $\text{Om}(z)$ is constant for the redshift. The slope of $\text{Om}(z)$, which is positive for the emerging $\text{Om}(z)$ and denotes phantom phase $(\omega<-1)$ and negative for quintessence region $(\omega > -1)$ also identifies the DE models.

The reconstructed $\text{Om}(z)$ parameter for the best-fit data is displayed in figure \ref{CH3_FIG8} as a function of redshift. Over redshift, it has been observed that the $\text{Om}(z)$ parameter decreases.
\begin{figure} [!htb]
\centering
\minipage{1\textwidth}
\includegraphics[width=\textwidth]{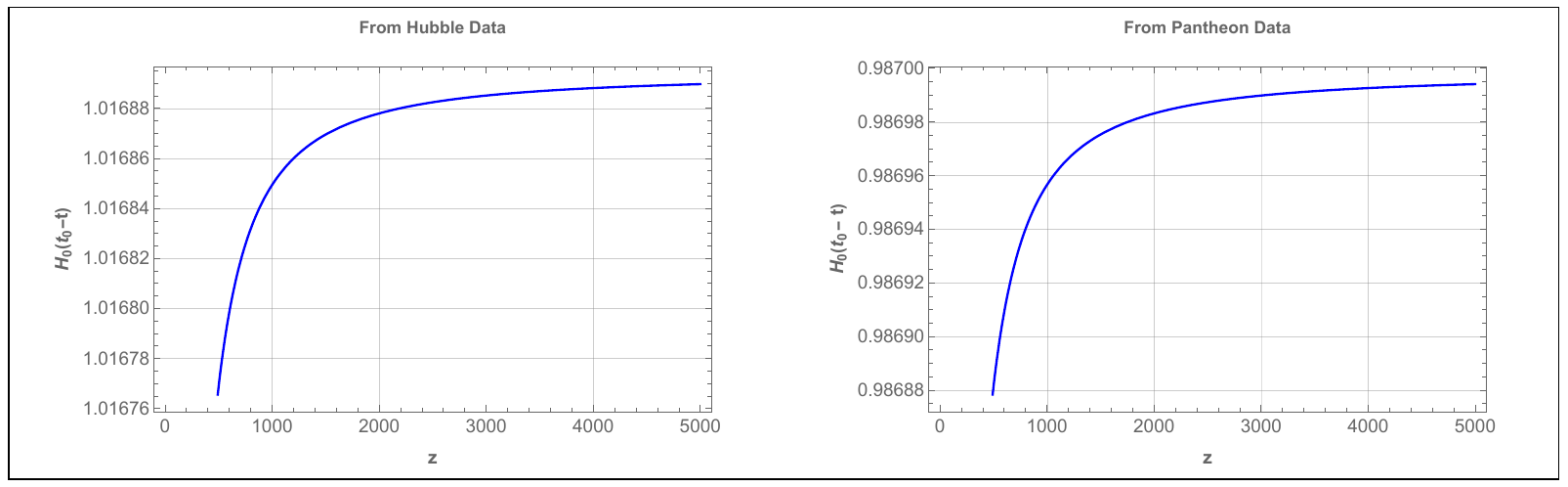}
\endminipage
\caption{Graphical behavior of time versus redshift with the constraint values of the coefficients obtained from figure \ref{CH3_FIG2} and \ref{CH3_FIG3} (The parameter scheme: Mean of parameter values).}
\label{CH3_FIG9}
\end{figure}
By figuring out the ages of the oldest objects in our galaxy, one can directly estimate the minimum age of the Universe. These are the stars in the Milky Way's galaxy that are metal-poor. The age of the Universe is computed as,
\begin{equation} \label{CH3_eq.22}
H_0 (t_0-t)=\int_{0}^{z} \frac{dx}{(1+x) E(x)}\,\,\, , \hspace{0.8cm} E(z)=\frac{H^2(z)}{H^2_0},
\end{equation}
where 
\begin{equation*}
H_0 t_0=\lim_{z\rightarrow \infty} \int_{0}^{z} \frac{dx}{(1+x) E(x)} \, .
\end{equation*}
    We may deduce from this straightforward observation that $1/H_0$ should indicate the current age of the Universe, possibly up to a multiplicative factor extremely near to one. The Universe is $13.8$ billion years old according to observations of the cosmic background radiation \cite{Gribbin_2015_book}. Figure \ref{CH3_FIG9} depicts the time behavior with a redshift. It is found that $H_0(t_0-t)$ converges to $1.01689$ and $0.9870$ for Hubble and Pantheon data, respectively, for infinitely large $z$. This translates to $t_0 = 1.01689 H^{-1}_0 \approx 14.01$ Gyrs and $t_0 = 0.987 H^{-1}_0 \approx 13.607$ Gyrs, which is the current age of the Universe and is very near to the age of the Universe calculated from Planck's findings, $t_0 = 13.786\pm 0.020$ Gyrs \cite{Aghanim_2020_641}.  It is well known that the age of the Universe at any redshift is inversely proportional to $H_0$. This requires the Universe to be older than the oldest objects it contains at any redshift, which will provide an upper limit on $H_0$. Assuming the $\Lambda$CDM model at late times, Vagnozzi et al. \cite{Vagnozzi_2022_36}  obtained the $95$ percent confidence level upper limit as, $H_0 < 73.2$ $\text{Km} \, \text{s}^{-1} \, \text{Mpc}^{-1}$.
\section{Conclusion} \label{CH3_SEC VI}
    In this chapter, we presented a class of modified $f(T, T_{\mathcal{G}})$ gravity models using cosmological data sets. We began by describing the fundamental features of a gravitational action that combines the torsion scalar and the Gauss--Bonnet invariant. The chosen function, $f(T, T_\mathcal{G}) = -T + \lambda_1 \sqrt{T^2 + \lambda_2 T_\mathcal{G}}$, simplifies to GR as the constant $\lambda_2$ approaches zero. This model is based on a well-motivated Hubble parameter within the $f(T, T_{\mathcal{G}})$ gravity framework. Using the parametrization method, we discussed the null, strong, weak, and dominant energy conditions for $f(T, T_\mathcal{G})$ gravity models. The coefficients for the Hubble parameter were constrained using the Hubble data set and the Pantheon SNe Ia data set.

    Energy conditions are crucial for evaluating the self-consistency of cosmological models. They help determine whether a novel cosmological model complies with the spacetime causal and geodesic structure. We outlined the major points of our work, including the testing of our cosmological solutions in section \ref{CH3_SEC III}. Table \ref{CH3_TABLE I} displays the values for the model parameters that best-fit with the data. According to these constrained values, the deceleration parameter $q$ indicates that the Universe transitions smoothly from a decelerated phase of expansion to an accelerated phase in our derived model, with a deceleration-acceleration redshift of $z_{\text{da}} = 0.87$ and $z_{\text{da}} = 0.77$ for the Hubble and Pantheon data, respectively. The EoS parameter suggests that the expansion of the Universe is accelerating, as it lies in the quintessence region. For the Hubble data and Pantheon samples, the EoS parameter at $z=0$ is $\omega_0=-0.77$ and $\omega_0=-0.755$, respectively.

    The determined values of cosmological parameters and their behavior indicate that the model is stable with the Hubble and Pantheon data sets and is a feasible method for understanding the late-time acceleration of the Universe in $f(T, T_\mathcal{G})$ gravity. The extracted value of the deceleration-acceleration redshift is consistent with current constraints. We examined the specific physical properties of the model and the evolution of physical parameters in combination with the energy conditions. It was observed that the NEC and DEC do not violate the model, but the SEC fails to fulfil it, producing a repulsive force and leading the Universe to jerk. As noted in \cite{Visser_1997_56}, the SEC violation in figure \ref{CH3_FIG6} and \ref{CH3_FIG7} demonstrates the viability of our model. The $\text{Om}(z)$ parameter reconstruction for the $f(T, T_\mathcal{G})$ model shows that it varies between positive prior values and high positive values at present. The behavior of the $\text{Om}(z)$ diagnostic indicates that our model potentially aligns with a quintessence-like evolution. Furthermore, we analyzed the variation of cosmic time with redshift, as illustrated in figure \ref{CH3_FIG9}. It was discovered that $H_0(t_0-t)$ converges to $1.01689$ and $0.9870$ for the Hubble and Pantheon data, respectively, for infinitely large $z$. This allows us to determine the age of the Universe at present as $t_0 = 1.01689 H_0^{-1} \approx 14.01$ Gyrs and $t_0 = 0.9870 H_0^{-1} \approx 13.607$ Gyrs, which is remarkably comparable to the age calculated using the Planck findings $t_0 = 13.786 \pm 0.020$ Gyrs. As a result, the model demonstrates the consistency of the accelerating evolutionary behavior of the Universe with the available data sets.

\chapter{Stability of \texorpdfstring{$f(Q, B)$}{} gravity via Dynamical System Approach: a Comprehensive Bayesian Statistical Analysis} 

\label{Chapter4} 

\lhead{Chapter 4. \emph{Stability of \texorpdfstring{$f(Q, B)$}{} gravity via Dynamical System Approach: a Comprehensive Bayesian Statistical Analysis}} 
\vspace{10 cm}
* The work in this chapter is covered by the following publication: 

\textbf{Santosh V. Lohakare} and B. Mishra, ``Stability of $f(Q, B)$ gravity via Dynamical System Approach: a Comprehensive Bayesian Statistical Analysis", \href{https://doi.org/10.3847/1538-4357/ad9602}{\color{blue}\textit{{The Astrophysical Journal}} \textbf{978} (2025) 26.}

\clearpage
  
\section{Introduction} \label{ch4_Sec 1}
    Novel classes of modified gravity theories have emerged, incorporating curvature, torsion, and nonmetricity scalars. These classes arise even though the fundamental theories are mathematically equivalent at the equation level. The key lies in the difference between the torsion scalar $T$ and the nonmetricity scalar $Q$, which deviates from the usual Levi--Civita Ricci scalar $R$ of GR due to additional terms: $R = - T + B$ and $R = Q + B$, respectively; where $B$ is the boundary term. The objects framed in GR can be identified by an over-circle symbol. A geometric trinity of gravity of second-order can be observed in $R$, $B-T$, $Q+B$, whereas $f(R)$, $f(B-T)$, $f(Q+B)$ can be regarded as a geometric trinity of gravity of fourth-order \cite{Jimenez_2019_5, Capozziello_2023_83, Capozziello_2022_82}. Consequently, arbitrary functions $f(R)$, $f(T)$, and $f(Q)$ no longer share a total derivative relationship. Furthermore, scalar fields can be introduced within this framework, leading to theories of scalar-tensor \cite{Felice_2011_84, Horndeski_1974_10}, scalar torsion \cite{Bahamonde_2019_100_064018, Bahamonde_2023_107, Geng_2011_704}, and scalar nonmetricity \cite{Runkla_2018_98, Jarv_2018_97}, each offering intriguing possibilities. Recently, Heisenberg \cite{Heisenberg_2023_1066_Review} reviewed various cosmological models in $f(Q)$ gravity. Considering energy conditions, Banerjee et al. \cite{Banerjee_2021_81} investigated wormhole geometry in $f(Q)$ gravity. Several $f(Q)$ parameterizations have been analyzed, including observational constraints and investigating compact objects beyond the standard maximum mass limit \cite{Maurya_2023_269, Maurya_2022_2022_003, Lohakare_2023_526, Maurya_2022_70}. In addition, Boehmer et al. \cite{Boehmer_2023_2303.04463}, Palianthanasis \cite{Palianthanasis_2024_43, Paliathanasis_2021_2402.02919} and Khyllep et al. \cite{Khyllep_2023_107} presented a dynamical system analysis in $f(Q)$ gravity with perturbations.

    In cosmology, the CC measurement is utilized to determine the age and expansion rate of the Universe. The CC technique consists of three primary components: (i) defining a sample of optimal CC tracers, (ii) determining the differential age, and (iii) assessing systematic effects \cite{Moresco_2022_25}. The Hubble parameter $H(z)$ is essential in determining the energy content of the Universe and its acceleration mechanism. The $H(z)$ estimation is mainly carried out at $z = 0$. Still, there are other methods to determine $H(z)$, such as the detection of BAO signal in the clustering of galaxies and quasars and analyzing SNe Ia observation \cite{Riess_2021_908, Font_Ribera_2014_2014_027, Raichoor_2020_500, Hou_2020_500, Riess_2018_853}. Pantheon$^+$ is an analysis that expands the original Pantheon framework to combine an even larger number of SNe Ia samples to understand the complete expansion history. In this study, we used the observational Hubble data (CC sample), Pantheon$^+$, and BAO data sets to investigate the expansion history of the Universe and the behavior of other geometrical parameters.
    
    This study investigates a specific subclass of the $f(Q, B)$ model to assess its potential as an alternative to the conventional cosmological framework. We have developed a numerical approach to predict the redshift behavior of the Hubble expansion rate. Our findings indicate that while the model can replicate the low-redshift behavior of the standard $\Lambda$CDM model, it exhibits notable differences at high redshifts. The $f(Q,B)$ model emerges as a viable candidate for explaining the current epochs and effectively captures the evolution of energy components over cosmic time, thereby supporting its validity as an alternative explanation for the observed acceleration of the Universe. We examined the background cosmological dynamics of the selected model and evaluated its feasibility using Bayesian analysis, supported by MCMC methods, applied to late-time cosmic observations, including Pantheon$^+$, CC and BAO data sets. Additionally, we introduced a dynamical system analysis to assess the stability of the model. A significant outcome of our analysis is the identification of a stable critical point within the dynamical system using center manifold theory. This critical point corresponds to the de Sitter phase, a well-established cosmological epoch characterized by accelerated expansion. The stability of this critical point suggests that, given certain initial conditions, the Universe will inevitably move towards and remain within the de Sitter phase. This finding aligns with current observations suggesting a late-time Universe dominated by DE and undergoing accelerated expansion.
    
    In teleparallel gravity, the boundary term $B$ can be incorporated into the Lagrangian, resulting in $f(T, B)$ theories that exhibit rich phenomenology \cite{Bahamonde_2015_92}. However, within the framework of nonmetricity gravity, the Lagrangian of symmetric teleparallel gravity does not account for the role of $B$. This has led to the development of the $f(Q, B)$ theory, which is currently of significant interest to cosmologists \cite{Capozziello_2023_83, De_2024_2024_50}. Our study explores the concept of an accelerating Universe by introducing a novel and straightforward parametrization for the Hubble parameter. This chapter is divided into five sections. Section \ref{ch4_Sec 2} presents the geometrical framework of symmetric teleparallelism, also formulating $f(Q,B)$ gravity and extracting the general metric and affine connection field equations. In section \ref{ch4_Sec 3}, we apply this formulation to a cosmological setup, resulting in $f(Q, B)$ cosmology with observational data sets. Building on the model presented in section \ref{ch4_Sec 4}, we performed a dynamical system analysis to investigate its long-term behavior and identify any stable or unstable states. Finally, section \ref{ch4_Sec 5} concludes the chapter with the results and discussions.

\section{Symmetric Teleparallel Gravity} \label{ch4_Sec 2}
        We examine a gravitational model defined by the four-dimensional metric tensor $g_{\mu\nu}$ and the covariant derivative $\nabla_{\mu}~$, which is constructed using the generic connection $\tilde{\Gamma}^\zeta{}_{\mu\nu}$. Within the framework of symmetric teleparallel GR, the connection $\tilde{\Gamma}^\zeta{}_{\mu\nu}$ is both flat and torsionless. Consequently, this results in $R^\zeta{}_{\eta\mu\nu} = 0$ and $\mathrm{T}^\eta{}_{\mu\nu}=0$. Furthermore, it retains the symmetries of the metric tensor $g_{\mu\nu}$. In teleparalllel theories the autoparallel \cite{Obukhov_2021_104} are defined as 
    \begin{equation}
        \frac{d^{2}x^{\mu}}{ds^{2}}+\tilde{\Gamma}^\mu{}_{\zeta\nu} \frac{dx^{\zeta}}{ds}\frac{dx^{\nu}}{ds}=0. \label{ch4_eq: 1}
    \end{equation}
    The Riemann tensor can be defined for the general connection
    \begin{equation}
        R^\zeta{}_{\eta\mu\nu} = \frac{\partial\tilde{\Gamma}^\zeta{}_{\eta\nu}%
        }{\partial x^{\mu}}-\frac{\partial \tilde{\Gamma}^\zeta{}_{\eta\mu}}{\partial x^{\nu}}+\tilde{\Gamma}^\sigma{}_{\eta\nu} \tilde{\Gamma}^\zeta{}_{\mu\sigma}%
        -\tilde{\Gamma}^\sigma{}_{\eta\mu} \tilde{\Gamma}^\zeta{}_{\nu\sigma}, \label{ch4_eq: 2}
    \end{equation}
        the torsion tensor
    \begin{equation}
    \mathrm{T}^\eta{}_{\mu\nu}=\tilde{\Gamma}^\eta{}_{\mu\nu}-\tilde{\Gamma}^\eta{}_{\nu\mu}, \label{ch4_eq: 3}
    \end{equation}
        and the nonmetricity tensor 
    \begin{eqnarray}
        Q_{\eta\mu\nu}= \nabla_{\eta} \ g_{\mu\nu}=\frac{\partial g_{\mu\nu}%
        }{\partial x^{\eta}}-\tilde{\Gamma}^\sigma{}_{\eta\mu} \ g_{\sigma\nu}%
        -\tilde{\Gamma}^\sigma{}_{\eta\nu} \ g_{\mu\sigma}. \label{ch4_eq: 4}
    \end{eqnarray}         
        In symmetric teleparallel theory, we can always choose a suitable diffeomorphism that vanishes the general affine connection $\tilde{\Gamma}^\zeta{}_{\mu\nu}$, known as the coincident gauge \cite{Jimenez_2018_98}. As a consequence, the covariant derivative reduces to the partial derivative, and the symmetric teleparallel postulates (i.e., vanishing curvature and torsion) enforce that the general connection becomes the Levi--Civita connection, which is symmetric by construction. In TEGR, instead of setting the curvature to zero like in symmetric teleparallel gravity, we set the Levi--Civita connection to zero and introduce the Weitzenb\"{o}ck connection, which has torsion but no curvature. Consequently, this leads to the vanishing of the Riemann curvature tensor $R^\zeta{}_{\eta\mu\nu} = 0$ and the nonmetricity tensor $Q_{\eta\mu\nu}=0$ ensuring a purely torsional spacetime in TEGR. In this context, the torsion scalar $T$ becomes the fundamental geometric object in teleparallel gravity.

        As a result the nonmetricity scalar $Q$ defined in \cite{Nester_1999} is introduced as
    \begin{equation}
        Q=Q_{\eta\mu\nu}P^{\eta\mu\nu} \, . \label{ch4_eq: 5}
    \end{equation}
        This statement represents the fundamental geometric quantity of gravity. The nonmetricity conjugate $P_{\;\mu\nu}^{\eta}$ is defined as
    \begin{equation}
        P_{\,\,\,\mu\nu}^{\eta}=-\frac{1}{4}Q_{\,\,\,\mu\nu}^{\eta}+\frac{1}{2}        Q_{(\mu\phantom{\eta}\nu)}^{\phantom{(\mu}\eta\phantom{\nu)}}+\frac
        {1}{4}\left(  Q^{\eta}-\tilde{Q}^{\eta}\right)  g_{\mu\nu}-\frac{1}
        {4}\delta_{\;(\mu}^{\eta}Q_{\nu)}, \label{ch4_eq: 6}
\end{equation}
        the traces $Q_{\mu}=Q_{\mu\nu}^{\phantom{\mu\nu}\nu}$ and $\tilde{Q}_{\mu}=Q_{\phantom{\nu}\mu\nu}^{\nu\phantom{\mu}\phantom{\mu}}$ are used in this context. 
        The boundary term is defined as
\begin{equation}
        B = R - Q = -\nabla_\mu(Q^\mu-\tilde Q^\mu) = -\frac1{\sqrt{-g}}\partial_\mu\left[\sqrt{-g}(Q^\mu-\tilde Q^\mu)\right].
\end{equation}
        The Ricci scalar $R$ corresponds to the Levi--Civita connection $\Gamma^\zeta{}_{\mu\nu}$ of the metric tensor $g_{\mu\nu}$. The nonmetricity scalar $Q$ for a symmetric and flat connection differs from $R$ by a boundary term $B$, which is defined as $B = R - Q$.

        The gravitational action integral for STEGR is expressed as follows
    \begin{equation}
        \int d^{4}x\sqrt{-g}Q\simeq\int d^{4}x\sqrt{-g}R - B \, , \label{ch4_eq: 7}
    \end{equation}
        this implies that symmetric teleparallel GR is dynamically equivalent to GR.
        However, the equivalence is lost when nonlinear components of the nonmetricity scalar $Q$ are introduced as in $f(Q)$-gravity in the gravitational action. Moreover, the corresponding gravitational theory has no longer dynamical equivalence with GR or its generalization, $f(R)$-gravity.
        The action integral for symmetric teleparallel $f(Q)$-gravity \cite{Jimenez_2018_98, Jimenez_2020_101} is expressed as follows
    \begin{equation}
        S_{f\left(  Q\right)  }=\int d^{4}x\sqrt{-g}f(Q). \label{ch4_eq: 8}
    \end{equation}
\subsection{\texorpdfstring{$f(Q,B)$}{} Cosmology} \label{ch4_Sec 2a}
        A recent extension of the $f(Q, B)$ theory \cite{Capozziello_2023_83, De_2024_2024_50, Palianthanasis_2024_43} incorporates a boundary term into the gravitational action integral. This generalization includes the gravitational action integral in the following manner
    \begin{equation} \label{ch4_eq: 9}
        S = \int d^{4}x \ \sqrt{-g} \left[\frac{1}{2 \kappa^2} f(Q, B) \right] \, , 
    \end{equation}

        To construct a realistic cosmological model, we consider a matter action $S_\text{m}$, associated with the energy-momentum tensor $\Theta_{\mu\nu}$. As shown in \cite{De_2024_2024_50}, varying the total action $S + S_\text{m}$ leads to the following Friedmann equations
    \begin{eqnarray}
        \kappa^2 T_{\mu\nu}&=&-\frac f2g_{\mu\nu}
        +\frac2{\sqrt{-g}}\partial_\eta \left(\sqrt{-g}f_Q P^\eta{}_{\mu\nu}\right)
        +(P_{\mu\alpha\beta}Q_\nu{}^{\alpha\beta}-2P_{\alpha\beta\nu}Q^{\alpha\beta}{}_\mu)
        f_Q
        \nonumber \\ & &+\left(\frac B2 g_{\mu\nu}- \nabla_{\mu} \nabla_{\nu}
    +g_{\mu\nu} \nabla^\alpha \nabla_\alpha-2P^\eta{}_{\mu\nu}\partial_\eta \right)f_B\,,
        \label{ch4_eqn:FE1-pre}
    \end{eqnarray}
        This can be expressed in a covariant manner
    \begin{eqnarray}
        \kappa^2 T_{\mu\nu}=-\frac f2g_{\mu\nu}+2P^\eta{}_{\mu\nu} \nabla_\eta(f_Q-f_B)
        + \left( G_{\mu\nu}+\frac Q2g_{\mu\nu}\right)f_Q
        +\left(\frac B2g_{\mu\nu}-\nabla_{\mu} \nabla_{\nu} +g_{\mu\nu} \nabla^\alpha \nabla_\alpha \right)f_B\,.\nonumber \\ \label{ch4_eq: 11}
    \end{eqnarray}
        A definition of the effective stress-energy tensor is as follows
    \begin{eqnarray} \label{ch4_T^eff}
        T^{\text{eff}}_{\mu\nu} =  T_{\mu\nu}+ \frac 1{\kappa^2}\left[\frac f2g_{\mu\nu}-2P^\eta{}_{\mu\nu} \nabla_\eta(f_Q-f_B)
        -\frac {Qf_Q}2g_{\mu\nu}
        -\left(\frac B2g_{\mu\nu}- \nabla_{\mu} \nabla_{\nu} +g_{\mu\nu} \nabla^\alpha \nabla_\alpha \right)f_B\right] \, , \nonumber \\
    \end{eqnarray}
        In order to produce an equation that is similar to that of GR
    \begin{align}
        G_{\mu\nu}=\frac{\kappa^2}{f_Q}T^{\text{eff}}_{\mu\nu}\,. \label{ch4_eq: 13}
    \end{align}
    In this section, we explore the application of $f(Q, B)$ gravity within a cosmological context and introduce $f(Q, B)$ cosmology. Our analysis considers a homogeneous and isotropic flat FLRW spacetime represented by its line element in Cartesian coordinates.
    
    Following this section, it has been demonstrated that within the context of $f(Q, B)$ gravity, an additional effective sector of geometrical origin can be obtained as shown in equation (\ref{ch4_T^eff}). Consequently, when considered in a cosmological context, this term can be interpreted as an effective dark-energy sector, which possesses an energy-momentum tensor
    \begin{eqnarray}
            T^{\text{DE}}_{\mu\nu}= \frac 1{f_Q}\left[\frac 
            f2g_{\mu\nu}-2P^\eta{}_{\mu\nu} \nabla_\eta(f_Q-f_B)
            -\frac {Qf_Q}2g_{\mu\nu} -\left(\frac B2g_{\mu\nu}-\nabla_{\mu} \nabla_{\nu} + g_{\mu\nu} \nabla^\alpha \nabla_\alpha \right)f_B\right] , \label{ch4_eq: 15}
    \end{eqnarray}
    \begin{align}
        R = & 6(2H^2 + \dot{H}), \quad Q = -6H^2,
        \quad B = 6(3H^2 + \dot{H}). \label{ch4_eq: 16}
    \end{align}
    In this case, we consider a vanishing affine connection $(\tilde{\Gamma}^\eta{}_{\mu \nu}=0)$, when fixing the coincident gauge. Our Friedmann-like equations can be derived from these data as follows
    \begin{eqnarray}
            3 H^2 &=&\kappa^2 \left(\rho + \rho_{\text{DE}}\right),  \label{ch4_first_field_equation}\\
            \left(2 \dot{H}+3 H^2\right)& =&-\kappa^2 \left(p + p_{\text{DE}}\right), \label{ch4_second_field_equation}
    \end{eqnarray}
    where $ \rho $ and $ p $ represent the energy density and pressure of the matter sector, respectively, treated as a perfect fluid. Additionally, we have defined the effective dark-energy density and pressure as follows
        \begin{eqnarray} 
            &\rho_{\text{DE}}=\frac{1}{\kappa^2}\left[3 H^2\left(1-2 f_Q\right)-\frac{f}{2}+\left(9 H^2+3 \dot{H}\right) f_B-3 H \dot{f}_B\right], \label{ch4_eq: 17} \\ 
            &p_{\text{DE}}=\frac{1}{\kappa^2}\Big[-2 \dot{H}\left(1-f_Q\right)-3 H^2\left(1-2 f_Q\right)+\frac{f}{2}+2 H \dot{f}_Q -\left(9 H^2+3 \dot{H}\right) f_B+\ddot{f}_B\Big]. \label{ch4_eq: 18}
        \end{eqnarray}
    Since standard matter is conserved independently, with $\dot{\rho}+3 H\left(\rho+p\right)=0$, it can be deduced from equations \eqref{ch4_eq: 17} and \eqref{ch4_eq: 18} that the DE density and pressure conform to the standard evolution equation
    \begin{eqnarray}
        \dot{\rho}_{\text{DE}}+3 H\left(\rho_{\text{DE}}+p_{\text{DE}}\right)=0 .
    \end{eqnarray}
    Finally, we can define the parameter for the DE EoS as
    \begin{eqnarray}
        \omega_{\text{DE}} = \frac{p_{\text{DE}}}{\rho_{\text{DE}}}.
    \end{eqnarray}
    In the $\Lambda$CDM limit, as expected, the EoS parameter $\omega_{\text{DE}} \to -1$.
\subsection{Power Law \texorpdfstring{$f(Q, B)$}{}} \label{ch4_SEC: 2b}
    In this study, we propose a specific mathematical form of $f(Q, B)$ to capture the characteristic power-law behaviors observed in different stages of the evolution history of the Universe, i.e., at different cosmological epochs. This form is inspired by the work of Bahamonde and Capozziello \cite{Bahamonde_2017_77_2}, which utilizes the Noether Symmetry approach. The proposed form is given by
    \begin{equation}
        f(Q, B) = f_0 Q^m B^n \, , \label{ch4_eq: fqb}
    \end{equation}    
    where $f_0$, $m$ and $n$ are arbitrary constants.

     To determine the theoretical values of the Hubble rate, we can numerically solve equation \eqref{ch4_first_field_equation}. Assuming matter behaves as a pressureless perfect fluid ($p_{\text{m}}=0$), the matter density can be expressed as $\rho_\text{m} = 3 H_0^2 \Omega_{\text{m}0} (1+z)^3$, where $z$ denotes the cosmological redshift and $\Omega_{\text{m}0}$ is the current matter density parameter. Consequently, for the specific model under consideration, the first Friedmann equation can be written as follows
    \begin{equation}
        \begin{aligned}
        H^{\prime \prime}(z) &=& \frac{-1}{f_0\, n (n-1) (1+z)^2 H(z)^3} \left[-9 f_0 (n-2 m-1) H(z)^4+f_0\, n (n-1) (1+z)\right. H(z)^3 H^{\prime}(z)\\
        & &-6 f_0 \left((n-1)^2+(2+n) m\right) (1+z) H(z)^3 H^{\prime}(z)+f_0\, n (n-1) (1+z)^2 H(z)^2 H^{\prime}(z)^2 \\
        & &+f_0 (1-n+2 m (1+n))(1+z)^2  H(z)^2 H^{\prime}(z)^2-6^{1-n-m} H_0^2(1+z)^3 \Omega_{\text{m} 0}\left(H(z)^2\right)^{-m} \\
        & & \left.\left(H(z)\left(3 H(z)-(1+z) H^{\prime}(z)\right)\right)^{2-n}\right] , \label{ch4_HZ_ode_QB}
        \end{aligned}
    \end{equation}
    where the prime $(')$ denotes differentiation with respect to $z$.
    Equation \eqref{ch4_HZ_ode_QB} is a second-order differential equation for the function $H(z)$. To solve this equation, we need to apply suitable boundary conditions. The first boundary condition is straightforward: $H(0) = H_0$, which sets the present value of the Hubble parameter. To satisfy the second boundary condition, it is essential to confirm that the current rate of change of the Hubble parameter aligns with the projections of the standard $\Lambda$CDM model. This model describes the expansion of the Universe and provides a specific expansion law that the derivative should follow. By aligning the first derivative of $H(z)$ with this expansion law, we can accurately determine the second initial condition needed to solve the differential equation
    \begin{eqnarray}
            H_{\Lambda \text{CDM}} = H_0 \sqrt{1 - \Omega_{\text{m}0} + \Omega_{\text{m}0} (1+z)^3} ,
    \end{eqnarray}
    after taking the derivative of the equation with respect to $z$, we can derive the second initial condition for equation \eqref{ch4_HZ_ode_QB} as $H'(0) = \frac{3}{2}H_0 \Omega_{\text{m}0}$. 
\section{Observational Data, Methodology and Constraints} \label{ch4_Sec 3}
    To model the Universe accurately, we require robust observational data and effective parameter estimation methodologies. Within this framework, we detail the observational data sets and methods used to constrain the model parameters $f_0$, $m$, and $n$. Our analysis includes a comprehensive array of data, such as CC, Pantheon$^+$ and BAO observations. By leveraging these diverse data sets, we effectively narrow down the model parameters, facilitating an in-depth exploration of the evolution of the Universe. Additionally, we explore $f(Q, B)$ gravity and its solutions involving the Hubble parameter. The CC data set, known for its reliability and model independence, provides Hubble parameters by measuring the age difference between two passively evolving galaxies. This method allows us to derive the Hubble function at various redshifts up to $z \approx 2$. The shape of $H(z)$ is further constrained by multiple sources, including 32 data points from Hubble data sets, BAO data from various sources, and CMB data from Planck 2018. The employed methodology, utilized data, and outcomes are detailed in subsequent sections.
\subsection{Cosmic Chronometers} \label{ch4_Sec 3a}
    To estimate the expansion rate of the Universe at redshift $z$, we use the widely used differential age (DA) method. In this way, it is possible to predict $H(z)$ using $(1+z) H(z)=-\frac{dz}{dt}$. The Hubble parameter is modeled on 32 data points (see \hyperref[Appendices]{Appendices}) for a redshift range of $0.07 \leq z \leq 1.965$ \cite{Moresco_2022_25, Lohakare_2023_40_CQG}. The mean value of the parameters $H_0$, $\Omega_{\text{m}0}$, $f_0$, $m$ and $n$ are determined by minimizing the chi-square value.    
\subsection{Supernovae Type Ia} \label{ch4_Sec 3b}
    We will also consider the Pantheon$^+$ SNe Ia data compilation, consisting of 1701 SNe Ia relative luminosity distance measurements spanning the redshift range of $0.00122 < z < 2.2613$ \cite{Brout_2022_938}. The Pantheon$^+$ data set contains distance moduli estimated from 1701 light curves of 1550 spectroscopically confirmed SNe Ia with a redshift range acquired from 18 distinct surveys. Notably, 77 of the 1701 light curves are associated with Cepheid-containing galaxies. The Pantheon$^+$ data set has the benefit of being able to constrain $H_0$ in addition to the model parameters. To fit the parameter of the model from the Pantheon$^+$ samples, we minimize the $\chi^2$ function.      
\subsection{Baryon Acoustic Oscillation} \label{ch4_Sec 3c}
    BAO data sets are essential for constraining cosmological parameters in standard and modified gravity theories. They measure large-scale structures by observing galaxy clustering, reflecting sound wave imprints from the early Universe. Derived from surveys like the Sloan Digital Sky Survey (SDSS) and the Dark Energy Survey (DES), BAO data provide precise measurements of cosmic distances and the comoving sound horizon. In modified gravity theories, BAO data sets allow for comparisons between theoretical predictions and observed structures, helping to evaluate deviations from standard $\Lambda$CDM models. This makes BAO a valuable tool in testing alternative gravity theories and exploring cosmic acceleration. 
    \begin{figure} [htbp]
        \centering
    \includegraphics[width=110mm]{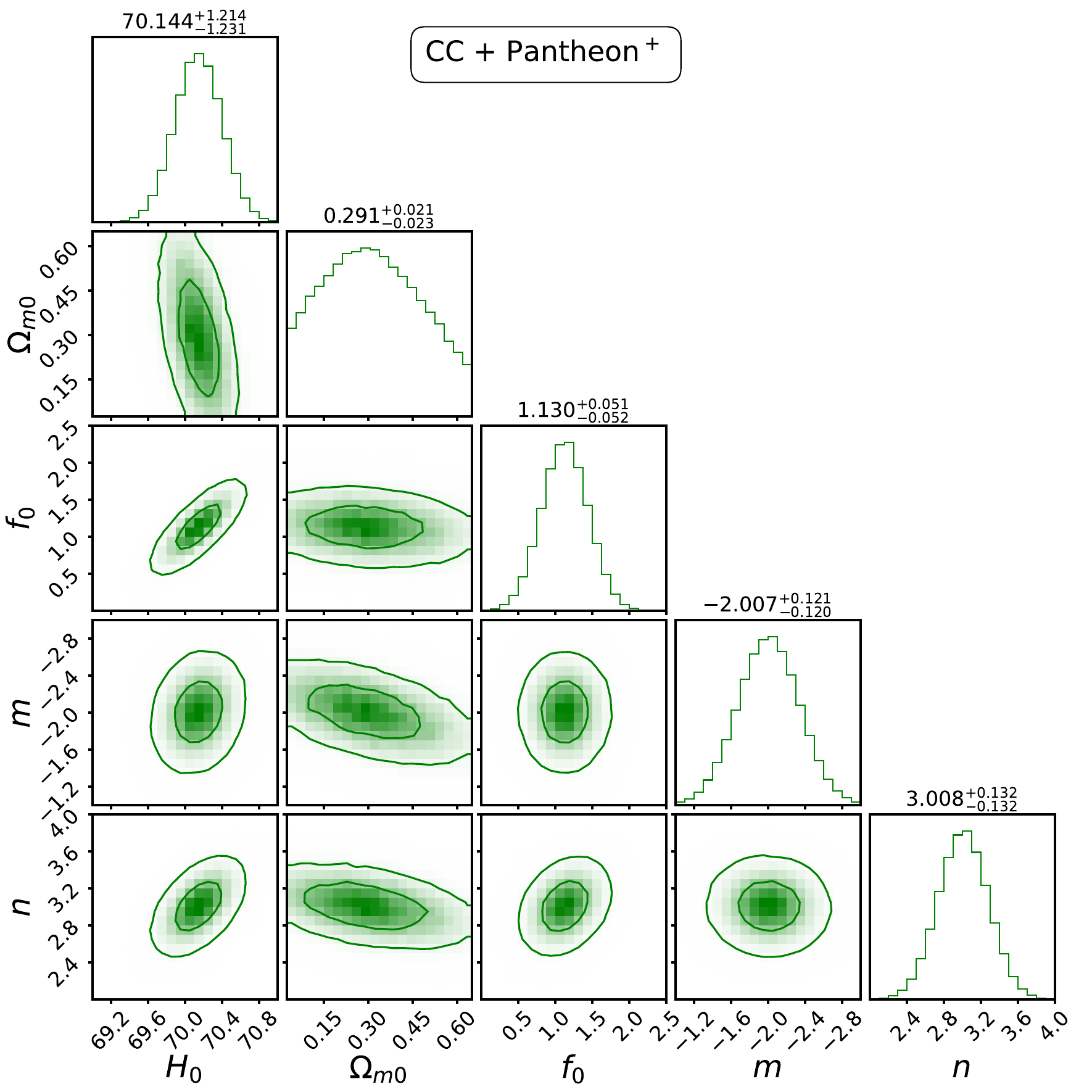}
        \caption{The contour plots display the $1\sigma$ and $2\sigma$ uncertainty regions for the model parameters $H_0$, $\Omega_{\text{m}0}$, $f_0$, $m$ and $n$. These contours are based on the combined CC + Pantheon$^+$ data sets.}
        \label{ch4_FIG1}
    \end{figure}
    \begin{figure} [htbp]
        \centering
    \includegraphics[width=110mm]{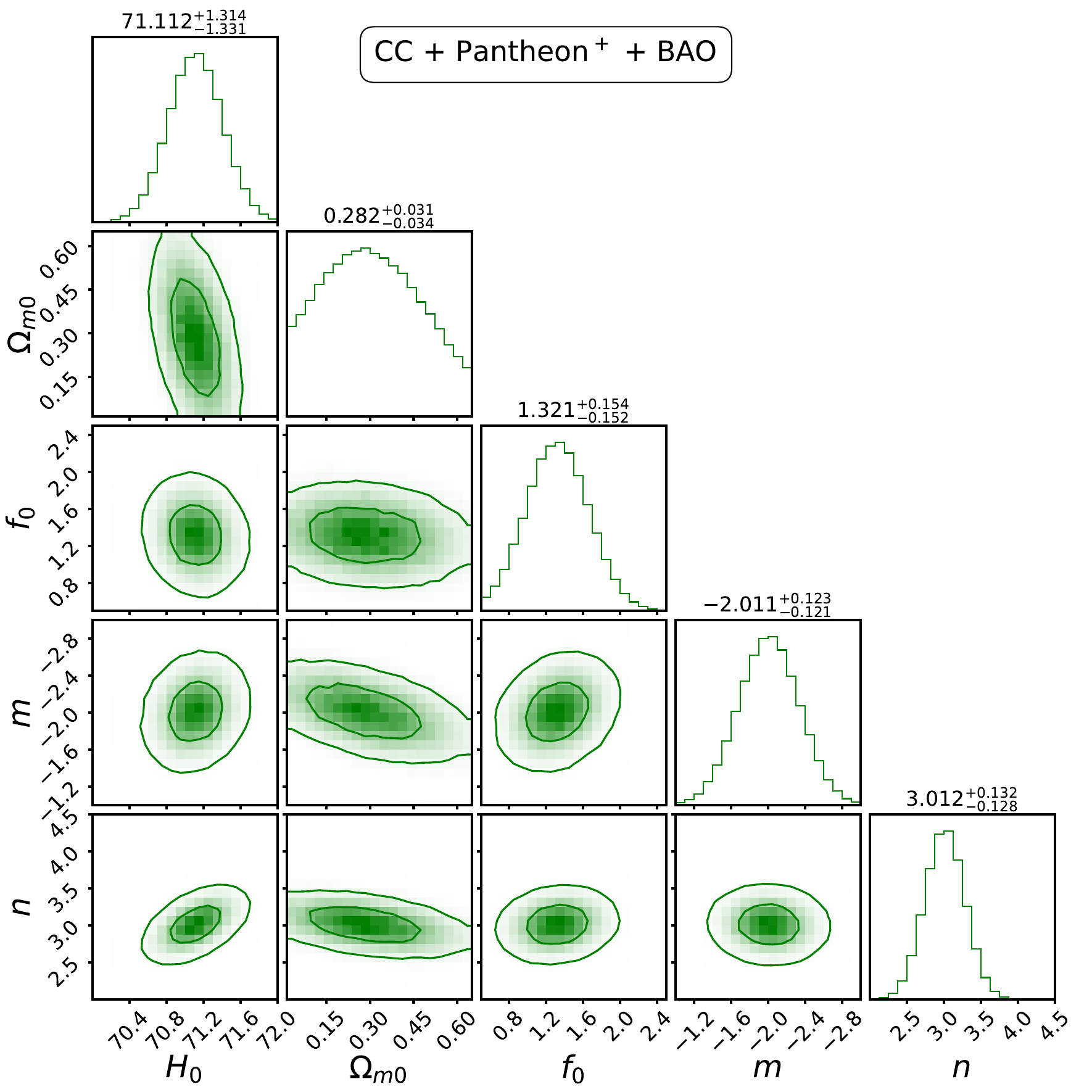}
        \caption{The contour plots display the $1\sigma$ and $2\sigma$ uncertainty regions for the model parameters $H_0$, $\Omega_{\text{m}0}$, $f_0$, $m$ and $n$. These contours are based on the combined CC + Pantheon$^+$ + BAO data sets.}
        \label{ch4_FIG2}
    \end{figure}
    \begin{figure} [htbp]
        \centering
    \includegraphics[width=.48\textwidth]{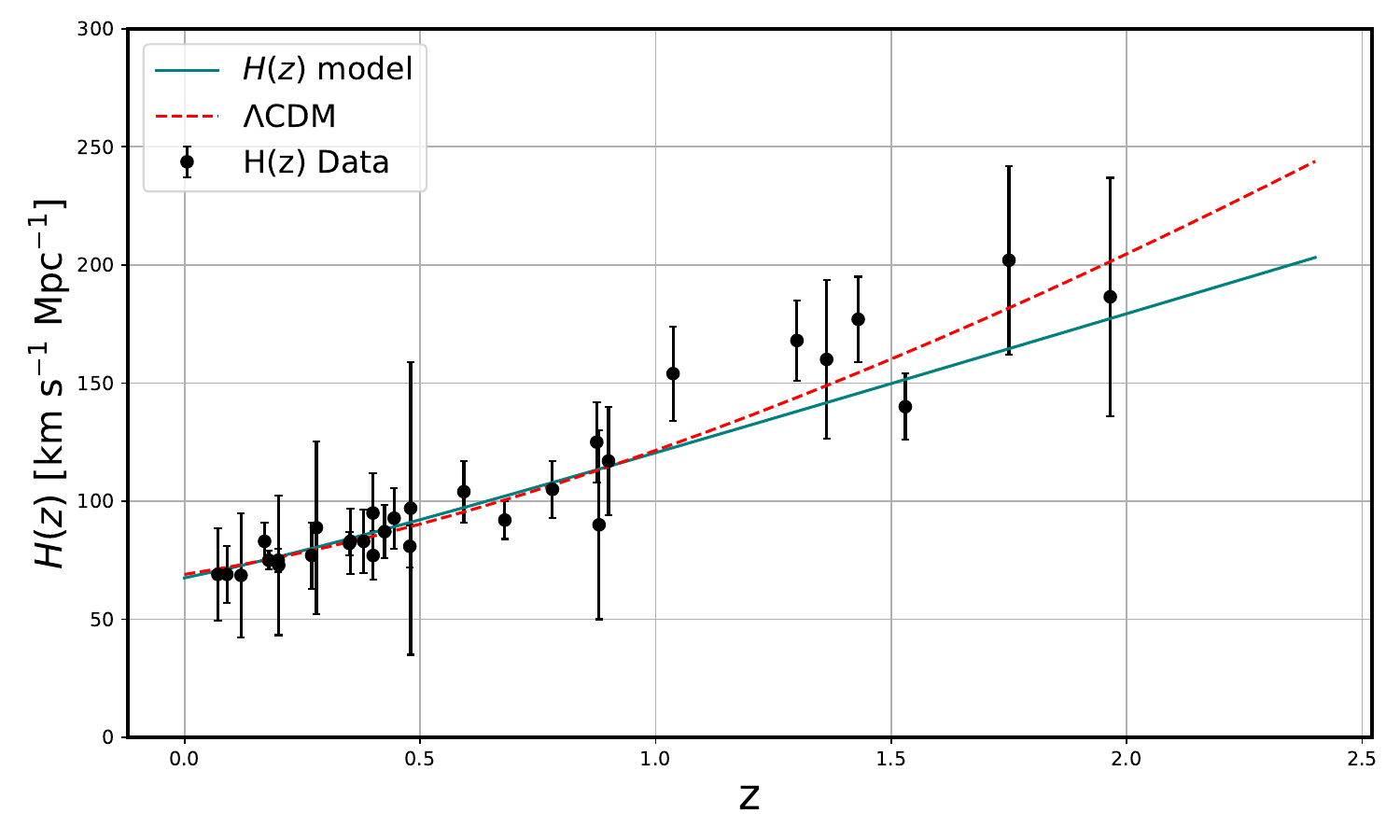}
        \includegraphics[width=.48\textwidth]{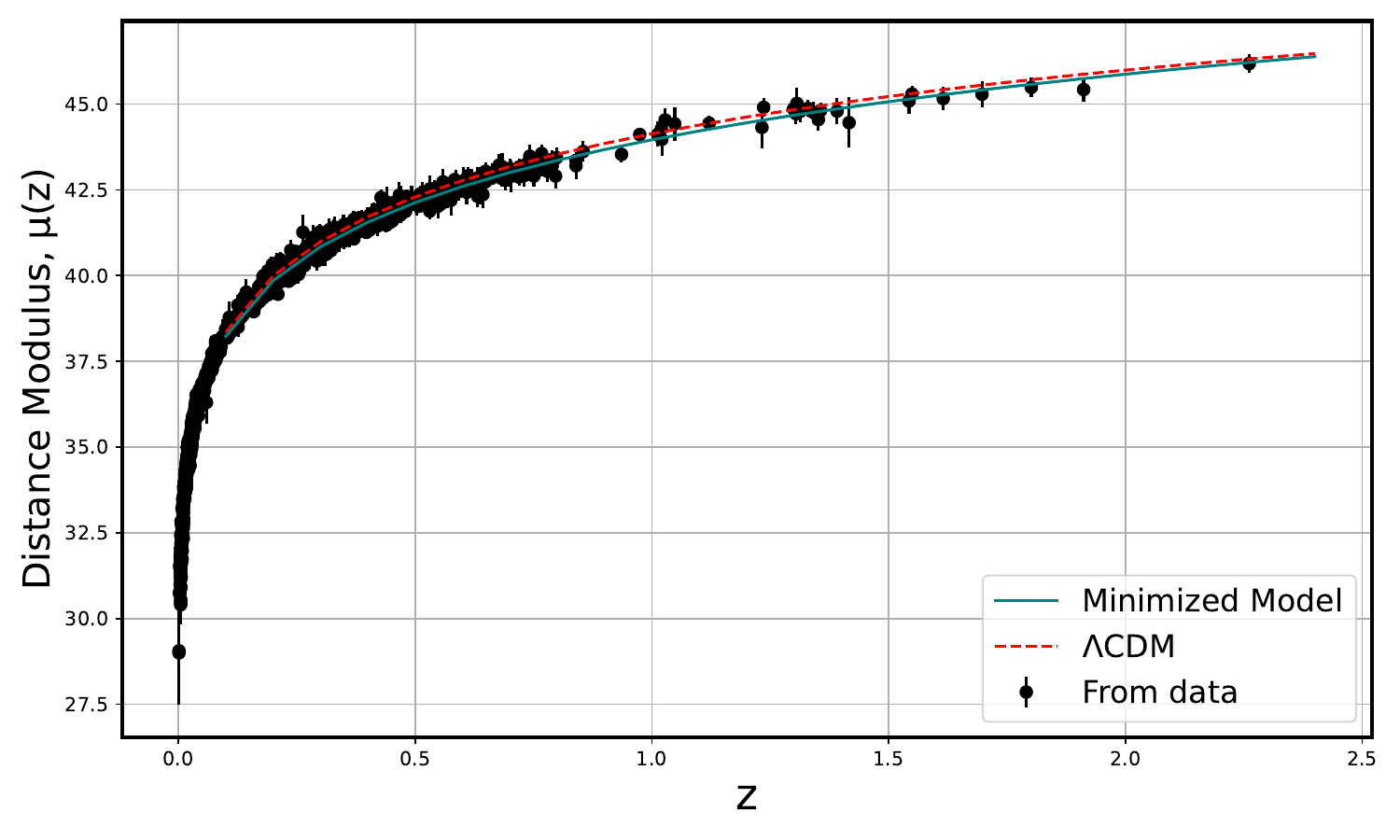}
        \caption{In the left panel, the black error bars represent the uncertainty associated with the 32 data points from the CC sample. The solid teal line corresponds to the model, while the dashed red line represents the $\Lambda$CDM. Moving to the right panel, we observe a red line that depicts the plot of the distance modulus of model $\mu(z)$ against redshift $z$. This teal line demonstrates a superior fit to the 1701 data points from the Pantheon$^+$ data set, including their associated error bars.}
        \label{ch4_FIG3}
    \end{figure}
    \begin{figure} [htbp]
        \centering
        \includegraphics[width=160mm]{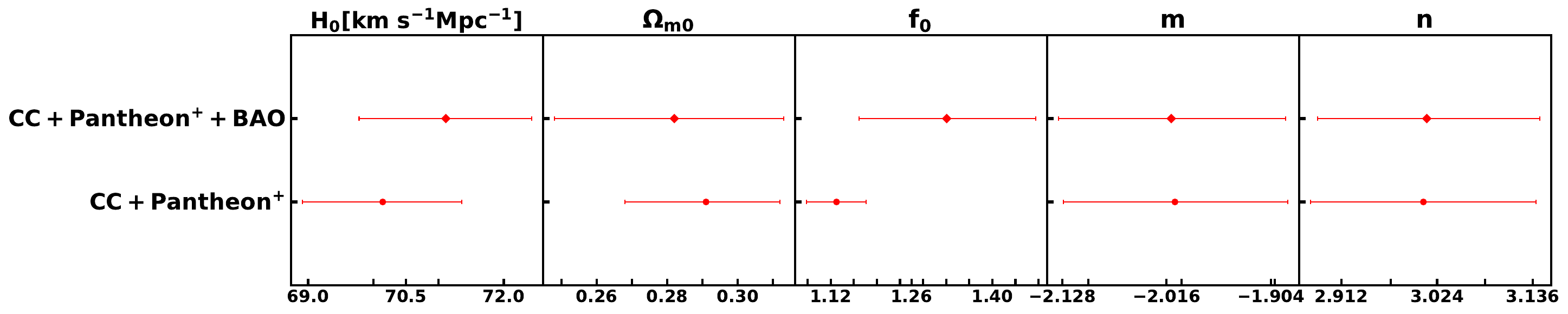}
        \caption{A whisker plot showing the model parameters $H_0$, $\Omega_{\text{m}0}$, $f_0$, $m$, and $n$, highlighting their discrepancies.}
        \label{ch4_FIG: whisker_plot}, 
    \end{figure}
    \begin{figure} [htbp]
    \vspace{0.5cm}
        \centering
        \includegraphics[width=.48\textwidth]{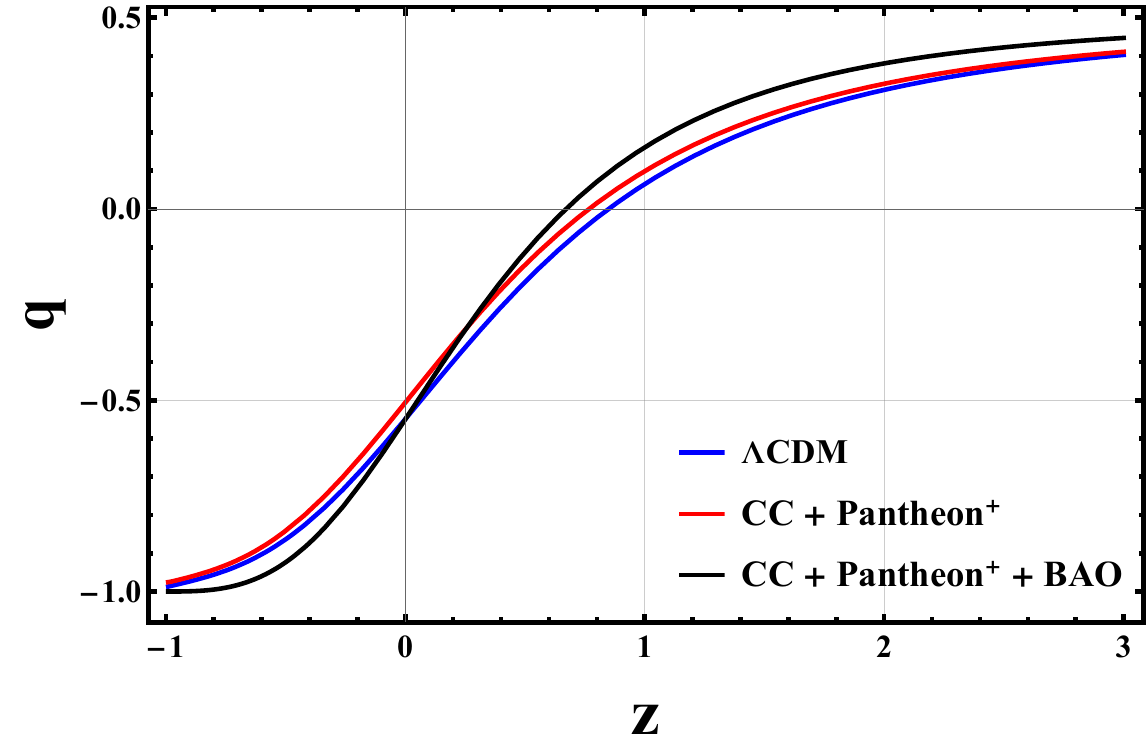}
        \includegraphics[width=.48\textwidth]{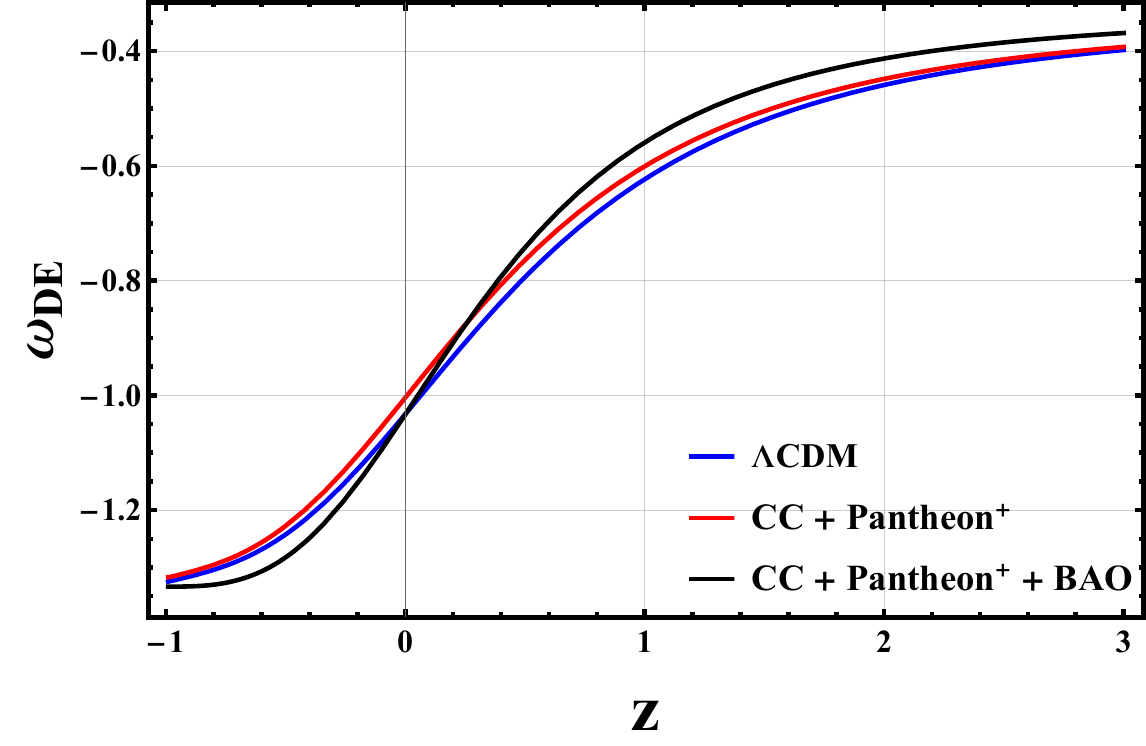}
        \caption{Behavior of the deceleration parameter (left panel) and EoS parameter (right panel) using the CC + Pantheon$^+$ and CC + Pantheon$^+$ + BAO data sets, with the mean values of parameters $ f_0 $, $ m $, and $ n $ as listed in Table \ref{ch4_TABLE I}.}
        \label{ch4_FIG4}
    \end{figure}
    \begin{table*} [htbp]
        \centering
        \scalebox{0.8}{
    \begin{tabular}{|*{6}{c|}}\hline
        {\centering  \textbf{Data sets}} & $H_0$ & $\Omega_{\text{m}0}$ & $f_0$ & $m$ & $n$ \\ [0.5ex]
    \hline \hline
        \parbox[c][0.7cm]{4cm}{\centering \textbf{CC + Pantheon$^+$}} & $70.144^{+1.214}_{-1.231}$ & $0.291^{+0.021}_{-0.023}$ & $1.130^{+0.051}_{-0.052}$ & $-2.007^{+0.121}_{-0.120}$ & $3.008^{+0.132}_{-0.132}$  \\
    \hline
        \parbox[c][0.7cm]{5cm}{\centering \textbf{CC + Pantheon$^+$ + BAO}} & $71.112^{+1.314}_{-1.331}$ & $0.282^{+0.031}_{-0.034}$ & $1.321^{+0.154}_{-0.152}$ & $-2.011^{+0.123}_{-0.121}$ & $3.012^{+0.132}_{-0.128}$ \\[0.5ex] 
    \hline
    \end{tabular}}
        \caption{Constrained values of model parameters based on the CC, Pantheon$^+$ samples, and BAO data sets.}
        \label{ch4_TABLE I}
    \end{table*}
        \begin{table*}[htbp]
        \centering
        \scalebox{0.53}{
        \begin{tabular}{|*{12}{c|} }
    \hline
        \parbox[c][0.7cm]{3.2cm}{\centering  \textbf{Data sets}}  &\multicolumn{2}{c|}{$\chi^2_{\text{min}}$} &\multicolumn{2}{c|}{AIC} &\multicolumn{2}{c|}{AIC$_\text{c}$} &\multicolumn{2}{c|}{BIC} & {$\Delta \text{AIC}$} & {$\Delta \text{AIC}_\text{c}$} & {$\Delta \text{BIC}$} \\
    \cline{2-9}
        \parbox[c][0.7cm]{1.5cm}& $f(Q, B)$ & $\Lambda$CDM & $f(Q, B)$ & $\Lambda$CDM & $f(Q, B)$ & $\Lambda$CDM & $f(Q, B)$ & $\Lambda$CDM & & &\\ 
    \hline \hline
        \parbox[c][0.7cm]{3.5cm}{\centering \textbf{CC + Pantheon$^+$}} & 1652.231 & 1654.270 & 1662.231 & 1658.270 & 1662.265 & 1658.277 & 1668.421 & 1660.746 & 3.961 & 3.988 & 7.675 \\
    \hline
        \parbox[c][0.7cm]{4.9cm}{\centering \textbf{CC + Pantheon$^+$ + BAO}} & 1659.321 & 1659.123 & 1669.321 & 1663.123 & 1669.355 & 1663.129 & 1675.521 & 1665.603 & 6.198 & 6.226 & 9.918 \\
    \hline
    \end{tabular}}
        \caption{The table presents the minimum $\chi^2$ values for the $f(Q, B)$ model, along with their corresponding AIC, AIC$_\text{c}$, and BIC values. It also includes a comparison of the differences with $\Lambda$CDM model in AIC, AIC$_\text{c}$, and BIC values.}
        \label{ch4_TABLE I b}
    \end{table*}
    
    Figures \ref{ch4_FIG1} and \ref{ch4_FIG2} provide the contour plots with $1\sigma$ and $2\sigma$ errors for the CC + Pantheon$^+$ and CC + Pantheon$^+$ + BAO data sets, respectively. In the left panel of figure \ref{ch4_FIG3}, the evolution of the Hubble parameter as a function of redshift is illustrated. This figure compares the predictions of two models: the $\Lambda$CDM model and the $H(z)$ model derived from the numerical approach proposed in this study (depicted by the teal line), alongside observational data. One line is included for the $\Lambda$CDM model to facilitate the comparison. Additionally, the dashed-red line (labeled as $\Lambda$CDM) is derived from the standard prediction of the $\Lambda$CDM model with the parameters $H_0$, and $\Omega_\text{m}$. In figure \ref{ch4_FIG3}, the right panel shows a comparison of the distance modulus using our $f(Q, B)$ model (teal line) and the $\Lambda$CDM model (dashed red line) predictions. Both models were considered with their respective parameters. The similarity between our model and $\Lambda$CDM prediction is evident. However, when the same shared parameter values were used, the models deviated from each other, mainly in the apparent magnitude prediction. The red dashed line in figure \ref{ch4_FIG3} shows that our model fits better with the $\Lambda$CDM. Also, the model accurately captures the behavior of the Hubble function, as shown by the consistency of error bars. In figure \ref{ch4_FIG3}, the observed distance modulus of the 1701 SNe Ia data set is depicted, along with the best-fit theoretical curves of the distance modulus function $\mu (z)$ shown as a teal line.

    Figure \ref{ch4_FIG: whisker_plot} displays the best-fit values and associated uncertainties for the model parameters $H_0$, $\Omega_{\text{m}0}$, $f_0$, $m$, and $n$, derived from our MCMC analysis. The plot visually represents the parameter ranges obtained from different data sets, including CC + Pantheon$^+$ and CC + Pantheon$^+$ + BAO. The Hubble constant $H_0$ values range from approximately 68.913 $\text{Km} \, \, \text{s}^{-1} \, \text{Mpc}^{-1}$ to 72.426 $\text{Km} \, \, \text{s}^{-1} \, \text{Mpc}^{-1}$, while the matter density parameter $\Omega_{\text{m}0}$ spans from 0.248 to 0.312. The parameters $f_0$, $m$, and $n$ exhibit ranges of 1.078 to 1.475, $-2.127$ to $-1.89$, and 2.876 to 3.144, respectively. These ranges highlight the variability and discrepancies in the parameter estimates, underscoring the robustness and reliability of the model fits to the observational data. The whisker plot effectively conveys the uncertainties inherent in the model parameters, providing a comprehensive overview of the results from the MCMC analysis.
 
    Figure \ref{ch4_FIG4} illustrates the significance of the deceleration parameter $q$, a crucial metric in cosmology that provides insights into the dynamics of the Universe. A positive $q$ indicates deceleration, while a negative $q$ signifies acceleration. Analysis of the CC + Pantheon$^+$ and CC + Pantheon$^+$ + BAO data sets reveals that $q$ transitions from positive in the past, indicating early deceleration, to negative in the present, indicating current acceleration, as depicted in figure \ref{ch4_FIG4}. At the current cosmic epoch, the deceleration parameter $q_0$ has been measured as $-0.506$ and $-0.549$ for the CC + Pantheon$^+$ and CC + Pantheon$^+$ + BAO data sets, respectively. These values are in good agreement with the range of $q_0 = -0.528^{+0.092}_{-0.088}$ determined by recent observations \cite{Christine_2014_89}. Current observations align with this deceleration parameter, and the derived model demonstrates a smooth transition from deceleration to acceleration at $z_t = 0.763$ and $ z_t = 0.67 $ for the CC + Pantheon$^+$ and CC + Pantheon$^+$ + BAO data sets, respectively. The recovered transition redshift $z_t$ is consistent with current constraints based on 11 $H(z)$ observations reported by Busca et al. \cite{Busca_2013_552} for redshifts $0.2 \leq z \leq 2.3$, $z_t = 0.74 \pm 0.5$ from Farooq and Rarta \cite{Farooq_2013_766}, $z_t = 0.7679^{+0.1831}_{-0.1829}$ by Capozziello et al. \cite{Capozziello_2014_90_044016}, and $z_t = 0.60^{+0.21}_{-0.12}$ by Yang and Gong \cite{Yang_2020_2020_059}. Similarly, the EoS parameter $(\omega_{\text{DE}})$ is integral to understanding the evolution of the Universe, as it correlates with the energy sources influencing this progression. The current EoS values for DE, represented by $\omega_{\text{DE}}(z = 0)$, are determined to be $-1.032$ and $-1.004$ for the CC + Pantheon$^+$ and CC + Pantheon$^+$ + BAO data sets, respectively. Various cosmological studies have also placed constraints on the EoS parameter. For instance, the Planck 2018 results yielded $\omega_{\text{DE}}=-1.03\pm 0.03$ \cite{Aghanim_2020_641}, and the WAMP + CMB analysis reported $\omega_{\text{DE}}=-1.079^{+0.090}_{-0.089}$ \cite{Hinshaw_2013_208}. By computing the associated energy density and pressure of DE, we can observe the fluctuations in the effective DE EoS, which are depicted in redshift [figure \ref{ch4_FIG4}].

    We evaluate the models against the standard $\Lambda$CDM model using the Akaike Information Criterion (AIC) and the Bayesian Information Criterion (BIC), in addition to $\chi^2_{\text{min}}$. Both AIC and BIC consider the goodness of fit of the model and its complexity, which depends on the number of parameters $(n)$. The AIC is calculated as
    \begin{eqnarray}
            \text{AIC} = \chi^2_{\text{min}} + 2 n \, .
    \end{eqnarray}
    In statistical modeling, a lower AIC value indicates a better fit to the data, accounting for model complexity. This penalizes models with more parameters, even if they fit the data better. The BIC is computed as
    \begin{eqnarray}
            \text{BIC} = \chi^2_{\text{min}} + n \, \text{ln} \,\mathcal{N} \, ,
    \end{eqnarray}
    where $ \mathcal{N} $ is the number of data samples used in the MCMC process. The corrected Akaike Information Criterion (AIC$_\text{c}$) is defined as
    \begin{eqnarray}
    \text{AIC}_\text{c} = \text{AIC} + \frac{2 n (n+1)}{\mathcal{N}-n-1} \, ,
    \end{eqnarray}
    for large sample sizes $ (\mathcal{N} \gg n) $, the correction term becomes negligible, making AIC$_\text{c}$ preferable over the original AIC.

    We compare the AIC and BIC values between the $ f(Q, B) $ model and the $\Lambda$CDM model to gain insights into how well each model aligns with the standard cosmological model. The differences in AIC, AIC$_\text{c}$ and BIC are expressed as $ \Delta \text{IC} = \text{IC}_{\text{Model}} - \text{IC}_{\Lambda\text{CDM}} $. Smaller $ \Delta $AIC and $ \Delta $BIC values indicate that a model, along with its selected data set, closely resembles the $\Lambda$CDM model, suggesting superior performance. To assess the effectiveness of our MCMC analysis, we computed the corresponding AIC, AIC$_\text{c}$, and BIC values, as shown in Table \ref{ch4_TABLE I b}. Our results strongly endorse the proposed $ f(Q, B) $ gravity models based on the analyzed data sets. Additionally, we observed that the $ f(Q, B) $ model exhibits higher precision when applied to the CC + Pantheon$^+$ data sets.
\section{Dynamical System Analysis} \label{ch4_Sec 4}   
    The methods of dynamical systems are valuable for analyzing the overall long-term dynamics of a particular cosmological model. This involves an equation, $x' = f(x)$, where $x$ is a column vector and $f(x)$ is the equivalent vector of autonomous equations. In this method, the prime symbol represents the derivative with respect to the number of e-folding, $N = \text{ln} \, a(t)$. The general form of the dynamical system for the modified FLRW equations defined by equation \eqref{ch4_first_field_equation} can be generated through this approach. Let us define a new variable 
    \begin{eqnarray}
        X=f_B, \hspace{0.3cm} Y=\frac{\dot{f}_B}{H}, \hspace{0.3cm} Z=\frac{\dot{H}}{H^2}, \hspace{0.3cm} V=\frac{\kappa^2 \, \rho_\text{r}} {3H^2}, \hspace{0.3cm} W=-\frac{f}{6 H^2}, \hspace{0.3cm} \Omega_{\text{m}}= \frac{\kappa^2 \, \rho_{\text{m}}}{3 H^2}, \hspace{0.3cm} \Omega_{\text{DE}}=\frac{\kappa^2 \, \rho_{\text{DE}}}{3 H^2} \, . \label{ch4_eq: 30} 
    \end{eqnarray}
    Thus, from equation \eqref{ch4_first_field_equation}, we have the algebraic identity
    \begin{eqnarray}
        \Omega_{\text{m}} + \Omega_{\text{r}} + \Omega_{\text{DE}}=1 \, , \label{ch4_eq: 31}
    \end{eqnarray}
    together with the density parameters
    \begin{eqnarray}
        & & \Omega_{\text{m}} = \frac{\kappa^2 \rho_\text{m}}{3 H^2}, \hspace{0.6cm} \Omega_{\text{r}} = V = \frac{\kappa^2 \rho_r}{3 H^2},\\ \hspace{0.3cm} & &\Omega_{\text{DE}}= 1-2 f_Q + W + 3 X + X Z - Y \, . \label{ch4_eq: density parameters}
    \end{eqnarray}
    So, taking the derivative of these variables with respect to $N$, we obtain the following dynamical system
    \begin{subequations} \label{ch4_eq: DS}   
    \begin{eqnarray} 
        \frac{dX}{dN} &=& Y \, ,\\
        \frac{dY}{dN} &=& 2 - 3V + \frac{2 Z}{3} - \frac{2 Z f_Q}{3} + 3 X - 2 f_Q + W -\frac{2 f_Q'}{3} + X Z - Y Z \, ,\\
        \frac{dZ}{dN} &=& \lambda - 2 Z^2 \, ,\\
        \frac{dV}{dN} &=& -4 V - 2 Z V \, ,
    \end{eqnarray}
    \end{subequations}
    where $\lambda = \frac{\ddot{H}}{H^3}$, we will concentrate on the scenario where $f(Q, B) = f_0 Q^m B^n$. The model for this scenario can be expressed using the dynamical variables.
    \begin{eqnarray}
        f_Q = m \, W \, ,
    \end{eqnarray}
    and we get the following dependency relation
    \begin{eqnarray}
        W &=& -\frac{X}{n} (Z+3) \, ,\\
        \Omega_{\text{m}} &=& -V-2 f_Q + W + 3 X + X Z - Y \, ,\\
        \Omega_{\text{DE}} &=& 1-2 f_Q + W + 3 X + X Z - Y \, ,
    \end{eqnarray}
    and
    \begin{eqnarray}
        \lambda = \frac{Y (Z + 3)-2 X Z \left(m (Z + 3) + 3 (n-1)\right)}{(n-1) X} \, .
    \end{eqnarray}
    It is possible to eliminate the equations for $W$, $\Omega_{\text{m}}$ and $\Omega_{\text{DE}}$ from our autonomous system using the relations mentioned and constraint \eqref{ch4_eq: 31}, resulting in a set of only four equations
    \begin{subequations} \label{ch4_eq: Autonomous_DS}   
    \begin{eqnarray} 
        \frac{dX}{dN} &=& Y \, ,\\
        \frac{dY}{dN} &=& \frac{1}{3 n (n-1)}\Big[n \big(-2 X (Z+3) (m (Z-3)+3) + 2 m Y (Z + 3) + 9 V + 3 Y Z - 2 (Z + 3)\big) \nonumber\\ & & - (2 m-1) X (Z + 3) (2 m Z + 3) + n^2 (-9 V + 3 X (Z + 3)-3 Y Z + 2 Z + 6)\Big] \, ,\nonumber\\ \\
        \frac{dZ}{dN} &=& \lambda - 2 Z^2 \, ,\\
        \frac{dV}{dN} &=& -4 V - 2 Z V \, .
    \end{eqnarray}
    \end{subequations}
        It is important to note that a dynamical system has a critical point and this point must be taken into account when analyzing the system
    \begin{eqnarray}
        \mathcal{P}_{\star}(X, Y, Z, V) = \Bigg( \frac{2 n}{3 (n-1)},\,  0,\, 0, \, 0\Bigg) \, ,
    \end{eqnarray}
        for existence condition $m = 1 - n$. We will now analyze the range of value for $n$, which will result in a stable critical point. While we will not explicitly mention the area of instability, whether it is saddle-like or repulsor-like. It is important to note that the critical point is a de Sitter acceleration phase, and therefore any kind of instability of the critical point is not supported by observations.

        The eigenvalue is given by
        \begin{eqnarray} \label{ch4_eigenvalues}
        \left\{0,-4,\frac{-n-\sqrt{-2 n^2-3 n}}{n},\frac{- n+\sqrt{-2 n^2-3 n}}{n}\right\} \, .
        \end{eqnarray}
        The phase portrait in figure \ref{ch4_FIG5} shows the behavior of a dynamical system near a stable critical point, $\mathcal{P}_{\star}$. As time progresses, trajectories in the phase space tend to move towards $\mathcal{P}_{\star}$, indicating that it is an attractor for the system. This convergence from various initial conditions signifies that small perturbations decay over time, returning the system to the stable state at $\mathcal{P}_{\star}$. The stability of $\mathcal{P}_{\star}$ can be analyzed using the Jacobian matrix evaluated at $\mathcal{P}_{\star}$. Stability of $\mathcal{P}_{\star}$ is ensured when all eigenvalues of the Jacobian possess negative real parts, which aligns with the observed behavior in the phase portrait. The overall dynamics of the system are governed by the differential equations defining it, and the phase portrait provides a graphical representation to visualize these dynamics and understand the long-term behavior of the system. The zero eigenvalues suggest the presence of a center manifold, which requires further analysis to determine overall stability.

\subsection{Stability Analysis for \texorpdfstring{$\mathcal{P}_{\star}$}{} by using Center Manifold Theory} \label{ch4_CMT}
    The Center manifold theory is presented in Section \ref{Intro: CMT}. The Jacobian matrix at the critical point $\mathcal{P}_{\star}$ for the autonomous system \eqref{ch4_eq: Autonomous_DS} is given below
        \begin{equation}
            J\left(\mathcal{P}_{\star}\right)=\left(
            \begin{array}{cccc}
            0 & 1 & 0 & 0 \\
            -3+\frac{3}{n} & -2 & -\frac{4}{3} & -3 \\
            0 & \frac{9}{2 n} & 0 & 0 \\
            0 & 0 & 0 & -4 \\
            \end{array}\right)
        \end{equation}
    The eigenvalues of the Jacobian matrix, as presented in (\ref{ch4_eigenvalues}), are $\lambda_1 = 0$, $\lambda_2 = -4$, $\lambda_3 = \frac{-n-\sqrt{-2 n^2-3 n}}{n}$, and $\lambda_4 = \frac{-n+\sqrt{-2 n^2-3 n}}{n}$. The corresponding eigenvectors are: $\Big[\frac{4 n}{9-9 n},0,1,0\Big]^T$, $\Big[-\frac{3 n}{11 n+3},\frac{12 n}{11 n+3},-\frac{27}{22 n+6},1\Big]^T$, $\Big[\frac{2 n}{9},\frac{2}{9} \left(\sqrt{-n (2 n+3)}-n\right),1,0\Big]^T$ and \\$\Big[\frac{2 n}{9},\frac{1}{9} (-2) \left(n+\sqrt{-n (2 n+3)}\right),1,0\Big]^T$.

    Using the center manifold theory, we examine the stability of the critical point $\mathcal{P}_{\star}$. By applying the transformation $X = x - \frac{2 n}{3 (n-1)}$, $Y = y$, $Z = z$, and $V = v$, we shift this critical point to the origin. The resulting equations in the new coordinate system are as follows
    \begin{equation}
    \left(\begin{array}{c}
    \dot{x} \\
    \dot{y} \\
    \dot{z} \\
    \dot{v} 
    \end{array}\right)=\left(\begin{array}{cccc}
     0 & 0 & 0 & 0 \\
     0 & -4 & 0 & 0 \\
    0 & 0 & \frac{-n-\sqrt{-2 n^2-3 n}}{n} & 0 \\
    0 & 0 & 0 & \frac{-n+\sqrt{-2 n^2-3 n}}{n}  
    \end{array}\right)\left(\begin{array}{l}
    x \\
    y \\
    z \\
    v 
    \end{array}\right)+\left(\begin{array}{c}
    \text { non } \\
    \text { linear } \\
    \text { term }
    \end{array}\right)
    \end{equation}
    Upon examining the diagonal matrix in relation to the standard form \eqref{ch4_CMT_system}, it is clear that the variables $y$, $z$ and $v$ remain stable, while $x$ acts as the central variable. At this critical point, matrices $A$ and $B$ take on the following form\\
        $$A=\left[0\right] , \quad
        B=\left[\begin{array}{ccc}
        -4 & 0 & 0  \\
        0 & \frac{-n-\sqrt{-2 n^2-3 n}}{n} & 0  \\
        0 & 0 & \frac{-n+\sqrt{-2 n^2-3 n}}{n} 
        \end{array}\right]  
        $$
    In the context of center manifold theory, the manifold is characterized by a continuous differential function. Assuming specific functions for the stable variables $y = g_1 (x)$, $z = g_2 (x)$, and $v = g_3 (x)$, we derived the zeroth approximation of the manifold functions using equation \eqref{local_center_manifold}
    \begin{eqnarray}
        & & N(g_1 (x)) = \left(-3+\frac{3}{n}\right) x+\mathcal{O}^2,\nonumber\\& & N(g_2 (x)) = 0 + \mathcal{O}^2 , \hspace{0.2cm} N(g_3 (x)) = 0 + \mathcal{O}^2,
    \end{eqnarray}
    where $\mathcal{O}^2$ term encompasses all terms that are proportional to the square or higher powers. The following expression gives the center manifold in this scenario
    \begin{eqnarray}
        \dot{x} = \left(-3+\frac{3}{n}\right) x + \mathcal{O}^2.
    \end{eqnarray}
    According to the center manifold theory, the critical point $\mathcal{P}_{\star}$ exhibits stable behavior for $(n < 0) \vee (n > 1)$.
    \begin{figure} [htbp]
        \centering
        \includegraphics[width=85mm]{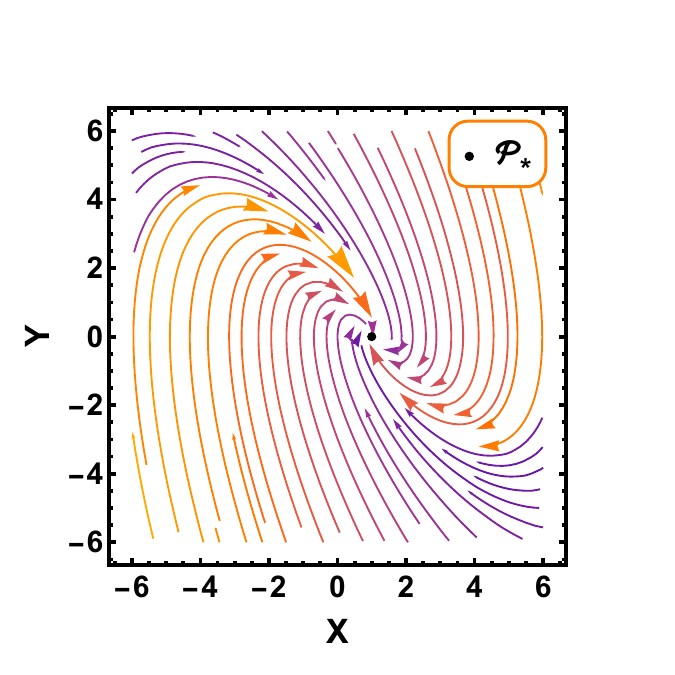}
        \caption{This graph shows the behavior of a four-dimensional system \eqref{ch4_eq: Autonomous_DS} simplified to two dimensions. The parameters are set as $Z = 0$, $V = 0$, and $n = 3.012$.}
        \label{ch4_FIG5}
    \end{figure}
    Figure \ref{ch4_FIG5} reveals that a specific point in this two-dimensional representation (the critical point $\mathcal{P}_{\star}$) attracts other points over time, suggesting its stable and attractive nature within the full four-dimensional system. Figure \ref{ch4_FIG5} shows the fascinating world of sink trajectories within a dynamical system, visualized through a phase portrait. This point signifies a location where trajectories tend to sink or converge. The critical point $\mathcal{P}_{\star}$ is non-hyperbolic due to the presence of zero eigenvalues and can describe the acceleration of the Universe. It is also an attractor solution, stable in the regions $( n < 0 ) \vee ( n > 1 )$ as determined by center manifold theory. The density parameters for radiation, matter, and DE are $\Omega_\text{r} = 0$, $\Omega_\text{m} = 2-\frac{4}{n-1}$, and $\Omega_\text{DE} = -1 + \frac{4}{n-1}$, respectively, satisfying the constraint in equation \eqref{ch4_eq: 31}. This scenario corresponds to a deceleration parameter $q = -1$ and a total EoS $\omega_\text{tot} =-1-\frac{2 \dot{H}}{3 H^2}= -1$, indicating a de Sitter phase and, consequently, an accelerating expansion of the Universe.
    \begin{figure} [htbp]
        \centering
        \includegraphics[width=100mm]{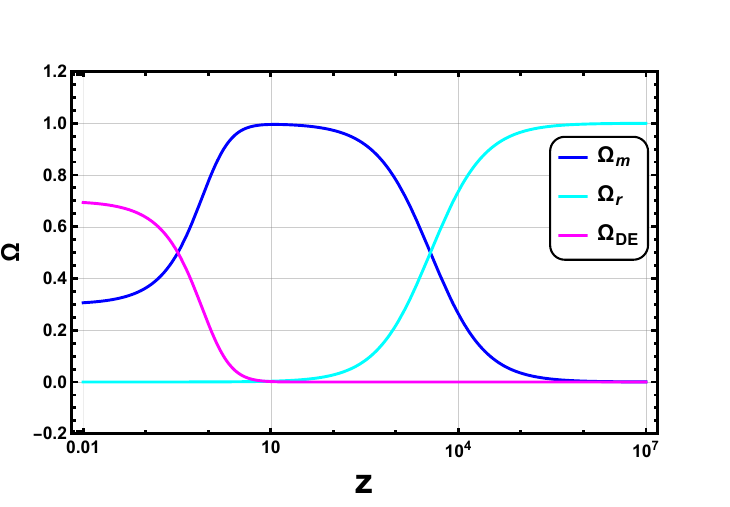}
        \caption{Evolution of density parameters DE (magenta line), matter (blue line), and radiation (cyan line) for the initial conditions: $X =10^{11}$, $Y = 3.29 \times 10^{14}$, $Z = 0.008$, and $V = 4.54 \times 10^{-5}$.}
        \label{ch4_FIG6}
    \end{figure}
    Figure \ref{ch4_FIG6} shows the evolution of density parameters for DE, matter, and radiation as a function of redshift. The model parameters $m = -2.011$ and $n = 3.012$, which are obtained from the parametrization method using MCMC analysis for CC + Pantheon$^+$ + BAO data sets, are used.
    \begin{itemize}
        \item The \textit{magenta line} represents the DE density parameter ($\Omega_{\text{DE}}$), which increases sharply at lower $z$, indicating the growing influence of DE in the accelerated expansion of the Universe.
        \item The \textit{blue line} shows the matter density parameter ($\Omega_{\text{m}}$), which decreases with increasing $z$, reflecting the dilution of matter as the Universe expands.
        \item The radiation density parameter ($\Omega_{\text{r}}$) is represented by the \textit{cyan line}, which remains almost constant at zero for small redshift values. This emphasizes its minimal contribution in the present epoch.
    \end{itemize}
    In figure \ref{ch4_FIG6}, we present the evolution of the density parameters for DE, matter, and radiation as a function of redshift ($z$), utilizing a model parameter value of $m=-2.011$ and $n = 3.012$ obtained from our MCMC analysis. The initial conditions for this plot are $X =10^{11}$, $Y = 3.29 \times 10^{14}$, $Z = 0.008$, and $V = 4.54 \times 10^{-5}$. The magenta line representing DE exhibits a significant increase at lower redshifts, indicating its dominance in the current epoch of the Universe. The blue line for matter density decreases as redshift decreases, reflecting the transition from a matter-dominated Universe at higher redshifts to a DE-dominated Universe at lower redshifts. The cyan line for radiation density is notably higher at early times (high redshifts) and diminishes rapidly as the Universe expands, consistent with the radiation-dominated era in the early Universe. This plot effectively captures the dynamic evolution of the energy components of the Universe, illustrating the transitions from radiation dominance to matter dominance and finally to DE dominance, providing valuable insights into the evolution of the Universe.

\section{Conclusion} \label{ch4_Sec 5}
    In this chapter, we have delved into the cosmological implications of a modified $f(Q, B) $ gravity model, which integrates both the nonmetricity scalar $ Q $ and the boundary term $ B $. Our approach adopted the coincident gauge, where the general affine connection vanishes, meaning the covariant derivative reduces to the partial derivative. We then applied Bayesian statistical analysis using MCMC techniques to constrain the model parameters. The analysis was grounded in observational data from CC measurement, the extended Pantheon$^+$ data set, and BAO measurements. Our results elucidate a smooth transition from a deceleration phase to an accelerating expansion phase in the evolution of the Universe. This transition is critical in understanding the dynamics of cosmic expansion and the role of DE. We developed a numerical approach to predict the redshift behavior of the Hubble expansion rate. This approach was instrumental in constraining the model parameters and understanding the kinematic evolution of the Universe. The $ f(Q, B)$ model has been compared with the standard $ \Lambda $CDM model, demonstrating its potential as a viable alternative cosmological framework. While the $ \Lambda $CDM model has been the cornerstone of modern cosmology, our findings suggest that the $ f(Q, B) $ model can replicate the low-redshift behavior of the $ \Lambda $CDM model and exhibits notable differences at high redshifts. Our findings align strongly with current cosmological observations of a late-time Universe dominated by DE and undergoing accelerated expansion. This supports the validity of the $ f(Q, B) $ model as an alternative explanation for the observed acceleration of the Universe.

    A dynamical system analysis framework has been introduced to assess the stability of the model. The identification of a stable critical point using center manifold theory underscores the robustness of the $ f(Q, B) $ model. A significant finding of our study is the identification of a stable critical point within the dynamical system of the model, corresponding to the de Sitter phase. The stability of this critical point implies that, given specific initial conditions, the Universe will inherently move towards and stay within the de Sitter phase. This observation aligns with current data indicating a Universe dominated by DE and undergoing late-time accelerated expansion. Future research could delve deeper into the specific initial conditions leading to the de Sitter phase and investigate the influence of the boundary term $ B $ on the dynamics of the system. The density parameter plot depicts the transition of the Universe from radiation dominance to matter dominance and ultimately to DE dominance. This offers valuable insights into the evolutionary dynamics of the Universe. Additionally, exploring the implications of this stable critical point for physical quantities like the Hubble parameter would offer valuable insights into the evolution of the Universe. In summary, the $ f(Q, B) $ gravity model not only aligns well with current cosmological observations but also provides a comprehensive framework for understanding the accelerated expansion of the Universe. The ability of the model to capture the transition from deceleration to acceleration, identify a stable critical point, and offer a viable alternative to the $\Lambda$CDM model makes it a promising candidate for further exploration in cosmological studies. Our study emphasizes the crucial role of modified gravity theories in understanding the expansion of the Universe and the nature of DE.

\chapter{Cosmology in \texorpdfstring{$f(\mathcal{G})$}{} Gravity: a Late Time Cosmic Phenomena} 

\label{Chapter5} 

\lhead{Chapter 5. \emph{Cosmology in \texorpdfstring{$f(\mathcal{G})$}{} Gravity: a Late Time Cosmic Phenomena}} 

\vspace{11 cm}
* The work in this chapter is covered by the following publication: 

\textbf{Santosh V. Lohakare}, Soumyadip Niyogi and B. Mishra, ``Cosmology in \texorpdfstring{$f(\mathcal{G})$}{} Gravity: a Late Time Cosmic Phenomena", \href{https://doi.org/10.1093/mnras/stae2302}{{\color{blue}\textit{Monthly Notices of the Royal Astronomical Society} \textbf{535} (2024) 1136.}}

\clearpage
\section{Introduction}        
    The $f(\mathcal{G})$ gravity theory modifies the Einstein--Hilbert action by introducing a function of the Gauss--Bonnet invariant, denoted $\mathcal{G}$, which is a combination of the Ricci scalar $R$, the Ricci tensor $R_{\mu \nu}$, and the Riemann tensor $R_{\mu \nu \sigma \gamma}$ \cite{Stelle_1978_9, Barth_1983_28, Felice_2009_6, Nojiri_2005_631, nojiri2006modified}. It belongs to an infinite class of curvature invariants known as the Lovelock scalars along with $R$. These do not introduce derivative terms greater than two into the equations of motion for the metric tensor. In four dimensions, the term $\sqrt{-g} \mathcal{G}$ is a total derivative, so the Gauss--Bonnet term contributes to the equations of motion only when coupled to something else, such as a scalar field $\phi$ with the form $f(\phi) \mathcal{G}$ coupling \cite{Tsujikawa_2002_66, Cartier_2000_2000_035}. A dilaton-graviton mixing term generates this kind of coupling in the low-energy effective action of string theory \cite{Gasperini_2003_373}. The interest in $f(\mathcal{G})$ gravity lies in its potential to explain the observed late-time cosmic acceleration in the Universe. This acceleration could be caused by a gravity modification rather than an unusual source of matter with negative pressure \cite{carroll2004dark}. In recent years, significant research has been conducted into modified gravity to understand the nature of DE \cite{Nojiri_2011_505}. Modified gravity models are particularly attractive because they align more closely with cosmological observations and local gravity experiments than models that rely on exotic matter sources \cite{joyce2015dark}. It is suggested that this theory can pass solar system tests \cite{Davis_2007_0709.4453, Felice_2009_80_063516} and may describe the most exciting features of late-time cosmology, such as the transition from deceleration to acceleration and the current acceleration of the Universe \cite{Nojiri_2005_631, Davis_2007_0709.4453, Felice_2009_80_063516, Cognola_2006_73}.

    In this chapter, we have explored a subclass of the $f(\mathcal{G})$ model to test its viability as an alternative to the standard cosmological paradigm. We have developed a numerical method to predict the redshift behaviour of the Hubble expansion rate, and our results suggest that the model can reproduce the low-redshift behaviour of the $\Lambda$CDM model but has significant differences at high redshifts. The $f(\mathcal{G})$ model is a feasible candidate for explaining the current epochs and effectively captures the evolution of energy components over cosmic time, supporting its validity as an alternative explanation for the observed acceleration of the Universe. We delved into the background cosmological dynamics of the chosen model and evaluated its feasibility using Bayesian analysis supported by MCMC methods applied to late-time cosmic observations, such as SNe Ia (Pantheon$^+$) and observational Hubble data (CC sample). We have introduced a dynamical system analysis framework to assess the stability of the model. Our research pinpointed critical points that illustrate different phases of the Universe and elucidated the evolutionary epochs. We have shown that the model effectively represents the changing energy components over cosmic time, supporting its credibility as an alternative explanation for the observed acceleration of the Universe.
    
    This chapter aims to establish constraints on $f(\mathcal{G})$ cosmology models using CC and Pantheon$^+$ data sets. The chapter comprehensively analyses $f(\mathcal{G})$ gravity and uses dynamical system analysis to investigate the stability of the model. The mathematical formalism of $f(\mathcal{G})$ gravity is detailed in section \ref{ch5_SEC II}. In section \ref{ch5_SEC III}, we use the MCMC method to establish correlations between the $f(\mathcal{G})$ gravity model and observational data to determine the best fits for the model parameters $H_0$, $\alpha$, $\beta$, and $m$. Additionally, we present plots of various cosmological parameters such as deceleration, effective EoS $\omega_{\text{eff}} = \frac{p_{\text{eff}}}{\rho_\text{eff}}$, Om diagnostic, and the $r-s$ parameter plot, which are essential for understanding the dynamical behaviour of the Universe under $f(\mathcal{G})$ gravity. Subsequently, in section \ref{ch5_SEC IV}, we construct a dynamical system framework to analyse the critical points of the $f(\mathcal{G})$ gravity model. This analysis is crucial for assessing the stability and viability of the model and its alignment with current cosmological observations. Finally, in section \ref{ch5_SEC V}, we present the conclusions of our results.
\section{\texorpdfstring{$R + f(\mathcal{G})$}{} Gravity} \label{ch5_SEC II}
    We consider an action that encompasses GR and a functional dependent on the Gauss--Bonnet term \cite{Nojiri_2005_631, Nojiri_2006_39}
    \begin{equation}
        S = \int d^4x \ \sqrt{-g} \left[ \frac{1}{2 \kappa^2} R + f(\mathcal{G}) + \mathcal{L}_\text{m} \right],
    \end{equation}
    The Gauss--Bonnet topological invariant is defined as
    \begin{equation}
        \mathcal{G} = R^2 - 4 R_{\mu \nu} R^{\mu \nu} + R_{\mu \nu \lambda \sigma} R^{\mu \nu \lambda \sigma}.
    \end{equation}
        By varying the action over $g_{\mu \nu}$, the following field equations are obtained
    \begin{eqnarray}
        0 &=& \frac{1}{2 \kappa^2} \left( - R^{\mu \nu} + \frac{1}{2} g^{\mu \nu} R \right) + T^{\mu \nu} + \frac{1}{2} g^{\mu \nu} f(\mathcal{G}) - 2 f_\mathcal{G} R R^{\mu \nu} + 4 f_\mathcal{G} R_\rho{}^{\mu} R^{\nu \rho} - 2 f_\mathcal{G} R^{\mu\rho\sigma\tau}R^\nu{}_{\rho\sigma\tau} \nonumber \\ & &- 4 f_\mathcal{G} R^{\mu\rho\sigma\nu} R_{\rho\sigma} + 2(\nabla^\mu \nabla^\nu f_\mathcal{G}) R - 2g^{\mu\nu}(\nabla^2f_\mathcal{G})R - 4 (\nabla_{\rho} \nabla^{\mu} f_\mathcal{G}) R^{\nu \rho} \nonumber \\ & &- 4 (\nabla_{\rho} \nabla^{\nu} f_\mathcal{G}) R^{\mu \rho} + 4 (\nabla^2 f_\mathcal{G}) R^{\mu \nu} + 4 g^{\mu \nu} (\nabla_{\rho} \nabla_{\sigma} f_\mathcal{G}) R^{\rho \sigma} - 4 (\nabla_{\rho} \nabla_{\sigma} f_\mathcal{G}) R^{\mu \rho \nu \sigma},
    \end{eqnarray}
        where we made the notations $f_\mathcal{G} = \frac{df}{d\mathcal{G}}$ and $f_{\mathcal{G}\mathcal{G}} = \frac{d^2 f}{d\mathcal{G}^2}$. By assuming a spatially flat FLRW Universe, we express the Ricci scalar $R$ and the Gauss--Bonnet invariant $\mathcal{G}$ as functions of the Hubble parameter as
    \begin{equation}
        \mathcal{G} = 24 \left( \dot{H} H^2 + H^4 \right), \quad R = 6 \left( \dot{H} + 2 H^2 \right).
    \end{equation}
        The field equations for the metric \eqref{ch2_flrw metric} yield the FLRW equations in the form
        \begin{eqnarray}
            3 H^2 &=&\kappa^2 \left(\rho_{\text{m}}+\rho_{\text{r}} + \rho_{\text{DE}}\right) = \kappa^2 \rho_{\text{eff}},  \label{ch5_first_field_equation}\\
            2 \dot{H}+3 H^2& =&-\kappa^2 \left(\frac{\rho_{\text{r}}}{3} + p_{\text{DE}}\right) = -\kappa^2 p_{\text{eff}}, \label{ch5_second_field_equation}
    \end{eqnarray}
    where $\rho_{\text{m}}$, $\rho_{\text{r}}$ and $\rho_{\text{DE}}$ denotes the matter density, radiation density and DE density, respectively. Furthermore, the effective DE density and pressure have been defined as follows
    \begin{subequations}
    \begin{eqnarray}
        & \rho_{\text{DE}} =\frac{1}{\kappa^2}\Big[ \mathcal{G} f_\mathcal{G} - f(\mathcal{G}) - 24 \dot{\mathcal{G}} H^3 f_{\mathcal{G}\mathcal{G}}\Big], \label{ch5_density field eq} \\
        &p_{\text{DE}} = \frac{1}{\kappa^2}\Big[ 8 H^2 \ddot{f}_\mathcal{G} + 16 H \left( \dot{H} + H^2 \right) \dot{f}_\mathcal{G} + f - \mathcal{G} f_\mathcal{G}\Big]\, . \label{ch5_pressure field eq}
    \end{eqnarray}
    \end{subequations}
    Without interactions between non-relativistic matter and radiation, these components independently follow their respective conservation laws $\dot{\rho}_{\text{m}}+3 H \rho_{\text{m}}=0$ and $\dot{\rho}_{\text{r}}+4 H \rho_{\text{r}}=0$. From Equations \eqref{ch5_density field eq} and \eqref{ch5_pressure field eq}, it can be concluded that the DE density and pressure follow the standard evolution equation
    \begin{eqnarray}
        \dot{\rho}_{\text{DE}}+3 H\left(\rho_{\text{DE}}+p_{\text{DE}}\right)=0.
    \end{eqnarray}
\subsection{Power-Law \texorpdfstring{$f(\mathcal{G})$}{} Model} \label{ch5_SEC II a}
        In light of the multiple cosmological data analyses and tests conducted within our solar system, all confirming the principles of GR, one can conclude that any departures from standard GR are expected to be negligible. Consequently, we led to consider the following $f(\mathcal{G})$ functional \cite{Davis_2007_0709.4453, Felice_2009_80_063516},
        \begin{eqnarray}
            f(\mathcal{G}) = \alpha \sqrt{\beta} \left(\frac{\mathcal{G}^2}{\beta^2}\right)^m \, ,
        \end{eqnarray}
        where $\alpha$, $\beta$ and $m$ are positive constants.
        
        In order to determine the theoretical values of the Hubble rate, we can calculate it by solving equation \eqref{ch5_density field eq} numerically. When we assume that matter behaves as a pressureless perfect fluid, we can express the matter density as $\rho_\text{m} = 3 H_0^2 \Omega_{\text{m}0} (1+z)^3$, with $z$ representing the cosmological redshift (defined as $\frac{a_0}{a} = 1 + z$, where $a_0$ represents the scale factor at present and $a$ denotes the scale factor when the light emitted) and $\Omega_{\text{m}0}$ representing the current value of the matter density parameter. Thus, for the particular model we are examining, the first Friedmann equation can be expressed as follows
        \begin{eqnarray} \label{ch5_HZ_ode}
            3 H^2 &=& 576^m (2 m-1) \frac{\alpha}{\beta ^{3/2}} H^6 \left(\frac{H^6 \left(H-(z+1) H'\right)^2}{\beta ^2}\right)^{m-1} \times \big[-2 m (z+1)^2 H H'' \nonumber\\ & & + (z+1) H' \big(2 (3 m-1) H -(6 m-1) (z+1) H'\big]+H^2\big) + 3 H_0^2 \Omega_{\text{m}0} (1+z)^3 \, ,
        \end{eqnarray}
        where the prime $(')$ indicates the derivative with respect to $z$.

        Equation \eqref{ch5_HZ_ode} represents a second-order differential equation for the function $H(z)$, which can be solved using appropriate boundary conditions. The first initial condition is simply $H(0) = H_0$. For the second initial condition to be determined, we can ensure that, at present, the first derivative of the Hubble parameter is consistent with the predictions of the standard $\Lambda$CDM model, which is characterized by the following expansion law
        \begin{eqnarray}
            H_{\Lambda \text{CDM}} = H_0 \sqrt{1 - \Omega_{\text{m}0} + \Omega_{\text{m}0} (1+z)^3}  \, ,
        \end{eqnarray}
        After differentiating the above equation with respect to $z$, the second initial condition for equation \eqref{ch5_HZ_ode} is obtained as $H'(0) = \frac{3}{2}H_0 \Omega_{\text{m}0}$.
\section{Observation with Numerical Solution} \label{ch5_SEC III}
        In this part, we will assess the observational feasibility of the model under examination by conducting a Bayesian analysis of the late-time cosmic data. Specifically, we will evaluate the data from the SNe Ia Pantheon$^+$ sample \cite{Brout_2022_938} and the CC derived from the observational Hubble data compiled in Ref. \cite{Moresco_2022_25}. We have not assumed that the Hubble and Pantheon$^+$ data sets are correlated. Rather, we will present our results independently for the CC and Pantheon$^+$ data sets. Utilizing these data sets for statistical analysis enables us to obtain reliable results that are not influenced by assumptions of any particular underlying reference model \cite{Agostino_2018_98, Agostino_2019_99, Lohakare_2023_40_CQG}. In the following subsections, we will outline the key characteristics of these measurements and the corresponding likelihood functions.
\subsection{Cosmic Chronometers} \label{ch5_SEC III a}
    To estimate the expansion rate of the Universe at redshift $z$, we use the widely used differential age (DA) method. In this way, it is possible to predict $H(z)$ using $(1+z) H(z)=-\frac{dz}{dt}$. The Hubble parameter is modeled on 32 data points (see \hyperref[Appendices]{Appendices}) for a redshift range of $0.07 \leq z \leq 1.965$ \cite{Moresco_2022_25}. 
\subsection{Supernovae Type Ia} \label{ch5_SEC III b}    
    We will incorporate the Pantheon$^+$ SNe Ia data set, which comprises 1701 measurements of the relative luminosity distance of SNe Ia across the redshift range of $0.00122 < z < 2.2613$ \cite{Brout_2022_938}. This compilation consists of distance moduli derived from 1701 light curves of 1550 spectroscopically confirmed SNe Ia sourced from 18 distinct surveys. Notably, 77 of these light curves pertain to galaxies harboring Cepheid variables. The Pantheon$^+$ data set offers the added advantage of constraining $H_0$ in addition to the model parameters. To estimate the model parameters using the Pantheon$^+$ samples, we will minimize the $\chi^2$ function. 
    \begin{figure}[H]
    \centering
        \includegraphics[width=110mm]{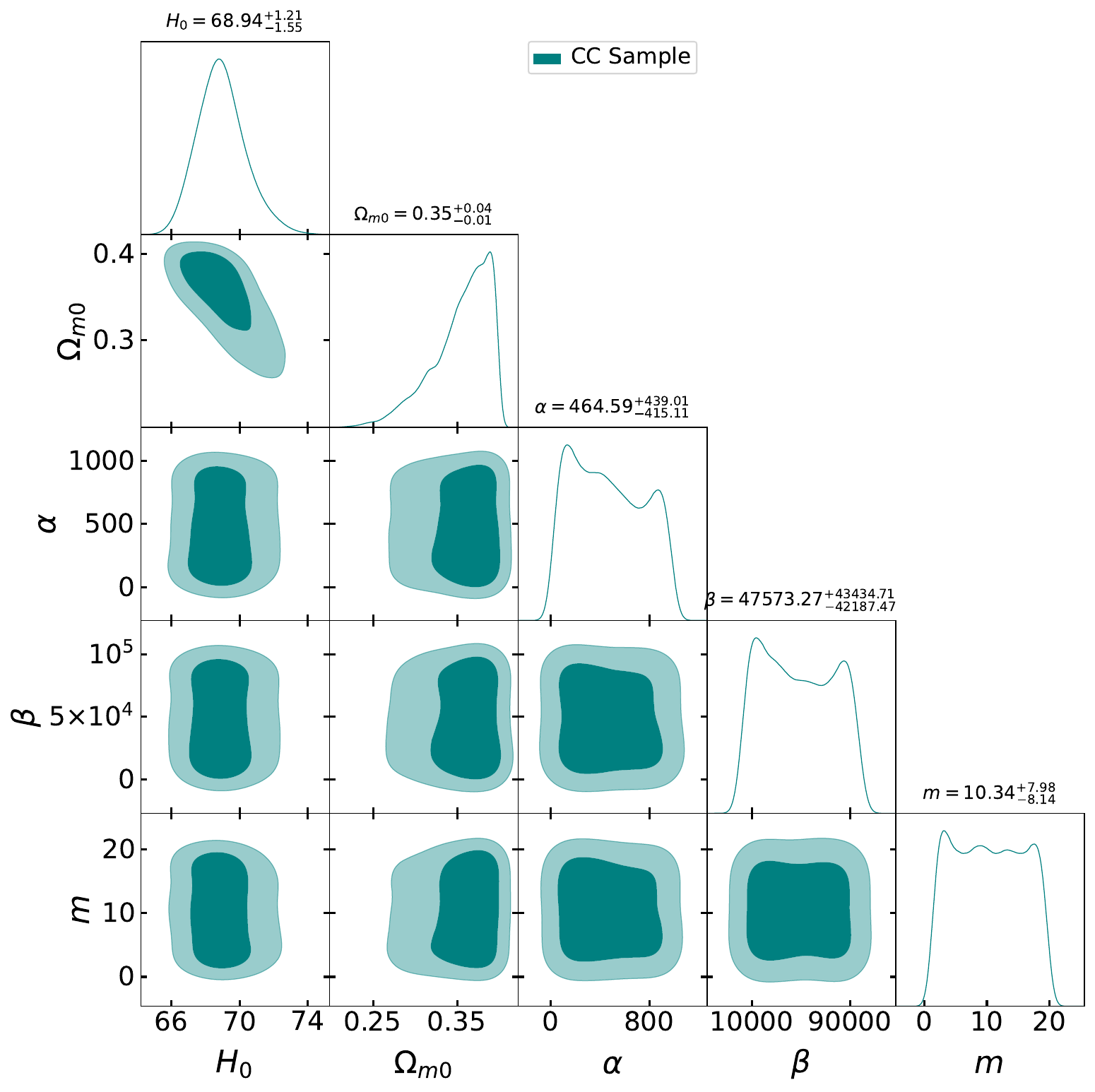}
        \includegraphics[width=110mm]{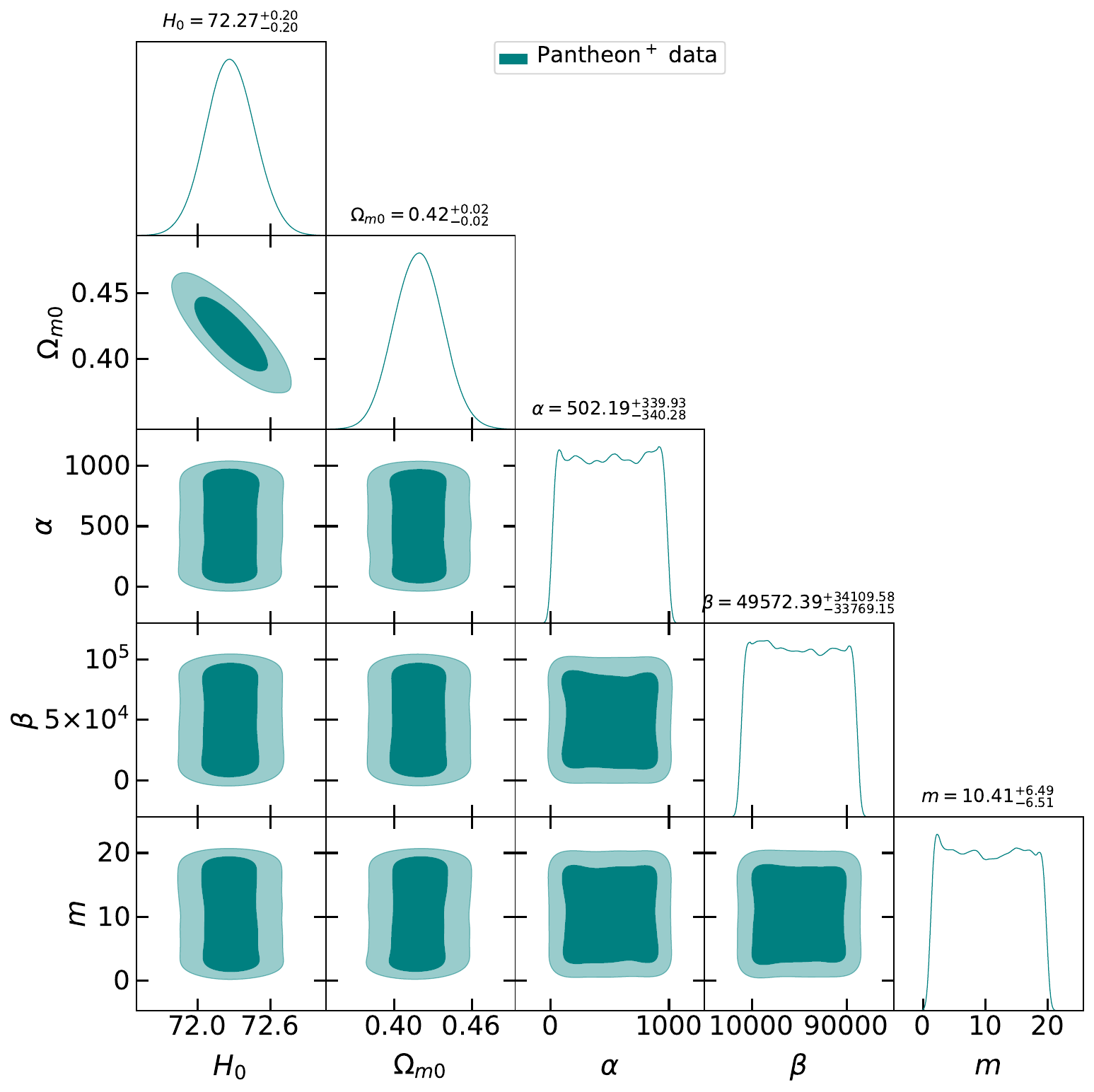}
    \caption{Contour plots show the $1\sigma$ and $2\sigma$ uncertainty regions for the variables $H_0$, $\Omega_{m0}$, $\alpha$, $\beta$, and $m$. These contours are derived from the CC sample (upper panel) and the Pantheon$^+$ data (lower panel).} 
    \label{ch5_FIG1}
    \end{figure} 
    \begin{figure}[H]
    \centering
        \includegraphics[width=100mm]{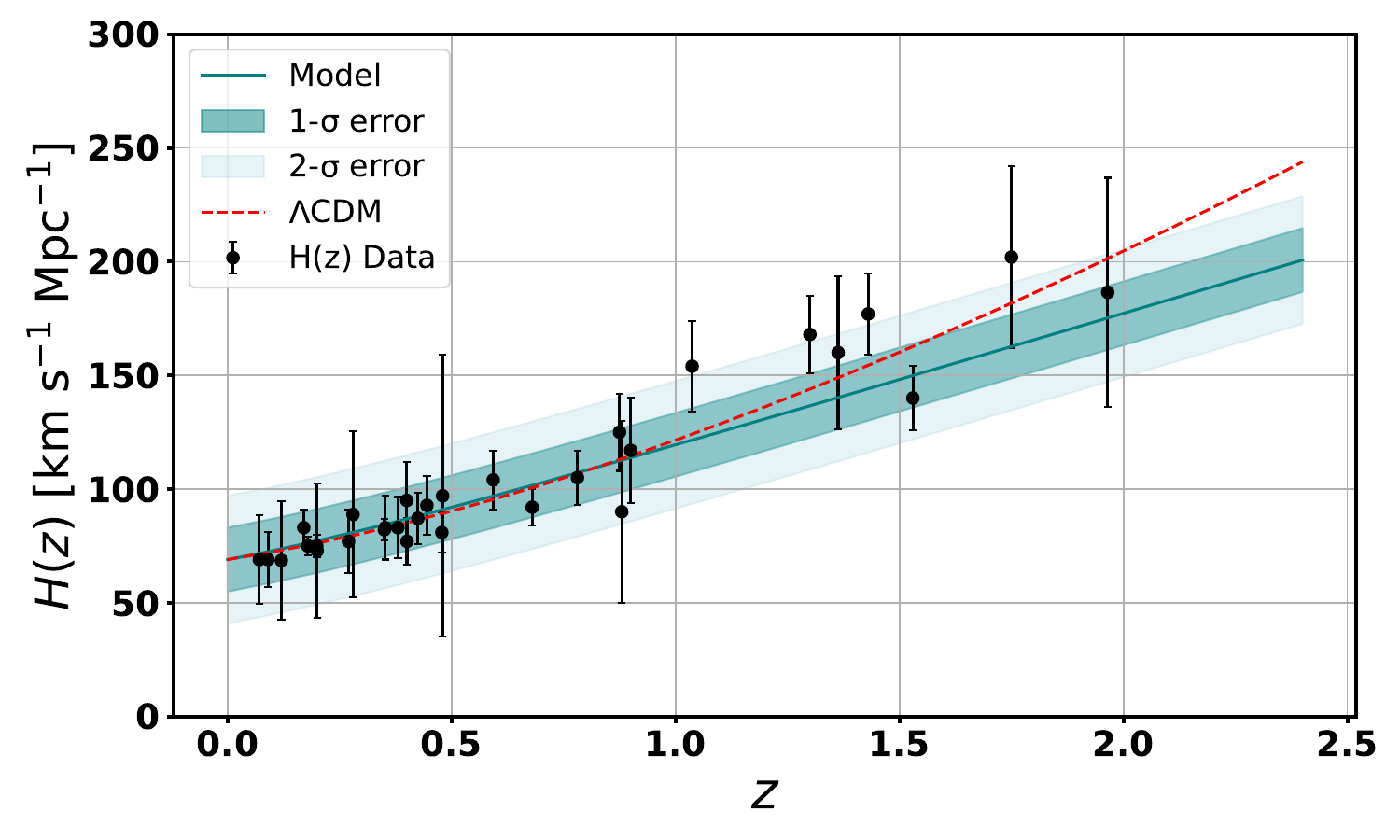}
        \includegraphics[width=100mm]{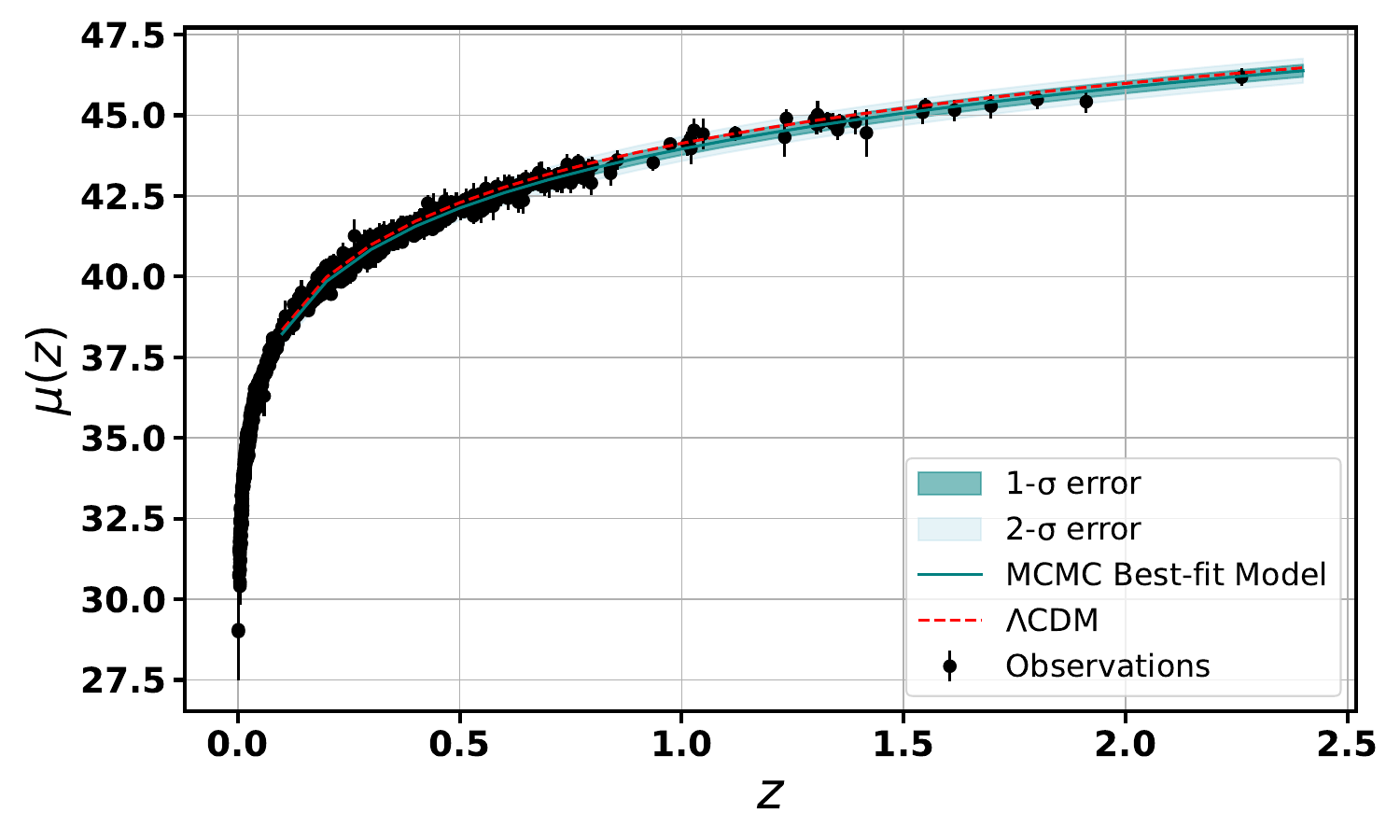}
    \caption{In the upper panel, the black error bars show uncertainty for 32 data points from the CC sample, with the solid teal line representing the model and the broken red line representing $\Lambda$CDM. In the lower panel, the solid teal line represents the distance modulus $\mu(z)$ of the model against redshift $z$, providing a superior fit to the 1701 data points from the Pantheon$^+$ data set with error bars.} 
    \label{ch5_FIG2}
    \end{figure} 
    We assess the models against the standard $\Lambda$CDM model using the Akaike Information Criterion (AIC) \cite{Akaike_1974_19} and the Bayesian Information Criterion (BIC) \cite{Schwarz_1978_6} in addition to $\chi^2_{\text{min}}$. Both AIC and BIC take into account the model’s goodness of fit as well as its complexity, which is influenced by the number of parameters $(n)$. The AIC is determined as
        \begin{eqnarray}
            \text{AIC} = \chi^2_{\text{min}} + 2 n \, .
        \end{eqnarray}
        In statistical modelling, a lower AIC value suggests a better fit to the data, taking into consideration the complexity of the model. This penalizes models with more parameters, even if they provide a better fit to the data. Alternatively, the BIC is calculated as
        \begin{eqnarray}
            \text{BIC} = \chi^2_{\text{min}} + n \, \text{ln} \, \mathcal{N},
        \end{eqnarray}
        where $\mathcal{N}$ represents the number of data samples used in the MCMC process. The corrected Akaike Information Criterion ($\text{AIC}_\text{c}$) is defined as
        \begin{eqnarray}
            \text{AIC}_\text{c} = \text{AIC} + \frac{2 n (n+1)}{\mathcal{N}-n-1} \, ,
        \end{eqnarray}
        given that the correction term becomes negligible for large sample sizes $(\mathcal{N} >> n)$, it is not restricted even in such cases. Therefore, it is always advantageous to employ $\text{AIC}_\text{c}$ over the original AIC.
        
        We evaluate the variances in AIC and BIC between the $f(\mathcal{G})$ model and the benchmark model, which is the $\Lambda$CDM model. As a result of this comparison, we can gain a deeper insight into how well each model matches the standard model of cosmology. The differences in AIC and BIC are expressed as $\Delta \text{AIC} = \Delta \chi^2_{\text{min}} + 2 \Delta n$, and $\Delta \text{BIC} = \Delta \chi^2_{\text{min}} + \Delta n \, \text{ln}\, m$, accordingly. A difference in $\text{AIC}_{\text{c}}$ between two competing models can be defined as $\Delta \text{AIC}_\text{c} = \text{AIC}_\text{c {$f(\mathcal{G})$}} - \text{AIC}_\text{c {$\Lambda \text{CDM}$}}$. These measures gauge how each model differs from the benchmark model, with smaller $\Delta$AIC and $\Delta$BIC values suggesting that a model, in conjunction with its selected data set, resembles the $\Lambda$CDM model more closely, indicating superior performance.
    \begin{table*} 
        \centering
        \scalebox{0.86}{
    \begin{tabular}{|*{6}{c|}}\hline
        {\centering  \textbf{Data sets}} & $H_0$ & $\Omega_{\text{m}0}$ & $\alpha$ & $\beta$ & $m$ \\ [0.5ex]
    \hline \hline
        {\centering \textbf{CC Sample}} & $68.944^{+1.210}_{-1.551}$ & $0.355^{+0.043}_{-0.014}$ & $464.593^{+439.011}_{-415.110}$ & $47573.273^{+43434.714}_{-42187.475}$ & $10.342^{+7.984}_{-8.144}$  \\
    \hline
        {\centering \textbf{Pantheon}$^+$} & $72.270^{+ 0.200}_{-0.201}$ & $0.420^{+0.020}_{-0.020}$ & $502.193^{+339.937}_{-340.285}$ & $49572.390^{+34109.581}_{-33769.153}$ & $10.412^{+6.491}_{-6.510}$ \\[0.5ex] 
    \hline
    \end{tabular}}
        \caption{The table presents an exploration of the parameters for the MCMC algorithm. It displays the best-fit values for the model parameters, including $H_0$, $\Omega_{\text{m}0}$, $\alpha$, $\beta$, and $m$, derived from the MCMC study using the CC and Pantheon$^+$ data sets.}
        \label{ch5_TABLE I a}
    \end{table*}
    \begin{table*}
        \centering
        \scalebox{0.6}{
        \begin{tabular}{| *{12}{c|} }
    \hline
        {Data sets}  &\multicolumn{2}{c|}{$\chi^2_{\text{min}}$} &\multicolumn{2}{c|}{AIC} &\multicolumn{2}{c|}{AIC$_\text{c}$} &\multicolumn{2}{c|}{BIC} & {$\Delta \text{AIC}$} & {$\Delta \text{AIC}_\text{c}$} & {$\Delta \text{BIC}$} \\
    \cline{2-9}
        & $f(\mathcal{G})$ & $\Lambda$CDM & $f(\mathcal{G})$ & $\Lambda$CDM & $f(\mathcal{G})$ & $\Lambda$CDM & $f(\mathcal{G})$ & $\Lambda$CDM & & &\\ 
    \hline \hline
        CC sample   & 26.132 & 29.046 & 36.132 & 33.046 & 38.439 & 33.459 & 33.682 & 32.066 & 3.086 & 4.98 & 1.616 \\
    \hline
        Pantheon$^+$  & 1618.774 & 1625.224 & 1628.774 & 1629.224 & 1628.809 & 1629.231 & 1634.924 & 1631.684 & -0.45 & -0.422 & 3.24 \\
    \hline
    \end{tabular}}
        \caption{The table provides minimum $\chi^2$ values for the $f(\mathcal{G})$ model, along with their corresponding AIC, AIC$_\text{c}$, and BIC values, and a comparison of AIC, AIC$_\text{c}$, and BIC differences between the model and $\Lambda$CDM.}
        \label{ch5_TABLE I b}
    \end{table*}
        The contour plots (see figure \ref{ch5_FIG1}) display the $1\sigma$ and $2\sigma$ uncertainty regions for the parameters $H_0$, $\Omega_{\text{m}0}$, $\alpha$, $\beta$, and $m$ using two data sets: the CC data and the Pantheon$^+$ data. These plots, derived using the MCMC method, illustrate the marginalized posterior distributions of parameter pairs, with inner and outer contours representing $68\%$ and $95\%$ confidence levels, respectively.

        The contour plot for the parameters $H_0$ and $\Omega_{\text{m}0}$ exhibits a more elliptical shape compared to the other parameter pairs, which display more square or elongated contours. This elliptical shape indicates that there is a relatively weak correlation between $H_0$ and $\Omega_{\text{m}0}$, suggesting that the data independently constrains these two parameters. This independence implies that the variations in one parameter do not significantly affect the value of the other, leading to a more symmetrical uncertainty region. In contrast, the square or elongated contours seen in other parameter pairs indicate stronger correlations, where changes in one parameter can be offset by adjustments in another to maintain a similar fit to the data. This strong coupling results in less symmetrical and more stretched uncertainty regions, reflecting the interdependence of these parameters within the $f(\mathcal{G})$ gravity model. The best-fit values derived from the MCMC analysis are presented in Table \ref{ch5_TABLE I a}. In order to evaluate the efficacy of our MCMC analysis, we calculated the associated AIC, $\text{AIC}_\text{c}$ and BIC values, which are presented in Table \ref{ch5_TABLE I b}. Our findings strongly support the assumed $f(\mathcal{G})$ gravity models when analysing the data sets. Moreover, we noted that the $f(\mathcal{G})$ model demonstrates greater precision when applied to the Pantheon$^+$ data sets.

        The upper panel of figure \ref{ch5_FIG2} shows that the $f(\mathcal{G})$ gravity model fits the observational $H(z)$ data well across the redshift range considered. Both the $f(\mathcal{G})$ gravity and $\Lambda$CDM models follow similar trends, but the $f(\mathcal{G})$ model predicts lower $H(z)$ values at higher redshifts ($z > 1$). This deviation suggests distinct underlying physics due to higher-order curvature terms in the $f(\mathcal{G})$ model, which also accounts for late-time cosmic acceleration without a cosmological constant ($\Lambda$). The lower panel of figure \ref{ch5_FIG2} illustrates that the $f(\mathcal{G})$ gravity model fits the distance modulus $\mu(z)$ data excellently across the redshift range. Both the $f(\mathcal{G})$ gravity and $\Lambda$CDM models align closely with the observational data, with minimal deviation between them. The strong agreement with observational data supports its viability as a competitive alternative to the $\Lambda$CDM model, with its natural incorporation of higher-order curvature terms making it an attractive option for future cosmological studies. 
    \begin{figure}[H]
        \centering
        \includegraphics[width=160mm]{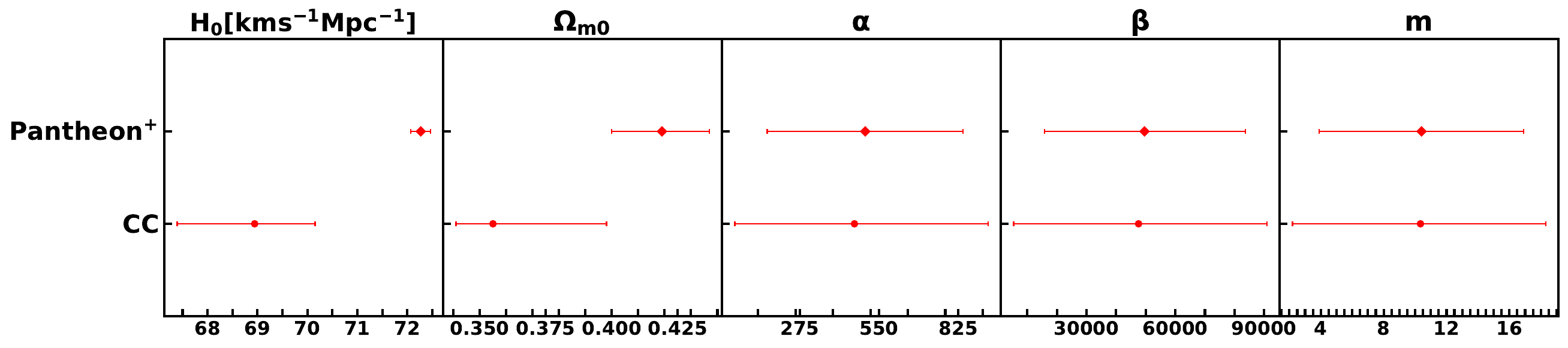}
        \caption{Whisker plot depicting the model parameters $H_0$, $\Omega_{\text{m}0}$, $\alpha$, $\beta$ and $m$, respectively, highlights their discrepancies.} 
        \label{ch5_FIG: whisker plot}
    \end{figure}
    Our analysis reveals a significant discrepancy between the Hubble constant $H_0$ values derived from the CC sample and the Pantheon$^+$ data sets, as shown in figure \ref{ch5_FIG: whisker plot}. This whisker plot highlights the ongoing $H_0$ tension in cosmology by presenting the model parameters $H_0$, $\Omega_{\text{m}0}$, $\alpha$, $\beta$, and $m$ along with their $1\sigma$ confidence intervals.
\subsection{Cosmological Parameter Evolution}\label{ch5_SEC III c}
    We analyse the evolution of crucial cosmological parameters, including the effective EoS, state-finder, and Om diagnostic parameters, by imposing constraints on model parameters using various observational data. We present a fully general expression for the deceleration parameter, $q = -\ddot{a}/aH^2$ as follows
        \begin{eqnarray}
            q(z) = -1 + \frac{H'(z)}{H(z)} (1+z)\, .
        \end{eqnarray}
        \begin{figure}[H]
        \centering
        \includegraphics[width=79mm]{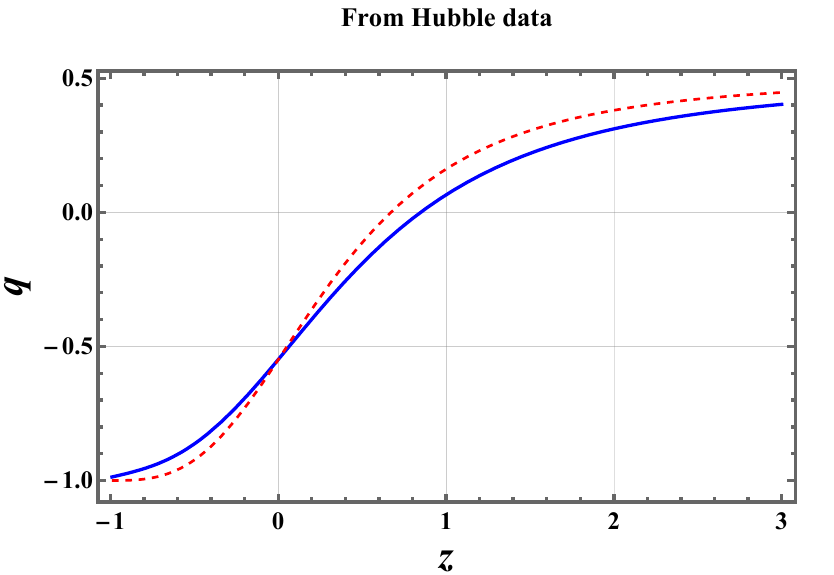}
        \includegraphics[width=79mm]{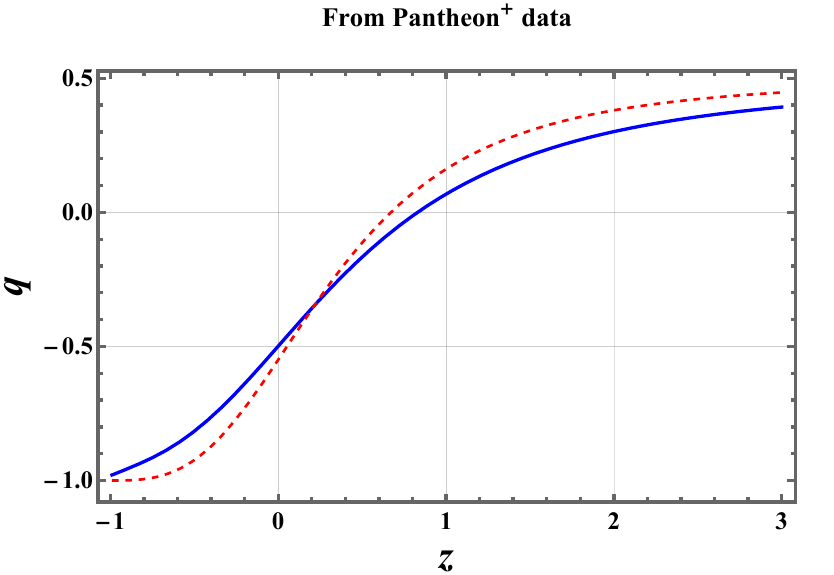}
        \caption{Graphical representation of the deceleration parameter versus redshift using the constrained coefficients from figure \ref{ch5_FIG1}. The thick line represents the behaviour of the deceleration parameter for the $f(\mathcal{G})$ model, while the dashed line shows the deceleration parameter for the $\Lambda$CDM model.}
        \label{ch5_FIG3}
        \end{figure}
    Figure \ref{ch5_FIG3} presents the deceleration parameter as a function of redshift $z$ for the $f(\mathcal{G})$ gravity model, derived from both Hubble data (upper panel) and Pantheon$^+$ data (lower panel). Figure \ref{ch5_FIG3} demonstrate that the restricted values of model parameters derived from the analysed CC and Pantheon$^+$ data sets indicate a transition of $q$ from positive (indicating early deceleration) in the past to negative (indicating current acceleration) in the present. The present value of deceleration parameter $q_0$ is measured to be $-0.527$ and $-0.499$ for the CC and Pantheon$^+$ data, respectively, at the current cosmic epoch, which aligns relatively well with the range of $q_0 = -0.528^{+0.092}_{-0.088}$ determined by recent observations \cite{Christine_2014_89}. Recent observations are consistent with this deceleration parameter, and the resulting model indicates a smooth transition from deceleration to acceleration at $z_t = 0.84$ and $z_t = 0.82$ for the CC and Pantheon$^+$ data sets, respectively. The derived transition redshift $z_t$ aligns with current constraints based on 11 $H(z)$ observations reported by Busca et al. \cite{Busca_2013_552} for redshifts $0.2 \leq z \leq 2.3$, $z_t = 0.74 \pm 0.5$ from Farooq and Ratra \cite{Farooq_2013_766}, $z_t = 0.7679^{+0.1831}_{-0.1829}$ by Capozziello et al. \cite{Capozziello_2014_90_044016}, and $z_t = 0.60^{+0.21}_{-0.12}$ by Yang and Gong \cite{Yang_2020_2020_059}. The consistency between the curves from Hubble and Pantheon$^+$ data underscores the robustness of the $f(\mathcal{G})$ model in capturing the expansion dynamics of the Universe.
    \begin{figure}[H]
        \centering
        \includegraphics[width=79mm]{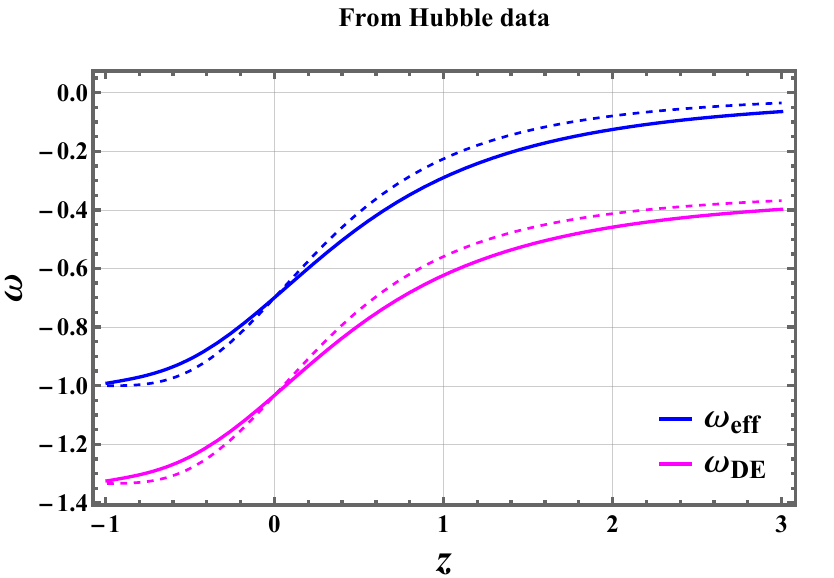}
        \includegraphics[width=79mm]{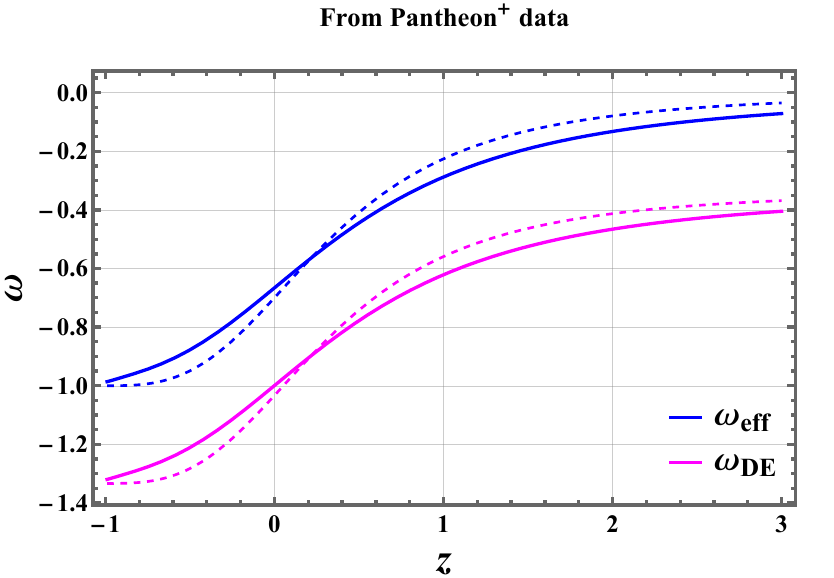}
    \caption{Graphical representation of the EoS parameter versus redshift using the constrained coefficients from figure \ref{ch5_FIG1}. The thick line represents the behaviour of the EoS parameter for the $f(\mathcal{G})$ model, while the dashed line shows the EoS parameter for the $\Lambda$CDM model.}
    \label{ch5_FIG4}
    \end{figure}
    The deceleration parameter is one of the key factors that characterize the behaviour of the Universe, determining whether it continuously decelerates, accelerates, or undergoes multiple phases of transition. Similarly, energy sources influence the evolution of the Universe through the EoS parameter, defined as $\omega_{\text{DE}}$ is shown in figure \ref{ch5_FIG4}. By calculating the energy density and pressure of DE, as depicted in figure \ref{ch5_FIG4}, we can observe the variations in the effective EoS of DE relative to the redshift variable. The current EoS values for DE, $\omega_{\text{DE}}(z = 0)$, are obtained as $-1.018$, $-0.999$ for the CC, Pantheon$^+$, data sets, respectively. These values indicate phantom behaviour (at $z < 0$) and a trend towards approximately $-1.32$ at late times. The present values of $\omega_{\text{eff}}$ are $-0.684$, $-0.666$ for the CC and Pantheon$^+$ data sets, respectively. Various cosmological studies have also constrained the EoS parameter, including the Supernovae Cosmology Project $\omega_{\text{DE}}=-1.035^{+0.055}_{-0.059}$ \cite{Amanullah_2010_716}, Planck 2018 $\omega_{\text{DE}}=-1.03\pm 0.03$ \cite{Aghanim_2020_641}, and WAMP+CMB $\omega_{\text{DE}}=-1.079^{+0.090}_{-0.089}$ \cite{Hinshaw_2013_208}.

    Figure \ref{ch5_FIG4} shows the effective EoS, $\omega_{\text{eff}}$ as a function of redshift $z$ for the $f(\mathcal{G})$ gravity model. At low redshifts ($z \approx 0$), $\omega_{\text{eff}}$ is close to $-1$, indicating DE dominance and accelerated expansion. As $z$ increases, $\omega_{\text{eff}}$ transitions from negative values to less negative values, reflecting a shift from a deceleration phase in the early Universe to an acceleration-dominated phase in the current epoch. The consistent behaviour of $\omega_{\text{DE}}$ across both data sets reinforces the capability of the $f(\mathcal{G})$ model to describe the expansion history of the Universe. This transition aligns with theoretical expectations of the $f(\mathcal{G})$ model, which incorporates higher-order curvature terms to account for cosmic dynamics without a cosmological constant. The observed $\omega_{\text{DE}}$ behaviour highlights the effectiveness of the model, supporting its viability as an alternative to the $\Lambda$CDM model.    
    \begin{figure}[H]
        \centering
        \includegraphics[width=79mm]{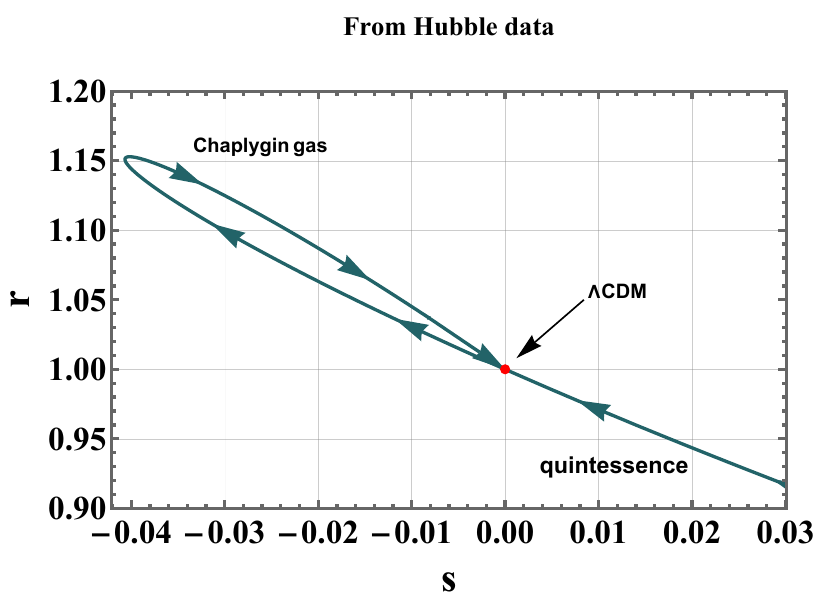}
        \includegraphics[width=79mm]{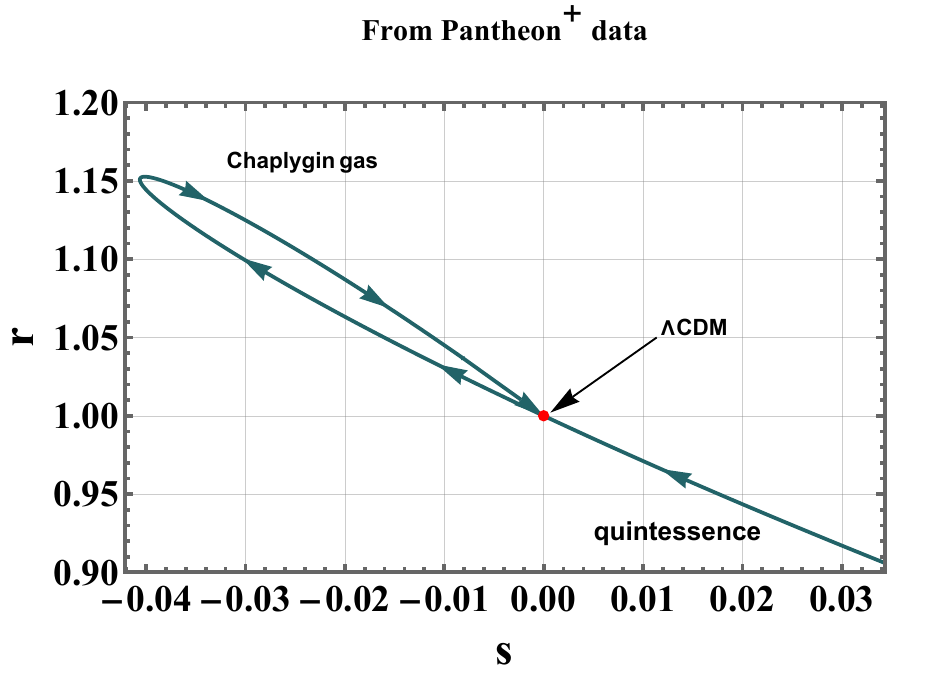}
    \caption{A plot showing the evolution of the given cosmological model in the $r - s$ plane using the constrained coefficients from figure \ref{ch5_FIG1}.} 
    \label{ch5_FIG5}
    \end{figure}
    The statefinder diagnostic proposed by V. Sahni \cite{Sahni_2003_77_201} provides a geometric method for discerning different DE models using statefinder parameters.
        \begin{eqnarray}
            r = \frac{\dot{\ddot{a}}}{a H^3}, \,\,\,\,\,\,\,\,\,\, s = \frac{r - 1}{3 (q-\frac{1}{2})},
        \end{eqnarray}
    The conditions $(r < 1, s > 0)$ correspond to the quintessence of DE, while the domain $(r > 1, s < 0)$ represents the phantom scenario. Additionally, the state $(r = 1, s = 0)$ reproduces the standard $\Lambda$CDM model.
        
    Figure \ref{ch5_FIG5} illustrates the $r-s$ parameter plot for the $f(\mathcal{G})$ gravity model. The trajectory in the $r-s$ plane highlights the evolutionary track of the expansion of the Universe. The $f(\mathcal{G})$ model passes through the region corresponding to the $\Lambda$CDM model, indicated by the red point. At lower values of $s$, the model aligns with quintessence characteristics, suggesting a dynamical DE component with $\omega > -1$. As $s$ increases, the trajectory moves towards regions associated with Chaplygin gas models, indicating a unified dark matter and DE scenario. The smooth transition observed in the $r-s$ parameter space demonstrates the flexibility of the $f(\mathcal{G})$ model in describing different cosmological behaviours. This capability allows the model to account for various dynamics, from quintessence-like to Chaplygin gas-like, providing a comprehensive description of the expansion of the Universe. 

    The $\text{Om}(z)$ diagnostic is a simple testing method that depends only on the first-order derivative of the cosmic scale factor. In particular, in DE theories, the $\text{Om}(z)$ parameter is followed as an additional effective diagnostic tool \cite{Sahni_2008_78_103502, Sahni_2003_77_201} and alternatively stated for simplification, when $\text{Om}(z_1, z_2) > 0$, it indicates quintessence, whereas when $\text{Om}(z_1,z_2) < 0$, it signifies phantom behaviour, where $(z_1 < z_2)$. The $\text{Om}(z)$ diagnostic in the $\Lambda$CDM model serves as a null test, as noted in \cite{Sahni_2008_78_103502}, and its sensitivity to the EoS parameter was further explored in subsequent data as seen in \cite{Qi_2018_18_066, Zheng_2016_825_17, Ding_2015_803_L22}. If $\text{Om}(z)$ remains constant for the redshift, the DE concept would be a cosmological constant. The DE concept will form a cosmological constant if $\text{Om}(z)$ is constant for the redshift. The slope of $\text{Om}(z)$, which is positive for the emerging $\text{Om}(z)$ and denotes phantom phase $(\omega<-1)$ and negative for quintessence region $(\omega > -1)$ also identifies the DE models.
    \begin{figure} [H]
        \centering
        \includegraphics[width=79mm]{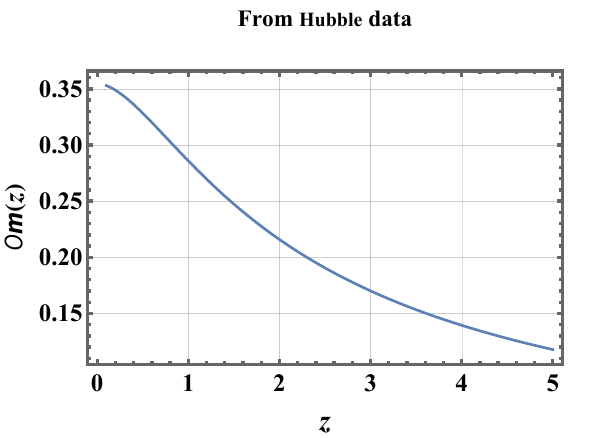}
        \includegraphics[width=79mm]{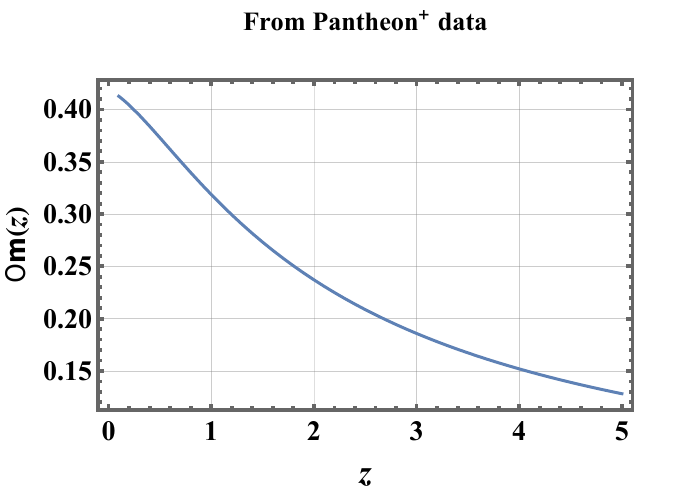}
    \caption{$\text{Om}(z)$ diagnostic parameter profile for the cosmological model using the constrained coefficients from figure \ref{ch5_FIG1}.} 
    \label{ch5_FIG6}
    \end{figure}
        The graph in figure \ref{ch5_FIG6} shows the reconstructed $\text{Om}(z)$ parameter based on the best-fitting data, plotted against redshift. It illustrates a decreasing trend in the $\text{Om}(z)$ parameter as redshift increases. At higher redshifts ($z > 1$), the Om diagnostic curve shows a significant decline, indicating the deviation of the $f(\mathcal{G})$ gravity model from the $\Lambda$CDM model. This behaviour suggests that the higher-order curvature terms in the $f(\mathcal{G})$ model influence the cosmic dynamics differently compared to the standard cosmological model. The shape of the Om diagnostic curve for the $f(\mathcal{G})$ gravity model highlights its potential to account for cosmic acceleration through modifications to gravity. This diagnostic tool effectively illustrates the differences between the $f(\mathcal{G})$ model and the $\Lambda$CDM model, reinforcing the former’s viability as an alternative explanation for the observed acceleration of the Universe.

\section{Stability Assessment via Dynamical Systems} \label{ch5_SEC IV}
    We used the autonomous dynamical system method to study the problem due to the complicated form of the equation \eqref{ch5_density field eq}. For a general $f(\mathcal{G})$ model, it will be helpful to introduce the following variables
    \begin{align}
        X = H^2 f_{\mathcal{G}} \,,\quad Y = H \dot{f}_{\mathcal{G}} \,,\quad Z=\frac{\dot{H}}{H^2} \,,\quad 
        W=-\frac{f}{6H^2}\,,\quad V=\frac{\kappa^2 \rho_r}{3H^2}\, .\label{ch5_dynamical variables}
    \end{align}
    alongside the density parameters
    \begin{align}
        \Omega_{\text{m}}=\frac{\kappa^2 \rho_{\text{m}}}{3H^2}, \quad \Omega_{\text{r}}=V=\frac{\kappa^2 \rho_{\text{r}}}{3H^2}, \quad \Omega_{\text{DE}}=\frac{\kappa^2 \rho_{\text{DE}}}{3H^2},
    \end{align}
        with the constraint
    \begin{align}   \Omega_{\text{m}}+\Omega_{\text{r}}+\Omega_{\text{DE}}=1.
    \end{align}
        In terms of dynamical variables, we have
    \begin{align} \label{ch5_constraint eq}
        \Omega_{\text{m}}+\Omega_{\text{r}} + 8 X Z + 8 X + 2 W - 8 Y=1,
    \end{align}
        and
    \begin{align}
        \Omega_{\text{DE}} = 8 X Z + 8 X + 2 W - 8 Y.
    \end{align}
        In order to study the time-dependent behaviour of the dynamical system, it is necessary to establish a dimensionless time parameter. In this study, we choose to use a time parameter expressed as the number of e-folds $N \equiv \text{ln}\, a/a_0$, where $a_0$ is a constant with the same units as $a$, and is typically set as $a_0 \equiv 1$. The evolution of each variable is then determined by its derivative with respect to $N$, which is expressed as follows
    \begin{subequations}\label{ch5_generaldynamicalsystem}
    \begin{eqnarray}
        \frac{dX}{dN}& = & 2XZ + Y\,, \\
        \frac{dY}{dN}& = & Y Z + (3 X - 2 Y) (Z + 1) + \frac{3 W}{4} - \frac{Z}{4} - \frac{3}{8} - \frac{V}{8}\,, \\
        \frac{dZ}{dN}& = & \lambda - 2 Z^2\,, \\
        \frac{dW}{dN}& = & - 4 X (4 Z + 2 Z^2 + \lambda)-2 W Z\,, \\
        \frac{dV}{dN}& = & -2 V (2 + Z) \,.
    \end{eqnarray}
    \end{subequations}
    \begin{table*}
    \centering
    \scalebox{0.72}{
    \begin{tabular}{|*{8}{c|}}\hline 
        {\textbf{Critical Point}} & $\textbf{X}$ & $\textbf{Y}$ & $\textbf{Z}$ & $\textbf{V}$ & \textbf{Existence} & $\omega_{\text{eff}}$ & $q$\\ [0.5ex]\hline \hline 
        {$\mathcal{P}_1= (x_1,y_1,z_1,v_1)$} & $x_1$ & $4 x_1$ & $-2$ & $1 + 24 x_1$ & \begin{tabular}{@{}c@{}}$x_{1} \neq 0$,\,\, $m = \frac{1}{4}$ \end{tabular} & $\frac{1}{3}$ & $1$ \\
    \hline
        {$\mathcal{P}_2= (x_2,y_2,z_2,v_2)$} & $x_{2}$ & $\frac{1}{4} (-1-8 x_2)$ & $1+ \frac{1}{8 x_2}$ & $0$ &  \begin{tabular}{@{}c@{}}$x_{2} \neq 0$,\,\, $m = \frac{1}{4}$ \end{tabular}  & $0$ & $\frac{1}{2}$ \\
    \hline
        {$\mathcal{P}_3 = (x_3,y_3,z_3,v_4)$} & $\frac{-1}{16}$ & $\frac{-1}{8}$ & $-1$ & $0$ &  \begin{tabular}{@{}c@{}}$ 2 m^2 - m \neq 0$ \end{tabular} & $- \frac{1}{3}$ & $0$ \\
    \hline
        {$\mathcal{P}_4 = (x_4,y_4,z_4,v_4)$} & $\frac{m}{-4+8 m}$ & $0$ & $0$ & $0$ & \begin{tabular}{@{}c@{}}$(-1+2 m) (-1+ 4 m) \neq 0$, $m \neq 0$ \end{tabular} & $-1$ & $-1$ \\
    \hline
    \end{tabular}}
        \caption{The critical points and physical characteristics of the system.}
    \label{ch5_TABLE-II}
    \end{table*}   
    \begin{table}
    \centering 
    \begin{tabular}{|*{8}{c|}}\hline 
        {\textbf{Critical Point}} & $\Omega_{\text{m}}$ & $\Omega_{\text{r}}$ & $\Omega_{\text{DE}}$ & $\text{Acceleration}$\\ [0.5ex]\hline \hline 
        {$\mathcal{P}_1$} & $0$ & $1+ 24 x_1$ & $-24 x_1$ & Never\\
    \hline
        {$\mathcal{P}_2$} & $1 + 20 x_{2}$ & $0$ & $-20 x_2$ & Never\\
    \hline
        {$\mathcal{P}_3$} & $0$ & $0$ & $1$ & Never\\
    \hline
        {$\mathcal{P}_4$} & $0$ & $0$ & $1$ & Always\\
    \hline
    \end{tabular}
        \caption{The density parameters associated with the critical point.}
    \label{ch5_TABLE-III}
    \end{table}
    \begin{table*}
    \centering
    \scalebox{0.77}{
    \begin{tabular}{|*{8}{c|}}\hline 
        {\textbf{Critical Point}} & $\textbf{Eigenvalues}$ & $\textbf{Stability}$ \\ [0.5ex]\hline \hline 
        {$\mathcal{P}_1$} & {$\left\{0, 1, \frac{-x_1+\sqrt{-x_1 (2+47 x_1)}}{2 x_1}, \frac{-x_1-\sqrt{-x_1 (2+47 x_1)}}{2 x_1} \right\}$} & Unstable \\
    \hline
        {$\mathcal{P}_2$} & {$\left\{0, -1, \frac{- 3 x_2+\sqrt{-4 x_2 - 71 {x_2}^2)}}{4 x_2}, \frac{- 3 x_2-\sqrt{-4 x_2 - 71 {x_2}^2)}}{4 x_2} \right\}$} & Stable for $-\frac{4}{71} \leq x_2 < -\frac{1}{20}$ \\
    \hline
        {$\mathcal{P}_3$} &  {$\left\{-2, -2, -1, 4+\frac{2}{-1+2 m} \right\}$} & Stable for $\frac{1}{4} < m <\frac{1}{2}$ \\
    \hline
        {$\mathcal{P}_4$} & {$\left\{-4, -3, \frac{3 m - 6 m^2 -\sqrt{(1-2 m)^2 (25 m -4) m}}{2 m (2 m - 1)}, \frac{3 m - 6 m^2 + \sqrt{(1-2 m)^2 (25 m -4) m}}{2 m (2 m - 1)} \right\}$} & Stable for $\frac{4}{25} \leq m < \frac{1}{4}$ \\
    \hline
    \end{tabular}}
    \caption{Eigenvalues and stability regime.}
    \label{ch5_TABLE-IV}
    \end{table*}
    Taking into account that $f(\mathcal{G}) = \alpha \sqrt{\beta} \left(\frac{\mathcal{G}^2}{\beta^2}\right)^m$, from equation \eqref{ch5_dynamical variables} we get
    \begin{eqnarray} \label{ch5_dependent_w}
        W = \frac{2 X (1+Z)}{-m} \, .
    \end{eqnarray}
    In order to obtain an expression for $\lambda = \frac{\ddot{H}}{H^3}$, we can use equation \eqref{ch5_dependent_w}
    \begin{eqnarray}
        \lambda = -\frac{m W (4 X Z + Y) + 8 m X^2 Z (Z + 2) + 4 X^2 Z^2}{2 (2 m-1) X^2} \, ,
    \end{eqnarray}
        As per the given relations and the constraint \eqref{ch5_constraint eq} and dependency relation \eqref{ch5_dependent_w}, we can remove the equations for $W$ from our autonomous system, leaving us with only a set of four equations
    \begin{subequations}\label{ch5_Autonomous DS}
        \begin{eqnarray}
            \frac{dX}{dN}& = & 2XZ + Y\,,\\
            \frac{dY}{dN}& = & -\frac{3 X Z}{2 m}-\frac{3 X}{2 m}-\frac{V}{8} + 3 X Z + 3 X - Y Z - 2  Y - \frac{Z}{4}-\frac{3}{8}\,, \\
            \frac{dZ}{dN}& = & \frac{(Z + 1) \big(Y + 4 (1-2 m) X Z \big)}{(2 m-1) X}\,, \\
            \frac{dV}{dN}& = & -2 V (2 + Z) \,.
        \end{eqnarray}
    \end{subequations}
    \begin{figure}[H]
    \centering
    \subfigure[]
        {\includegraphics[width=50mm]{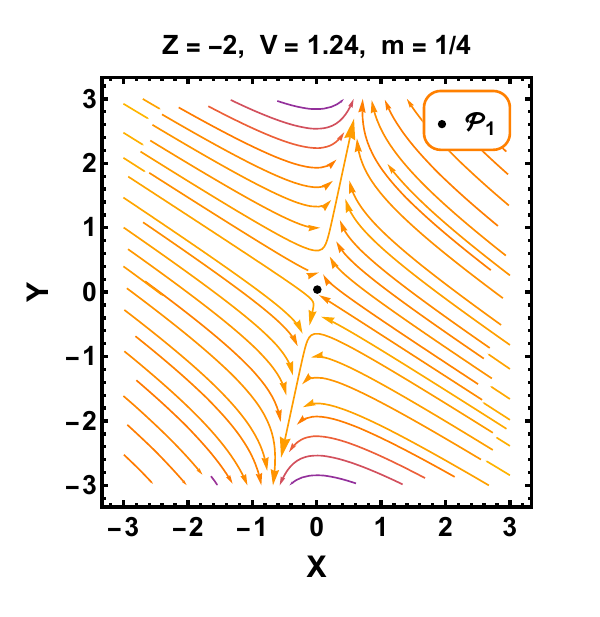} \label{ch5_portrait_P1}} 
    \subfigure[]
        {\includegraphics[width=50mm]{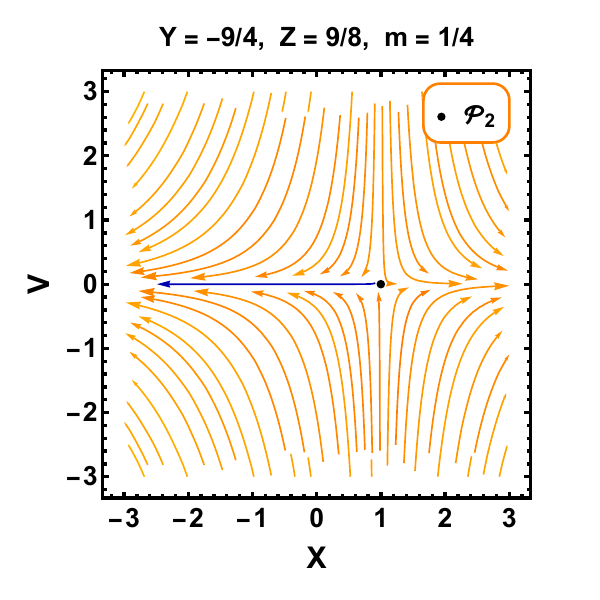} \label{ch5_portrait_P2}} \\
    \subfigure[]
        {\includegraphics[width=50mm]{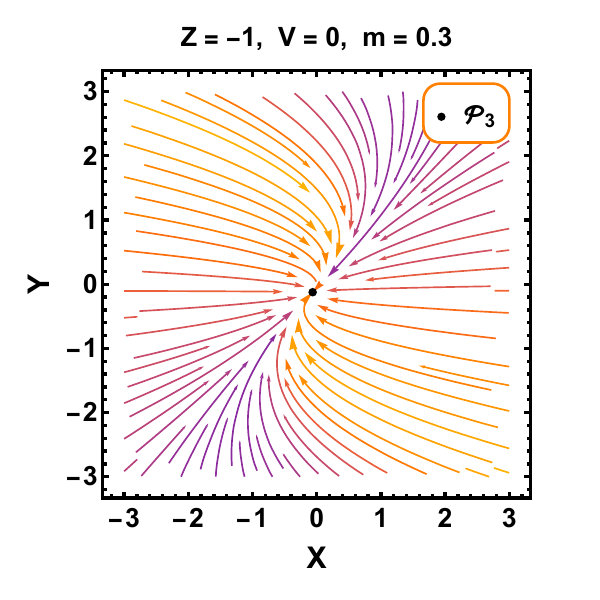} \label{ch5_portrait_P4}}
    \subfigure[]
        {\includegraphics[width=50mm]{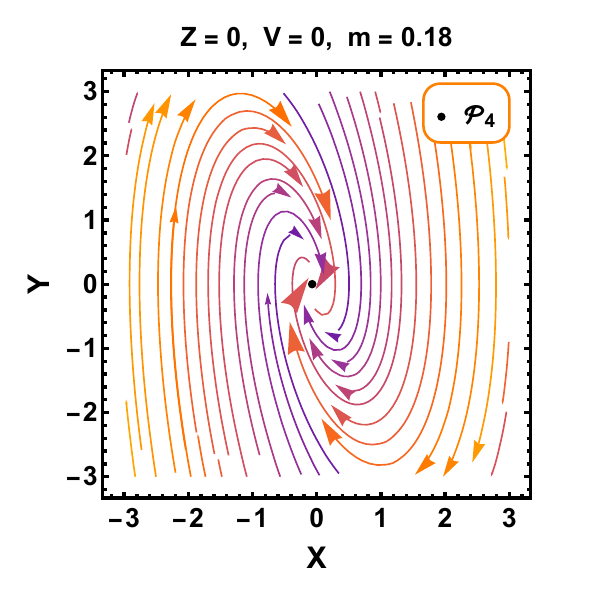} \label{ch5_portrait_P5}}
    \caption{Two-dimensional phase portrait for the dynamical system.} 
    \label{ch5_FIG7}
    \end{figure}
    The two-dimensional phase portraits shown in figure \ref{ch5_FIG7} depict the dynamics of the system for $m = \frac{1}{4}, 0.3, 0.18$ by mapping the trajectories onto the $XY$ and $XV$ planes. These visualizations provide insights into the stability and nature of the critical points $\mathcal{P}_1$ to $\mathcal{P}_4$.
    \begin{itemize}
        \item {\bf Critical point $\mathcal{P}_1$:} The critical point $\mathcal{P}_1$ is identified on the $XY$ plane in figure \ref{ch5_portrait_P1}. The phase portrait indicates an unstable node. The trajectories are seen diverging away from the critical point in all directions, confirming the instability of $\mathcal{P}_1$ as suggested by its eigenvalues. This divergence implies that small perturbations will cause the system to evolve away from $\mathcal{P}_1$, never allowing it to settle into a steady state.
    \item {\bf Critical point $\mathcal{P}_2$:}
        The critical point $\mathcal{P}_2$ is depicted in the $XV$ plane in figure \ref{ch5_portrait_P2}. The critical point $\mathcal{P}_2$ corresponds to a non-standard CDM-dominated epoch in which the density of DE is negligible ($\Omega_{\text{DE}} = -20 x_2$). When $x_2 = 0$, this critical point reflects a standard cold dark matter-dominated era. The phase portrait shows trajectories approaching $\mathcal{P}_2$ along specific paths, forming a saddle point structure. This behaviour indicates that $\mathcal{P}_2$ is a saddle point, with some trajectories being attracted towards it along stable manifolds and repelled away along unstable manifolds. This behaviour is consistent with the mixed stability eigenvalues obtained for $\mathcal{P}_2$.
    \item {\bf Critical point $\mathcal{P}_3$:}
        The critical point $\mathcal{P}_3$ is illustrated on the $XY$ plane in figure \ref{ch5_portrait_P4}. It is important to note that the critical point $\mathcal{P}_3$, where $q = 0$ and $\omega = \frac{-1}{3}$, does not depict accelerating expansion. This critical point occurs at $2 m^2 - m \neq 0$ and is stable when $\frac{1}{4} < m < \frac{1}{2}$. The deceleration parameter solely relies on the variable $Z = \frac{\dot{H}}{H^2}$. In this context, when $Z=-1$, it signifies that the contribution from the Gauss--Bonnet invariant $\mathcal{G} = 24 H^4 (Z+1)$ term vanishes, which could explain why it does not demonstrate the transition phase, as depicted in figure \ref{ch5_portrait_P4}. The phase portrait shows trajectories spiralling inwards towards $\mathcal{P}_3$, indicating that it is a stable spiral. This suggests that small perturbations will cause the system to oscillate while eventually converging to $\mathcal{P}_3$. The eigenvalues, consisting of negative real parts, confirm the stable nature of this critical point.
    \item {\bf Critical point $\mathcal{P}_4$:}
        The critical point $\mathcal{P}_4$ is depicted on the $XY$ plane in figure \ref{ch5_portrait_P5}. The critical point $\mathcal{P}_4$ represents the de Sitter solutions with $\Omega_{\text{DE}}=1$, $\Omega_{\text{m}}=0$, $\Omega_{\text{r}}=0$, indicating the current accelerated expansion of the Universe. This de Sitter solution is valid at the critical point $\mathcal{P}_4$ within the parameter range $(-1+2 m) (-1+ 4 m) \neq 0$ and $m \neq 0$. Consequently, the value of $\omega_{\text{eff}}=q=-1$ highlights the significance of this critical point in describing the current dynamics of the Universe. The trajectories show a clear spiral structure converging towards $\mathcal{P}_4$, indicating a stable focus. This implies that the system will exhibit damped oscillations as it approaches $\mathcal{P}_4$. The stability analysis of $\mathcal{P}_4$ supports this observation, showing that the real parts of the eigenvalues are negative, thus confirming stability.
    \end{itemize}
        Understanding the stability and characteristics of critical points is essential for gaining insight into the long-term dynamics and behaviour of the specified dynamical system.
    \begin{figure}[H]
    \centering
        \includegraphics[width=100mm]{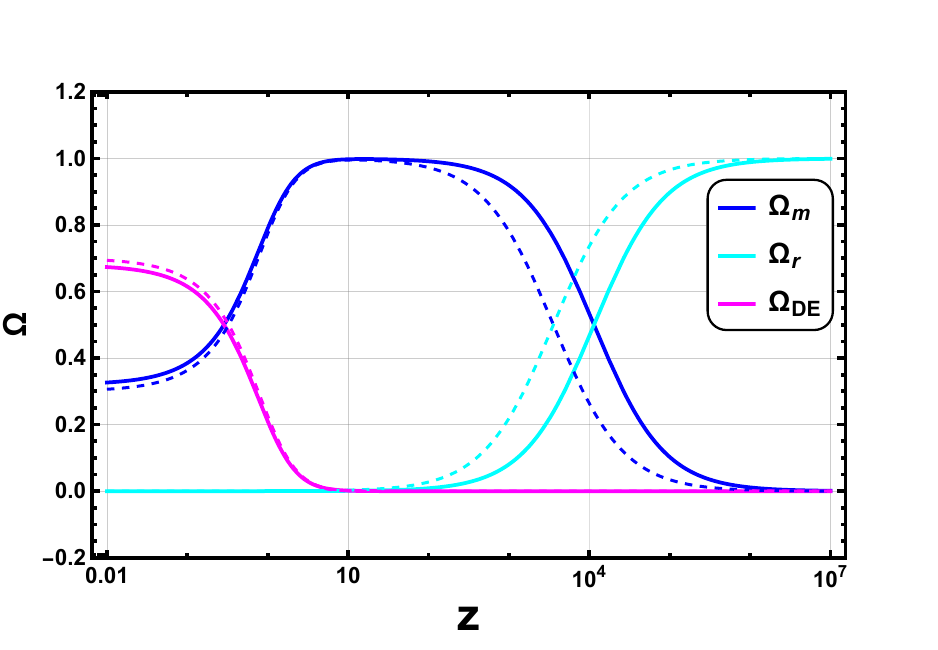}
        \caption{Evolution of the relative energy densities of dark matter $\Omega_{\text{m}}$, radiation $\Omega_{\text{r}}$, and DE $\Omega_{\text{DE}}$. The thick line represents the evolution of the density parameter for the $f(\mathcal{G})$ model, while the dashed line shows the evolution for the $\Lambda$CDM model.} 
    \label{ch5_FIG8}
    \end{figure}
        In the system described by equations \eqref{ch5_Autonomous DS}, it is possible to perform numerical integration with suitable initial conditions to capture the complete cosmological evolution across different epochs. Current measurements of cosmological parameters \cite{Aghanim_2020_641} suggest that the Universe is almost flat. For this specific example, we use the initial conditions $X = 10^{14}$, $Y = -1.2 \times 10^{14}$, $Z = 0.005$, and $V = 8.2 \times 10^{-5}$ and model parameter $m = 0.18$. The behaviour observed aligns with current cosmic observations regarding the evolution of density parameters. By integrating equation \eqref{ch5_generaldynamicalsystem} using the summarized initial conditions, we obtain numerical solutions for the density parameters $\Omega_{\text{m}}$, $\Omega_{\text{r}}$, and $\Omega_{\text{DE}}$, as shown in figure \ref{ch5_FIG8}. These results reveal that the Universe evolves through a radiation-dominated phase at early times $(q = 1)$. Subsequently, it transitioned into a matter-dominated phase with a deceleration parameter $\frac{1}{2}$. Currently, it is moving into an exponentially accelerating epoch with a deceleration parameter of $-1$. The $f(\mathcal{G})$ model represents a cosmological scenario in which the Universe undergoes successive eras of radiation domination, matter domination, and currently, DE domination. Our results indicate that the point where matter and radiation contribute equally is slightly higher than in the $\Lambda$CDM model. The behaviour of the model is consistent with current cosmic observations on the evolution of density parameters. The current densities are approximately $\Omega_{\text{m}} \approx 0.3$, $\Omega_{\text{DE}} \approx 0.7$, and $\Omega_{\text{r}} \approx 10^{-4}$. Similar behaviour in the evolution of density parameters has been noted in the literature \cite{Granda_2020_80}.

\section{Conclusion} \label{ch5_SEC V}
        In this chapter, we explored the cosmological properties of a specific modified Gauss--Bonnet gravity model. Initially, we discussed the main features of a gravitational action, which includes a general combination of the Ricci scalar and the Gauss--Bonnet invariant. Then, assuming a flat FLRW cosmological background, we derived the point-like Lagrangian of the theory and the corresponding equations of motion. The specific function we focused on, $f(\mathcal{G}) = \alpha \sqrt{\beta}(\frac{\mathcal{G}^2}{\beta^2})^m$, approaches GR as the real constant $\alpha$ gets closer to zero. However, our study does not explicitly converge to the cosmological constant case, making it particularly interesting as a potential alternative to the standard $\Lambda$CDM model. This model shows the ability to replicate DE behaviour while avoiding the conceptual issues associated with $\Lambda$. Importantly, we showed that the right-hand sides of the modified Friedmann equations can be understood as effective energy density and pressure resulting from curvature.

        We investigated the cosmological properties of the $f(\mathcal{G})$ model in the presence of matter fields. We assumed non-relativistic pressureless matter and neglected the late-time contribution of the radiation fluid. By numerically solving the first Friedmann equation, we determined the redshift behaviour of the Hubble parameter. We used the $\Lambda$CDM model to establish appropriate initial conditions for $H(z)$ and its derivatives. Subsequently, we utilized the most recent low-redshift observations to compare our theory directly with the model-independent predictions of the cosmic expansion. Specifically, we used a Bayesian analysis with the MCMC method, analysing using the Pantheon$^+$ and CC data sets separately. By assuming uniform prior distributions, we obtained constraints on the free parameters of the model at the $1\sigma$ and $2\sigma$ confidence levels. This enabled us to reconstruct the cosmological evolution of the Hubble expansion rate and the total effective EoS parameter. Our analysis indicates that the $f(\mathcal{G})$ model effectively accounts for the current acceleration of the Universe without the need for $\Lambda$. However, upon closer analysis and comparison with the predictions of the standard cosmological scenario, it becomes evident that the $f(\mathcal{G})$ model exhibits significant deviations from $\Lambda$CDM as the redshift increases, demonstrating its inability to describe a standard matter-dominated era. Our analysis identifies a notable discrepancy between the $H_0$ values obtained from the CC sample and the Pantheon$^+$ data sets, as seen in figure \ref{ch5_FIG: whisker plot}, highlighting the ongoing $H_0$ tension in cosmology. This result emphasizes the necessity for further research into possible systematic errors or new physics to resolve this issue. Addressing these discrepancies is essential for enhancing our understanding of the expansion rate of the Universe.

        In the second phase of our study, we conducted a dynamical system analysis, focusing on the type of $f(\mathcal{G})$ function under consideration. This analysis has enabled us to assess the global behaviour and stability of the cosmological model. It provided insights into the critical points associated with the model and their characteristics, which could be relevant to observable cosmology and the evolution of the Universe. Table \ref{ch5_TABLE-IV} presents the eigenvalues of the critical points along with their corresponding stability conditions. Our findings revealed stable critical points describing the late-time cosmic accelerated phase. This indicates non-standard matter and radiation-dominated eras of the Universe. Interestingly, our results align with the standard quintessence model on $z>0$. We made notable preliminary discoveries regarding the finite phase space of a power-law class of the Gauss--Bonnet gravity model. In addition, the equations describing the dynamical system for the power law $f(\mathcal{G})$ gravity model are provided in Equations \eqref{ch5_Autonomous DS}. Furthermore, Table \ref{ch5_TABLE-II} includes critical points, existing conditions, effective EoS, and deceleration parameters for the autonomous system. In contrast, Table \ref{ch5_TABLE-III} presents density parameter values for the acceleration phase. In total, we identified four critical points, three being stable ($\mathcal{P}_2$, $\mathcal{P}_3$, $\mathcal{P}_4$) and one unstable ($\mathcal{P}_1$). Finally, figure \ref{ch5_FIG8} shows the proficiency of the model in depicting the evolution of dark matter, radiation, and DE densities, effectively capturing the transitions through various cosmic epochs and reinforcing the robustness of the model in explaining late-time cosmic phenomena.

\chapter{Concluding Remarks and Future
Perspectives} 

\label{Chapter6} 

\lhead{Chapter 6. \emph{Concluding Remarks and Future
Perspectives}} 

This thesis extensively investigated the cosmological consequences of different modified gravity models, offering valuable insights into the accelerated expansion of the Universe and the influence of DE. Through rigorous analyses and comparisons with observational data, we have demonstrated the potential of these models as viable alternatives to the standard $\Lambda$CDM model.

Chapter \ref{Chapter2} introduced a modified Gauss--Bonnet gravity model, focusing on the gravitational action involving the Ricci scalar and Gauss--Bonnet invariant. Using various data sets (CC, Pantheon$^+$, BAO), we have parameterized the Hubble and other geometrical parameters, finding best-fit values for the coefficients. The model shows a smooth transition from deceleration to accelerated expansion, with transition redshifts ranging from $0.636$ to $0.74$. The DE EoS parameter indicates an increase in the expansion of the Universe, remaining within the phantom region for certain redshifts. In the second phase, a dynamical system analysis of the $f(R, \mathcal{G})$ function reveals five critical points, with stable points during the de Sitter phase and unstable points during the radiation-dominated phase. The model confirms the accelerating behavior with specific EoS and deceleration parameters, aligning with recent cosmological observations. The density parameters obtained are consistent across different approaches, supporting the stable accelerating behavior of the model.

In Chapter \ref{Chapter3}, we introduced modified $f(T, T_{\mathcal{G}})$ gravity models using cosmological data sets. The models combine the torsion scalar and Gauss--Bonnet invariant, simplifying to GR as a constant approaches zero. The Hubble parameter in this framework was constrained using Hubble and Pantheon SNe Ia data sets. Energy conditions were evaluated to ensure model consistency. The deceleration parameter indicates a transition from deceleration to acceleration in the expansion of the Universe. The EoS parameter suggests an accelerating expansion in the quintessence region. The model aligns well with current data, showing stability and feasibility for late-time acceleration. The age of the Universe was estimated to be consistent with Planck's findings. The study concludes that the model is viable at the background level with potential for further analysis.

Chapter \ref{Chapter4} delved into the cosmological implications of a modified $f(Q, B)$ gravity model, integrating the nonmetricity scalar $Q$ and the boundary term $B$. We constrained the model parameters by using Bayesian statistical analysis with MCMC techniques and observational data from CC measurements, the extended Pantheon$^+$ data set, and BAO measurements. Our results revealed a smooth transition from deceleration to acceleration, which is essential to understanding the dynamics of cosmic expansion and the role of DE. We constrained the model parameters for the comparison of the model with the data set and BAO measurements. Our results revealed a smooth transition from deceleration to acceleration, which is critical for understanding cosmic expansion dynamics and the role of DE. The comparison of the model with the $\Lambda$CDM framework highlighted its ability to replicate low-redshift behavior while exhibiting notable differences at high redshifts. The dynamical system analysis identified a stable critical point corresponding to the de Sitter phase, reinforcing the robustness of the model in explaining late-time cosmic phenomena.

In chapter \ref{Chapter5}, we investigated the cosmological implications of a modified Gauss--Bonnet gravity model, focusing on a specific function, $f(\mathcal{G}) = \alpha \sqrt{\beta}(\frac{\mathcal{G}^2}{\beta^2})^m $, that approaches GR as $\alpha$ tends to zero. The model does not converge to the cosmological constant case, offering an alternative to the standard $\Lambda$CDM model by explaining DE without needing $\Lambda$. The modified Friedmann equations are interpreted as effective energy density and pressure. The study examines the Hubble parameter's redshift behavior by solving the first Friedmann equation and comparing the results with low-redshift observations using Pantheon$^+$ and CC data. Bayesian analysis with MCMC methods revealed deviations from $\Lambda$CDM, particularly at higher redshifts, and highlighted discrepancies in the $H_0$ values between data sets, reflecting the ongoing Hubble tension. In the second part of the chapter \ref{Chapter5}, dynamical system analysis assessed the stability of the model, identifying stable critical points corresponding to the late-time accelerated phase. Four critical points were found, three of which were stable, showing the ability of the model to describe non-standard cosmic epochs. The findings align with the quintessence model at $z>0$ and demonstrate the effectiveness of model in describing cosmic evolution across dark matter, radiation, and DE densities. Further research is needed to address systematic errors and reconcile the $H_0$ tension.

In summary, this thesis has demonstrated that modified gravity models, such as those based on Gauss--Bonnet, $f(T, T_{\mathcal{G}})$, and $f(Q, B)$ frameworks, offer promising alternatives to the $\Lambda$CDM model. The aforementioned models not only demonstrate strong agreement with present-day cosmological observations but also offer a robust framework for comprehending the phenomenon of accelerated expansion in the Universe. The ability to capture the transition from deceleration to acceleration, identify stable critical points, and offer viable alternatives to the $\Lambda$CDM model underscores the crucial role of modified gravity theories in advancing our understanding of the expansion of the Universe and the nature of DE. 

In the future, this analysis can be extended to strong gravity regimes, providing a platform to investigate how these models influence the propagation of gravitational waves. By bridging theoretical predictions with observational data, researchers may uncover deviations from General Relativity, offering a gateway to finding new physics in the rapidly advancing field of gravitational wave astronomy. Continued exploration of these models is essential, as it enables the resolution of potential discrepancies and contributes to refining our understanding of cosmic evolution. This thesis highlights the pivotal role of modified gravity theories in deepening our comprehension of the fundamental mechanisms underlying the accelerated expansion of the Universe.

\addtocontents{toc}{\vspace{1em}} 

\backmatter


\label{References}

\lhead{\emph{References}}



\providecommand{\href}[2]{#2}\begingroup\raggedright\endgroup

\cleardoublepage
\clearpage

\lhead{\emph{Appendices}}
\chapter{Appendices} \label{Appendices}
\section*{Cosmological data sets}
\subsection*{A) Hubble data with 32 data points:}
\begin{table}[H]
\centering
\scalebox{0.85}{
\begin{tabular}{|ccccc||ccccc|}
\hline
No. & Redshift & $H(z)$ & $\sigma_{H(z)}$ & Ref. & No. & Redshift & $H(z)$ & $\sigma_{H(z)}$ & Ref. \\ [0.5ex]
\hline \hline
1. & 0.070 & 69.00 & 19.6 & \cite{Zhang_2014_14} & 17. & 0.4783 & 80.90 & 9.00 & \cite{Moresco_2016_2016_014} \\
2. & 0.090 & 69.00 & 12.0 & \cite{Simon_2005_71} & 18. & 0.480 & 97.00 & 62.00 & \cite{Stern_2010_2010_008} \\
3. & 0.120 & 68.60 & 26.2 & \cite{Zhang_2014_14} & 19. & 0.593 & 104.00 & 13.00 & \cite{Moresco_2012_2012_006} \\
4. & 0.170 & 83.00 & 8.00 & \cite{Simon_2005_71} & 20. & 0.680 & 92.00 & 8.00 & \cite{Moresco_2012_2012_006} \\
5. & 0.179 & 75.00 & 4.00 & \cite{Moresco_2012_2012_006} & 21. & 0.750 & 98.80 & 33.60 & \cite{Borghi_2022_928} \\
6. & 0.199 & 75.00 & 5.00 & \cite{Moresco_2012_2012_006} & 22. & 0.781 & 105.00 & 12.00 & \cite{Moresco_2012_2012_006} \\
7. & 0.200 & 72.90 & 29.60 & \cite{Zhang_2014_14} & 23. & 0.875 & 125.00 & 17.00 & \cite{Moresco_2012_2012_006} \\
8. & 0.270 & 77.00 & 14.00 & \cite{Simon_2005_71} & 24. & 0.880 & 90.00 & 40.00 & \cite{Stern_2010_2010_008} \\
9. & 0.280 & 88.80 & 36.60 & \cite{Zhang_2014_14} & 25. & 0.900 & 117.00 & 23.00 & \cite{Simon_2005_71} \\
10. & 0.352 & 83.00 & 14.00 & \cite{Moresco_2012_2012_006} & 26. & 1.037 & 154.00 & 20.00 & \cite{Moresco_2012_2012_006} \\
11. & 0.380 & 83.00 & 13.50 & \cite{Moresco_2016_2016_014} & 27. & 1.300 & 168.00 & 17.00 & \cite{Simon_2005_71} \\
12. & 0.400 & 95.00 & 17.00 & \cite{Simon_2005_71} & 28. & 1.363 & 160.00 & 33.60 & \cite{Moresco_2015_450} \\
13. & 0.4004 & 77.00 & 10.20 & \cite{Moresco_2016_2016_014} & 29. & 1.430 & 177.00 & 18.00 & \cite{Simon_2005_71} \\
14. & 0.425 & 87.10 & 11.20 & \cite{Moresco_2016_2016_014} & 30. & 1.530 & 140.00 & 14.00 & \cite{Simon_2005_71} \\
15. & 0.445 & 92.80 & 12.90 & \cite{Moresco_2016_2016_014} & 31. & 1.750 & 202.00 & 40.00 & \cite{Simon_2005_71} \\
16. & 0.470 & 89.00 & 49.60 & \cite{Ratsimbazafy_2017_467} & 32. & 1.965 & 186.50 & 50.40 & \cite{Simon_2005_71} \\ [0.5ex]
\hline
\end{tabular}}
\caption{$H(z)$ measurements were made using the CC technique, expressed in [km s$^{-1}$ Mpc$^{-1}$] units, along with the corresponding errors.}
\label{table: Table II}
\end{table}

\subsection*{B) Hubble data with 55 data points:}
    \begin{table}[H]
    \centering 
    \scalebox{0.77}{
    \begin{tabular}{|c c c c c || c c c c c|} 
        \hline 
    No. & $z$ & H(z) & $\sigma_{H}$ & Ref. & No. & $z$ & H(z) & $\sigma_{H}$ & Ref. \\ [0.5ex] 
        \hline \hline 
    1. & 0.070 & 69.00 & 19.6 & \cite{Zhang_2014_14} & 29. & 0.480 & 87.79 & 2.03 & \cite{Wang_2017_469} \\
    2. & 0.090 & 69.00 & 12.0 & \cite{Jimenez_2003_593} & 30. & 0.480 & 97.00 & 62.00 & \cite{Stern_2010_2010_008} \\
    3. & 0.120 & 68.60 & 26.2 & \cite{Zhang_2014_14} & 31. & 0.510 & 90.40 & 1.90 & \cite{Alam_2017_470} \\
    4. & 0.170 & 83.00 & 8.00 & \cite{Simon_2005_71} & 32. & 0.520 & 94.35 & 2.64 & \cite{Wang_2017_469} \\
    5. & 0.179 & 75.00 & 4.00 & \cite{Moresco_2012_2012_006} & 33. & 0.560 & 93.34 & 2.30 & \cite{Wang_2017_469} \\
    6. & 0.199 & 75.00 & 5.00 & \cite{Moresco_2012_2012_006} & 34. & 0.590 & 98.48 & 3.18 & \cite{Wang_2017_469} \\
    7. & 0.200 & 72.90 & 29.60 & \cite{Zhang_2014_14} & 35. & 0.593 & 104.0 & 13.00 & \cite{Moresco_2012_2012_006} \\
    8. & 0.240 & 79.69 & 3.32 & \cite{Gaztanaga_2009_399} & 36. & 0.600 & 87.90 & 6.10 & \cite{Blake_2012_425} \\
    9. & 0.270 & 77.00 & 14.00 & \cite{Simon_2005_71} & 37. & 0.610 & 97.30 & 2.10 & \cite{Alam_2017_470} \\
    10. & 0.280 & 88.80 & 36.60 & \cite{Zhang_2014_14} & 38. & 0.640 & 98.02 & 2.98 & \cite{Wang_2017_469} \\
    11. & 0.300 & 81.70 & 5.00 & \cite{Oka_2014_439} & 39. & 0.680 & 92.00 & 8.00 & \cite{Moresco_2012_2012_006} \\
    12. & 0.310 & 78.18 & 4.74 & \cite{Wang_2017_469} & 40. & 0.730 & 97.30 & 7.00 & \cite{Blake_2012_425} \\
    13. & 0.340 & 83.80 & 2.96 & \cite{Gaztanaga_2009_399} & 41. & 0.781 & 105.0 & 12.00 & \cite{Moresco_2012_2012_006} \\
    14. & 0.350 & 82.70 & 9.10 & \cite{Chuang_2013_435} & 42. & 0.875 & 125.0 & 17.00 & \cite{Moresco_2012_2012_006} \\
    15. & 0.352 & 83.00 & 14.00 & \cite{Moresco_2012_2012_006} & 43. & 0.880 & 90.00 & 40.00 & \cite{Stern_2010_2010_008} \\
    16. & 0.360 & 79.94 & 3.38 & \cite{Wang_2017_469} & 44. & 0.900 & 117.0 & 23.00 & \cite{Simon_2005_71} \\
    17. & 0.380 & 81.50 & 1.90 & \cite{Alam_2017_470} & 45. & 1.037 & 154.0 & 20.00 & \cite{Moresco_2012_2012_006} \\
    18. & 0.3802 & 83.00 & 13.50 & \cite{Moresco_2016_2016_014} & 46. & 1.300 & 168.0 & 17.00 & \cite{Simon_2005_71} \\
    19. & 0.400 & 95.00 & 17.00 & \cite{Simon_2005_71} & 47. & 1.363 & 160.0 & 33.60 & \cite{Moresco_2015_450} \\
    20. & 0.400 & 82.04 & 2.03 & \cite{Wang_2017_469} & 48. & 1.430 & 177.0 & 18.00 & \cite{Simon_2005_71} \\
    21. & 0.4004 & 77.00 & 10.20 & \cite{Moresco_2016_2016_014} & 49. & 1.530 & 140.0 & 14.00 & \cite{Simon_2005_71} \\
    22. & 0.4247 & 87.10 & 11.20 & \cite{Moresco_2016_2016_014} & 50. & 1.750 & 202.0 & 40.00 & \cite{Simon_2005_71} \\
    23. & 0.430 & 86.45 & 3.27 & \cite{Gaztanaga_2009_399} & 51. & 1.965 & 186.5 & 50.40 & \cite{Moresco_2015_450} \\
    24. & 0.440 & 82.60 & 7.80 & \cite{Blake_2012_425} & 52. & 2.300 & 224.0 & 8.60 & \cite{Busca_2013_552} \\
    25. & 0.440 & 84.81 & 1.83 & \cite{Wang_2017_469} & 53. & 2.330 & 224.0 & 8.00 & \cite{Bautista_2017_603} \\
    26. & 0.4497 & 92.80 & 12.90 & \cite{Moresco_2016_2016_014} & 54. & 2.340 & 222.0 & 7.00 & \cite{Delubac_2015_574} \\
    27. & 0.470 & 89.00 & 34.00 & \cite{Ratsimbazafy_2017_467} & 55. & 2.360 & 226.0 & 8.00 & \cite{Font_Ribera_2014_2014_027} \\
    28. & 0.4783 & 80.90 & 9.00 & \cite{Moresco_2016_2016_014} & & & & & \\ [1ex] 
        \hline 
    \end{tabular}}
        \caption{The observational Hubble data set that was used in this paper}
    \label{CH3_table: Table II} 
    \end{table}
    

\cleardoublepage
\pagestyle{fancy}
\label{Publications}
\lhead{\emph{List of Publications and Presentations}}
\chapter{List of Publications and Presentations}
\section*{Thesis Publications}
\begin{enumerate}
    \item \textbf{Santosh V. Lohakare}, K. Rathore, and B. Mishra, ``Observational constrained $f (R, \mathcal{G})$ gravity cosmological model and the dynamical system analysis", \href{https://iopscience.iop.org/article/10.1088/1361-6382/acfc0f/meta}{\color{blue}\textit{Classical and Quantum Gravity} \textbf{40} (2023) 215009.}

    \item \textbf{Santosh V. Lohakare}, B. Mishra, S.K. Maurya, and Ksh. N. Singh, ``Analyzing the geometrical and dynamical parameters of modified Teleparallel--Gauss--Bonnet model", \href{https://doi.org/10.1016/j.dark.2022.101164}{\color{blue}\textit{Physics of the Dark Universe} \textbf{39} (2023) 101164.}

    \item \textbf{Santosh V. Lohakare}, and B. Mishra, ``Exploring Stability of $f(Q, B)$ gravity via Dynamical System Approach: a Comprehensive Bayesian Statistical Analysis", \href{https://doi.org/10.3847/1538-4357/ad9602}{\color{blue}{\em The Astrophysical Journal} \textbf{978} (2025) 26.}

    \item \textbf{Santosh V. Lohakare}, S. Niyogi, and B. Mishra, ``Cosmology in Modified $f(\mathcal{G})$ Gravity: a Late Time Cosmic Phenomena", \href{https://doi.org/10.1093/mnras/stae2302}{\color{blue}\textit{Monthly Notices of the Royal Astronomical Society} \textbf{535} (2024) 1136.}
\end{enumerate}

\section*{Other Publications}
\begin{enumerate}
    \item S. A. Narawade, \textbf{Santosh V. Lohakare}, and B. Mishra \textit{“Cosmological reconstruction and stability in covariant $f(Q)$ gravity,"} \href{https://doi.org/10.1016/j.aop.2024.169913}{\color{blue}\textit{Annals of Physics} \textbf{474} (2025) 169913.}

    \item S. A. Kadam, \textbf{Santosh V. Lohakare}, and B. Mishra, ``Dynamical complexity in teleparallel Gauss–Bonnet gravity", \href{https://doi.org/10.1016/j.aop.2023.169563}{\color{blue}\textit{Annals of Physics} \textbf{460} (2024) 169563.}

    \item L. K. Duchaniya, \textbf{Santosh V. Lohakare}, B. Mishra, ``Cosmological models in $f(T,\mathcal{T})$ gravity and the dynamical system analysis", \href{https://doi.org/10.1016/j.dark.2023.101402}{\color{blue}\textit{Physics of the Dark Universe} \textbf{43} (2024) 101402.}

    \item \textbf{Santosh V. Lohakare}, S. K. Maurya, Ksh. N. Singh, and B. Mishra, ``Influence of three parameters on maximum mass and stability of strange star under linear $f(Q)$ - action", \href{https://doi.org/10.1093/mnras/stad2861}{\color{blue}\textit{Monthly Notices of the Royal Astronomical Society} \textbf{526} (2023) 3796.}

    \item \textbf{Santosh V. Lohakare}, B. Mishra, F. Tello-Ortiz, and S. K. Tripathy, ``The Fate of the Universe Evolution in the Quadratic Form of Ricci–Gauss–Bonnet Cosmology", \href{https://doi.org/10.1134/S0202289323040138}{\color{blue}\textit{Gravitation and Cosmology} \textbf{29} (2023) 443.}

    \item \textbf{Santosh V. Lohakare}, F. Tello-Ortiz, S. K. Tripathy, and  B. Mishra, ``Bouncing Cosmology in Modified Gravity with Higher-Order Gauss–Bonnet Curvature Term", \href{https://doi.org/10.3390/universe8120636}{\color{blue}\textit{Universe} \textbf{8} (2022) 636.}

    \item L. K. Duchaniya, \textbf{Santosh V. Lohakare}, and B. Mishra, ``Dynamical stability analysis of accelerating $f(T)$ gravity models", \href{https://doi.org/10.1140/epjc/s10052-022-10406-w}{\color{blue}\textit{The European Physical Journal C} \textbf{82} (2022) 448.}

    \item S. K. Maurya and Ksh. N. Singh, \textbf{Santosh V. Lohakare} and B. Mishra, ``Anisotropic Strange Star Model Beyond Standard Maximum Mass Limit by Gravitational Decoupling in $f(Q)$ Gravity", \href{https://doi.org/10.1002/prop.202200061}{\color{blue}\textit{Fortschritte der Physik} \textbf{70} (2022) 2200061.}

    \item \textbf{Santosh V. Lohakare}, S. K. Tripathy, and B. Mishra, ``Cosmological model with time varying deceleration parameter in $f(R, \mathcal{G})$ gravity", \href{https://iopscience.iop.org/article/10.1088/1402-4896/ac40d6/meta}{\color{blue}\textit{Physica Scripta} \textbf{96} (2021) 125039.}    
\end{enumerate}

\section*{Conferences and Presentations}
\begin{enumerate}
		\vspace{1mm}
            \item Presented a paper entitled \emph{“Stability of $f(Q, B)$ Gravity via Dynamical System Approach: a Comprehensive Bayesian Statistical Analysis,"} in \textit{II International Scientific Conference Space. Time. Civilization. (STC -- 2024)} organized by BITS Pilani, Hyderabad Campus, India (November 02 – 07, 2024).
            \item Presented a paper entitled \emph{“$f(R, G)$ Gravity cosmological model with variable deceleration parameter,"} in \textit{International conference on Gravitation-Theory (Gravitex-2021)} organized by University of KwaZulu Natal, Durban, South Africa, (August 9 - 12, 2021).
            \item Presented a paper entitled \emph{“Constraining the cosmological parameters of modified Teleparallel-Gauss--Bonnet model"} in the \textit{32nd meeting of Indian Association for General Relativity and Gravitation (IAGRG32)} organized by IISER, Kolkata, India (December 19 – 21, 2022). 
            \item Presented a paper entitled \emph{“Cosmological model with time-varying deceleration parameter in $f(R,G)$ gravity,"} in the \textit{Prof.~P.C.~Vaidya National Conference on Mathematical Sciences} organized by Sardar Patel University, Vallabh Vidyanagar, Gujarat (March 15 - 16, 2022).
            \item Presented a paper entitled \emph{“Observational constrained $f(R,\mathcal{G})$ gravity cosmological model and the dynamical system analysis"} in the \textit{Physical Interpretations of Relativity Theory (PIRT-2023)} organized by Bauman Moscow State Technical University, Moscow, Russia (July 3 - 6, 2023).
            \item Presented a paper entitled \emph{“Matter-bounce scenario in $f(R,G)$ gravity"} in the \textit{27th International Conference of International Academy of Physical Sciences on Advances in Relativity and Cosmology (PARC-2021)} organized by BITS Pilani, Hyderabad Campus, India (October 26 – 28, 2021).
            \item Presented a paper entitled \emph{“Cosmological model with time-varying deceleration parameter in $f(R,\mathcal{G})$ gravity"} in the \textit{The Metric-Affine Frameworks for Gravity 2022 conference} organized by Laboratory of Theoretical Physics, Institute of Physics, University of Tartu, Estonia (July 27 – 01, 2022).
    \vspace{2mm}
            \item Presented a paper entitled \emph{“Cosmological model with time-varying deceleration parameter in $f(R,\mathcal{G})$ gravity"} in the \textit{23rd International Conference on General Relativity and Gravitation} organized by Institute of Theoretical Physics, Chinese Academy of Sciences, Liyang, China (July 3 - 8, 2022).
    \vspace{2mm}
            \item Presented a paper entitled \emph{“Matter-bounce scenario in $f(R, \mathcal{G})$ gravity"} in the \textit{International Conference on Mathematical Sciences and its Applications (ICMSA-2022)} organized by Swami Ramanand Teerth Marathwada University, Nanded (July 28 – 30, 2022).
    \vspace{2mm}
            \item Presented a paper entitled \emph{“Late-time cosmic acceleration in Ricci--Gauss--Bonnet gravity"} in the \textit{Recent Advances in Science and Technology (RAST-2021)} organized by IGIT, Sarang, Odisha (July 28, 2021).
    \vspace{2mm}
            \item Presented a paper entitled \emph{“Dynamical behavior of accelerating cosmological model $f(R,\mathcal{G})$ Gravity"} in the \textit{Cosmology from Home-2021}, (July 05 - 16, 2021).
    \vspace{2mm}
            \item Presented a paper entitled \emph{“Observational constrained $f(R,\mathcal{G})$ gravity cosmological model and the dynamical system analysis"} in the \textit{The Metric-Affine Gravity - 2024} organized by Laboratory of Theoretical Physics, Institute of Physics, University of Tartu, Estonia (June 17 – 21, 2024).
    \vspace{2mm}
            \item Participated in the international webinar on \emph{Recent Advances in Science and Technology (RAST-2021)} organized by IGIT, Sarang, Odisha (November 08 – 09, 2020). 
    \vspace{2mm}
            \item Participated in the international webinar on \emph{Recent Developments in Modified Gravity and Cosmology (RDCM - 2021)} organized by BITS Pilani, Hyderabad Campus, India (November 09 – 11, 2021). 
    \vspace{2mm}
            \item Participated in the international workshop on \emph{Mathematical Modelling, Cosmology and Data Science (IWMMCDS - 2024)} organized by Woxsen University, Hyderabad, India (October 25 – 29, 2024).
    \vspace{2mm}
            \item Participated in the \emph{CA21136 CosmoVerse: Training Series - 2024} organized by Cosmoverse COST (European Cooperation in Science and Technology) (September 26 – October 23, 2024).
    \vspace{2mm}
            \item Participated in the \emph{Teacher’s Enrichment Workshop} organized by BITS-Pilani Hyderabad Campus (January 09 - 14, 2023).
	\end{enumerate}

\paragraph{International conference / Research visit (In person)} 
\begin{enumerate}
    \item Presented a paper entitled \emph{“Observational constrained $f(R, \mathcal{G})$ gravity cosmological model and the dynamical system analysis"} in the \textit{International Conference on Particle Physics and Cosmology (Rubakov-2023)} organized by Yerevan State University, Yerevan, Armenia (October 2-7, 2023).
    \item A research visit to the \emph{Inter-University Centre for Astronomy and Astrophysics (IUCAA)} at Pune, Maharashtra (31 March – 6 April, 2024).
\end{enumerate}
\cleardoublepage
\pagestyle{fancy}
\lhead{\emph{Biography}}
\setstretch{1.13}
\chapter{Biography}
\textbf{\Large Brief Biography of the Candidate:}\\
    \textbf{Mr. Lohakare Santosh Vijay} received his Master's degree from N.~E.~S.~Science College in Nanded, Maharashtra, in 2017. He successfully cleared the Council of Scientific and Industrial Research (CSIR) National Eligibility Test (NET) for Assistant Professor and Junior Research Fellow (JRF) in December 2019, the State Eligibility Test (SET) Maharashtra in 2021 and the Graduate Aptitude Test in Engineering (GATE) in 2020. His academic contributions include multiple publications in esteemed national and international journals, along with presentations at various national and international conferences. Notably, he has been invited to present at Yerevan Technical University in Armenia from December 2 to 7, 2023. Additionally, he has secured two international travel grants from the DST-SERB under the ITS scheme and received a one-time travel grant from CSIR.\\

\textbf{\Large Brief Biography of the Supervisor:}\\
    \textbf{Prof.~Bivudutta Mishra} received his Ph.D.~degree from Sambalpur University, Odisha, India, in 2003. His main research areas are Geometrically Modified Theories of Gravity, Theoretical Aspects of Dark Energy and Wormhole Geometry. He has published over 170 research papers in national and international journals, presented papers at conferences in India and abroad, supervised four Ph.D.~students, and is currently guiding eight more. He has also organized academic and scientific events in the department. He has become a member of the scientific advisory committee of national and international academic events. He has successfully completed multiple sponsored projects funded by Government Funding agencies and is at present working on three projects funded by CSIR, SERB-DST (MATRICS), and SERB-DST (CRG-ANRF). He is also an awardee of DAAD-RISE, 2019, 2022. He has also reviewed several research papers in highly reputed journals, is a Ph.D. examiner, and is a BoS member of several universities. He has been invited by many foreign universities to share his research in scientific events, some of which are Canada, Germany, Republic of China, Russia, Australia, Switzerland, Japan, United Kingdom, Poland, etc. As an academic administrator, he was Head of the Department of Mathematics from September 2012 to October 2016 and was Associate Dean of International Programmes and Collaborations from August 2018 to September 2024. He is also a visiting professor at Bauman Moscow State Technical University, Moscow, a visiting associate at Inter-University Centre for Astronomy and Astrophysics, Pune, a Fellow of the Royal Astronomical Society, UK, and a Fellow of the Institute of Mathematics and Applications, UK. Foreign member of the Russian Gravitational Society, Moscow. He has been listed among the top $2\%$ of scientists according to the Stanford University author database of standardized citation indicators.
                        
\end{document}